\documentclass[review]{elsarticle}

\usepackage{amssymb}
\usepackage{epsfig}
\usepackage{amsmath}

\usepackage{amsbsy}
\newcommand{\be}{\begin{equation}}
\newcommand{\ee}{\end{equation}}
\newcommand{\bea}{\begin{eqnarray}}
\newcommand{\eea}{\end{eqnarray}}

\def\Journal#1#2#3#4{{#1} {#2} (#4) #3 }

\def\PRO{{\em Prog. Theor. Phys.}}

\def\PRL{\em Phys. Rev. Lett.}
\def\PREV{\em Phys. Rev.}

\def\ANNP{\em Ann. Phys. (N.Y.)}

\def\la{\langle}\def\ra{\rangle}
\def\bi{\bibitem}
\def\be{\begin{eqnarray}}\def\ee{\end{eqnarray}}
\def\lsim{\mathrel{\rlap{\lower3pt\hbox{\hskip1pt$\sim$}}
     \raise1pt\hbox{$<$}}} 
\def\gsim{\mathrel{\rlap{\lower3pt\hbox{\hskip1pt$\sim$}}
     \raise1pt\hbox{$>$}}} 
\def\le{ \begin{array}{ll}}\def\re{\end{array}}

\def\lear{ \left( \begin{array}{cc}}\def\rear{\end{array} \right)}
\def\tr{\textnormal{tr}}

\def\le{ \left( \begin{array}{cc}}\def\re{\end{array} \right)}
\def\tr{\textnormal{tr}}

\def\bi{\bibitem}

\def\tr{{\rm tr}}
\def\Tr{{\rm Tr}}
\def\ln{{\rm ln}}
\def\del{\partial}
\def\O{{\cal O}}

\usepackage[usenames]{color}

\journal{Physics Reports}

\begin{document}

\begin{frontmatter}



\title{Chiral symmetry and effective field theories for \\ hadronic, nuclear and stellar matter}


\author[label1]{Jeremy W. Holt}
\author[label2,label3]{Mannque Rho}
\author[label4,label5]{Wolfram Weise}

\address[label1]{Department of Physics, University of Washington, Seattle, 98195, USA}
\address[label2]{Department of Physics, Hanyang University, Seoul 133-791, Korea}
\address[label3]{Institut de Physique Th\'{e}orique, CEA Saclay, 91191 Gif-sur-Yvette, France}
\address[label4]{Physik Department, Technische Universit\"{a}t M\"{u}nchen, D-85747 Garching, Germany}
\address[label5]{ECT$^{\, *}$, Villa Tambosi, I-38123 Villazzano (TN), Italy}

\begin{abstract}
Chiral symmetry, first entering in nuclear physics in the 1970's for which Gerry Brown  played a seminal role, has led to a stunningly successful framework for describing strongly-correlated nuclear dynamics both in finite and infinite systems. We review how the early, germinal idea conceived with the soft-pion theorems in the pre-QCD era has evolved into a highly predictive theoretical framework for nuclear physics, aptly assessed by Steven Weinberg: ``it (chiral effective field theory) allows one to show in a fairly convincing way that what they  (nuclear physicists) have been doing all along ... is the correct first step in a consistent approximation scheme." Our review recounts both how the theory presently fares in confronting Nature and how one can understand its extremely intricate workings in terms of the multifaceted aspects of chiral symmetry, namely, chiral perturbation theory, skyrmions, Landau Fermi-liquid theory, the Cheshire cat phenomenon, and
hidden local and mended symmetries.
\end{abstract}


\end{frontmatter}



\vspace{.4in}

\tableofcontents

\vspace{.4in}

\section{Prologue}

This review covers a broad spectrum of mutually related topics, from the symmetry breaking pattern of QCD and the structure of hadrons to various facets of the nuclear many-body problem and to the physics of dense, strongly interacting matter as it is realized in the core of neutron stars. It is written as a tribute to Gerry Brown who pioneered and shaped these fields of research in his own very special style. Each one of the authors benefited greatly from multiple exchanges with Gerry, as senior collaborator over decades (M.R.), as postdoctoral fellow and frequent visitor to Stony Brook (W.W.), and as Gerry's former PhD student in his later years (J.W.H.), thus representing three generations of researchers inspired by Gerry Brown's intuitive way of thinking.

Chiral symmetry and its realizations in the physics of hadrons and nuclei was one of the persistent guiding themes throughout four decades of Gerry Brown's scientific life. He pioneered, with MR's participation, implementing S.~Weinberg's principal framework of chiral effective field theory (EFT) into nuclear physics - thus building a bridge between QCD and nuclei. An instructive early account is given in a 1979 article discussing the role of chiral symmetry in the nucleon-nucleon interaction \cite{brown79}.

This is a suitable moment for one of us (M.R.) to inject some personal recollections. An article \cite{comments} written in 1981 presaged the impressive modern development of the chiral expansion first put forward for nuclear application following Weinberg's EFT paper \cite{Wei79} and Witten's large-N paper \cite{witten79} that both appeared in 1979. In Erice in 1981 we (G.E.B. and M.R.) were both giving lectures, on nuclear forces and nuclear electroweak currents respectively, describing how to perform the chiral perturbation expansion in this context. These steps were on the way to formulate (possibly) the full chiral expansion but we were diverted by the rediscovery of the skyrmion model in 1983. It was only after Weinberg's seminal papers \cite{weinbergseminal} appeared in the early 1990's that we went back to that problem. We communicated with Weinberg pointing out that we were doing just what he was proposing then. This is what Weinberg meant in his field theory book -- which one can read in a more general context in his talk given at the 1996 conference on ``Conceptual Foundations of Quantum Field Theory" \cite{wtheorem} -- when he made his statement recalled in the Abstract above.

In the decades following these early developments, applications of chiral effective field theory to nuclear few- and many-body problems have deepened and broadened in many ways, both conceptually and in terms of a growing number of calculations with direct contact to observables. It is in this spirit that the present review is composed. It reflects the interplay between concepts, applications and phenomenology, viewed from different angles but under the common frame of chiral symmetry as the principle that rules low-energy QCD
at the interface with nuclear physics. We mention that part of the materials have been adapted and updated from an earlier recent review \cite{hkwreview}.


\section{Introductory survey}
\label{intro}

The fundamental theory of strongly interacting matter is quantum chromodynamics (QCD).
There are two limiting cases in which QCD can be accessed with ``controlled"
approximations: at momentum scales exceeding several GeV (corresponding to short
distances on the scale of $r < 0.1$ fm), QCD is a theory of weakly-coupled quarks and
gluons. At low momentum scales well below 1 GeV (corresponding to long distances on the
order of $r > 1$ fm), QCD is governed by color confinement and a non-trivial vacuum, namely the ground state of QCD hosts strong condensates of gluons and quark-antiquark pairs. Color confinement arises at nearly the same energy scale as the dynamical (spontaneous) breaking of chiral symmetry, a global symmetry of QCD that in the limit of massless quarks is an exact symmetry and conformal symmetry that is explicitly broken by a quantum anomaly known as trace anomaly.
The spontaneous breaking of chiral symmetry, which is explicitly broken by quark masses generated at the Higgs scale, implies the emergence of pseudoscalar Nambu-Goldstone bosons. For two quark flavors ($u$ and $d$) with nearly zero mass, these Goldstone bosons of chiral symmetry\footnote{By Goldstone bosons, unless otherwise noted, we will refer to both Goldstone and pseudo-Goldstone bosons.} are identified with the isospin triplet of pions. The spontaneous breaking of scale symmetry, known to be possible only in the presence of an explicit breaking of scale symmetry (triggered in the case of QCD by the trace anomaly), generates an isoscalar dilaton. Low-energy QCD in the matter-free vacuum is
therefore realized as a (chiral) effective field theory of the active, light degrees of freedom:
the pions as pseudo-Goldstone bosons. In the low-energy, long-wavelength limit,
Goldstone bosons have the property that they interact weakly with one another and with any
massive hadron, dictated mainly by the derivative coupling. In this limit a perturbation theory based
on the systematic expansion in powers of a ``small" parameter can be performed \cite{Wei66/67,GL84}.

The nonlinear realization of chiral symmetry that renders feasible the systematic chiral expansion at low energies and long-wavelengths allows an extension of the theory to a higher energy scale given by the mass of the next mesonic excitations, the vector mesons $\rho$ and $\omega$.\footnote{In what follows, we will refer to the $\rho(770)$ and $\omega(782)$ mesons simply as ``vector mesons," generically denoted as $V\in U(2)$. Later we will encounter higher-mass vector mesons in the infinite tower that appears in a holographic description of hidden local symmetry.}  As will be explained below in somewhat more detail, this extension follows from the redundancy inherent in the nonlinear chiral field that can be elevated to a local gauge symmetry. The resultant Lagrangian, referred to as a ``hidden local symmetry" (HLS for short) Lagrangian~\cite{HLS}, is gauge-equivalent to the nonlinear sigma model. In the low-energy regime where the nonlinear sigma model is applicable, there is no significant power in hidden local symmetry. However if the vector meson can be considered ``light" in some sense, then the theory has an advantage -- in predictiveness -- over the nonlinear sigma model~\cite{georgi,HY:PR}: one can perform a systematic chiral expansion in which such low-energy properties as vector dominance, KSRF relations, etc.\ appear naturally. This means that the HLS approach can in principle handle the condition where the vector meson mass becomes as light as the pion mass, a situation that may arise in the approach to chiral symmetry restoration.

In the low-energy and low-density domain accessed by currently available experimental probes, chiral effective field theory as outlined above in terms of Goldstone boson fields {\it alone}
provides a sufficiently powerful framework for a highly successful description of nuclear interactions \cite{evgenireview, hammerreview,
machleidtreview}, and it is the starting point for a systematic approach to nuclear
many-body dynamics and thermodynamics at densities and temperatures well within the
confined phase of QCD \cite{hkwreview}. An alternative effective field theory incorporating hidden local symmetry, as well as scale symmetry and mended symmetry, could have the potential to provide a more direct interface between QCD and nuclear physics and go beyond the standard chiral effective field theory. The various aspects of chiral symmetry, as it is manifested in the low-energy effective field theory of strong interactions and as it is extrapolated to high temperatures and densities, will be the subject of this review.

\subsection{Low-energy QCD and chiral symmetry}

Let us recall how the special role of the pion emerges through the Nambu-Goldstone mechanism of spontaneous chiral symmetry breaking in QCD.

\subsubsection{\it Chiral symmetry and the pion}

Historically, our understanding of the pion as a Nambu-Goldstone boson
emerged \cite{G61,NJL61} in the pre-QCD era of the 1960's and culminated in the current algebra approach \cite{AD68} (combined with the PCAC relation for the pion). Inspired by the BCS theory of superconductivity, Nambu and Jona-Lasinio (NJL) \cite{NJL61} developed a model that helped clarify the dynamics that drives spontaneous chiral symmetry breaking and the formation of pions as pseudo-Goldstone bosons. The NJL model in the SM (standard model) context is known to be equivalent to the nonlinear sigma model. So one could just as well discuss what is given below in terms of a generalized nonlinear sigma model but we find it more transparent to follow the logic of the NJL which gives a hint at the link to QCD degrees of freedom. In the next section, we will address the connection between QCD variables and hadronic variables with the help of topology closely tied to quantum anomalies.

Starting from the color current of quarks, ${\bf J}_\mu^a = \bar{q}
\gamma_\mu {\bf t}^a q$, where $q$ denotes the quark fields with $4N_c N_f$ components
representing their spin, color and flavor degrees freedom $(N_c=3, N_f=2)$ and
${\bf t}^a$ ($a = 1, ... ,8$) are the generators of the
$SU(3)_c$ color gauge group, one can make the additional ansatz that the distance over which color
propagates is restricted to a short correlation length $l_c$. Then the interaction between
low-momentum quarks, mediated by the coupling of the quark color current to the gluon fields,
can be schematically viewed as a local coupling between their color currents:
\begin{equation}
{\cal L}_{\rm int} = -G_c\,{\bf J}_\mu^a(x)\,{\bf J}^\mu_a(x)\,\, ,
\label{eq:Lint1}
\end{equation}
where $G_c \sim \bar{g}^2\, l_c^2$ represents an effective coupling strength proportional to the square of the QCD coupling, $g$, averaged over the relevant distance
scales, in combination with the correlation length, $l_c$, squared.

Given the local interaction in Eq.\ (\ref{eq:Lint1}), the model Lagrangian for the quark fields
\begin{equation}
{\cal L} = \bar{q}(x)(i\gamma^\mu\partial_\mu - m_q)q(x) + {\cal L}_{\rm int}(\bar{q},q)
\label{eq:NJL}
\end{equation}
arises  by ``integrating out" the gluon degrees of freedom and absorbing them in the local four-fermion interaction ${\cal L}_{\rm int}$. In this way the local $SU(3)_c$ gauge
symmetry of QCD is replaced by a global one. Confinement is lost but all
symmetries of QCD are maintained. In Eq.\ (\ref{eq:NJL}) the mass matrix $m_q$
incorporates the small ``bare" quark masses, and in the chiral limit ($m_q \rightarrow 0$) the Lagrangian in Eq.\ (\ref{eq:NJL}) has a chiral symmetry of left- and right-handed quarks,
$SU(N_f)_L\times SU(N_f)_R$, analogous to that of the original QCD Lagrangian for
$N_f$ massless quark flavors.

Fierz transforming the color current-current interaction in Eq.\ (\ref{eq:Lint1}) produces a
set of exchange terms acting in quark-antiquark channels. For the case of $N_f = 2$:
\begin{equation}
{\cal L}_{\rm int} \rightarrow {G\over 2}\left[(\bar{q}q)^2 + (\bar{q}\,i\gamma_5\,
\vec\tau \,q)^2\right] + ... \,\, , \label{eq:Lint2}
\end{equation}
where $\vec \tau = (\tau_1,\tau_2,\tau_3)$ represents the vector of isospin Pauli matrices.
Included in Eq.\ (\ref{eq:Lint2}) but not shown explicitly are a series of terms with
vector and axial vector currents, both in color-singlet and color-octet channels. The new constant
$G$ is proportional to the color coupling strength $G_c$, and their ratio is uniquely
determined by the number of colors and flavors. This derivation can be viewed as a
contemporary way of introducing the time-honored NJL model \cite{NJL61}, which has
been applied \cite{VW91,HK94} to a variety of problems in hadronic physics. The virtue of
the model is its simplicity in illustrating the basic mechanism behind spontaneous chiral
symmetry breaking, as we now outline.

In the mean-field approximation, the equation of motion associated with the Lagrangian in
Eq.\ (\ref{eq:NJL}) leads to a gap equation
\begin{equation}
M_q = m_q - G\langle0|\bar{q}q|0\rangle\,\, ,
\end{equation}
which connects the dynamically generated constituent quark mass $M_q$
to the appearance of the chiral quark condensate
\begin{equation}
\langle 0|\bar{q}q|0\rangle= -{\rm tr}\lim_{\,x\rightarrow 0}\langle0| {\cal T}q(0)
\bar{q}(x)|0\rangle = -2iN_fN_c\int {d^4p\over (2\pi)^4}{M_q\,\theta(\Lambda -
|\vec{p}\,|)\over p^2 - M_q^2 + i\epsilon}\,\, .
\label{gapeq}
\end{equation}
The chiral condensate plays the role of an order parameter for the spontaneous breaking
of chiral symmetry. In the chiral limit, where $m_q = 0$, a non-zero constituent quark
mass $M_q$ develops dynamically, together with a non-vanishing chiral condensate
$\langle 0|\bar{q}q|0\rangle$, provided that $G$ exceeds a critical value on the order of
$G_{\rm crit} \simeq 10$ GeV$^{-2}$. The integral in Eq.\ (\ref{gapeq}) requires a
momentum-space cutoff $\Lambda \simeq 2M_q$ beyond which the interaction is
``turned off". The strong non-perturbative interactions polarize the
vacuum and generate a condensate of quark-antiquark pairs, thereby turning an
initially point-like quark with its small bare mass $m_q$ into a dressed quasi-particle
with a size on the order of $(2M_q)^{-1}$.

\subsubsection{\it Pseudoscalar meson spectrum}

By solving the Bethe-Salpeter equations in the color-singlet quark-antiquark channels,
the lightest mesons are generated as quark-antiquark excitations of the correlated QCD
ground state (with its condensate structure). Several calculations have been
performed previously in the three-flavor NJL model \cite{VW91,HK94,KLVW90}. This
model has an undesired $U(3)_L\times U(3)_R$ symmetry, but due to the axial
$U(1)_A$ anomaly of QCD this symmetry\footnote{The axial $U(1)$ transformation
acting on quark fields, $ q \to \exp(i\alpha \gamma_5)q $, constitutes a symmetry only
of classical chromodynamics, not its quantum version.} is reduced to $SU(3)_L\times
SU(3)_R\times
U(1)_V$. In the three-flavor NJL model, instanton-driven interactions are incorporated
in a flavor determinant \cite{tH76} $\det[\bar{q}_i(1 \pm \gamma_5)q_j]$. This
interaction necessarily involves all three quark flavors $u, d, s$ simultaneously in a genuine
three-body contact term.

 \begin{figure}
       \centerline{\includegraphics[width=8.3cm] {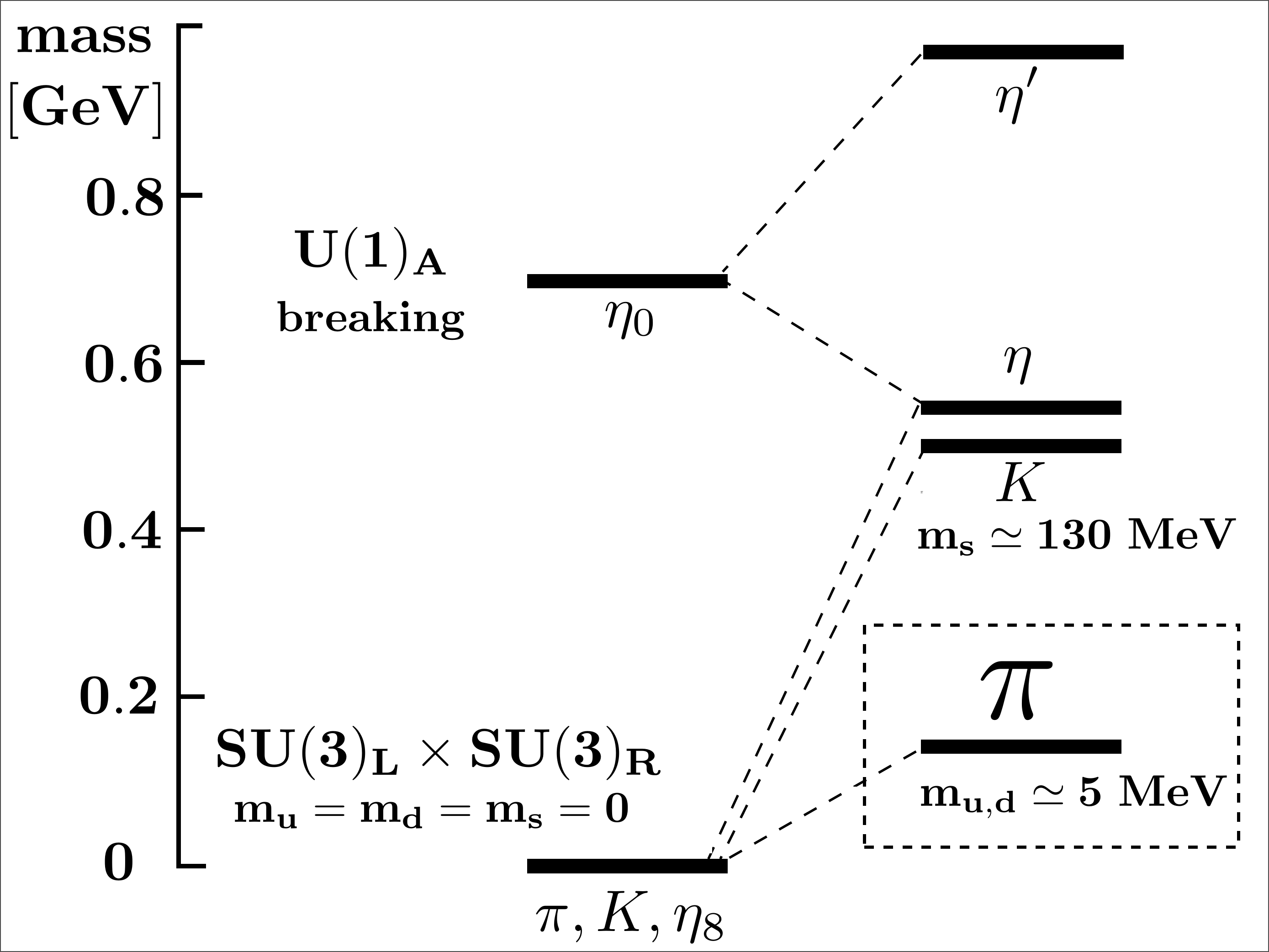}}
\caption{Symmetry breaking pattern in the pseudoscalar meson nonet calculated in the
massless three-flavor NJL model \cite{KLVW90}.}
   \label{fig:2}
 \end{figure}

In Fig.\ \ref{fig:2} we show the symmetry breaking pattern resulting from such a calculation
of the pseudoscalar meson spectrum. Assuming massless $u, d$ and
$s$ quarks, the pseudoscalar octet emerges as a set of massless Goldstone bosons of the
spontaneously broken $SU(3)_L\times SU(3)_R$ symmetry, while the anomalously broken
$U(1)_A$ symmetry drives the singlet $\eta_0$ away from the Goldstone boson sector.
The inclusion of finite quark masses that explicitly break chiral symmetry shifts the
pseudoscalar ($J^\pi = 0^-$) nonet into its empirically observed hierarchy, including
$\eta_0$-$\eta_8$ mixing.

The pion mass is related to the scalar quark condensate and bare quark masses through
the famous Gell-Mann-Oakes-Renner relation \cite{GOR68}:
\begin{equation}
m^2_{\pi}\,f_{\pi}^2 = - {1\over 2}(m_u + m_d) \langle 0| \bar{q} q |0\rangle +
{\cal O}(m^2_{u,d} \,\ln (m_{u,d})).
\label{eq:GOR}
\end{equation}
derived from current algebra and PCAC.
It involves additionally the pion decay constant, $f_\pi = 92.4$\,MeV, which is defined
by the matrix element connecting the pion state with the QCD vacuum via the isovector
axial vector current, $A^\mu_a = \bar{q}\gamma^\mu\gamma_5 {\tau_a \over 2} q$:
\begin{equation}
\langle 0 | A^{\mu}_a (0) | \pi_b (p) \rangle = i \delta_{ab}\, p^{\mu} f _\pi\, .
\label{eq:fpi}
\end{equation}
The pion decay constant, like the chiral condensate $\langle 0| \bar{q} q |0\rangle$,
is a measure of spontaneous chiral symmetry breaking associated with the
scale $\Lambda_\chi \sim 4\pi f_\pi \sim 1$ GeV. The non-zero pion mass,
$m_\pi \simeq 138$\,MeV\,$\ll \Lambda_\chi$, is a reflection of the explicit chiral symmetry
breaking by the small quark masses, with $m^2_{\pi} \sim m_q$. The quark masses
$m_{u,d}$ as well as the scalar quark condensate $\langle 0|\bar{q} q |0\rangle$
are scale-dependent quantities, and only their product is renormalization group
invariant. At the renormalization scale of 2 GeV, a typical average quark mass
${1\over 2}(m_u + m_d) \simeq 3.5$ MeV leads to a condensate $\langle 0 |\bar{q} q
|0 \rangle \simeq -(0.36$\,GeV$)^3$.

The quark masses set the primary scales in QCD, and their
classification into ``light" and ``heavy" quarks determines very different
physics phenomena. The heavy quarks (e.g., the $t$, $b$ and -- within limits --
the $c$ quarks), whose reciprocal masses offer a natural ``small parameter", can be
treated in non-relativistic approximations (that is, expansions of observables in powers of
$1/m_{t,b,c}$). The sector of the light quarks (e.g., the $u$, $d$ quarks and -- to some extent -- the
$s$ quark), however, is governed by quite different principles and rules. Evidently, the light quark
masses themselves are now ``small parameters" to be compared with a
``large" scale of dynamical origin. This large scale is characterized by a
mass gap of about 1 GeV separating the QCD vacuum from nearly all of its
excitations, with the exception of the pseudoscalar meson octet shown in Fig.~\ref{fig:2}.
This mass gap is comparable to $4\pi f_\pi$, the energy scale associated with
the spontaneous breaking of chiral symmetry in QCD.

\subsection{Chiral effective field theory}

The basic premise on which our discussion will be anchored is Weinberg's ``folk theorem"~\cite{Wei79} which underlies the rich variety of concepts involved in formulating effective field theories in physics. The ``theorem" that we shall often refer to as WFT (standing for Weinberg folk theorem) states:
``When you use quantum
field theory to study low-energy phenomena, then according to the folk theorem
you're not really making any assumption that could be wrong, unless
of course Lorentz invariance or quantum mechanics or cluster decomposition
is wrong, provided you don't say specifically what the Lagrangian is. As
long as you let it be the most general possible Lagrangian consistent with
the symmetries of the theory, you're simply writing down the most general
theory you could possibly write down. This point of view has been used in
the last fifteen years or so to justify the use of effective field theories, not just
in the tree approximation where they had been used for some time earlier,
but also including loop diagrams. Effective field theory was first used in this
way to calculate processes involving soft $\pi$ mesons, that is, $\pi$ mesons with
energy less than about $2\pi F_\pi\approx 1200$ MeV. The use of effective quantum
field theories has been extended more recently to nuclear physics, where
although nucleons are not soft they never get far from their mass shell, and
for that reason can be also treated by similar methods as the soft pions.
Nuclear physicists have adopted this point of view, and I gather that they
are happy about using this new language because it allows one to show in a
fairly convincing way that what they've been doing all along (using two-body
potentials only, including one-pion exchange and a hard core) is the correct
first step in a consistent approximation scheme.''\cite{wtheorem}

Low-energy QCD is the physics of systems of light quarks at energies and momenta
smaller than 1 GeV, the scale that provides a natural separation between ``light"
and ``heavy" (or ``fast" and ``slow") degrees of freedom. The basic
idea of effective field theories is, in the spirit of the WFT, to
introduce the active light particles as dynamical degrees of freedom, while the
heavy particles are integrated out or  ``frozen'' and treated as (almost) static sources. The dynamics
is described by the most general effective Lagrangian that incorporates all relevant
symmetries of the underlying fundamental theory and whose terms are organized
in powers of the light scale over the heavy scale. In low-energy QCD, confinement and spontaneous
chiral symmetry breaking imply that the light degrees of freedom are the
Nambu-Goldstone bosons. In the following, we restrict ourselves to $N_f = 2$.

\subsubsection{\it The Nambu-Goldstone boson sector}

In this section we briefly summarize the steps \cite{Wei66/67,GL84} for constructing the
chiral Lagrangian in the pure meson sector (baryon number $B= 0$). A chiral
field is introduced as
\begin{equation}
U(x) = \exp\bigg({i \over f_\pi}\, \vec \tau \cdot \vec \pi(x) \bigg) \in SU(2)~,
\end{equation}
where $f_\pi$ is the the pion decay constant in the chiral limit ($m_\pi \rightarrow 0$).
The physics of QCD is then phrased in terms of an
effective Lagrangian involving the chiral field $U(x)$ and its derivatives:
\begin{equation}
{\cal L}_{QCD} \to {\cal L}_{eff} (U, \partial^\mu U, ...) .
\label{eq:Leff}
\end{equation}
Since Goldstone bosons interact only when they carry non-zero four-momenta, the low-energy
expansion of Eq.~(\ref{eq:Leff}) allows for an ordering in powers of $\partial^{\mu} U$. From Lorentz
invariance only even numbers of derivatives are permitted\footnote{Note that for $N_f=3$, there is the Wess-Zumino term with three derivatives.}. One writes for the first few terms of the
chiral Lagrangian:
\begin{equation}
{\cal L}_{\rm eff} = {\cal L}_{\pi\pi}^{(2)} + {\cal L}_{\pi\pi}^{(4)} + ...\,,
\end{equation}
where the leading term (encoding the current algebras, leading order in the non-linear sigma model) involves two derivatives:
\begin{equation}
{\cal L}^{(2)}_{\pi\pi} = {f_\pi^2 \over 4} {\rm tr} ( \partial^{\mu} U \partial_{\mu}
U^{\dagger} )\, .\end{equation}
At fourth order the additional terms permitted by symmetries are
\begin{equation}
{\cal L}^{(4)}_{\pi\pi} = {\ell_1 \over 4} \Big[ {\rm tr}(\partial^{\mu} U \partial_{\mu}
U^{\dagger})\Big]^2 + {\ell_2 \over 4}  {\rm tr} (\partial_{\mu} U \partial_{\nu} U^{\dagger})
\, {\rm tr} (\partial^{\mu} U \partial^{\nu} U^{\dagger}) +\dots \, ,\end{equation}
where further contributions involving the light quark mass $m_q$ and external fields are not shown.
The constants $\ell_1, \ell_2$ absorb loop
divergences and their finite scale-dependent parts must be fixed by matching to experiment.

The symmetry-breaking mass term is small, so that it too can be handled perturbatively.
The leading contribution introduces a term that is linear in the quark mass matrix $m_q = {\rm diag}
(m_u,m_d)$:
\begin{equation}
{\cal L}^{(2)}_{\pi\pi} = {f_\pi^2 \over 4} {\rm tr}(\partial_{\mu} U \partial^{\mu}
U^{\dagger}) + {f_\pi^2 \over 4} m_\pi^2 \, {\rm tr}(U + U^{\dagger}) \,,
\end{equation}
where $m_\pi^2 \sim (m_u+m_d)$. The fourth-order Lagrangian ${\cal L}^{(4)}_{\pi\pi}$ also
receives explicit chiral symmetry breaking contributions (proportional to $m_q$ and
$m_q^2$)\footnote{Higher-order quark mass terms that can be systematically brought in using the spurion field ${\cal M}$ that transforms like the chiral field $U$ with $\la{\cal M}\ra=m_q$ will figure in the skyrmion matter description of dense matter described below.} with the introduction of additional low-energy constants $\ell_3$ and $\ell_4$.

Provided that the effective Lagrangian includes all terms allowed by the symmetries
of QCD, chiral effective field theory is the low-energy equivalent \cite{Wei79,L94}
of QCD. The framework for systematic perturbative calculations of (on-shell) $S$-matrix elements involving
Nambu-Goldstone bosons, chiral perturbation theory (ChPT), is then defined by the following
rules: Collect all Feynman diagrams generated by ${\cal L}_{eff}$. Classify
individual terms according to powers of the small quantity $p/(4\pi f_{\pi})$,
where $p$ stands generically for three-momenta of Goldstone bosons or for the pion mass $m_{\pi}$. Loops are evaluated in
dimensional regularization and are renormalized by appropriate chiral counter terms.

How to incorporate massive mesons, i.e., the vector mesons $V$ and the dilaton $\chi$,  as needed for certain processes in consistency with chiral symmetry, hidden local symmetry and spontaneously broken scale symmetry, will be described in detail below.

\subsubsection{\it The baryon sector}

In a systematic large $N_c$ expansion to access QCD, the effective Lagrangian is  given by a weakly coupled mesonic Lagrangian of the nonlinear sigma model form described above or hidden-local-symmetrized form to be described below. The baryon arises as a toplogical soliton with the topology lodged in the chiral field $U$~\cite{witten}. This is the famous skyrmion~\cite{skyrme}. It will be seen in  Section \ref{cshs} that the skyrmion structure of the baryon characterized by topology can be ``smoothly" transformed into the quark-bag structure of QCD. It thereby provides the ``bridge" between the hadronic degrees of freedom and the QCD degrees of freedom. Furthermore when looked at in terms of holographic QCD that comes from string theory, the baryon appears as an instanton in 5D Yang-Mills Lagrangian, point-like in certain limits (large $N_C$ and large 't Hooft $\lambda\equiv N_c g_s^2$ limit)~\cite{hongetal}. It makes sense to take the baryon, the mass scale of which is much greater than that of fluctuating mesonic fields, $\pi$, $V$ and dilaton, as a point-like local field. Thus in the spirit of the Weinberg theorem, we may simply take the nucleon as a local field as in \cite{hongetal}. In nuclear physics  a local nucleon field is simply introduced as a matter field subject to the proper symmetries involved. Certain subtle properties of the skyrmion structure in dense matter that is most likely inaccessible by the local baryon field description will be discussed.

\subsubsection{\it Chiral pion-nucleon effective Lagrangian}

The prominent role played by the pion as a pseudo-Goldstone boson of spontaneously
broken chiral symmetry impacts as well the low-energy structure
and dynamics of nucleons \cite{TW01}. When probing the nucleon with a
long-wavelength electroweak field, a substantial part of the response comes from
the pion cloud, comprising the ``soft'' surface of the nucleon.
The calculational framework for this, baryon chiral perturbation theory
\cite{EM96,BKM95} has been applied successfully to diverse low-energy
processes (such as low-energy pion-nucleon scattering, threshold pion photo- and
electro-production and Compton scattering on the nucleon).

We now consider the physics of the pion-nucleon system, the sector with baryon number $B = 1$.
The nucleon is represented by a local isospin-$\frac{1}{2}$ doublet field
$\Psi= \left(\!\!\begin{array}{c} p \\ n\end{array}\!\!\right)$ of protons and neutrons with free Lagrangian
\begin{equation}
{\cal L}_N^{\rm free} = \bar{\Psi}(i\gamma_{\mu}\partial^{\mu} - M_0)\Psi
\end{equation}
where $M_0$ is the nucleon mass in the chiral limit.
Unlike the pion, the nucleon has a large mass of the same order as the chiral symmetry breaking
scale $4\pi f_\pi$, even in the limit of vanishing quark masses.
The additional term in the chiral Lagrangian involving the nucleon, denoted
by ${\cal L}_{\pi N}$, is again dictated by chiral symmetry and expanded in powers of the quark masses and
derivatives of the Goldstone boson field:
\begin{equation}
{\cal L}_{\pi N} = {\cal L}_{\pi N}^{(1)} + {\cal L}_{\pi N}^{(2)} + \dots
\end{equation}
In the leading term, ${\cal L}_{\pi N}^{(1)}$, there is a vector current coupling between pions and
the nucleon (arising from the replacement of $\partial^{\mu}$
by a chiral covariant derivative) as well as an axial vector coupling:
\begin{equation}
{\cal L}_{\pi N}^{(1)} =  \bar{\Psi}\Big[i\gamma_{\mu}(\partial^{\mu} +\Gamma^{\mu})- M_0 +
g_A \gamma_{\mu}\gamma_5\, u^{\mu}\Big]\Psi \, .\label{eq:LeffN}
\end{equation}
The vector and axial vector quantities involve the pion fields via $\xi = \sqrt{U}$ in the form
\begin{eqnarray}
\Gamma^{\mu} & = & {1\over 2}[\xi^{\dagger},\partial^{\mu}\xi] = {i\over 4f_\pi^2}
 \, \vec \tau \cdot (\vec \pi \times \partial^{\mu}\vec \pi) + ...~~, \\
u^{\mu} & = & {i\over 2}\{\xi^{\dagger},\partial^{\mu}\xi\}= - {1\over 2f_\pi}\, \vec \tau
\cdot \partial^{\mu}\vec \pi + ...~~,
\end{eqnarray}
where the last steps result from expanding $\Gamma^{\mu}$ and $u^{\mu}$ to
leading order in the pion fields. Up to this point the only parameters that enter are the
nucleon mass $M_0$, the pion decay constant $f_\pi$, and the nucleon axial vector
coupling constant $g_A$, all three taken in the chiral limit.

The next-to-leading order pion-nucleon Lagrangian, ${\cal L}_{\pi N}^{(2)}$, contains
the chiral symmetry breaking
quark mass term, which shifts the nucleon mass to its physical value. The nucleon sigma term
\begin{equation}
\sigma_N = m_q\frac{\partial M_N}{\partial m_q} =
\langle N | m_q(\bar{u}u + \bar{d}d) |N\rangle
\end{equation}
measures the contribution of the non-vanishing quark mass to the nucleon mass $M_N$.
Its empirical value, deduced \cite{GLS91} from low-energy pion-nucleon data, is in the range
$\sigma_N \simeq (45 \pm 8)$ MeV.
Up to next-to-leading order, the $\pi N$ effective Lagrangian has the form
\begin{eqnarray}
{\cal L}_{eff}^{N} & = & \bar{\Psi}(i\gamma_{\mu}\partial^{\mu} - M_N)\Psi -
{g_A \over 2f_{\pi}} \bar{\Psi}\gamma_{\mu}\gamma_5\vec\tau\,\Psi \cdot\partial^{\mu}
\vec \pi  \nonumber \\ && -{1 \over 4f_{\pi}^2} \bar{\Psi}\gamma_{\mu} \vec \tau\,\Psi
\cdot (\vec \pi\times \partial^{\mu} \vec \pi\,) +{\sigma_N\over  2f_\pi^2}\,\bar{\Psi}
\Psi\,\vec \pi^{\,2} + ...~~, \end{eqnarray}
where we have not shown additional terms involving
$(\partial^{\mu} \vec\pi)^2$ that arise from the complete Lagrangian ${\cal L}_{\pi N}^{(2)}$.
These terms come with further low-energy constants $c_3$ and $c_4$ that encode physics at smaller
distance scales and that need to be fitted to experimental data,
e.g., from pion-nucleon scattering.



\section{Chiral symmetry and hadron structure}
\label{cshs}

In this section, the notion that there is a duality between hadronic interactions -- involving mesons and baryons -- and QCD interactions -- involving quarks and gluons -- will be developed. The point is that  at low energies, one should be able to go ``smoothly" from the description in hadronic language to a description in quark-gluon language. This means that the nuclear dynamics that has been developed since decades can capture what can be given in terms of QCD. In fact this has been the underlying theme dating back to 1970's as discussed in Section \ref{MEC}. This continuity between hadrons and quarks and gluons is implied in the principle named ``Cheshire Cat." The crucial point here is that the quark-gluon degrees of freedom of QCD, the ``smile of the Cheshire Cat," can be traded in for topology -- and vice-versa -- for low-energy hadronic physics, with the boundary conditions providing the mediation.

The chiral bag model that marries chiral symmetry with confinement, the key elements of QCD, has played the principal role in the development of the structure of elementary nucleons and many-body nuclear systems including compressed baryonic matter that is relevant to the core of neutron stars. Given that what is involved is the trading between QCD degrees of freedom and hadron degrees of freedom via topology, we might call this ``quark-topology duality."  We will encounter a similar phenomenon in Section \ref{skyrm} where a dense matter consisting of half-skyrmions resembles strong-coupling quark matter.

We will pick two cases that illustrate the essential points. One is how the baryon charge can be shared between the quarks and the pions carrying topological charge. The other is how the flavor singlet axial charge is shared between the quarks and $\eta^\prime$ and the gluons. In both cases, it is the quantum anomalies, the first in the vector current and the second in the axial current,  that play the key role.

\subsection{From little bag to chiral bag}
Before quantum chromodynamics was put forward, nucleons were treated as point-like fermionic particles or local fermionic fields in phenomenological Lagrangians. The physical sizes of the nucleons as seen by external electroweak fields were
then described in terms of the form factors suitably parameterized by the meson degrees of freedom considered to be involved, such as the pion or the vector mesons $\rho$ and $\omega$. This had been the starting point of many-body approaches to nuclear structure and interactions, generically the whole edifice of nuclear physics. The advent of QCD brought a drastic conceptual change to this ``old" picture.

The basic premise of QCD, the nonabelian gauge theory, namely, asymptotic freedom, implies that the microscopic constituents of hadrons, colored quarks and gluons, are confined. A simple and highly successful model that encapsulates this feature is the MIT bag model. In this model~\cite{MITbag}, the nucleon is described by three quarks -- corresponding to $N_c=3$ -- of flavor $u$ and $d$ confined in a bag of size $R\sim 1$ fm, equivalent roughly to the size of the nucleon. The nearly massless quarks are confined and interact only weakly, via gluon exchange, inside the bag. In this picture which has enjoyed a large success in the physics of light-quark hadrons, there is no distinction between mesons that contain a quark and an antiquark and baryons that contain three quarks. It is still widely applied to describing what happens in baryonic matter at high density as in compact stars.

A naive implementation of the MIT bag model to nuclear structure got into conflict with the size of the bag $R\approx 1$ fm, roughly of the baryon size seen by EM probes, in nuclear medium vs. the independent particle picture of nucleons in heavy nuclei such as the well-studied Pb nucleus. Colloquially speaking, the big MIT bag in a big nucleus looked like a ``grapefruit in a salad bowl," too big to be consistent with what is known experimentally of nuclear structure. As a way-out, the ``Little Bag" was proposed by exploiting that restoring chiral symmetry -- which is missing in the MIT bag model -- would require the presence of pions outside of the bag and the pressure asserted by the pions would squeeze the bag to a smaller size~\cite{littlebag}. It turns out that this picture was incorrect, and what emerged with the rediscovery of the skyrmion model was the chiral bag picture leading  to the Cheshire Cat principle that made the bag (confinement) size unphysical~\cite{magicangle}. In this section, this development is described in terms of a simple model that illustrates how the Cheshire Cat works and how the hadron-quark continuity emerges.

\subsection{Encoding chiral symmetry and confinement in the chiral bag}
In low-energy dynamics, one works, in accordance with the WFT, with color-singlet effective fields for baryons and mesons without invoking quarks and gluons. In fact, all light-quark hadrons can be described in terms of local meson fields, with the baryons arising as skyrmions from the meson Lagrangian. There confinement is implicit. If, however, one wants to incorporate both chiral symmetry and confinement on the same footing in an explicit form, the chiral bag model offers a simple approach to the problem.

\begin{figure}[h]
\begin{center}
\includegraphics[width=8cm]{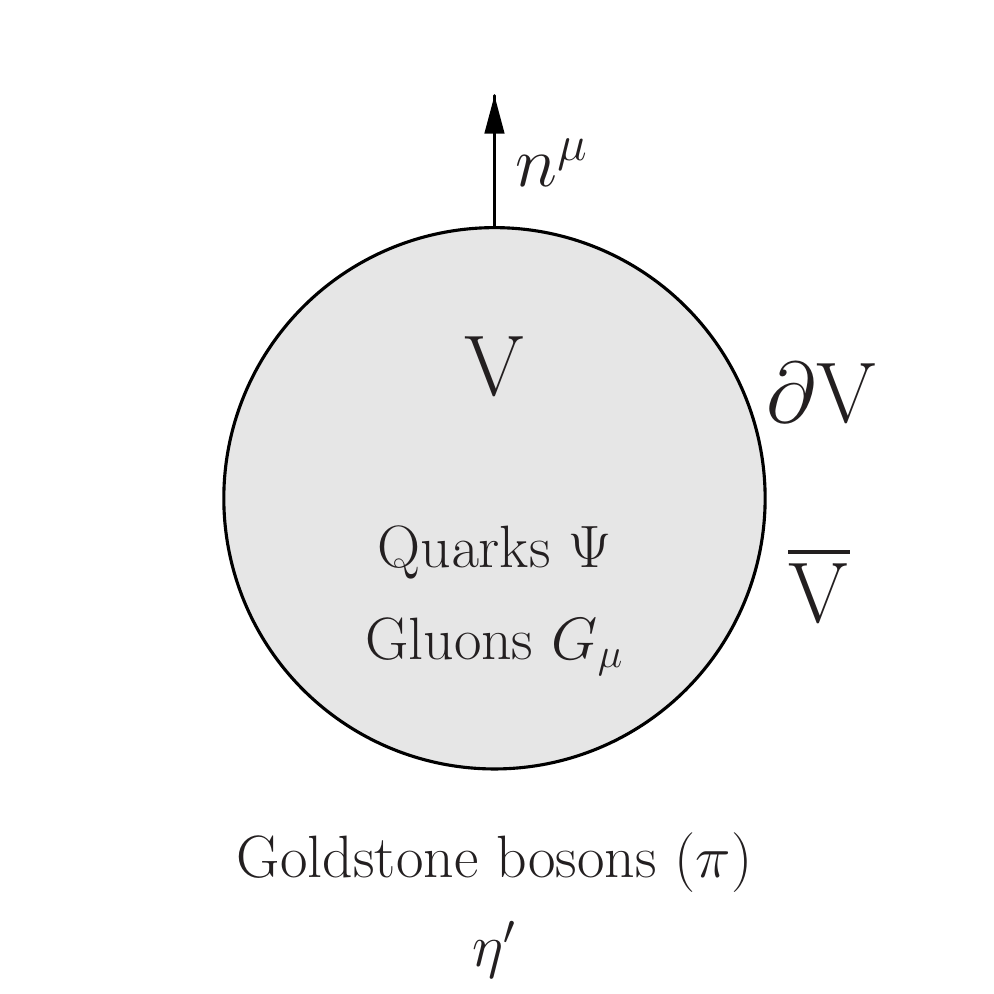}
\caption{The chiral bag with quarks and gluons inside the bag of volume $V$, taken to be spherical and Goldstone bosons plus $\eta^\prime$ (9th pseudoscalar meson) outside the bag of volume $\bar{V}$,
}
\label{chiral-bag}
\end{center}
\end{figure}

The model, caricatured in Fig.~\ref{chiral-bag}, is given in the simplest form\footnote{In this section, the subscript $s$ will be used for QCD constants, i.e., $g_s$ and $\alpha_s$, to be distinguished from other constants, such as the hidden gauge coupling denoted as $g$ and the effective vector coupling $g_V$, that appear in other sections.}
\be
S&=& S_V+S_{\tilde{V}}+S_{\del V},\label{cheshire}\\
S_V&=& \int_V d^4x \left(\bar{\psi}i\not\!\!{D}\psi -\frac{1}{2}
{\rm Tr}\ G_{\mu\nu}G^{\mu\nu}\right)+\cdots\nonumber\\
S_{\tilde{V}}&=&\frac{f^2}{4}\int_{\tilde{V}} d^4x \left(\tr\
\del_\mu U^\dagger \del^\mu U +\frac{1}{4N_f}
m^2_{\eta^\prime}({\rm ln}U-{\rm ln}U^\dagger)^2
\right) +\cdots  + S_{WZW},\nonumber\\
S_{\del V}&=&
\frac{1}{2}\int_{\del V} d\Sigma^\mu\left\{(n_\mu \bar{\psi} U^{\gamma_5}\psi)
+i\frac{g^2}{16\pi^2}{K_5}_\mu
(\tr\ {\rm ln} U^\dagger-\tr\ {\rm ln} U)\right\}.\nonumber
\ee

Here $\psi$ is the quark field $\psi^T=(uds)$, $G_\mu$ the gluon field, $g_c$ the color gauge coupling and $K_5^\mu$ is what is known as Chern-Simons current, $U=e^{i\eta^\prime/f_0} e^{2i\pi/f} $ is the chiral field and $U^{\gamma_5}=e^{i\eta^\prime\gamma_5/f_0} e^{2i\pi\gamma_5/f}$. The constant $f$ can be identified with the pion decay constant $f_\pi$ within the approximation we are applying. Here we use $\Tr$ for color trace in contrast to $\tr$ reserved for flavor trace. The ellipses stand for other terms that can figure.
This action encapsulates the long-wavelength properties of baryons in the simplest form. Implementing with vector mesons as hidden gauge fields discussed in Section \ref{him} would make the action applicable at higher energy scales or at higher densities as in our case. For what is described below, this simple action is sufficient.

The action in Eq.~(\ref{cheshire}) that we will refer generically to as the ``chiral-bag action" possesses two properties which are highly non-trivial. The volume actions both in $V$ and $\bar{V}$ are standard. What is not standard is the boundary action. As it stands at the classical level, the boundary action  $S_{\del V}$ confines quarks inside the bag, but the color charge is accumulated at the boundary, so the gauge invariance is broken. Quantum mechanically, however, the baryon charge carried by the quarks is not confined inside the bag but leaks out, and the gauge non-invariant surface term is precisely cancelled by a quantum anomaly term, thereby restoring gauge invariance. We will discuss these two features, the former involving the vector anomaly and the latter involving the axial vector anomaly. These capture the subtleties of baryons in QCD.

\subsubsection{\it Leakage of baryon charge}\label{cc-leakage}
The baryon in the MIT bag has its baryon charge entirely inside the bag. Classically the baryon charge is confined inside. It turns out that when the pions are coupled to the quarks in the chiral bag, this is no longer the case. This was noted a long time ago when the pion field outside the bag had a hedgehog  configuration~\cite{rgb} at the ``magic angle"~\cite{magicangle} $\theta=\pi/2$,
\be
U=e^{i\tau\cdot\hat{r}\theta}.
\ee
In analogy to a fermion coupled to a soliton (e.g., magnetic monopole) involving a zero mode for the fermion which turns out to have a 1/2 fermion charge~\cite{jackiw-rebbi}, the bag containing 3 quarks was found to have a 1/2 baryon charge. The rediscovery of the skyrmion model for the nucleon in 1983 immediately led to the identification of the missing 1/2  baryon charge in the pion cloud~\cite{rgb,magicangle}. This was for the magic angle $\theta=\pi/2$. That the partition of one unit of baryon charge into the inside and outside of the bag takes place for any chiral angle was then shown by Goldstone and Jaffe~\cite{goldstone-jaffe} using the same chiral bag model. This meant that when the bag radius goes to $\infty$, the baryon charge is entirely inside the bag whereas when the bag radius shrinks to a point, the charge gets lodged entirely in the pion cloud.

Why and how this ``leakage of baryon charge" occurs is a story of anomaly caused by infinities, often referred to as ``infinite-hotel phenomenon."~\cite{CC,CNDII} A simple way of understanding this phenomenon is to recognize that the boundary condition resulting from imposing the axial-current conservation, i.e., chiral invariance, renders the $U(1)_V$ vector current to be violated at the surface and hence the baryon charge leaks out. This is because the chiral invariance generates at the boundary an abelian axial vector field and this leads to the vector anomaly. The chiral bag model is constructed in such a way that this leaking baryon charge is picked up precisely by the topology lodged in the pion field.

What this implies is that as far as the baryon charge is concerned, the bag size has no physical meaning. The statement that this phenomenon applies to {\em all} physical processes is the Cheshire Cat principle  (CCP for short)~\cite{CC}, presaged in \cite{magicangle}. The boundary conditions connect the baryon, a fermion, to the pion, a boson. In (1+1) dimensions, the exact bosonization technique is available to establish the CCP. There is no exact bosonization in (3+1)D, hence the idea of CC can be at best approximate, and the chiral bag described by the action in Eq.~(\ref{cheshire}) is expected to work at low energy, at best, approximately. As developed elsewhere, one can make the model applicable at higher-energy scale by incorporating hidden local fields.

The upshot of the reasoning developed here is that QCD dynamics can be traded in for hadron dynamics via topology. It is the boundary that does the trading.\footnote{Such a singular role of boundaries is fairly well recognized in condensed matter physics.}  Although this model has not been fully explored due to such subtleties as Casimir effects and associated $1/N_c$ corrections that are extremely difficult to treat reliably, what it suggests is that there is a {\it hadron-quark duality} in nuclear interactions. This feature will be an underlying principle of this article in a variety of nuclear processes at low energy here as well as in other sections.

\subsubsection{\it Flavor-singlet axial charge $g_A^0$}
The CCP for the baryon charge we saw above is exact  thanks to the topological invariance,  independently of the dimensions of the space-time and of specific dynamics involved. We now consider a case where the topology does not impose an obvious direct condition, and yet the CCP, nonetheless, holds, albeit approximately. A subtle nontrivial manifestation of CCP is in the role the color degree of freedom of QCD plays in hadron properties. A highly intriguing example is the flavor-singlet axial charge $g_A^0$ of the proton. This object has figured importantly in the so-called ``proton spin problem." The way  $g_A^0$ enters into the proton spin is quite intricate and remains still controversial, and we won't go into it.

The physically relevant object involved is  the sum rule for the first moment of the proton, i.e.,
$$
\Gamma^P_1(Q^2) \equiv \int_0^1dx g_1^p(x,Q^2) =
\frac{1}{12}C_1^{NS}(\alpha_s(Q^2))\left(a^3
+ \frac{1}{3} a^8\right) $$
\be
+\frac{1}{9}C_1^S(\alpha_S(Q^2))a^0(Q^2) .
\ee
Here $C_1(\alpha_s)$'s are  first
moments of the Wilson coefficients of the singlet ($S$) and nonsinglet
($NS$) axial currents and $\alpha_s$ the perturbatively running $QCD$ coupling
constant.
$a^3$, $a^8$ and $a^0(Q^2)$ are the form factors in the
forward proton matrix elements of the renormalized axial current,
i.e.,
$$
\langle p,s|A^3_\mu|p,s \rangle =s_\mu \frac{1}{2} a^3, \;\;\;\;\;
\langle p,s|A^8_\mu|p,s \rangle =s_\mu
\frac{1}{2\sqrt{3}} a^8, $$
and
\be
\langle p,s|A^0_\mu|p,s \rangle =s_\mu  a^0 ,\label{fsac}
\ee
where $p_\mu$ and $s_\mu$ are the momentum and the polarization
vector of the proton. $a^3$ and $a^8$ can be chosen $Q^2$
independent and may be determined from the $\frac{G_A}{G_B}$ and
$\frac{F}{D}$ ratios. $a^0(Q^2)$ evolves due to the anomaly.
Naive models or the $OZI$ approximation to $QCD$ lead, at low energies, to
\be
a^0 \approx a^8 \approx 0.69 \pm 0.06 .
\ee
Experimentally~\cite{EMC}\footnote{This discrepancy between the constituent quark model and the experiment, with the FSAC interpreted in terms of the proton spin, led to the proton spin crisis.}
\be
a^0 = 0.12 (17) \ {\rm or}\ 0.29 (6)\,,
\ee
and lattice QCD gives $a_0=0.20 (12)$.

As with other physical quantities considered above, we partition the axial current that we are interested in into a component inside and a component outside as
\be
A^\mu =A^\mu_B \Theta_B + A^\mu_M \Theta_M\label{current}
\ee
where $\Theta_B=\theta (R-r)$ and $\Theta_M=\theta (r-R)$ with $R$ being the radius of the bag which may be taken spherical.  Since we are dealing only with the flavor-singlet axial current, we  omit the flavor index in the current and understand in what follows that $A_\mu$ stands for the flavor-singlet axial current (FSAC for short).

Due to the $U(1)_A$ anomaly, the FSAC is not conserved. The divergence of the FSAC is taken to be given by\footnote{In what follows, $\eta$ will stand for $\eta^\prime$.}
\be
\partial_\mu A^\mu =
\frac{\alpha_s N_F}{2\pi}\sum_a \vec{E}^a \cdot \vec{B}^a \Theta_{B}+
f m_\eta^2 \eta \Theta_{M}.\label{abj}
\ee
This formula should be subject to boundary conditions at the bag boundary
\be
n_\mu A^\mu_B=n_\mu A^\mu_M\label{bc}
\ee
where $n^\mu$ is the outward normal to the bag
surface with $n^2=-1$.

Due to the $U(1)_A$ anomaly, this boundary condition becomes a lot subtler than above. Since $\eta^\prime$ is present in addition to the pion field which accounts for spontaneously broken chiral symmetry, the vacuum fluctuations inside the bag that induce the baryon charge leakage into the skyrmion outside also induce a color leakage due to coupling to a pseudoscalar~\cite{nrwz}. This is a quantum effect associated with the chiral anomaly. In order to preserve color invariance, this leakage of color charge has to be prevented and this can be done by putting into the CBM Lagrangian a counter term of the form (for $N_c=3$)
\be {\cal
L}_{CT}=i\frac{g_s^2}{32\pi^2}\oint_{\Sigma} d\beta K^\mu n_\mu
({\tr}\,{ \ln}\, U^\dagger -{\tr}\, {\ln}\, U)\label{lct}
\ee
where $\beta$ is a point
on a surface $\Sigma$, $U$ is the $U(N_F)$ matrix-valued field written as
$U=e^{i\pi/f} e^{i\eta/f}$ and $K^\mu$ the properly regularized
Chern-Simons current $K^\mu=\epsilon^{\mu\nu\alpha\beta} (G_\nu^a
G_{\alpha\beta}^a -\frac 23 f^{abc} g_s G_\nu^a G_\alpha^b
G_\beta^c)$ given in terms of the color gauge field $G^a_\mu$.
Note that Eq.~(\ref{lct}) manifestly breaks color gauge invariance
(both large and small, the latter due to the bag), so the action
of the chiral bag model with this term is not gauge invariant at
the classical level but  when quantum fluctuations are calculated~\cite{nrwz}, there appears an induced anomaly term on the surface which is exactly canceled by Eq.~(\ref{lct}).  Gauge
invariance is preserved, not at the classical level, i.e., in the Lagrangian, but at the quantum level.

For numerically evaluating the matrix elements, it is found to be convenient to rewrite the current in Eq.~(\ref{current}) as
\be
A^\mu=A_{B_{Q}}^\mu + A_{B_{G}}^\mu + A_\eta^\mu\label{sep}
\ee
such that Eq.~(\ref{abj}) is in the form
\be
\partial_\mu (A_{B_{Q}}^\mu + A_\eta^\mu) &=& f m_\eta^2 \eta
\Theta_{M},\label{Dbag}\\
\partial_\mu A_{B_G}^\mu &=&
\frac{\alpha_s N_F}{2\pi}\sum_a \vec{E}^a \cdot \vec{B}^a
\Theta_{B}\label{Dmeson} \ee
where the subindices Q and G imply that these currents are written in terms of quark and gluon fields respectively. It should be stressed that since one is dealing with
an interacting theory, there is no unique way to separate the
different contributions from the gluon, quark and $\eta$
components. In particular, the separation into Eqs.~(\ref{Dbag})
and (\ref{Dmeson}) is neither unique nor physically meaningful, although the sum has no ambiguity. Separately, they may not necessarily be gauge-invariant. The merit of this separation is a natural partition of the contributions in the framework of the bag description for the confinement mechanism that we are using here.

\begin{figure}[ht]
\begin{center}
\includegraphics[width=12cm]{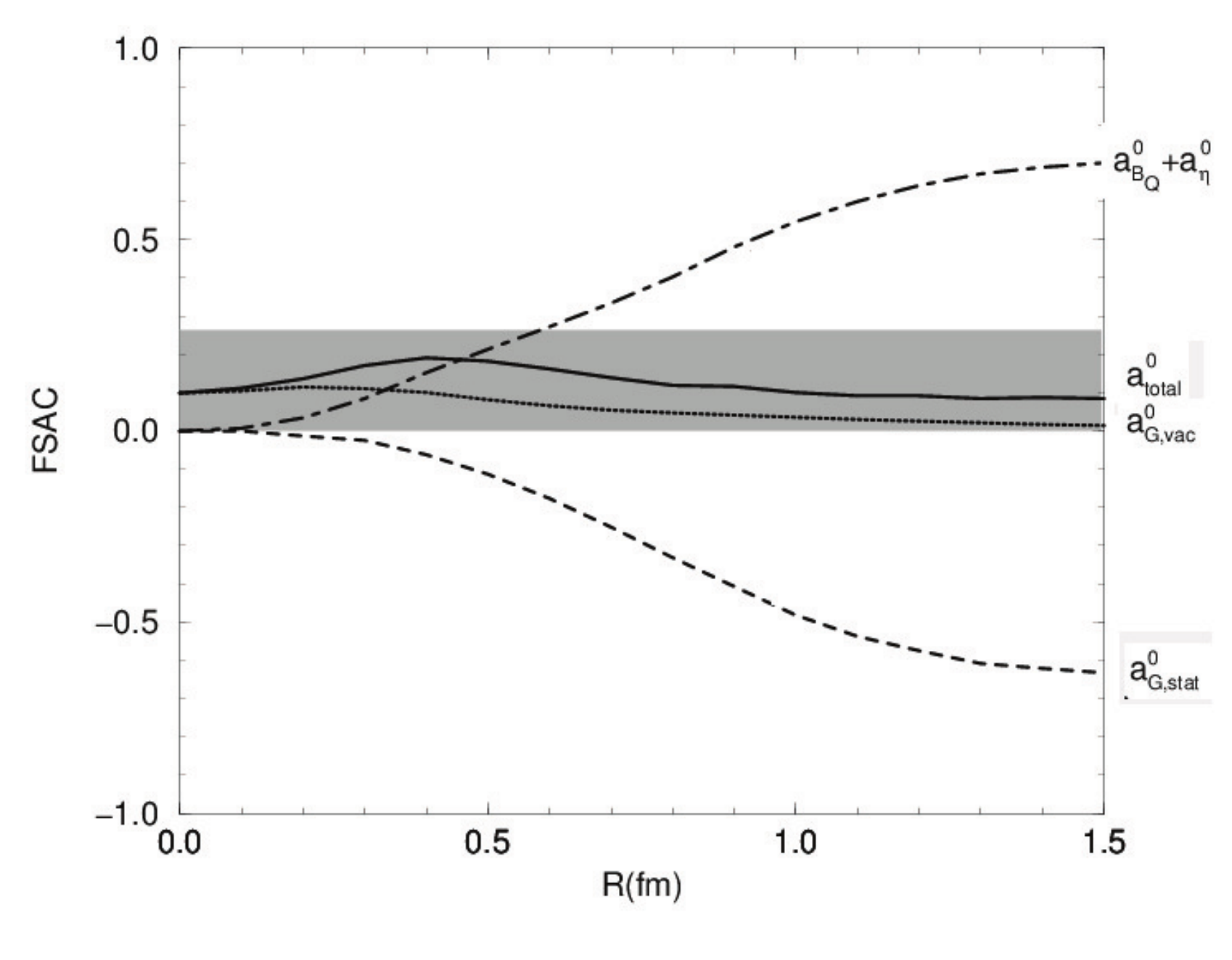}
\caption{``Quarkish" and ``gluonish" FSAC given in the chiral bag model that illustrates the working of the CCP. Details of the calculation and the parameters that enter in the calculation are found in \cite{LPRV}.
}
\label{FSAC}
\end{center}
\end{figure}

The flavor singlet matrix element $a^0\equiv g_A^0$ of Eq.~(\ref{fsac}) is comprised of two terms, ``quarkish" ($Q$) and ``gluonish" ($G$):
\be
s^\mu a^0_Q&=&\langle p,s|(A_{B_{Q}}^\mu + A_\eta^\mu)|p,s \rangle,\\
s^\mu a^0_G&=&\langle p,s|A_{B_{G}}^\mu |p,s \rangle.
\ee
The boundary condition Eq.~(\ref{bc}) allows one to express the $\eta$ current contribution in terms of the quark-bag contribution,
\be
a^0_Q=(1+c(m_\eta R))a^0_{B_Q}\label{quarkishi}
\ee
with the coefficient $c$ that can be readily calculated with the (MIT) bag wavefunction. It is found that $c=1/2$ for $m_\eta R=0$ and $c=0$ for $m_\eta =\infty$. This shows that $a^0_Q$, zero for $R=0$, will increase as $R$ increases as one can see in Fig.~\ref{FSAC}. The $U(1)_A$ anomaly plays no role here.

On the contrary, the gluonish component of the FSAC is controlled strongly by the axial anomaly with the boundary term Eq.~(\ref{lct}) playing a crucial role. The charge $a^0_G$ consists also of two terms. One, denoted $a^0_{G,stat}$, arises from the mixing of light quarks with the anomaly $\vec{E}^a \cdot \vec{B}^a$. For large bag size, this term contributes importantly with the sign opposite to the quarkish component. This is seen in Fig.~\ref{FSAC}. At $R=0$, both the quarkish charge and  $a^0_{G,stat}$ vanish identically. This is what one would expect in the skyrmion result in the chiral limit. For $R\neq 0$, the two contributions largely cancel leaving a small positive value.

The other gluonish component is the Casimir effect due to vacuum fluctuations in the gluon sector. As in all Casimir problems, it is difficult to evaluate this term reliably. Fortunately, however, suppressed in the large $N_c$ limit, it is expected to be negligible except for $R\sim 0$.  What is given in Fig.~\ref{FSAC} has a large uncertainty but confirms this expectation. What is notable is that it is nonzero for $R\rightarrow 0$ and deviates from the pure skyrmion for $R=0$.

The message from this exercise is that although numerical values may change with more sophisticated treatments, the qualitative feature exhibiting the smallness of the FSAC is robust. While the baryon charge anomaly induced at the boundary gives rise to the baryon charge leakage from the bag assuring the exact CC for the baryon charge, here it is the color anomaly lodged at the boundary that assures the CC for the FSAC. Note that the principal action is at the boundary.

\subsubsection{\it Vector dominance}
Another case where the Cheshire Cat principle is operative and shown to work in a surprising way is the vector dominance (VD) in the EM form factors of light-quark hadrons, in particular, the pion and the nucleon. Here the infinite tower of hidden gauge fields described in Section \ref{him} play an essential role.

The key development on this matter is the holographic QCD modeled by Sakai and Sugimoto mentioned in Section \ref{him}.
What transpires from the development of the gravity-gauge duality in hadron physics at low energy is that the EM form factors of the pion~\cite{SS} and nucleon~\cite{HRYY:VD} are given entirely by the infinite tower of the isovector vector mesons $\rho, \rho^\prime, \rho^{\prime\prime}, ...$ and the isoscalar vector mesons $\omega, \omega^\prime, \omega^{\prime\prime}, ...$. That the pionic form factor is described by the infinite sum of the isosvector vector mesons is perhaps not surprising. Indeed, it has been known -- as Sakurai vector dominance -- for a long time that the dominance of the lowest-lying $\rho$ works well for the pion or more generally light-quark mesons. One can, therefore, think of the holographic result as an improvement over the Sakurai vector dominance. That the nucleon form factor is also given entirely by the sum of the infinite tower of vector mesons is highly non-trivial and significant.

It has been known since many years that in stark contrast to the pion, the Sakurai VD failed to work for the nucleon. One possible remedy for this defect was found in the hybrid structure of the nucleon with the quark bag and the cloud of the $\rho$ and $\omega$ mesons. In Ref.~\cite{BRW} was shown that a chiral bag in which the baryon number is partitioned equally into the interior and exterior of the bag, so-called chiral bag at magic angle $\theta=\pi/2$, can fairly well explain the nucleon form factors at low momentum transfers. What the holographic structure is indicating is that the physics of quark-bag with the explicit quark degree of freedom,  identified pictorially as  ``Cheshire Cat smile," can be {\em entirely} captured by the infinite tower of vector mesons. How the Cheshire Cat smile can ``hide" in the infinite tower of vector meson cloud is explained by Zahed in the  holographic picture of QCD~\cite{zahed}. The crucial difference in structure between mesons and baryons is the role of the Cheshire Cat: It is in the baryons but absent in the mesons. In this connection, an interesting possibility is that this difference in structure between the baryons and mesons, though not visible in other models such as constituent quark models, could be in QCD as suggested by Kaplan~\cite{kaplan}. Another intriguing theoretical question is whether there is a process in which the Cheshire Cat smile cannot be completely  ``hidden." A possible case was suggested in the proton decay~\cite{protondecay} where the ``smile" may provide a potentially important nonperturbative mechanism for suppressing the decay matrix element. This issue will be settled in future high-energy experiments.


\section{Chiral symmetry in nuclear many-body systems}
\label{nceft}

In this and the following sections, unless otherwise noted, the discussion will be confined to ChEFT
as given through a chiral effective Lagrangian with Goldstone bosons coupled to the massive
nucleon field treated as a matter field. Other (massive) degrees of freedom are considered to
be integrated out, with their effects lodged in the parameters of the chiral expansion. The
parameters of the bare Lagrangian will have no density dependence as they are constants fixed
at zero density. We will discuss in Section \ref{hadrons-in-matter} how they can pick up density
dependence when the effective theory is matched to QCD at an appropriate matching scale. As
an effective theory, ChEFT is bound to break down at a higher-mass scale at which new
(hadronic) degrees of freedom intervene. The strategy here is to develop the theory to as high
a mass scale as feasible without explicitly introducing other massive degrees of freedom. How
and where the effective theory may break down will be discussed in the later part of this review.

\subsection{Meson exchanges}

\begin{figure}
      \centerline{\includegraphics[width=10cm] {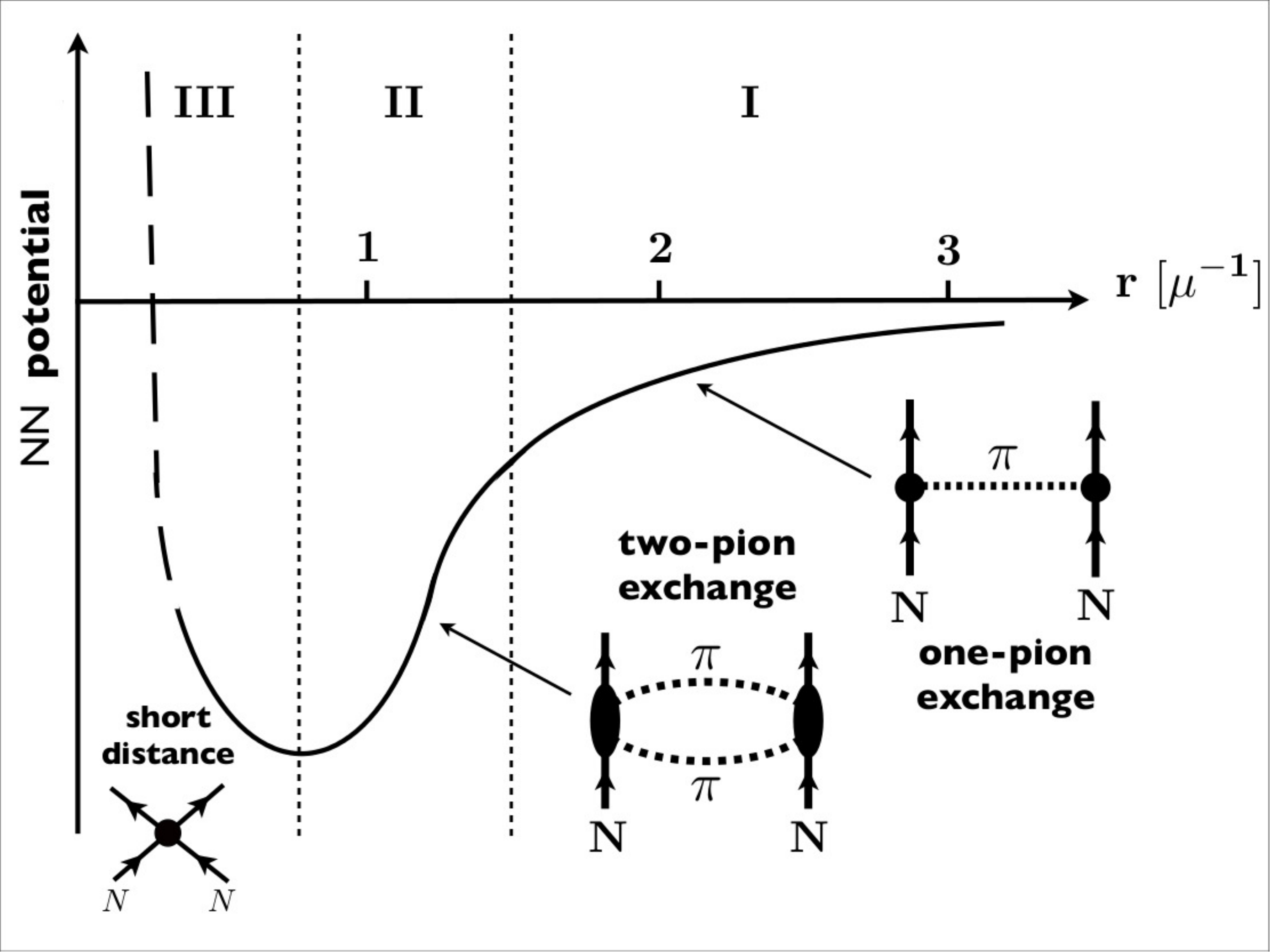}}
   \caption{Inward-bound organization of the nucleon-nucleon interaction (adapted from Taketani \cite{Tak56}).
   The distance $r$ is given in terms of the pion Compton wavelength $\mu^{-1} \simeq 1.4$ fm.}
   \label{fig:1}
 \end{figure}

Eight decades ago, Yukawa's pioneering article \cite{Yuk35} introduced the
framework for studying nuclear interactions based on the
exchange of a boson identified later as the pion. The next generation of
Japanese theorists, in particular Taketani, developed an inward-bound approach \cite{TNS51} to the nucleon-nucleon
interaction, sketched in Fig.\ \ref{fig:1}, whereby the long-distance part is governed by
one-pion exchange, the intermediate-range part is dominated by two-pion exchange processes
and the short-range dynamics remain unresolved at the low energies characteristic of
nuclear physics. At very short distance scales, the nuclear potential is given in a parameterized
form with strengths fit to scattering data. Taketani's program turned out to be immensely
useful and was further developed in
the late fifties of the twentieth century.  One example is the calculation of the
two-pion exchange potential \cite{Tak52,KMO57} (using dispersion relation techniques)
and early focus \cite{FM50} on the resonant pion-nucleon amplitude which
anticipated the $\Delta$-isobar models of later decades \cite{BW75}. Today this strategy is the one
pursued by modern effective field theory approaches.

One salient feature missing from these early models of nuclear forces based on pion exchange
was chiral symmetry. The pion-nucleon coupling was assumed to be renormalizable and
pseudoscalar in nature, but this resulted in unrealistically large pion-nucleon scattering lengths.
Some of the first attempts to incorporate constraints from chiral symmetry, which at the time was being developed in the particle-physics community, in nuclear interaction models were due to Gerry Brown \cite{brown68,brown70,brown79}. However, it was only when chiral symmetry was included in the context of the effective field theory framework that a systematic path to constructing the nuclear force was paved. This process was initiated in the form of soft-pion theorems in meson-exchange currents~\cite{comments} and culminated in chiral perturbation approach~\cite{weinbergseminal,vankolck-ordonez}.

\subsection{Chiral symmetry in meson-exchange currents}\label{MEC}

Historically the modern nuclear effective field theory anchored on chiral symmetry of QCD
germinated in the 1970's from the work on meson-exchange currents in nuclei. The basic
philosophy behind the initial step was that whatever degrees of freedom could be uncovered
in nuclear physics were to be probed in response functions to the electroweak (EW) external
fields. This attitude resembles on a deeper level ``seeing" QED at work in the Lamb shift. In
fact, ``seeing" chiral symmetry at work in nuclei -- and in nuclear matter -- is tantamount to
seeing the role of the pions in the response to the EW fields.

The long-standing problem in nuclear physics on the role of meson exchange currents in nuclei, purported to see explicitly how Yukawa's meson theory works in nuclei, began to see its solution in 1972 in the extremely simple but prescient calculation by Riska and Brown~\cite{riska-brown} of the $\sim 9.5\%$ discrepancy between experiment and theory of the process
\be
n+p\rightarrow d+\gamma\label{np}.
\ee
In doing so, it paved the road to the development of chiral perturbation theory as an effective
nuclear field theory, now widely accepted in the nuclear physics community as is described in this
article and elsewhere in this volume. This story is not as widely recognized as it deserves to be, so we highlight the issue in this review.

While the Yukawa theory of meson exchange in nuclear interactions was widely recognized and accepted, the explicit evidence for meson exchanges in nuclear systems had eluded theoretical efforts up to the 1970's due to lack of controlled and systematic methods for addressing strong interactions in nuclear processes. Compounded with a wide range of length scales involved in nuclear interactions, nuclear potentials probed by nuclear scattering and excitation spectra -- apart from the fact that the potential is not a unique physical quantity on its own -- did not provide a direct ``snapshot" of the appearance of mesons. The effort was therefore focused on singling out mesonic imprints in nuclear responses to external electroweak fields. The standard approach then was to construct the current operators and take matrix elements with appropriate wavefunctions obtained with phenomenological potentials fit to the free-space nucleon-nucleon scattering data. Assuming that the single-nucleon currents are known from free space, the problem then was how to compute corrections to the single-nucleon responses, often referred to as ``impulse approximations."  In the absence of a fundamental theory for strong interactions, multi-body currents thought to arise from the exchange of mesons known at the time, such as pions and heavier $\rho$ and $\omega$ mesons etc., were computed and added arbitrarily and without any systematics. Lacking organizational schemes, each diagram involving strong coupling would enter with a widely varying strength and sign,  the validity and reliability of which could not be gauged. In one word, it was a mess.

A way out of this conundrum was found in exploiting what is now recognized as the working  of chiral symmetry in nuclear dynamics. It was recognized~\cite{chemtob-rho} that in organizing multi-body currents the pion, which was known to figure prominently in nuclear forces, played a special role and the soft-pion theorems based on the current algebras -- being developed then prior to the advent of QCD -- could play an important role in nuclear electroweak (EW) processes at low momentum transfers~\cite{KDR}. We now know, following the formulation of chiral perturbation theory~\cite{Wei79}, that the soft-pion theorems constitute the leading terms in the chiral expansion of the currents~\cite{mr91} and the soft-pion terms in the currents~\cite{chemtob-rho, KDR} are the leading contributions to the two-body exchange currents, dominant in the magnetic dipole transition that governs the process in Eq.\ (\ref{np}) and also in nuclear axial-charge transitions of the type~\cite{warburton}
\be
A(J^\pm)\rightarrow A^\prime (J^\mp)+e^-(e^+)+\bar{\nu} (\nu), \hspace{.25in} \Delta T=1 .
\ee

This apparent dominance of soft-pion terms in certain specific processes was dubbed for lack of a more imaginative name the ``chiral filter process."  It is now clear that this soft-pion process stands for a ``smoking gun" evidence, not only for the role of pion but also for chiral symmetry in action in nuclei.

Soon after Weinberg's 79 paper~\cite{Wei79}, the chiral counting rule written down in that paper was applied~\cite{mr-erice,comments} to the irreducible two-body exchange current diagrams, showing that the soft-pion exchange terms were the leading-order terms in the chiral expansion and that in the chiral-filter process, the corrections appear at next-to-next-to-leading order, so highly suppressed\footnote{In terms of the chiral Lagrangian consisting of baryons and pions, the relevant physical amplitudes are expanded in power series in $(p/\Lambda)^\nu$,
\be
\nu&=&n_B -2 +2L+\sum_i \nu_i,\label{counting}\\
\nu_i&=&d_i +e +\frac{n_i}{2} -2
\ee
where $n_B$ is the number of nucleons involved in the irreducible diagrams, $L$ is the number of loops, $e$ the number of external fields -- 0 for nuclear forces and 1 for electroweak current-- and $d_i$ and $n_i$, respectively, are the number of derivatives and internal nucleon lines entering the $i$-th vertex. In \cite{mr-erice}, where two-body currents were considered and restricted to irreducible graphs, no internal nucleon lines were involved, so $n_i=0$, and since one slowly varying external field was considered, $e$ did not figure explicitly. In the modern development, the same counting rule is used for nuclear forces for which $e=0$.}. This accounts for the accurate calculation of the exchange currents in the process and explains how the Riska-Brown calculation succeeded to give a firm prediction with a few simple diagrams.

There are two points to stress. First is that given the current matrix elements and the rules for calculating higher-order diagrams, the link to nuclear forces is almost immediate. For instance one can resort to the conserved vector current (CVC) hypothesis for constraints on the potential. It is in this sense that the development of exchange currents presaged the modern structure of nuclear chiral effective field theory. Secondly not only were these calculations performed at leading-order but also the corrections to the operators turned out to be suppressed by two chiral orders~\cite{mr91}. In a recent calculation~\cite{pmr-prl} at next-to-next-to-next-to leading order (N$^3$LO) in the chiral counting, it has been possible to reproduce the experimental value with the theoretical error pinned down to $\lsim 1\%$. With the technique developed up to date, the error could be reduced to $\lsim 1/2 \%$. This is one of the most accurate nuclear physics prediction in the literature.

It should, however, be noted that this calculation is of hybrid nature in that the currents are constructed in chiral perturbation theory to the highest order feasible, while the wave functions are computed with the ``most sophisticated" phenomenological potential fitted accurately to free-space scattering data. We shall refer to this approach as ``EFT$^*$" -- with an asterisk to differentiate it from ``EFT" standing for bona-fide effective field theory. Ad noted, this approach presaged the modern development of {\it precision calculations} in effective field theory anchored on chiral symmetry in nuclear physics~\cite{eftstar}. A fully consistent chiral perturbation theory should of course consist of both the potential and the currents computed with the same chiral expansion. This approach -- ``EFT" -- is what is being actively pursued currently in the literature.

On a purely conceptual ground, as applied to EW matrix elements, EFT$^*$ is as consistent as EFT. Both use two different counting schemes, ``irreducible" for the current and the potential, and  ``reducible" for the wave functions. The difference between EFT$^*$ and EFT is that for EFT$^*$, while the currents are computed with irreducible graphs, the wave functions are obtained with an ``accurate phenomenological potential"  whereas in EFT, both the currents and the potential for the wave functions are computed to the same (irreducible) chiral order N$^n$LO for some $n$.
A pertinent practical question is: How does the EFT$^*$ fare with the fully consistent chiral perturbative approach when compared with the same $n$? Since in EFT$^*$, the currents are computed to $n$-th order while the potentials are fitted to experiments up to a certain momentum scale, corresponding to, say, $m$-th order with $m\gg n$, there is a mismatch in the chiral order. This would in principle entail, among others, certain off-shell ambiguities absent in EFT. So how serious is this mismatch in making predictions?

As of today, this question remains unanswered. Given that various numerical approximations are made in each approach, it is difficult to make a numerical estimate of errors due to mismatch in EFT$^*$ as compared with EFT. One can, however, say with confidence that there is no unambiguous evidence of ``failure" of EFT$^*$ in making accurate predictions. Furthermore there is no obvious reason to expect that the hybrid approach is inferior in accuracy to the consistent chiral counting.

Since nearly two decades after the early calculation of \cite{pmr-prl,eftstar}, there have been extensive new developments in doing EFT and EFT$^*$ calculations~\cite{recent results}.  To gain some idea where things stand as of today, we compare the recent developments with the old results. For this, let us look at a transition that is not dominated by soft-pion terms, namely, the Gamow-Teller transition. When soft-pion terms enter at the leading order, because of the two chiral-order suppression mentioned before, the mismatch in the chiral counting is expected to be unimportant. On the contrary, the multi-body terms contributing to the Gamow-Teller operator appear at N$^2$LO. It would be interesting to consider the ``hep" process $p+ ^3{\rm  He} \rightarrow ^4{\rm He} + e^+ +\nu_e$ which is relevant to the solar neutrino problem. This process was computed in EFT$^*$ in \cite{pp-park}, a genuine prediction, the precision of which has remained unmatched by other calculations. This process has not yet been treated in EFT so a comparison cannot be made.  We consider,  therefore, the $pp$ fusion process to which both approaches have been applied
\be
p+p\rightarrow d+e^+ +\nu_e.
\ee
This process is dominated by the Gamow-Teller operator, and, as predicted a long time ago~\cite{KDR},  the exchange current contribution, receiving no soft-pion corrections in one-pion exchange, should be suppressed relative to the single-particle contribution. It turns out that both EFT$^*$ and EFT -- with some fine-tuning such as EM corrections to the potential and more sophistication in the wave functions in the latter --  give nearly the same $S$ factor when calculated to N$^3$LO~\cite{pp-park,pp-marcucci}. The difference comes to $\lsim 1\%$, with a theoretical uncertainty for both of $\lsim 0.1\%$.  Even more significantly, the potential model approach in which the meson-exchange current is calculated with phenomenological Lagrangians, unconstrained by chiral symmetry (except for the pion)~\cite{chemtob-rho}, gives essentially the same result.

It is worth stressing that the discrepancy,  if any,  between EFT and EFT$^*$ should be in soft-pion suppressed processes such as the pp and hep processes and in large-momentum transfer charge form factors. On the other hand, in the soft-pion enhanced processes such as  magnetic dipole and axial-charge transitions~\cite{KDR},  EFT has not yet matched the accuracy of EFT$^*$. Some examples of axial charge transitions in nuclei will be given in Section \ref{him}.

\subsection{Nuclear forces}
\subsubsection{\it Two-nucleon forces}

\begin{figure}
      \centerline{\includegraphics[width=18cm] {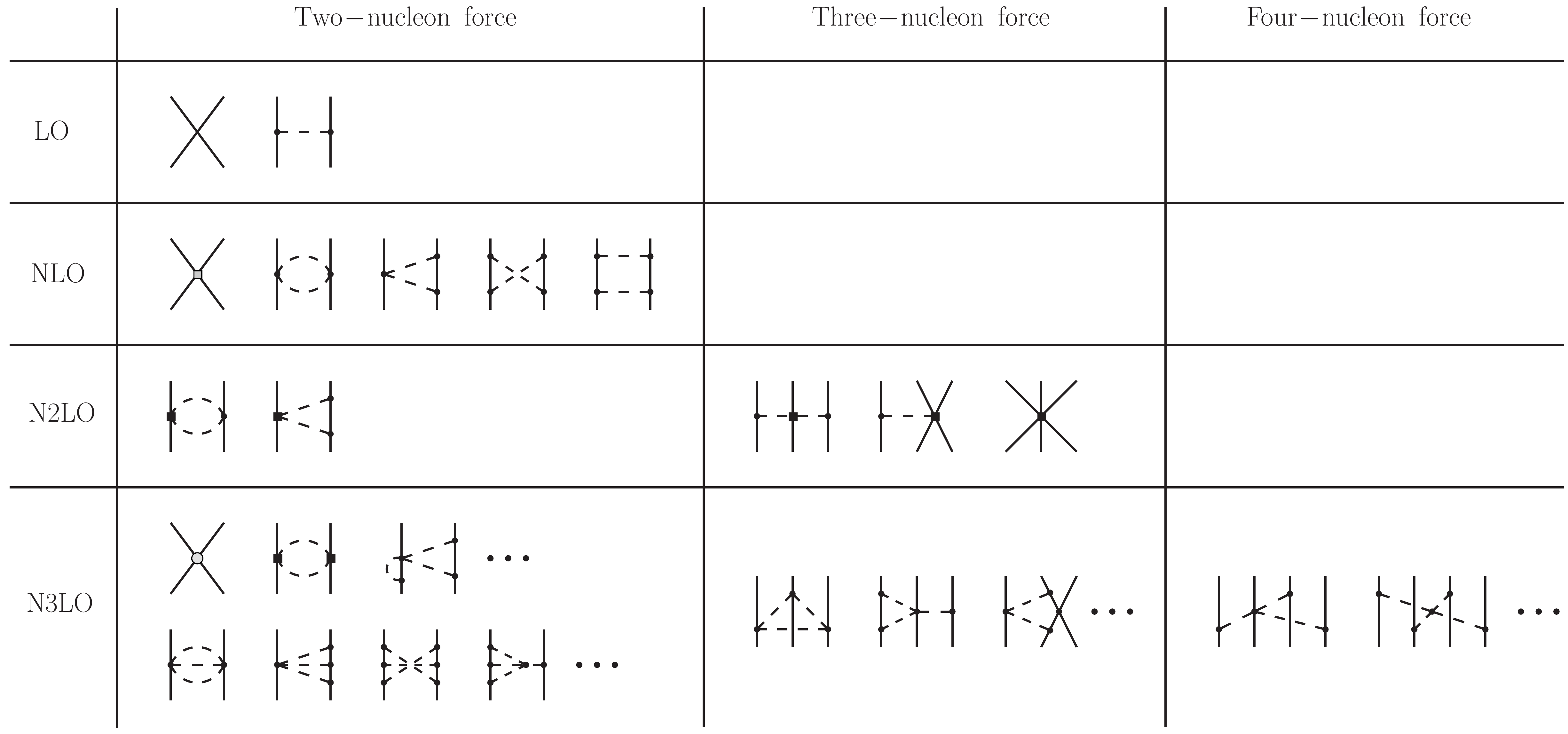}}
\caption{Diagrammatic hierarchy of two-, three- and four-body chiral nuclear forces up to fourth order in powers
of the light scale divided by the heavy scale.}
\label{fignforce}
 \end{figure}

In chiral effective field theory (ChEFT), the nuclear force is given in terms of explicit one-pion and
multi-pion exchange process constrained by chiral symmetry as well as a complete set of contact-terms
encoding unresolved short-distance dynamics (see Fig.\ \ref{fignforce}). Diagrammatic
contributions are organized in powers of small external momenta
(or the pion mass) over the chiral symmetry breaking scale $\Lambda_\chi \sim 1$\,GeV
 \cite{evgenireview,hammerreview,machleidtreview}.
Two-body forces enter first at leading-order (LO), corresponding to $(q/\Lambda)^0$, and include the
one-pion exchange contribution together with two contact terms acting in relative $S$-waves:
\begin{equation}
V_{NN}^{(\rm LO)} = -{g_A^2\over 4 f_\pi^2} \, {\vec \sigma_1 \cdot \vec q\,\,
\vec \sigma_2 \cdot \vec q \over m_\pi^2+q^2}\,\vec \tau_1\cdot \vec \tau_2 + C_S + C_T \,
\vec \sigma_1 \cdot\vec \sigma_2 \,.
\end{equation}
Here, $\vec \sigma_{1,2}$ and $\vec \tau_{1,2}$ denote the spin- and isospin operators of
the two nucleons, and $\vec q$ is the momentum transfer.
Next-to-leading order (NLO) terms proportional to $(q/\Lambda)^2$ include two-pion exchange diagrams
generated by the vertices of the chiral $\pi N$-Lagrangian ${\cal L}_{\pi N}^{(1)}$.
All loop integrals can be performed analytically, and the resulting expression reads
\begin{eqnarray}
V_{NN}^{(\rm 2\pi-NLO)} &=& {1\over 384 \pi^2 f_\pi^4} \bigg\{ 4m_\pi^2
(1+4g_A^2-5g_A^4) +q^2(1+10g_A^2-23g_A^4) - {48g_A^4 m_\pi^4 \over 4m_\pi^2+q^2} \bigg\} L(q)\,\,
\vec \tau_1\cdot \vec \tau_2 \nonumber \\ && +{3g_A^4 \over 64\pi^2 f_\pi^4}  L(q) \,
( \vec \sigma_1 \cdot\vec \sigma_2 \, \vec q^{\,2} -  \vec \sigma_1 \cdot \vec q\,\,
\vec \sigma_2 \cdot \vec q\,)\,, \end{eqnarray}
which introduces a nontrivial logarithmic loop function
\begin{equation}
L(q) = {\sqrt{4m_\pi^2+q^2} \over q} \ln{ q+\sqrt{4m_\pi^2+q^2} \over 2m_\pi} \,.
\end{equation}

The NLO contributions generate only isovector central, isoscalar
spin-spin and isoscalar tensor forces. Additional polynomial pieces generated
by the pion-loops have been left out in the above expression and can be
absorbed into the most general contact-term at NLO:
\begin{eqnarray} V_{NN}^{(\rm ct-NLO)} &=& C_1 \, \vec q^{\,2} +  C_2 \, \vec k^{\,2} +  (C_3 \,
\vec q^{\,2} +  C_4 \, \vec k^{\,2})\, \vec \sigma_1 \cdot\vec \sigma_2 + {i \over 2 }C_5 \,
( \vec \sigma_1 +\vec \sigma_2) \cdot (\vec q \times \vec k\,) \nonumber \\
&& + C_6\,\vec \sigma_1 \cdot \vec q\,\,\vec \sigma_2 \cdot \vec q\, +  C_7\,\vec \sigma_1
\cdot \vec k\,\,\vec \sigma_2 \cdot \vec k\, , \end{eqnarray}
with $\vec k = (\vec p + \vec p\,')/2$ the half-sum of initial $\vec p$ and final $\vec p\,'$ nucleon momenta.
The low-energy constants $C_1, \dots, C_7$ are adjusted to
fit empirical NN scattering phase shifts. The coefficient $C_5$ parameterizes the strength of the
short-distance spin-orbit interaction, which dominates over the finite-range contributions to the spin-orbit
force arising at higher orders. Consequences for the nuclear
energy density functional, where $C_5$ determines the self-consistent
single-particle spin-orbit potential in finite nuclei proportional to the density
gradient, will be discussed in Section \ref{dfm}.

The most important pieces from chiral two-pion exchange that generate attraction in the
isoscalar central channel and reduce the strong $1\pi$-exchange isovector tensor force
are still absent at NLO. These terms arise \cite{nnpap1} from subleading $2\pi$-exchange
through the chiral $\pi\pi NN$ contact couplings $c_{1,3,4}$ in ${\cal L}_{\pi N}^{(2)}$ (or
from the inclusion of explicit $\Delta(1232)$-isobar degrees of freedom):
\begin{eqnarray}
V_{NN}^{(\rm 2\pi-N^2LO)} &=& {3g_A^2 \over 16\pi f_\pi^4} \Big[ c_3\,
q^2+2m_\pi^2(c_3-2c_1) \Big] (2m_\pi^2+q^2) A(q) \nonumber \\ && + {g_A^2 c_4\over 32\pi
f_\pi^4} (4m_\pi^2+q^2) A(q)\, ( \vec \sigma_1 \cdot\vec \sigma_2 \, \vec q^{\,2} -
\vec \sigma_1 \cdot \vec q\,\,\vec \sigma_2 \cdot \vec q\,)\,\vec \tau_1\cdot \vec \tau_2 \, ,
\end{eqnarray}
where the arctan loop function
\begin{equation}
A(q) = {1\over 2 q} \arctan{q \over 2m_\pi} \,.
\end{equation}
In addition there are relativistic $1/M_N$-corrections to $2\pi$-exchange whose
explicit form \cite{nnpap1,evgenireview} depends on the precise definition of the
nucleon-nucleon potential $V_{NN}$, which itself is not an observable. Current state of the art
NN potentials have been constructed up to order N$^3$LO, $(q/\Lambda)^4$, which includes two-loop $2\pi$-exchange
processes, $3\pi$-exchange terms and contact forces quartic in the momenta parameterized by 15 additional low-energy
constants $D_1, \dots, D_{15}$. When solving the Lippmann-Schwinger equation the
chiral NN potential is multiplied by an exponential regulator function with a cutoff scale
$\Lambda = 400 - 700\,$MeV in order to
restrict the potential to the low-momentum region where chiral effective field theory is applicable.

At order N$^3$LO the chiral NN potential reaches the quality of a
``high-precision'' ($\chi^2/DOF \sim 1$) potential in reproducing empirical NN scattering
phase shifts and deuteron properties. At the same time it provides the foundation for systematic
nuclear structure studies of few- and many-body systems. \vskip 0.2cm

\subsubsection{\it Nuclear many-body forces}\label{NMF}

An advantage of the organization scheme provided by chiral effective field theory is that
nuclear many-body forces are generated consistently with the nucleon-nucleon potential.
Three-body forces arise first at N$^2$LO and contain three different topologies. The two-pion
exchange component contains terms proportional to the
low-energy constants $c_1$, $c_3$, and $c_4$ and has the form
\begin{equation}
V_{3N}^{(2\pi)} = \sum_{i\neq j\neq k} \frac{g_A^2}{8f_\pi^4}
\frac{\vec{\sigma}_i \cdot \vec{q}_i \, \vec{\sigma}_j \cdot
\vec{q}_j}{(\vec{q_i}^2 + m_\pi^2)(\vec{q_j}^2+m_\pi^2)}
F_{ijk}^{\alpha \beta}\tau_i^\alpha \tau_j^\beta,
\label{3n1}
\end{equation}
where $\vec{q}_i$ is the difference between the final and initial momenta of
nucleon $i$, and the isospin tensor $F_{ijk}^{\alpha \beta}$ is given by
\begin{equation}
F_{ijk}^{\alpha \beta} = \delta^{\alpha \beta}\left (-4c_1m_\pi^2
 + 2c_3 \,\vec{q}_i \cdot \vec{q}_j \right ) +
c_4 \,\epsilon^{\alpha \beta \gamma} \tau_k^\gamma \,\vec{\sigma}_k
\cdot \left ( \vec{q}_i \times \vec{q}_j \right ).
\label{3n4}
\end{equation}
The low-energy constants $c_1$, $c_3$, and $c_4$ can be fit to empirical pion-nucleon
\cite{buttiker00} or nucleon-nucleon scattering \cite{rentmeester03,entem03}.

The one-pion exchange term proportional to the low-energy constant $c_D$ has the form
\begin{equation}
V_{3N}^{(1\pi)} = -\sum_{i\neq j\neq k} \frac{g_A c_D}{8f_\pi^4\, \Lambda}
\,\frac{\vec{\sigma}_j \cdot \vec{q}_j}{\vec{q_j}^2+m_\pi^2}\, \vec{\sigma}_i \cdot
\vec{q}_j \, {\vec \tau}_i \cdot {\vec \tau}_j \, ,
\label{3n2}
\end{equation}
and the three-nucleon contact force proportional to $c_E$ is written:
\begin{equation}
V_{3N}^{(\rm ct)} = \sum_{i\neq j\neq k} \frac{c_E}{2f_\pi^4\, \Lambda}
\,{\vec \tau}_i \cdot {\vec \tau}_j\, ,
\label{3n3}
\end{equation}
where $\Lambda = 700$ MeV sets a natural scale.
Ideally the low-energy constants $c_D$ and $c_E$ are fit to properties of three-body
systems only, and the triton binding energy and lifetime provide a pair of weakly-correlated
observables for this purpose \cite{gardestig06,gazit09}.

At order N$^3$LO in the chiral power counting,
additional three- and four-nucleon forces arise without any additional undetermined low-energy
constants. The N$^3$LO three-body force is written schematically as
\be
V_{3N}^{(4)} = V_{2\pi}^{(4)} + V_{1\pi-\rm cont.}^{(4)} + V_{2\pi-1\pi}^{(4)}+V_{\rm ring}^{(4)}
+ V_{2\pi-\rm cont.}^{(4)} + V_{1/m}^{(4)},
\ee
corresponding to the $2\pi$ and $1\pi-{\rm contact}$ topologies (present already at order N$2$LO)
as well as the $2\pi-1\pi$, ring, $2\pi-{\rm contact}$ and $1/m$ topologies. All contributions have been
worked out and presented in Refs.\ \cite{ishikawa07,bernard08,bernard11}.
The N$^3$LO four-nucleon force has been calculated \cite{epelbaum06} and the resulting expressions
are considerably simpler than the N$^3$LO three-nucleon force.

\subsubsection{\it In-medium effective nucleon-nucleon interactions}\label{dripline}

To facilitate the implementation of chiral three-nucleon forces in studies of finite
nuclei and infinite nuclear matter, a strategy pursued in many recent works is
to replace the full chiral three-body force with a normal-ordered two-body
approximation \cite{bogner05,hagen07,holt09,holt10,hebeler10,otsuka10}.
Then the three-body force in second-quantized notation
\be
V_{3N} = \frac{1}{36} \sum_{123456} \langle 1 2 3 | \bar V | 4 5 6 \rangle
\hat a^\dagger_1 \hat a^\dagger_2 \hat a^\dagger_3 \hat a_6 \hat a_5 \hat a_4
\ee
with antisymmetrized matrix elements $\langle 1 2 3 | \bar V | 4 5 6 \rangle$, is written
\bea
V_{3N} &=& \frac{1}{6} \sum_{ijk}\langle ijk | \bar V_{3N} | ijk \rangle
+ \frac{1}{2} \sum_{ij1 4} \langle ij1 | \bar V_{3N} | ij4 \rangle :\!\hat a^\dagger_1 \hat a_4\!:
+ \frac{1}{4}\sum_{i1245}\langle i12 | \bar V_{3N} | i 4 5 \rangle :\!\hat a^\dagger_1 \hat a^\dagger_2
\hat a_5 \hat a_4\!:  \nonumber \\
&& + {1\over 36} \sum_{123456} \langle 1 2 3 | \bar V_{3N} | 4 5 6 \rangle
:\!\hat a^\dagger_1 \hat a^\dagger_2 \hat a^\dagger_3 \hat a_6 \hat a_5 \hat a_4\!:,
\label{nord}
\eea
where the indices $i,j,k$ represent filled orbitals in the chosen reference state $|\Omega \rangle$,
and $:\,\,:$ denotes the normal-ordered product of operators that satisfies
\be
:\! \hat a^\dagger_1\dots \hat a_n \!: |\Omega \rangle= 0.
\ee

\begin{figure}
\begin{center}
\includegraphics[scale=0.65,clip]{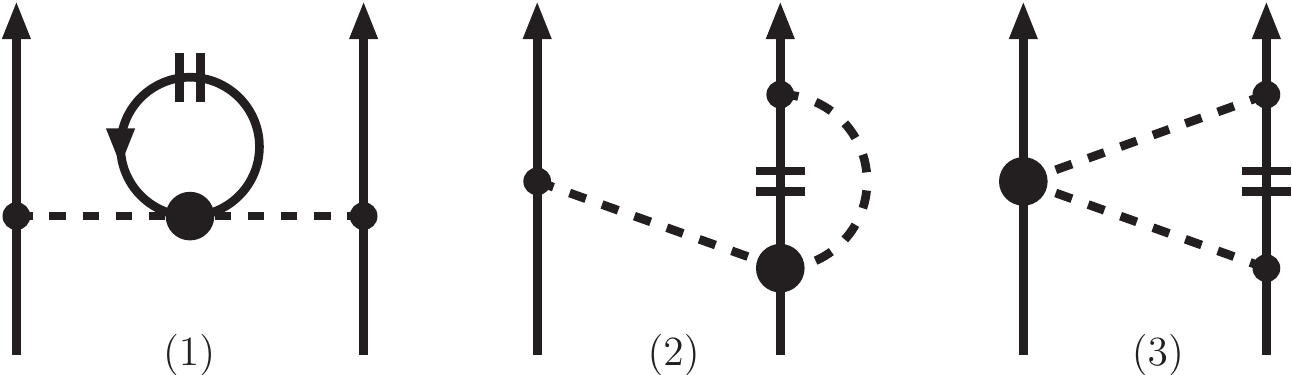}\hspace{.4in}
\includegraphics[scale=0.65,clip]{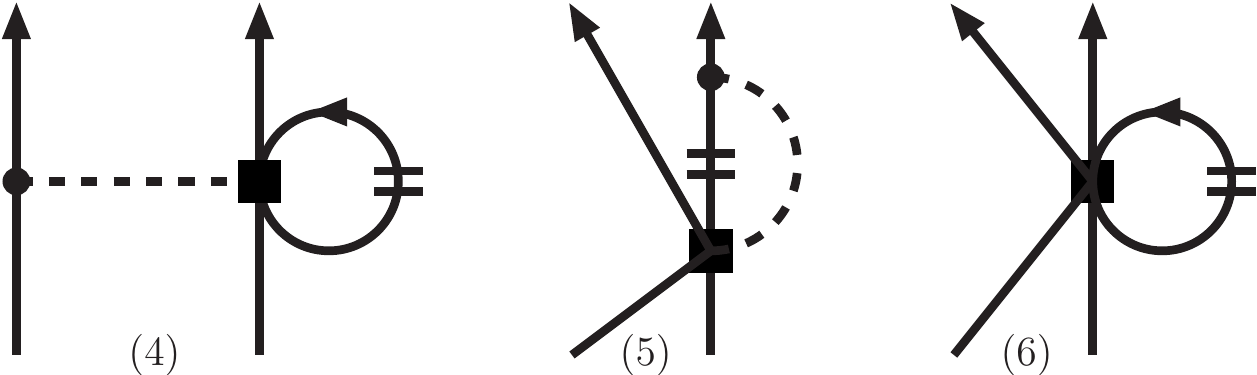}
\end{center}
\vspace{-.5cm}
\caption{In-medium nucleon-nucleon interaction from the leading chiral three-nucleon force.
The short double-line denotes the ``medium insertion'' representing summation over the filled
Fermi sea of nucleons. Reflected diagrams of (2), (3), (4), and (5) are not explicitly shown.}
\label{mfig1}
\end{figure}

In Refs.\ \cite{holt09,holt10} the reference state was chosen to be the ground state of a
noninteracting homogeneous gas of nucleons characterized by density and isospin
asymmetry parameter $\delta_{np} = (\rho_n-\rho_p)/(\rho_n+\rho_p)$. Imposing the
additional kinematical constraint that the scattering takes place in the rest
frame of the medium, the on-shell scattering amplitude has the same
form as the free-space nucleon-nucleon interaction:
\bea
V_{NN}^{\rm med} &=& V_C + W_C\, \vec \tau_1 \cdot \vec \tau_2
+ \left [ V_S + W_S \,\vec \tau_1 \cdot \vec \tau_2  \right ] \vec \sigma_1 \cdot \vec \sigma_2
+\left [ V_T + W_T\, \vec \tau_1 \cdot \vec \tau_2 \right ]
  \vec \sigma_1 \cdot \vec q \, \vec \sigma_2 \cdot \vec q  \nonumber \\
&+&\left [ V_{LS}+ W_{LS}\, \vec \tau_1 \cdot \vec \tau_2  \right ]
  i (\vec \sigma_1 + \vec \sigma_2)\cdot ( \vec q \times \vec p\,)
+\left [ V_Q +W_Q\, \vec \tau_1 \cdot \vec \tau_2 \right ]
  \vec \sigma_1 \cdot (\vec q \times \vec p\,) \vec \sigma_2 \cdot (\vec q \times \vec p\,),
  \label{fsint}
\eea
where the scalar functions $V_C,\dots, V_Q$ and $W_I,\dots,W_Q$ depend on $p$ and $q$, the
initial relative momentum and the momentum transfer, respectively. The resulting in-medium
two-body force can then be implemented (with appropriate symmetry factors \cite{hebeler10}) in
the many-body method of choice.

\begin{figure}
\begin{center}
\includegraphics[angle=270,scale=0.325,clip]{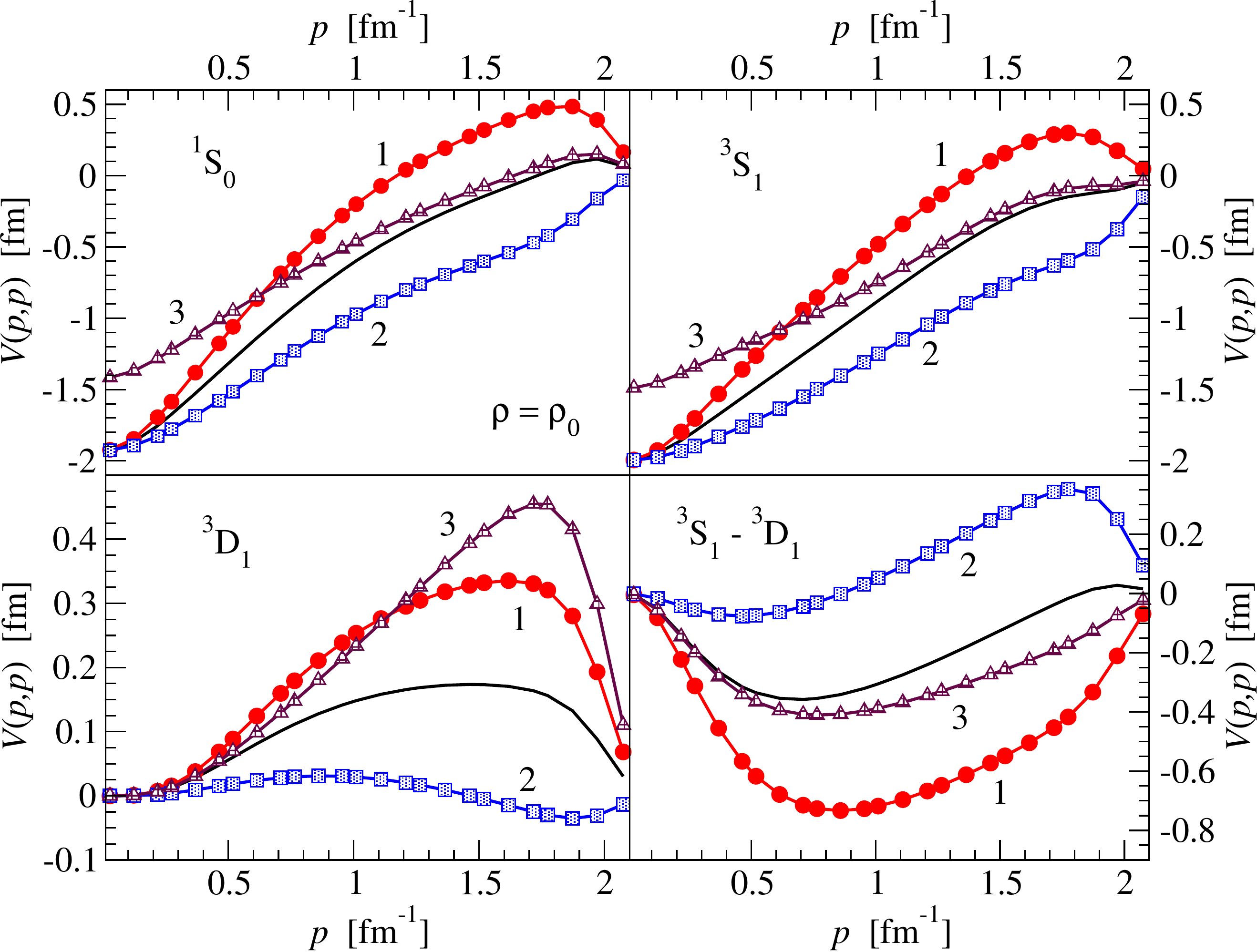} \hspace{.1in}
\includegraphics[angle=270,scale=0.325,clip]{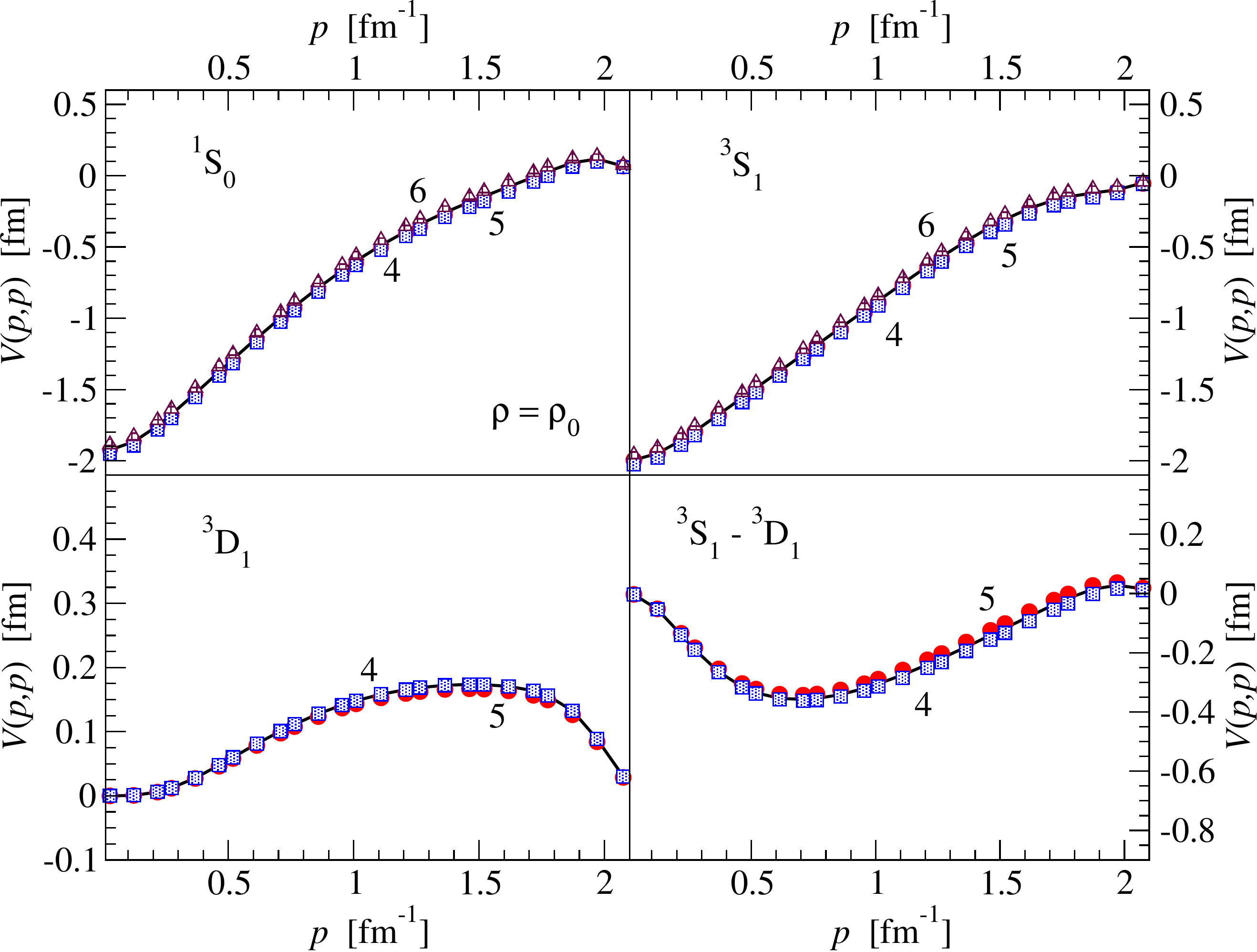}
\end{center}
\vspace{-.5cm}
\caption{Contributions to the on-shell $S$- and $D$-wave matrix elements of the
chiral nucleon-nucleon potential with $\Lambda = 414$\,MeV from the
density-dependent nucleon-nucleon interaction derived in Refs.\ \cite{holt09,holt10}. The
bare NN potential is shown by the the solid black line.}
\label{sdwaves}
\end{figure}

For the N$^2$LO chiral three-body force, the six topologically distinct diagrams
contributing to the in-medium NN interaction are shown in Fig.\ \ref{mfig1}. The
contributions labeled (1), (2), and (4) modify normal one-pion exchange, while
contributions (5) and (6) affect the strength of the NN contact interaction. The
Pauli-blocked two-pion exchange diagram, labeled (3) in Fig.\ \ref{mfig1}, gives rise
to many of the different terms in Eq.\ (\ref{fsint}). The on-shell matrix elements of the
lowest partial waves of the chiral NN potential with $\Lambda = 414$\,MeV are shown
as the solid black line in Fig.\ \ref{sdwaves}. In addition we show separately the modifications
arising from each of the six different topologies shown in Fig.\ \ref{mfig1}. The values of the
low-energy constants associated with the N$^2$LO chiral three-body force are
$c_1 = -0.81$\,GeV$^{-1}$, $c_3 = -3.0$\,GeV$^{-1}$, $c_4 = 3.4$\,GeV$^{-1}$, $c_D = -0.4$,
and $c_E = -0.071$ \cite{coraggio13,coraggio14}. The dominant effects arise from two-pion exchange diagrams,
but in all partial waves the pion self-energy correction Fig.\ \ref{mfig1}(1) and
the pion-exchange vertex correction Fig.\ \ref{mfig1}(2) approximately cancel. This leaves
Pauli-blocking effects in two-pion exchange, represented by Fig.\ \ref{mfig1}(3)
as the dominant effects from three-body forces. At this cutoff scale, the three-body contact term
is quite small, but at higher resolution scales, this term can give rise to substantial additional
repulsion \cite{holt09} in the singlet $S$-waves.

\begin{figure}
\begin{center}
\includegraphics[angle=270,scale=0.35,clip]{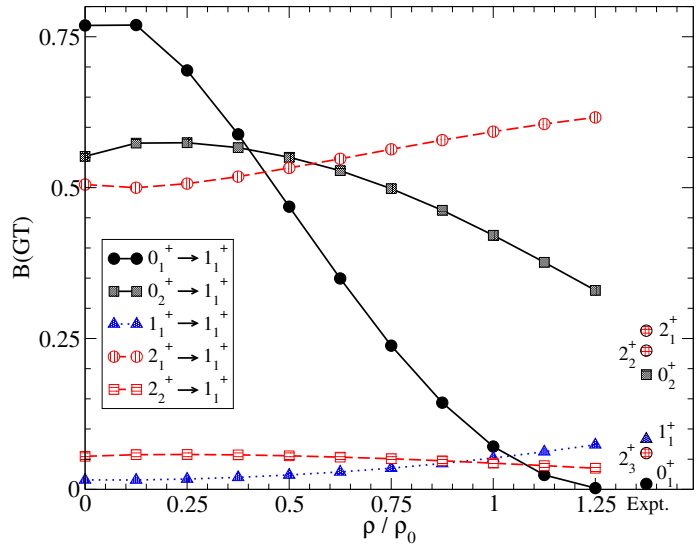}
\end{center}
\vspace{-.5cm}
\caption{Gamow-Teller strengths for transitions from the low-energy even-parity states
of $^{14}$C to the ground state of $^{14}$N as a function of the nuclear density.}
\label{bgtc14}
\end{figure}

With this normal-ordered two-body approximation, three-nucleon forces were shown to provide
a microscopic description for the anomalously-long beta-decay lifetime of $^{14}$C \cite{holt09}, which
later no-core shell model calculations \cite{maris11} confirmed and clarified. We show in Fig.\
\ref{bgtc14} the Gamow-Teller strengths from low-lying states in $^{14}$C
to the ground state of $^{14}$N computed with a density-dependent NN interaction as a function
of the average nuclear density experienced
by valence nucleons in $^{14}$C and $^{14}$N. Only the ground state to ground state transition receives
significant medium modifications, which results in a strong enhancement of the $^{14}$C lifetime.
Further work has employed in-medium two-body interactions to study three-nucleon force
effects on the oxygen neutron drip-line \cite{otsuka10,jdholt13}. Benchmark studies of the accuracy of the
two-body normal ordering approximation have been presented in Refs.\ \cite{hagen07,roth12}.

Density-dependent nucleon-nucleon interactions have also been employed extensively in calculations of the
equation of state of nuclear and neutron matter \cite{hebeler10,hebeler11,tews12,coraggio13,coraggio14,carbone14}.
This has allowed for the inclusion of three-nucleon forces beyond the Hartree-Fock approximation,
but certain topologies are necessarily omitted in such an approximation. The full set of second-order contributions
to the energy per particle of infinite matter from a three-body contact force has been carried out in Ref.\ \cite{kaiser12}.
There it was found that the diagrams neglected in the two-body normal ordering approximation give a
contribution to the energy per particle that is roughly half that from the density-dependent NN interaction.
In most previous applications, the third particle is averaged over an uncorrelated reference state, and recent work
\cite{carbone14} has verified the accuracy of such an approximation through detailed studies including
correlations in the reference state of infinite nuclear and neutron matter.

\subsubsection{\it Role of the $\Delta$(1232) isobars}\label{delta-sigma}
\label{delt}

The ``effectiveness" of an effective field theory is due in part on the proper
identification of the relevant low-energy degrees of freedom. Since pion-nucleon scattering is
dominated by the $p$-wave $\Delta(1232)$ resonance lying only 293 MeV above the
nucleon mass, the baryon number $B=1$ chiral effective Lagrangian is often extended
\cite{HHK97,pasca1,pasca2} to include the $\Delta(1232)$ isobar as an explicit degree of freedom.
If the physics of the $\Delta(1232)$ is absorbed in the low-energy constants, $c_3$ and $c_4$,
of the effective theory that includes pions and nucleons only, the limit of applicability is narrowed down to an
energy-momentum range small compared to the $\Delta$-nucleon mass difference.

\begin{figure}
\centerline{\includegraphics[width=14cm] {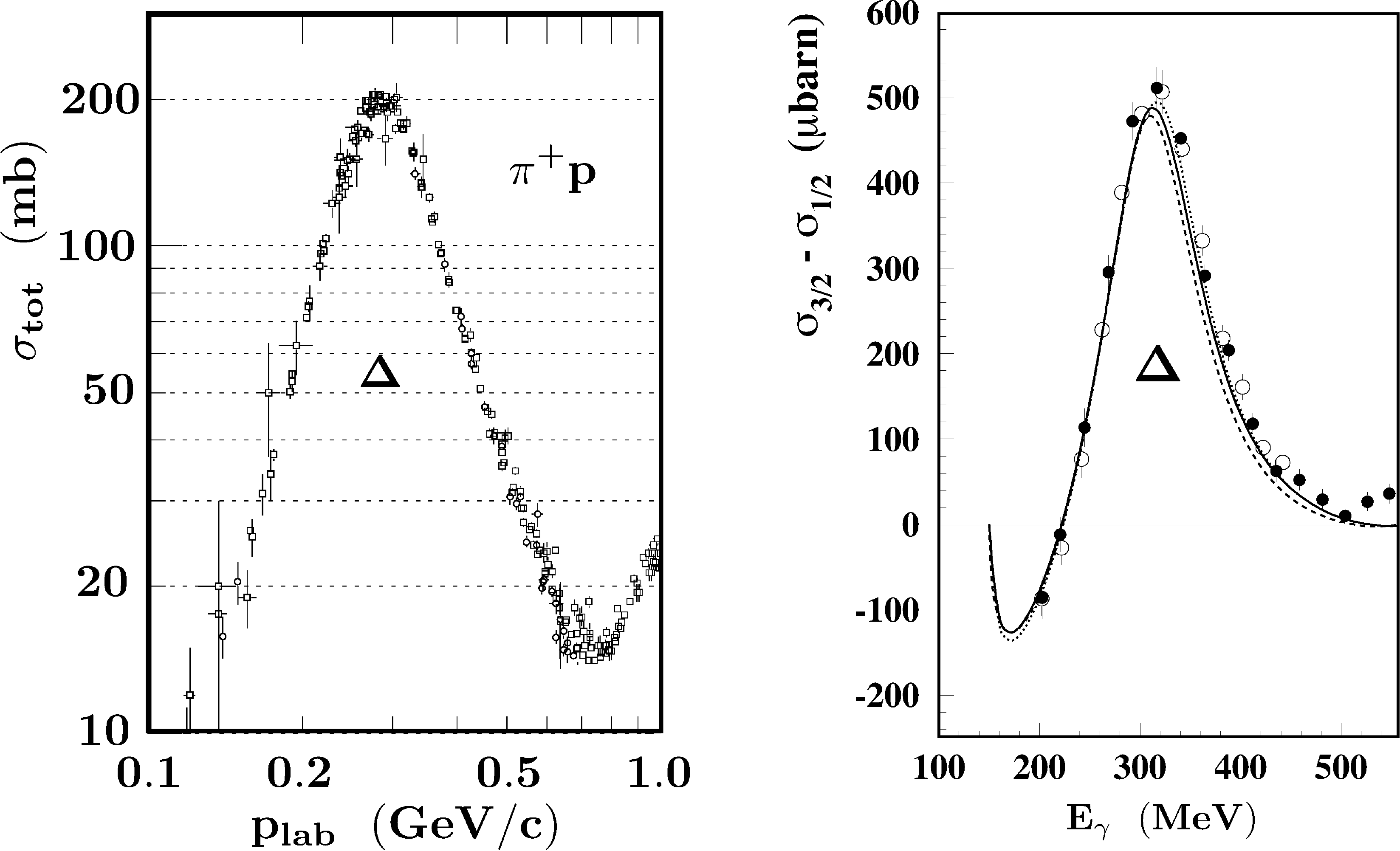}}
\caption{Left: Total cross section for $\pi^+p$ scattering in the region of the $\Delta(1232)$ resonance
(adapted from \cite{pdg2012}). Right: Difference of polarized Compton scattering cross sections \cite{ahrens2001}
of the proton in the total angular momentum channels $\sigma_{3/2}$ and $\sigma_{1/2}$. Curves represent
dispersion relation and multipole analysis cited in Ref.\ \cite{ahrens2001}.}
\label{Delta}
\end{figure}

In fact, the $\Delta$ isobar is the dominant feature in the total cross section for pion-nucleon scattering
and polarized Compton scattering, shown in the left and right panels respectively of Fig.\ \ref{Delta}.
In low-energy pion-nucleon scattering the strong spin-isospin response of the nucleon is encoded in a
large ``axial'' polarizability:
\begin{equation}
\alpha_A^{(\Delta)} = {g_A^2\over f_\pi^2 (M_\Delta - M_N)} \simeq 5\, {\rm fm}^3~~.
\end{equation}
which we note is several times the volume of the nucleon itself.

\begin{figure}
\begin{center}
\includegraphics[width=3cm]{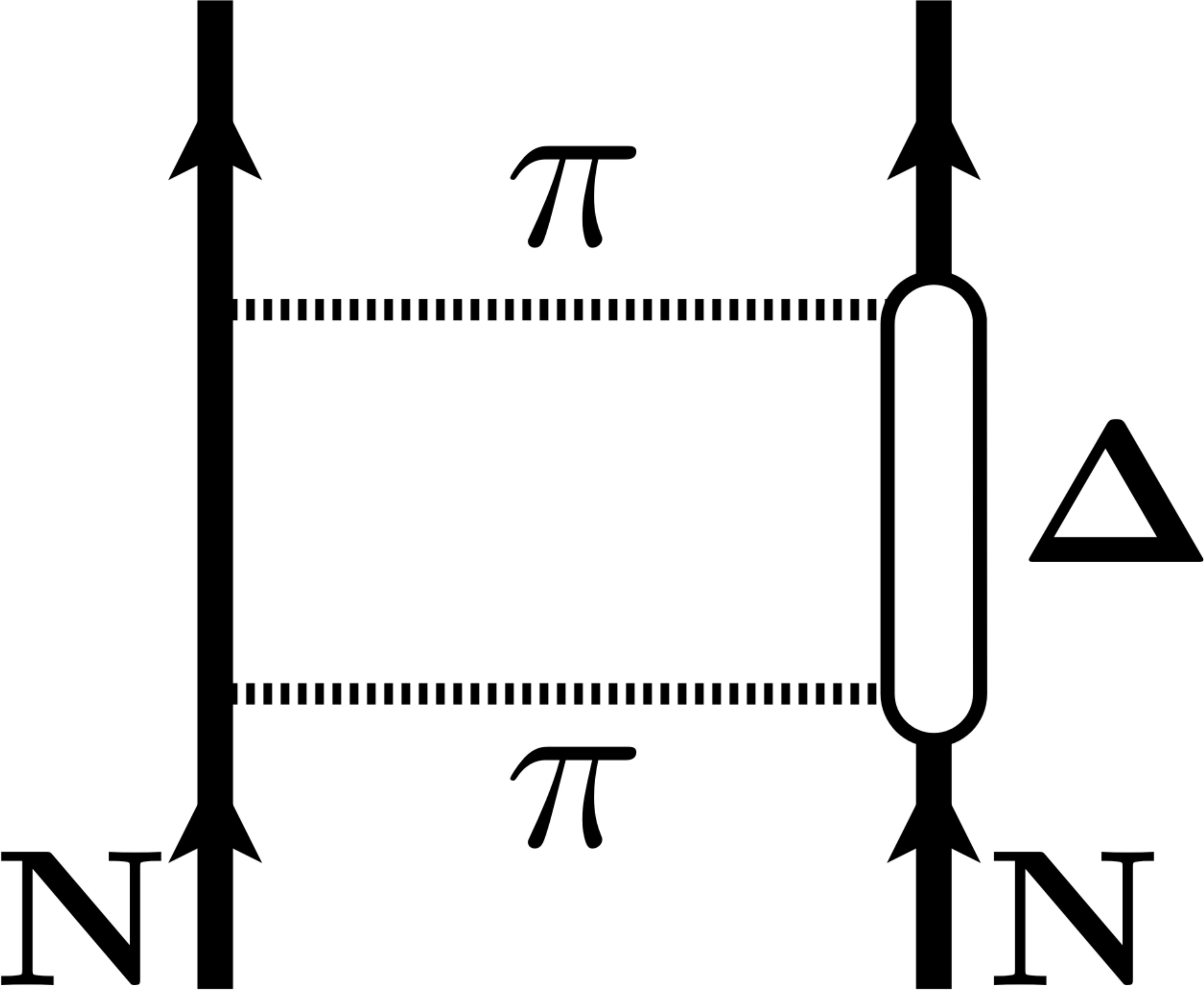}\hspace{.4in}
\includegraphics[height=2.3cm]{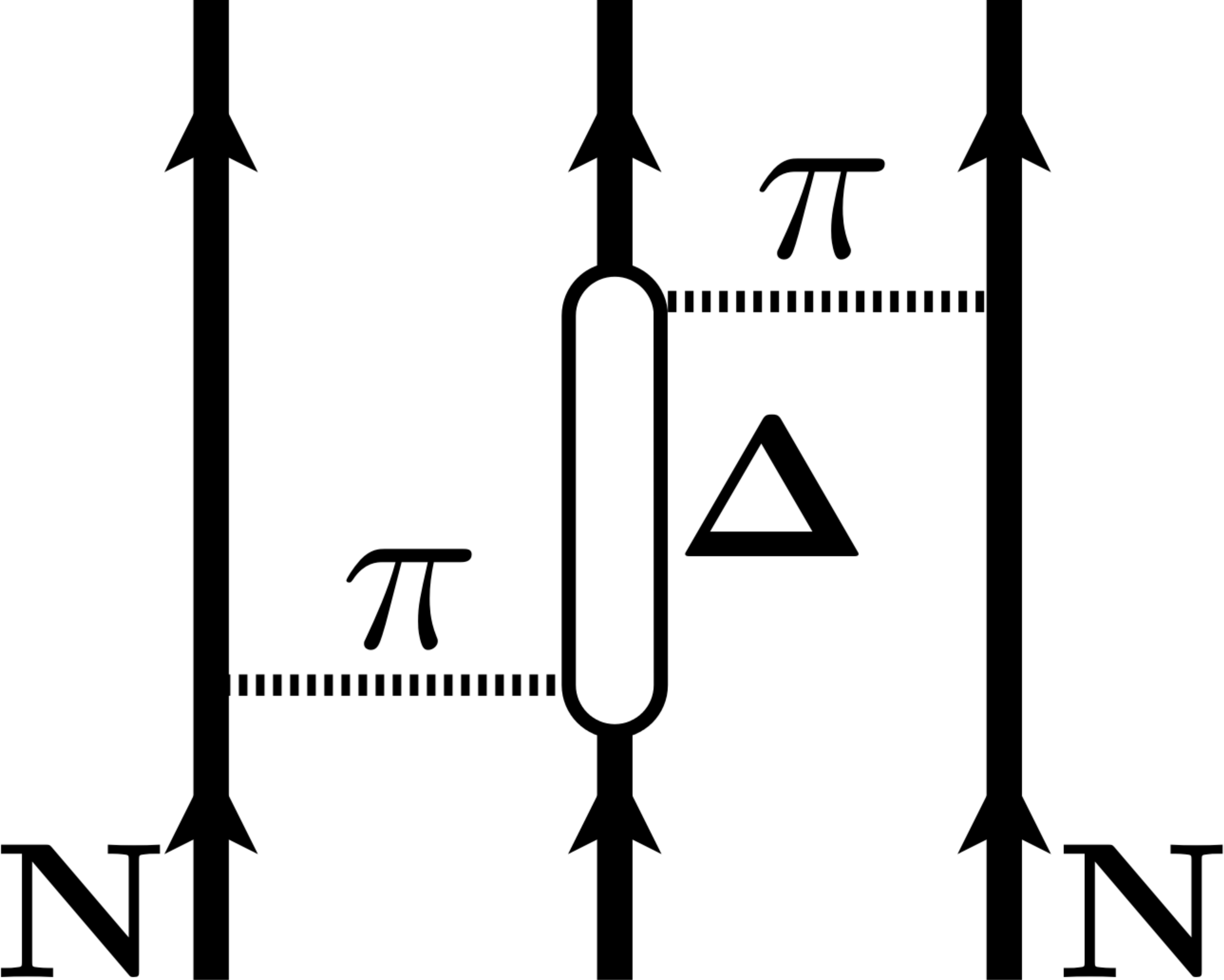}
\end{center}
\vspace{-.5cm}
\caption{Left panel: two-pion exchange involving a virtual $N\rightarrow\Delta\rightarrow N$ transition. Right panel:
three-nucleon interaction generated by the same mechanism.}
\label{2piDelta}
\end{figure}

The $\Delta$ isobar plays a similarly important role for nuclear interactions, where
two-pion exchange processes, such as the one shown in Fig.\ \ref{2piDelta} (left),
contribute significantly to the attractive isoscalar central NN interaction \cite{fkw2}. In one-boson
exchange models of the nucleon-nucleon interaction, such effects are parametrized in terms of
a fictitious ``sigma" boson. A parameter-free calculation of the isoscalar central potential generated by
single and double $\Delta$ excitation \cite{gerstendorfer} agrees almost perfectly with phenomenological
``$\sigma$" exchange at distances r $>$ 2 fm.
The behavior of the $2\pi$-exchange isoscalar central potential with
virtual excitation of a single $\Delta$ is reminiscent of a van der Waals potential:
\begin{equation}
V_C^{N\Delta}(r) = - {3g_A^2\,\alpha_A^{(\Delta)}\over (8\pi f_\pi)^2}\,{e^{-2m_\pi r}\over r^6} P(m_\pi r)~~,
\end{equation}
where $P(x)=6+12x+10x^2+4x^3+x^4$ is a fourth-order polynomial in $x = m_\pi r$. In the chiral limit
the familiar $r^{-6}$ dependence of the van der Waals interaction emerges naturally.

The $\Delta$ degrees of freedom also gives rise to an important effective three-nucleon interaction,
Fig.\ \ref{2piDelta} (right), which was suggested already more than half a century ago by Fujita and Miyazawa
\cite{fujita}. In chiral effective field theory with explicit $\Delta$ isobars, the low-energy constants $c_3$ and $c_4$ in
Eq.\ (\ref{3n4}), related to $p$-wave pion-nucleon scattering, are readjusted and reduced in magnitude since then
they have to account only for the remaining non-resonant background. Then important physics of the $\Delta$
are actually promoted from N$^2$LO to NLO in the chiral hierarchy of the NN interaction \cite{evgenireview},
leading to improved convergence.

One of the principal questions at the starting point of the chiral effective field theory approach is
how much information about the intrinsic structure of the nucleon is actually needed in order for the theory to work efficiently? The above considerations indicate that the $\Delta(1232)$ is indeed the dominant feature of nucleon
structure at low energy scales, while other less prominent excitations, such as baryon resonances appearing at higher energies are conveniently absorbed into low-energy constants.

\subsection{Infinite nuclear matter}
\label{inm}

Efforts to understand the properties of infinite nuclear matter from chiral effective field theory generally fall
into two classes. In one approach free-space two- and three-body nuclear potentials, whose low-energy
constants are fit to NN scattering phase shifts and properties of bound two- and three-body systems,
are combined with a many-body method of choice to compute the energy per particle. The unresolved
short-distance dynamics is therefore fixed at the few-body level. In the second approach, in-medium
chiral perturbation theory, the energy per particle is constructed as a diagrammatic expansion in the
number of loops involving explicit pion-exchange corresponding to an expansion in small external
momenta and the pion mass. Here the short-distance dynamics is fine-tuned to globally
reproduce the bulk properties of infinite matter. In nuclear matter the Fermi momentum $k_f$, related
to the nucleon density by $\rho=2k_f^3/3\pi^2$, sets the relevant momentum scale. At nuclear matter
saturation density $\rho_0 \simeq 0.16\,$fm$^{-3}$ the Fermi momentum $k_{f0}$ and the pion mass
are of comparable magnitude: $k_{f0} \simeq 2 m_\pi$. This implies that pions should be included as
explicit degrees of freedom. With these two small scales, $k_f$ and $m_\pi$, the nuclear matter equation
of state will be given by an expansion in powers of the Fermi momentum.
The expansion coefficients are non-trivial functions of the dimensionless parameter $u=k_f/m_\pi$, the
ratio of the two relevant low-energy scales in the problem. Two-nucleon reducible
diagrams modify the standard power counting \cite{weinbergseminal} in which higher loop contributions
are suppressed by powers of $k_f^2/(4\pi f_\pi)^2$. As we show explicitly below one of the small scales
gets replaced by the large nucleon mass $M_N$, giving rise to the scaling factor
$k_f M_N/(4\pi f_\pi)^2$.

The new feature in nuclear many-body calculations (compared to
scattering processes in the vacuum) is the in-medium nucleon propagator. For a
non-relativistic nucleon with the four-momentum $p^\mu =(p_0, \vec p\,)$ it reads:
\begin{equation}
G(p_0, \vec p\,) = { i \over p_0 -\vec p^{\,2}/2M_N + i \epsilon} -
2\pi \delta(p_0 -\vec p^{\,2}/2M_N) \, \theta(k_f -|\vec p\,|)\, .
\label{imp}
\end{equation}
The first term is the normal free-space propagator, while the second term is the so-called
``medium insertion'' and accounts for the filled
Fermi sea of nucleons. We note that this expression for $G(p_0, \vec p\,)$ can be rewritten as the sum
particle and hole propagators, a form more commonly used in non-relativistic many-body perturbation
theory. With the decomposition in Eq.\ (\ref{imp}), closed
multi-loop diagrams representing the energy density can be organized
systematically in the number of medium insertions.

\begin{figure}
\centerline{\includegraphics[width=10cm] {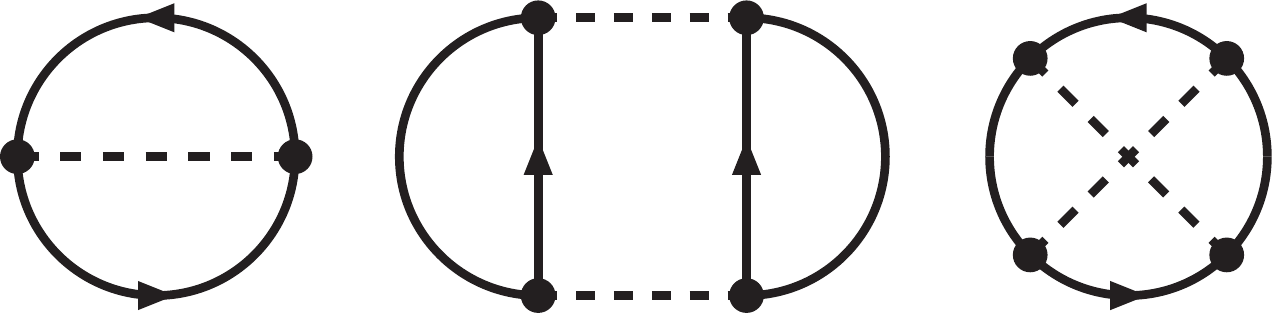}}
\caption{One-pion exchange Fock diagram and iterated one-pion exchange Hartree (left) and Fock (right)
diagrams contributing to the energy density $\rho \bar E(k_f)$ of isospin-symmetric nuclear matter.}
\label{pionfigs}
 \end{figure}

The leading contributions \cite{kfw1} to the energy per particle $\bar E(k_f)$ of
isospin-symmetric spin-saturated nuclear matter include the relativistically improved
kinetic energy
\begin{equation}  \bar E(k_f)^{\rm (kin)} = {3k_f^2 \over 10M_N}- {3k_f^4 \over 56M_N^3}\,
\end{equation}
together with the one-pion exchange Fock term
\begin{equation}
\bar E(k_f)^{(1\pi)} = {3g_A^2m_\pi^3 \over(4\pi f_\pi)^2}
\bigg\{{u^3\over 3} +{1\over 8u} -{3u\over 4}+\arctan 2u -\Big( {3\over
8u}+{1\over 32u^3}\Big) \ln(1+4u^2)\bigg\}\, ,
\end{equation}
where $u= k_f/m_\pi$. Diagrammatically this contribution is represented by the leftmost
term in Fig.\ \ref{pionfigs}.
The next terms in the $k_f$-expansion are the second-order $1\pi$-exchange
Hartree and Fock diagrams (shown as the middle and rightmost terms in Fig.\ \ref{pionfigs}).
The two-particle-reducible part of the planar $2\pi$-exchange box diagram gets
enhanced by a small energy denominator proportional to differences of nucleon kinetic
energies. The Hartree term, which can be computed analytically, reads:
\begin{equation}
\bar E(k_f)^{(H2)} = {3g_A^4M_N m_\pi^4 \over 5(8\pi)^3f_\pi^4}\bigg\{
{9\over 2u}-59 u+(60+32u^2) \arctan 2u -\Big( {9\over 8u^3} +{35\over 2u}
\Big) \ln(1+4u^2) \bigg\} \,
\end{equation}
and the Fock term is given in terms of a remaining one-dimensional integral:
\begin{equation}
\bar E(k_f)^{(F2)} = {g_A^4 M_N m_\pi^4 \over (4\pi)^3f_\pi^4}\bigg\{
{u^3\over 2} + \int_0^u \!dx {3x (u-x)^2(2u+x) \over 2u^3(1+2x^2)} \Big[
(1+8x^2+8x^4) \arctan x-(1+4x^2)\arctan2x\Big] \bigg\}\,.
\end{equation}
The above expressions for the second-order terms do not include a linear divergence
$\int_0^\infty dl\, 1$ coming from the momentum-space loop integral. In dimensional
regularization such a linear divergence is set to zero, in contrast to cutoff regularization
which gives a contribution linear in the momentum cutoff $\Lambda$. In fact, contributions
that are generally expected to be attractive in second-order perturbation theory here reveal only
their finite repulsive parts. The attractive component can be restored via a term linear in the
cutoff and the density
\begin{equation}  \bar E(k_f)^{(\Lambda)} = -{10 g_A^4 M_N \over (4\pi f_\pi)^4}\, \Lambda\, k_f^3 .
\,, \end{equation}

\begin{figure}
      \centerline{\includegraphics[width=10cm] {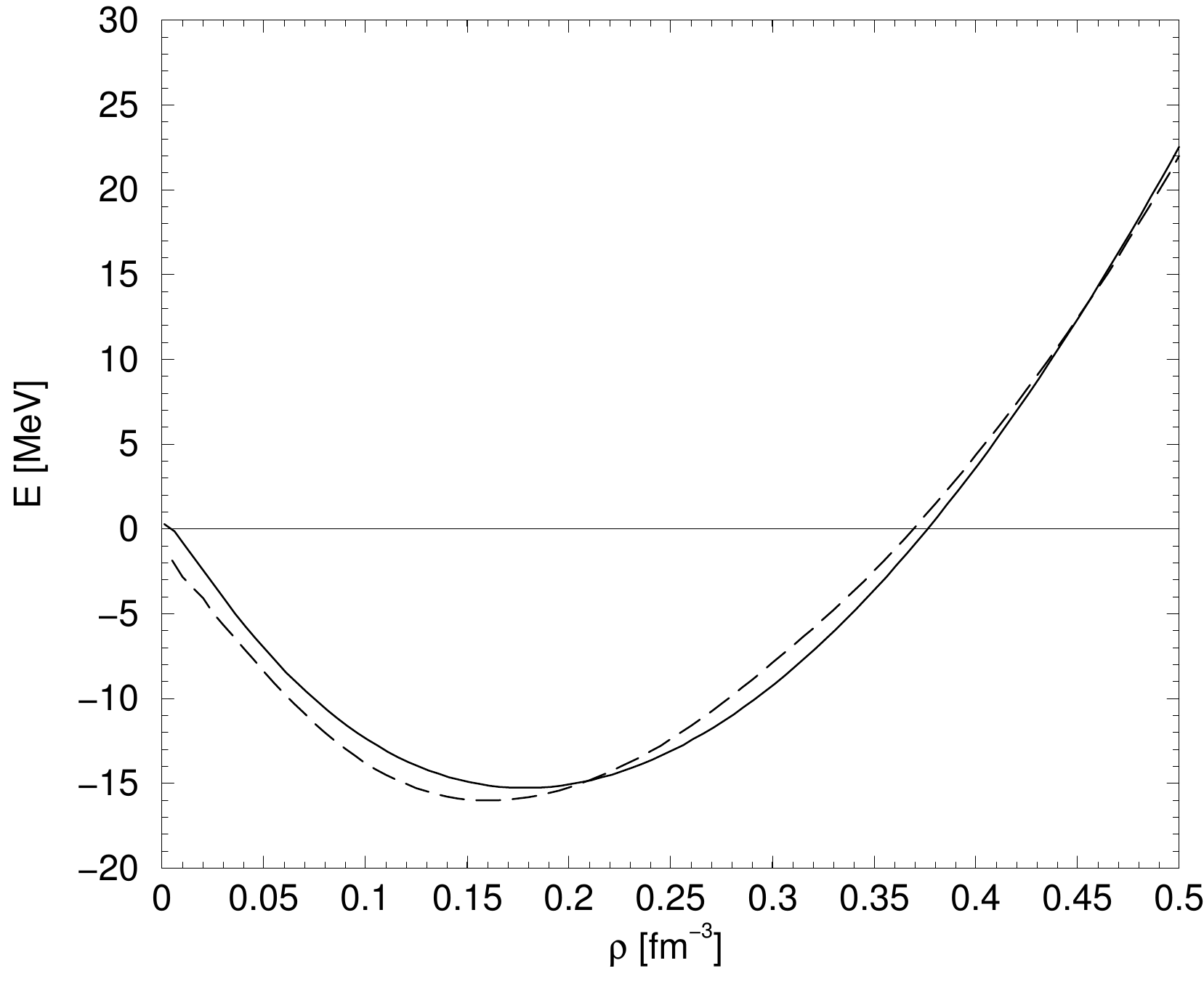}}
\caption{Equation of state of isospin-symmetric nuclear matter obtained from $1\pi$-exchange
and iterated $1\pi$-exchange together with a fine-tuned short-distance term linear in the density.
The dashed line stems from the many-body calculation of ref.\ \cite{friedpand}.}
\label{ebarpion}
\end{figure}

Replacing the free propagators with medium insertions in the above second-order diagrams
gives rise to Pauli-blocking corrections due to the nuclear medium (for explicit expressions
see Ref.\ \cite{kfw1}). Summing the above contributions to the energy per particle and
treating the cutoff $\Lambda$ as an adjustable fine-tuning parameter to incorporate
unresolved short-distance dynamics, one obtains the equation of state
of nuclear matter shown in Fig.\ \ref{ebarpion}. With a cutoff scale of
$\Lambda=611\,$MeV, the saturation density and energy per particle are
$\rho_0 =0.173\,$fm$^{-3}$ and $\bar E(k_{f0}) = -15.3\,$MeV. As a nontrivial check,
the nuclear matter compressibility comes out to be ${\cal K} = k_{f0}^2 \bar E''(k_{f0}) = 252\,$MeV,
in good agreement with the presently accepted empirical value ${\cal K} = (250\pm 50)\,$MeV
\cite{blaizot,youngblood,stoitsov}.

Let us now explain how the saturation of nuclear matter arises in the
framework of in-medium chiral perturbation theory. We
consider the following simple parameterization of the energy per particle \cite{kfw1}:
\begin{equation}  \bar E(k_f) = {3k_f^2 \over 10 M_N} - \alpha {k_f^3 \over M_N^2} + \beta
{k_f^4 \over M_N^3} \,, \end{equation}
which includes an attractive $k_f^3$-term and a repulsive $k_f^4$-term. Generically, this
family of curves has a saturation minimum provided that $\alpha, \beta>0$, and interestingly
once the two parameters are adjusted ($\alpha = 5.27$ and $\beta =12.22$ ) to the empirical
saturation point \{$\rho_0 = 0.16\,$fm$^{-3}$, $\bar E_0 = -16\,$\,MeV\}, the compressibility
${\cal K} \simeq 240$\,MeV comes out
correctly. Moreover, this simple parametrization for $\bar E(k_f)$ reproduces the equation
of state from sophisticated many-body calculations \cite{friedpand} using realistic nuclear forces
up to quite high densities $\rho \simeq 1\,$fm$^{-3}$.

The connection between this simple model and in-medium chiral perturbation theory is that in the
chiral limit, the leading interaction contributions calculated from $1\pi$- and
iterated $1\pi$-exchange turn into exactly such a two-parameter form where the coefficient
$\beta$ is given by \cite{kfw1}:
\begin{equation}
\beta = {3 \over 70}\bigg({g_A M_N \over 4\pi f_\pi}\bigg)^4(4\pi^2+237-24\, \ln \,2) = 13.6\, .
\end{equation}
This value of $\beta$ is quite close to that ($\beta =12.22$) extracted from a realistic nuclear
matter equation of state. The mechanism for nuclear matter saturation can be summarized
roughly as follows: while pion-exchange processes up to second order generate the necessary
attraction, Pauli-blocking effects due to the nuclear medium counteract this attraction in the
form of a repulsive contribution with a stronger density dependence (a $k_f^4$-term). This
interpretation is similar to previous analyses \cite{machleidt89} with high-precision NN
interaction models that found Pauli-blocking of the second-order tensor force to be the driving
force behind nuclear matter saturation. However, without the inclusion of three-body forces (or their
equivalent), such microscopic nuclear forces cannot achieve simultaneously the correct
saturation energy and density.

\begin{figure}
      \centerline{\includegraphics[width=10cm] {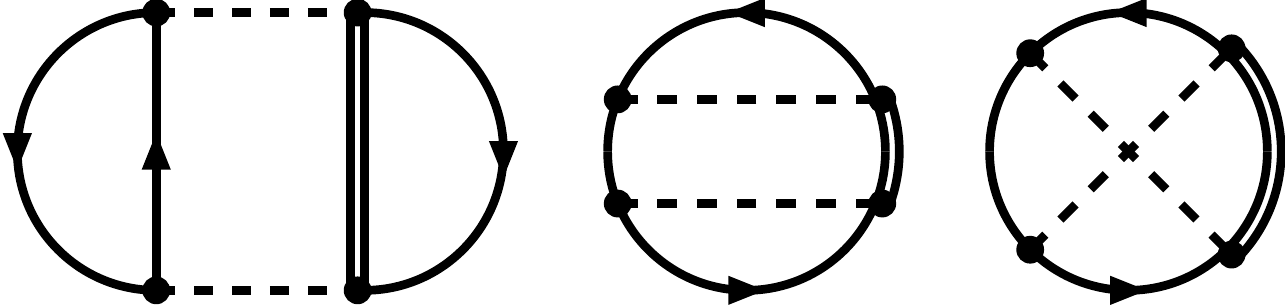}}
\caption{Hartree and Fock three-body diagrams related to $2\pi$-exchange with virtual
$\Delta(1232)$-isobar excitation. They represent long-range interactions between three
nucleons in the Fermi sea.}
\label{figdelt}
 \end{figure}

The study of nuclear matter from in-medium chiral perturbation theory has been extended
further by including the irreducible two-pion exchange contributions in the medium \cite{fkw2}.
A compact form for the associated Fock term is given in terms of a subtracted spectral
function representation:
\begin{eqnarray}
\bar E(k_f)^{(2\pi F)}&=& {1 \over 8\pi^3} \int_{2m_\pi}^{\infty}
\!\! d\mu\,{\rm Im}(V_C+3W_C+2\mu^2V_T+6\mu^2W_T) \bigg\{ 3\mu k_f -{4k_f^3
\over 3\mu }\nonumber \\ &&+{8k_f^5 \over 5\mu^3 } -{\mu^3 \over 2k_f}-4\mu^2
\arctan{2k_f\over\mu} +{\mu^3 \over 8k_f^3}(12k_f^2+\mu^2) \ln\bigg( 1+{4k_f^2
\over \mu^2} \bigg) \bigg\} \,, \label{specrep}
\end{eqnarray}
where Im$V_C$, Im$W_C$, Im$V_T$ and Im$W_T$ are the imaginary parts
(or spectral functions) of the isoscalar and isovector central and tensor NN amplitudes.
At this order also the $2\pi$-exchange diagrams with virtual $\Delta(1232)$-isobar
intermediate states come into play. As mentioned previously these terms
provide a dominant mechanism for intermediate-range attraction in the isoscalar
central channel \cite{gerstendorfer}. Furthermore, Pauli-blocked diagrams with three
medium insertions and a single $\Delta(1232)$ excitation
are equivalent to contributions from genuine long-range three-nucleon forces.
The two-ring Hartree diagram in Fig.\ \ref{figdelt} for instance leads to the following
contribution to the energy per particle \cite{fkw2}:
 \begin{equation}
 \bar E(k_f)^{(\Delta H3)}={g_A^4 m_\pi^6 \over \Delta(2\pi f_\pi)^4}
\bigg[ {2\over3}u^6 +u^2-3u^4+5u^3 \arctan2u-{1\over 4}(1+9u^2)
\ln(1+4u^2) \bigg] \,,
\label{fkw2h}
\end{equation}
where $\Delta = 293\,$MeV is the delta-nucleon mass splitting. The total
contribution of the one-ring Fock diagrams in Fig.\ \ref{figdelt} are given by
\begin{equation}
\bar E(k_f)^{(\Delta F3)}=-{3g_A^4 m_\pi^6 u^{-3}\over 4\Delta(4\pi
f_\pi)^4 }\int_0^u \!\! dx \Big[ 2G^2_S(x,u)+G^2_T(x,u)\Big] \,,
\label{fkw2f}
\end{equation}
where the explicit forms for the two auxiliary functions $G_S$ and $G_T$ can be
found in Ref.\ \cite{fkw2}. Evaluation of the terms in Eqs.\ (\ref{fkw2h}-\ref{fkw2f})
reveals that the three-body Hartree term is repulsive while the Fock term is weakly
attractive.

The subtraction constants associated with the spectral function
representation in Eq.\ (\ref{specrep}) result in short-distance contributions of the form
\begin{equation}
\bar E(k_f)^{(sd)} = B_3 {k_f^3 \over M_N^2}+ B_5 {k_f^5 \over M_N^4}+  B_6 {k_f^6
\over M_N^5}\,,
\end{equation}
The parameters $B_3,B_5,B_6$ are assumed to encode all unresolved short-distance NN
dynamics relevant for symmetric nuclear matter up to moderate densities. They are adjusted
to the bulk properties of nuclear matter, and thereafter the model is able to predict more detailed
quantities like the single-particle potential or the thermodynamic properties of nuclear matter
at finite temperatures. As we discuss in more detail in Section \ref{nct}, in this approach the
pion-mass dependence of all interaction terms is explicitly known. Differentiation of the nuclear
matter equation of state with respect to the pion mass then provides access to the density- and
temperature-dependent chiral condensate $\langle \bar q q \rangle(\rho,T)$.

\subsubsection{\it Fermi liquid description of nuclear matter}
\label{flsnm}

Landau's theory of normal Fermi liquids \cite{landau57} is a standard framework for
describing the low-energy, long-wavelength excitations of strongly-interacting Fermi
systems at low temperatures. The theory treats on an equal footing quasiparticle
excitations carrying the quantum numbers of single-particle states, as well as
collective modes of bosonic character. In addition the theory directly links the interaction
between quasiparticles to properties of the interacting ground state. The microscopic
foundation for Fermi liquid theory is grounded in many-body perturbation theory
\cite{abrikosov59}, where it has been shown that in the vicinity of the Fermi surface,
the one-body Green's function can be conveniently decomposed into quasiparticle and
background contributions, the effects of the latter are then absorbed into the effective
coupling between quasiparticles. In the original works \cite{migdal63,migdal67} of
Migdal applying Fermi liquid theory to finite nuclei, the quasiparticle couplings were adjusted
to empirical nuclear data, leading to a phenomenological theory that was capable of linking
different phenomena. The microscopic approach to nuclear Fermi liquid theory
based upon realistic models of the strong nuclear force was initiated by Gerry Brown and
collaborators \cite{brown71} and has remained a favorite topic of Gerry's even up until in
recent years \cite{schwenk02,holt07}. The use of chiral effective field theory
methods to study the quasiparticle interaction in nuclear and neutron matter has been
realized in Refs.\ \cite{fermiliq,holt12,holt13}. In this section we focus on symmetric
nuclear matter, while in Section \ref{flt} we discuss application of the theory to neutron matter
in the context of neutron-star dynamics.

The starting point of the microscopic approach to Fermi liquid theory is Landau's assumption
that the total energy $E$ of a weakly excited many-fermion system at zero temperature is a
functional of the quasiparticle distribution $n_p$. Taylor expanding to second order
in small variations of the distribution function then yields
\be
\delta {\cal E} = \sum_{\vec p s t} \epsilon^{(0)}_{\vec p}\, \delta
n_{\vec p s t} + \frac{1}{2} \sum_{\substack{{\vec p}_1 s_1 t_1 \\
{\vec p}_2 s_2 t_2}}{\cal F}({\vec p}_1 s_1 t_1;
{\vec p}_2 s_2 t_2)\, \delta n_{{\vec p}_1 s_1 t_1}
\,\delta n_{{\vec p}_2 s_2 t_2} + \cdots,
\label{deltae}
\ee
where $s_i$ and $t_i$ label the spin and isospin quantum numbers and ${\cal F}$ is the
quasiparticle interaction arising as the second functional derivative of the energy density
with respect to the quasiparticle distribution function. In diagrammatic perturbation theory
the first- and second-order contributions to the quasiparticle interaction are shown in
Fig.\ \ref{pphhph}. Including as well the leading-order corrections from chiral three-nucleon
forces \cite{holt12}, the bulk properties of symmetric nuclear matter around the saturation
density are found to be well reproduced. These properties are related to specific ``Landau
parameters'' ($F_L, G_L, \dots$) coming from a Legendre polynomial decomposition of the
central quasiparticle interaction:
\begin{figure}
\begin{center}
\includegraphics[width=18cm]{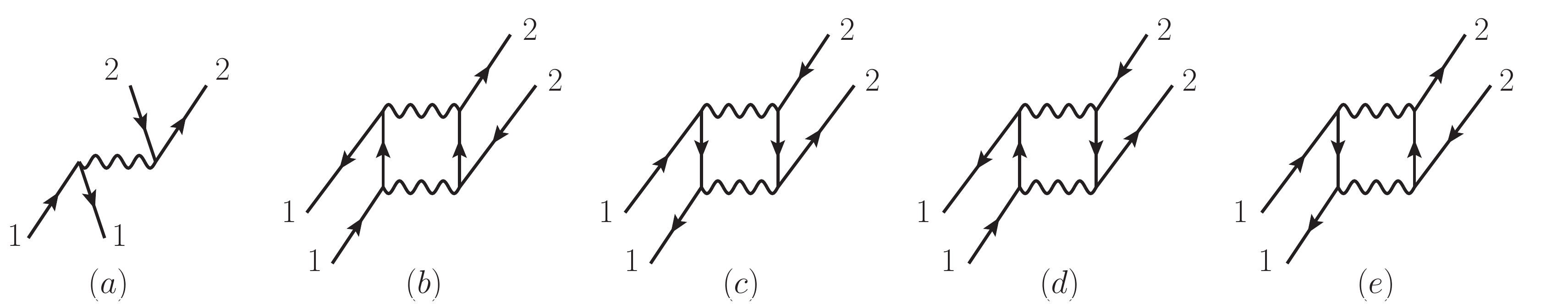}
\end{center}
\vspace{-.5cm}
\caption{Diagrams contributing to the first- and second-order quasiparticle interaction (wavy
lines represent antisymmetrized interactions): (a) first-order diagram, (b) particle-particle diagram,
(c) hole-hole diagram, and (d)+(e) particle-hole diagrams.}
\label{pphhph}
\end{figure}
\begin{equation}
{\cal F}_{\rm cent}({\vec p}_1, {\vec p}_2)=\frac{1}{N_0}\sum_{L=0}^\infty \left [
F_L + F^\prime_L \,\vec \tau_1 \cdot
\vec \tau_2 + (G_L + G^\prime_L \,\vec \tau_1 \cdot \vec \tau_2)\, \vec \sigma_1
\cdot \vec \sigma_2 \right ]P_L({\rm cos}\, \theta),
\label{ffunction2}
\end{equation}
where $N_0 = 2M^* k_f/\pi^2$ is the density of states at the Fermi surface, and we
have restricted the quasiparticle momenta to lie on the Fermi surface: $|\vec p_1| = |\vec p_2| =k_f$.
Below we will sometimes refer to the dimensionful Landau parameters defined by $f_L = F_L / N_0$.
The noncentral components of the quasiparticle interaction, which are more relevant for
neutron matter, will be discussed in the following Section \ref{flt}. The connection between
the Landau parameters and nuclear observables is well known \cite{migdal67}:
\begin{eqnarray}
{\rm Quasiparticle \, \, effective \, \, mass:} \hspace{.2in} \frac{M^*}{M_N} &=& 1+F_1/3,
\nonumber \\
{\rm Compression \, \, modulus:} \hspace{.2in} {\cal K}&=&\frac{3k_F^2}{M^*}
\left (1+F_0\right ), \nonumber \\
{\rm Isospin\, \,  asymmetry \, \, energy:} \hspace{.2in} \beta &=&
\frac{k_F^2}{6M^*}(1+F_0^\prime), \nonumber \\
{\rm Orbital \, \,} g\mbox{-factor:} \hspace{.2in}
g_l &=& \frac{1+\tau_3}{2} + \frac{F^\prime_1-F_1}{6(1+F_1/3)}\tau_3 , \nonumber \\
{\rm Spin\mbox{-}isospin\, \, response:} \hspace{.2in} g_{NN}^\prime &=&
\frac{4M_N^2}{g_{\pi N}^2N_0} G_0^\prime.
\label{obs}
\end{eqnarray}

In Table \ref{n3loc} we show the values of the Landau parameters at different orders
of perturbation theory employing the high-precision chiral nuclear force with momentum-space
cutoff of $\Lambda = 500$\,MeV whose low-energy constants are fit in Refs.\
\cite{entem03,gazit09}. Only the leading N$^2$LO part of the chiral three-nucleon force
is included in addition with the full N$^3$LO two-body potential. When two-body forces
alone are considered (first row of Table \ref{n3loc}), the quasiparticle interaction shares many features of previous
calculations \cite{sjoberg73a,sjoberg73b,backman85} employing the $G$-matrix effective
interaction. Specifically, the quasiparticle effective mass $M^*/M_N = 0.70$ is well below unity,
the nuclear symmetry energy $\beta = 21$\,MeV is small compared to the experimental values
$\beta_{emp} \simeq 30-35$\,MeV, and the compression modulus ${\cal K} = -44$\,MeV is
negative, leading to an instability with respect to density fluctuations. This last feature was
particularly troubling in early studies using microscopic two-body forces, and Gerry Brown had
the insight to include screening effects (the Babu-Brown induced interaction \cite{babu73}) in
the quasiparticle interaction coming from virtual collective excitations that cut down the strong
attraction in the scalar-isoscalar channel. In the chiral effective field theory approach, we
will see that three-nucleon forces provide very strong repulsion in this channel.

\setlength{\tabcolsep}{.07in}
\begin{table}
\begin{center}
\begin{tabular}{|c||c|c|c|c||c|c|c|c|} \hline
\multicolumn{9}{|c|}{Chiral 2N + 3N interactions at $k_F=1.33$ fm$^{-1}$} \\ \hline
 & $f_0$ [fm$^2$] & $g_0$ [fm$^2$] & $f^\prime_0$ [fm$^2$] & $g^\prime_0$ [fm$^2$] &
$f_1$ [fm$^2$] & $g_1$ [fm$^2$] & $f^\prime_1$ [fm$^2$] & $g^\prime_1$ [fm$^2$] \\ \hline
$V_{2N}^{(1)}$ & $-$1.274 & 0.298 & 0.200 & 0.955 & $-$1.018 & 0.529 & 0.230 & 0.090   \\ \hline \hline
$V_{2N}^{(2-pp)}$ & $-$1.461 & 0.023 & 0.686 & 0.255 & 0.041 & $-$0.059 & 0.334 & 0.254 \\ \hline
$V_{2N}^{(2-hh)}$ & $-$0.271 & 0.018 & 0.120 & 0.041 & 0.276 & 0.041 & $-$0.144 & $-$0.009 \\ \hline
$V_{2N}^{(2-ph)}$ & 1.642 & $-$0.057 & 0.429 & 0.162 & 0.889 & $-$0.143 & 0.130 & 0.142  \\ \hline \hline
$V_{3N}^{(1)}$ & 1.218 & 0.009 & 0.009 & $-$0.295 & $-$0.073 & $-$0.232 & $-$0.232 & $-$0.179 \\ \hline
\end{tabular}
\caption{$L = 0,1$ Fermi liquid parameters for symmetric nuclear matter at
saturation density from the N$^3$LO chiral nucleon-nucleon potential
as well as from the leading N$^2$LO chiral three-body force.}
\label{n3loc}
\end{center}
\end{table}

In rows 3-5 of Table \ref{n3loc} we show the second-order contributions
to the Landau parameters from the N$^3$LO chiral two-body force.
The particle-particle term (represented diagrammatically Fig.\
\ref{pphhph}(b)) enhances the attraction in the spin- and
isospin-independent interaction, while the particle-hole terms (Fig.\ \ref{pphhph}(d)
and (e)) are strongly repulsive. Together the particle-particle and particle-hole contributions
largely cancel in the calculation of the $F_0$ Landau parameter. The second-order
particle-hole diagram accounts for much of the full Babu-Brown induced interaction, and
therefore the mechanism to stabilize nuclear matter for chiral nuclear interactions
must be qualitatively different than in the case of the $G$-matrix effective interactions. As
seen in Table \ref{n3loc} both the $f_0^\prime$ and $f_1$ Landau parameters
receive coherent contributions from all second-order terms. In particular, this results in
a quasiparticle effective mass at saturation density that is nearly equal to the
free-space nucleon mass. By itself, this effect would decrease the
isospin asymmetry energy. However, the second-order contributions strongly
enhance the isospin-dependent parts of the quasiparticle interaction, and
at nuclear matter saturation density the resulting isospin asymmetry energy is
$\beta = 34$\,MeV, in very good agreement with the empirical value.

\begin{figure}[t]
\begin{center}
\includegraphics[height=9cm,angle=270]{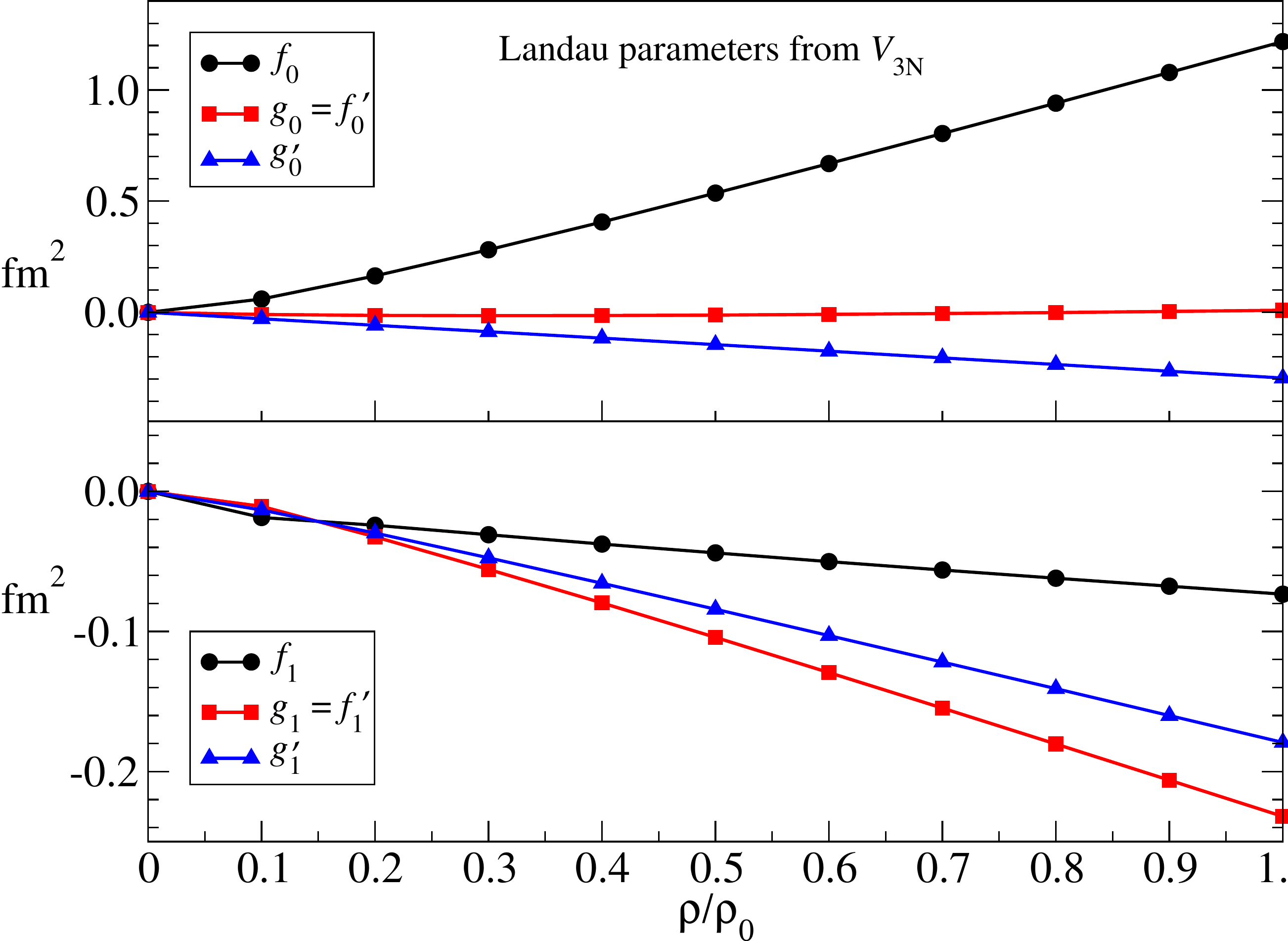}
\end{center}
\vspace{-.5cm}
\caption{Density-dependent $L=0,1$ Fermi liquid parameters from the
leading chiral three-nucleon force with low-energy constants given in ref.\ \cite{gazit09}.}
\label{DFLPn3lo}
\end{figure}

We now discuss the role of three-body forces. We first note that three-quasiparticle
interactions, omitted in the expansion in Eq.\ (\ref{deltae}), are expected to be
small because excitations of the many-body system result in a dilute gas of
quasiparticles (see, for instance, the discussion in Ref.\ [20]). However, genuine
three-body forces contribute to ${\cal F}$ in the form
of a density-dependent interaction. The first-order contributions from the
leading chiral three-nucleon force are shown diagrammatically in Fig.\ \ref{mfig1}
and were computed analytically for the first time in Ref.\ \cite{holt12}. The
surprising feature was the substantial repulsion
present in the spin- and isospin-independent quasiparticle interaction resulting
from two-pion exchange processes and to a lesser extent the repulsive chiral three-nucleon
contact interaction. We show in Fig.\ \ref{DFLPn3lo} the density-dependent $L=0,1$ Landau
parameters from the leading chiral three-nucleon force alone. In particular, $f_0$
grows strongly with the density and provides sufficient repulsion to stabilize
nuclear matter at the empirical saturation density. Summing all contributions discussed so
far and including self-consistent single-particle energies in the second-order diagrams, the
compression modulus comes out to be ${\cal K} = 200$\,MeV.
The magnitude of $f_0$ relative to the other Landau parameters (see Fig.\
\ref{DFLPn3lo} and the last row of Table \ref{n3loc}) is unexpectedly large. However, the
momentum dependence of this scalar-isoscalar interaction, given to first order by the
Landau parameter $f_1$, is especially weak and yields only a very small decrease in the
quasiparticle effective mass at saturation density. The scalar-isovector and vector-isoscalar
quasiparticle interactions from the chiral three-nucleon force are identical and nearly zero
when averaged over all momenta. Three-nucleon forces therefore play only a minor role in
determining, e.g.\ the asymmetry energy $\beta$ of nuclear matter.
Although most properties of symmetric nuclear matter at saturation density are well reproduced,
the anomalous orbital $g$-factor
\be
\delta g_l \simeq 0.07 \pm 0.02\label{chpt-deltagl}
\ee
is much weaker than the value $\delta g_{l,\rm exp} \simeq 0.23 \pm 0.03$ extracted
from an experimental analysis of the isovector giant dipole sum rule \cite{brown80}:
\begin{equation}
\int_0^{\omega_{\rm max}} d\omega \, \sigma(E1) = \frac{2\pi \alpha}{M_N} \frac{NZ}{A} (1+2\delta g_l)\, ,
\end{equation}
where $\alpha = 1/137$.\footnote{It will be seen in Section \ref{orbital} that this result can be greatly improved when the vector mesons $V=\rho, \omega$ are explicitly taken into account in the effective Lagrangian as in $hls$EFT.}
Overall, however, the first-order (free-space) contribution from two-body forces
together with the leading Pauli-blocking and core polarization effects from two- and
three-body forces has been shown to provide a very successful description of
isospin-symmetric nuclear matter in the vicinity of the empirical
saturation density.

\subsubsection{\it Nuclear mean field and the optical potential}

The nuclear single-particle potential is a central
concept in nuclear structure and reaction theory. For negative energy bound states,
the single-particle potential is associated with the shell model potential, while for
positive-energy scattering states it is identified with the nuclear optical model potential.
In this section we focus on optical model potential as it arises in many-body perturbation
theory with chiral nuclear forces, while in Section \ref{dfm} we discuss the nuclear mean
field based on the energy density functional formalism.

While phenomenological optical potentials \cite{becchetti69,varner91,koning03} are
widely employed to describe reactions on target nuclei close to the valley of stability,
microscopic optical potentials have no adjustable parameters and therefore can access
regions of the nuclear chart where there is little available empirical data. In microscopic
many-body perturbation theory, the optical model potential is identified with the
nucleon self energy \cite{bell59}, a complex-valued quantity whose imaginary part in the
context of nuclear scattering accounts for the strongly absorptive features of the
nucleon-nucleus interaction. Numerous calculations of the optical potential have been
carried out with realistic nuclear forces using a variety of many-body methods with
differing degrees of reliability
\cite{jeukenne76,grange87,haider88,arnold81,haar87,dickhoff04,waldecker11}.
Here we review the recent calculations \cite{kaiser02,fkw2,holt13b} that have been
performed in the framework of chiral effective theory.

\begin{figure}[t]
\begin{center}
\includegraphics[height=16cm,angle=270]{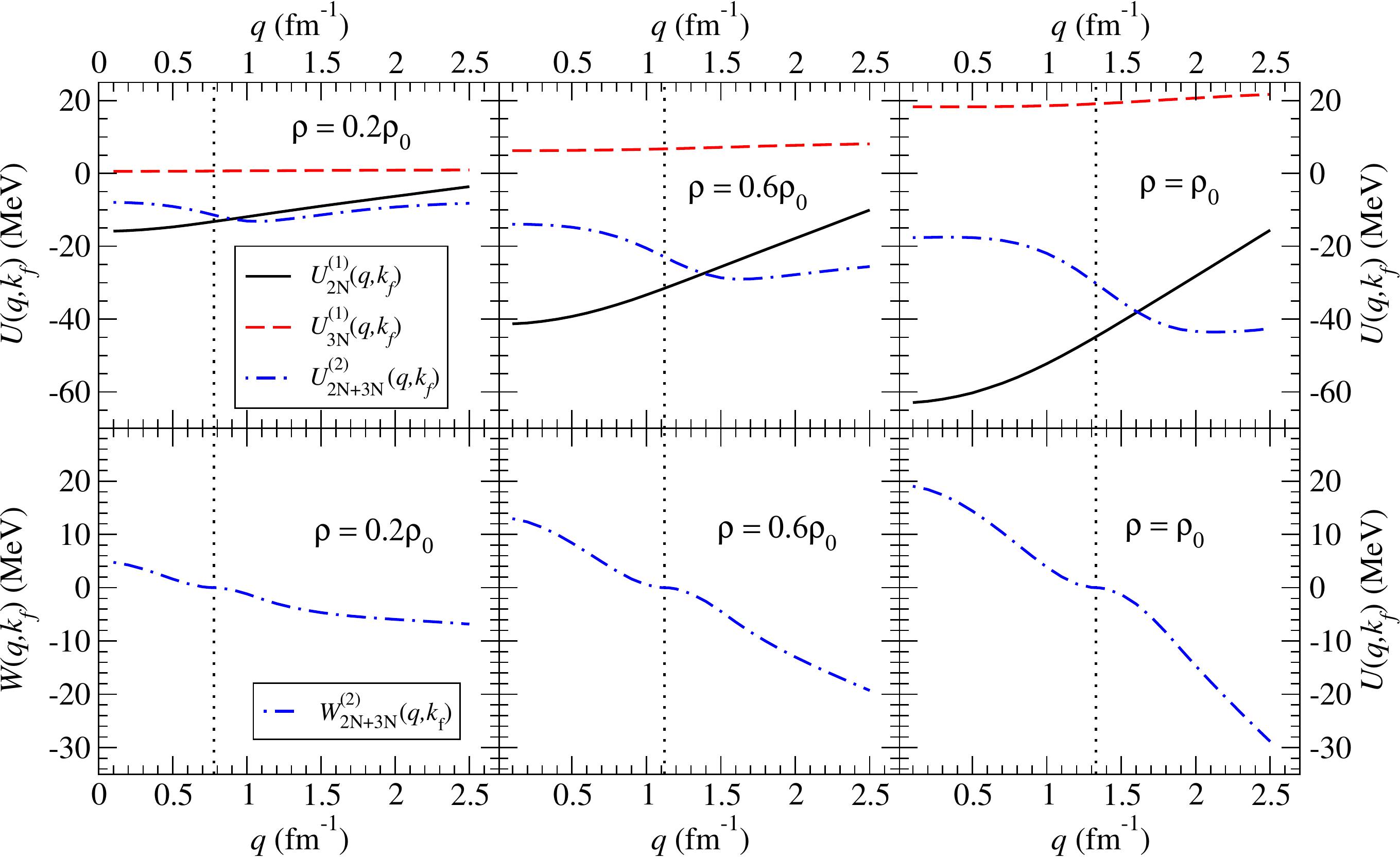}
\end{center}
\vspace{-.5cm}
\caption{Contributions to the real and imaginary parts of the on-shell ($\omega = q^2/(2M_N)$)
momentum- and density-dependent optical potential from chiral two- and three-body forces. The
vertical dotted line denotes the Fermi momentum.}
\label{ddopr}
\end{figure}

We start from the high-precision N$^3$LO chiral two-nucleon potential with a cutoff
of $\Lambda = 500$\,MeV supplemented with the leading chiral three-nucleon force
whose low-energy constants ($c_1=-0.81$\,GeV$^{-1}$, $c_3=-3.2$\,GeV$^{-1}$,
$c_4=5.4$\,GeV$^{-1}$, $c_D=-0.2$, and $c_E=-0.205$) have been fit to nucleon-nucleon
scattering phase shifts and the binding energy and lifetime of the triton \cite{entem03,gazit09}.
The nucleon self-energy function is in general complex, non-local and energy dependent:
$\Sigma(\vec r, \vec r^{\, \prime}, E) = U(\vec r, \vec r^{\, \prime}, E) + i
W(\vec r, \vec r^{\, \prime}, E)$. In a medium of homogeneous matter,
the self-energy can be more simply written as a function of the momentum, energy, and density (or
Fermi momentum) $\Sigma(q,\omega;k_f)$.
The leading two perturbative contributions from two-body forces are given by
\begin{eqnarray}
\Sigma^{(1)}_{2N}(q,\omega;k_f) &=& \sum_{1} \langle \vec q \, \vec h_1 s s_1 t t_1 | \bar V_{2N} | \vec q \,
\vec h_1 s s_1 t t_1 \rangle n_1, \nonumber \\
\Sigma^{(2)}_{2N}(q,\omega;k_f) &=& \frac{1}{2}\sum_{123} \frac{| \langle \vec p_1 \vec p_3 s_1 s_3 t_1
t_3 | \bar V_{2N} | \vec q \, \vec h_2 s s_2 t t_2 \rangle |^2}{\omega + \epsilon_2 - \epsilon_1
-\epsilon_3 + i \eta} \bar n_1 n_2 \bar n_3 (2\pi)^3 \delta(\vec p_1 + \vec p_3 - \vec q - \vec h_2), \nonumber \\
&+& \frac{1}{2}\sum_{123} \frac{| \langle \vec h_1 \vec h_3 s_1 s_3 t_1
t_3 | \bar V_{2N} | \vec q \, \vec p_2 s s_2 t t_2 \rangle |^2}{\omega + \epsilon_2 - \epsilon_1
- \epsilon_3 - i \eta} n_1 \bar n_2 n_3 (2\pi)^3 \delta(\vec h_1 + \vec h_3 - \vec q - \vec p_2),
\label{se1}
\end{eqnarray}
where sums are taken over the momentum, spin and isospin of the intermediate states, $\bar V_{2N}$
is the antisymmetrized two-body potential, $n_i = \theta(k_f-|\vec k_i|)$ is the zero-temperature Fermi
distribution, and $\bar n_i = 1-n_i$. The first-order contribution from three-body
forces is written
\begin{equation}
\Sigma^{(1)}_{3N}(q,\omega;k_f) = \frac{1}{2} \sum_{12} \langle \vec q \, \vec h_1\vec h_2; s s_1s_2; t t_1t_2 |
 \bar V_{3N} | \vec q \, \vec h_1\vec h_2; s s_1s_2; t t_1t_2 \rangle n_1 n_2,
\label{se31}
\end{equation}
where $\bar V_{3N}$ is the properly antisymmetrized three-body interaction. Including three-nucleon
forces at second order in perturbation theory can be achieved through the use of density-dependent
two-nucleon interactions \cite{holt09,holt10}.

\begin{figure}
\begin{centering}
\includegraphics[height=9cm,angle=270]{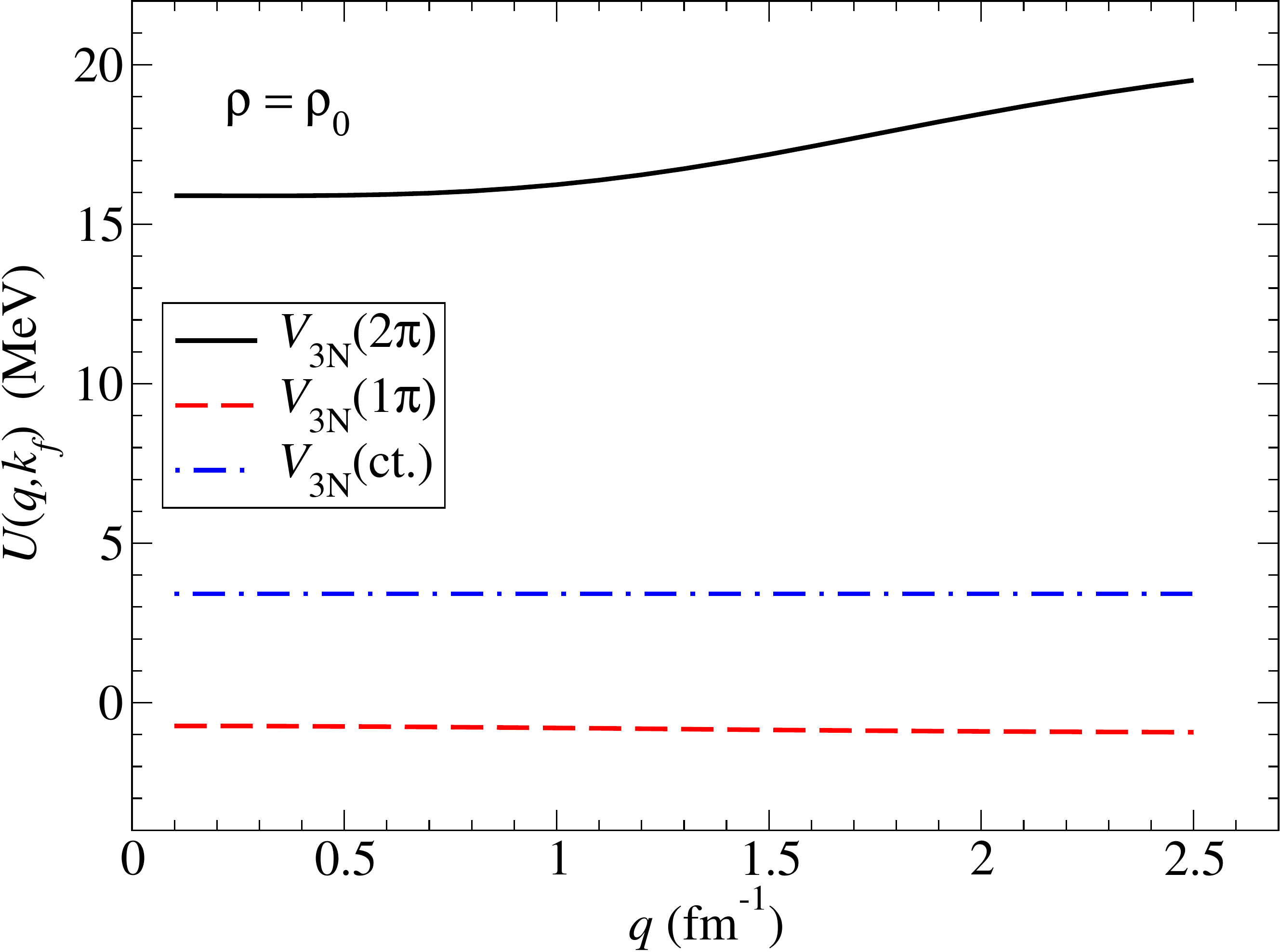}
\caption{Three-nucleon force contributions to the momentum-dependent real part of the nuclear
optical potential in the Hartree-Fock approximation. Individual results for the $2\pi$-exchange,
$1\pi$-exchange and contact three-nucleon forces are shown for symmetric nuclear at saturation
density.}
\end{centering}
\label{opt3n}
\end{figure}

In Fig.\ \ref{ddopr} we show the real and imaginary parts of the self-energy
as a function of the density and momentum. The first-order
Hartree-Fock contributions, which are real and explicitly energy independent, from two-
and three-body forces are denoted respectively by the solid and dashed lines in the
upper set of panels. The Hartree-Fock contribution from two-body forces is attractive up
to the maximum momentum considered and decreases in magnitude as the
momentum increases. In contrast, the N$^2$LO chiral three-nucleon force gives rise
to a repulsive contribution that increases nearly linearly with the density but which
is largely momentum independent. As shown in Fig.\ \ref{opt3n}
the two-pion exchange
component of the N$^2$LO chiral three-nucleon force (encoding the physics of the $\Delta$
isobar) provides nearly 80\% of the total repulsion and is responsible for the momentum
dependence of the three-body Hartree-Fock mean field.
\begin{figure}
\includegraphics[height=9cm,angle=270]{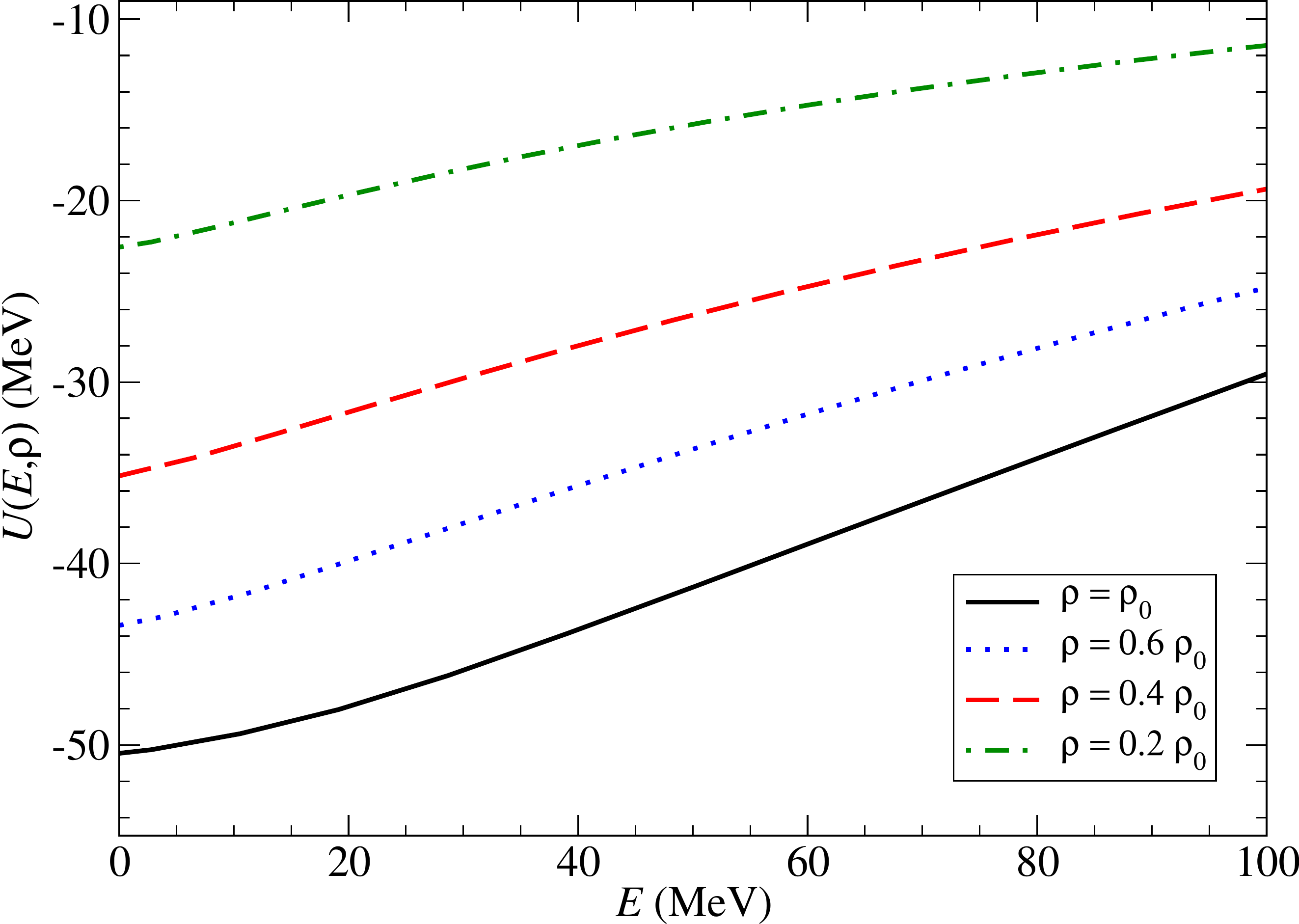}\hspace{.1in}
\includegraphics[height=9cm,angle=270]{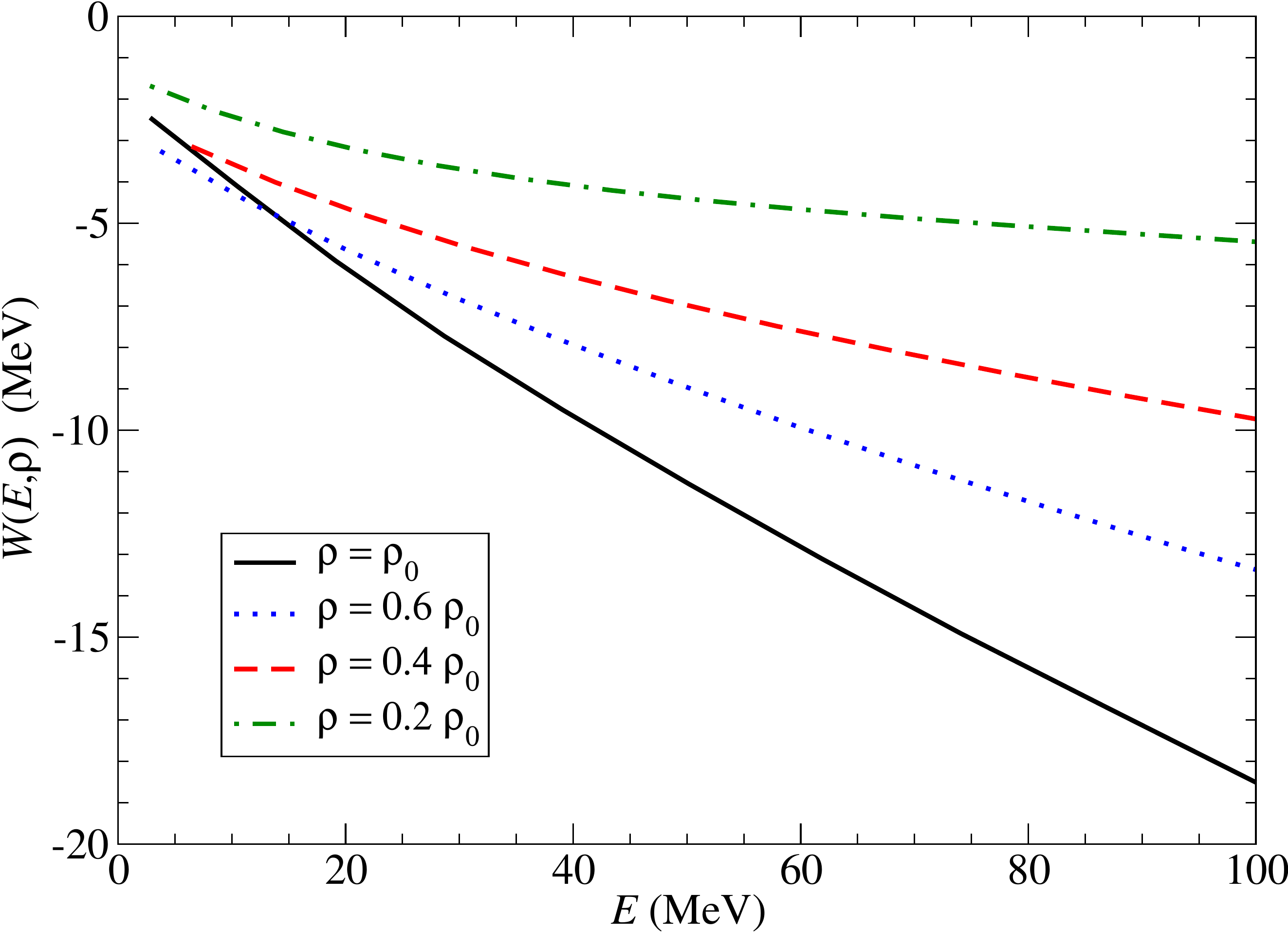}
\caption{Left panel: Three-body Hartree-Fock contributions to the real part of the nuclear optical
potential. The individual single-particle potentials from the three components of the N$^2$LO chiral
three-nucleon force are shown separately and plotted as a function of the momentum at nuclear
matter saturation density.
Right panel: Density dependence of the real part of the momentum-dependent optical potential
at second order in perturbation theory from chiral two- and three-body forces.}
\label{ddop3}
\end{figure}
Since the sum of the contact and $1\pi$-exchange three-body forces is nearly momentum-independent
as seen in Fig.\ \ref{opt3n}, variations in the low-energy constants $c_D$ and $c_E$ along the line
\be
c_E = \alpha \cdot c_D + {\rm const}
\ee
give nearly equivalent descriptions of the mean field. The constant of proportionality has the
value $\alpha \simeq 0.21\pm 0.02$, where the uncertainty results from a weak dependence
of the value on the momentum and density. In the chiral limit the single-particle potential
coming from the one-pion exchange three-body force has a particularly simple form
\cite{holt13b} which yields the correlation coefficient $\alpha = g_A/4 \simeq 0.3$.

Second-order corrections to the single-particle potential generate both real and imaginary
components as well as explicit energy dependence. We consider first the on-shell self-energy
where $\omega = q^2 / (2M_N)$. In Fig.\ \ref{ddopr} we show with dashed-dotted lines the
second-order contributions from two- and three-body forces (taken together). Below the Fermi
surface, the momentum dependence of the real part of the second-order self-energy is nearly
opposite to that of the first-order Hartree-Fock contributions. The result is an effective nucleon
mass at the Fermi surface that is very close to the free-space mass.
At positive energies (with respect to the Fermi energy) the imaginary part of the optical potential
is negative and absorptive. In the vicinity of the Fermi surface, the imaginary part vanishes
quadratically close to the Fermi surface, in agreement with Luttinger's theorem
\cite{luttinger61}. Moreover, the imaginary part is approximately inversion-symmetric about the
Fermi momentum, $W(q,k_f) \simeq -W(2k_f-q,k_f)$. Although not shown explicitly here, as
discussed in Ref.\ \cite{holt13b} the second-order effects from three-body forces are rather small
for both the real and imaginary parts, resulting in additional attraction at both low and
high momenta.

Solving the self-consistent Dyson equation for the momentum as a function of the energy
\cite{jeukenne77,negele81,fantoni81}:
\begin{equation}
E(q) = \frac{q^2}{2M_N} + {\rm Re}\, \Sigma(q,E(q);k_f)
\end{equation}
we obtain the real and imaginary parts of the optical potential relevant for comparisons
with phenomenology:
\begin{eqnarray}
U(E, \rho) &=& {\rm Re}\, \Sigma(q(E),E; k_f) \\
W(E, \rho) &=& \left (1+\frac{M_N}{q}\frac{\partial U}{\partial q}\right )^{-1} {\rm Im} \, \Sigma(q(E),E; k_f).\nonumber
\label{sce}
\end{eqnarray}
Self-consistency reduces the second-order contributions due to the larger energy denominators.
For convenience we employ the effective mass plus energy shift parameterization
$\epsilon_q = \frac{q^2}{2M^*} + U$, where $U$ is momentum independent, in computing the
second-order terms in Eq.\ (\ref{se1}). The resulting real and imaginary optical potentials
are shown in the left and right panels of Fig.\ \ref{ddop3} for several nuclear densities as a
function of the energy $E$. We see that the well depth of the real potential for a scattering
state at zero incident energy is approximately $50$\,MeV, in good agreement with the
phenomenological depth of $50-55$\,MeV \cite{koning03}, however, the energy dependence
is slightly weaker than in phenomenological models.
The absorptive strength of the imaginary potential at saturation density for an intermediate
scattering energy of $E \simeq 100$\,MeV is approximately
$18$\,MeV, which indeed is large compared to the empirical value, $|W| \simeq 10-12$\,MeV
\cite{koning03}. This rather large imaginary optical potential is a feature common to many
microscopic calculations. Second-order one-pion exchange alone gives rise to quantitatively
similar results (see Fig.\ 6 in Ref.\ \cite{fkw2}). In finite nuclei the absorptive strength
$|W|$ is reduced due to the characteristic gap in the single-particle energy spectrum around
the Fermi energy. The fact that nuclear matter does not feature such a gap increases the
phase space open for absorptive processes, leading to an overestimate of $|W|$.

\subsection{Finite nuclei from density functional methods}
\label{dfm}

In this section we discuss how chiral low-momentum interactions can be employed
to describe the properties of finite nuclei \cite{dmeimprov,microefun,stoitsov2010,
platter,efun,isofun} in density functional theory, which is the many-body method of
choice to compute the masses and certain excited states of medium-mass and
heavy nuclei \cite{reinhard,stone}. Both non-relativistic Skyrme functionals \cite{sly,pearson}
and relativistic mean-field models \cite{serot,ring} have been highly successful in
self-consistent mean-field calculations of wide-ranging nuclear properties. An alternative
and complementary approach \cite{lesinski,drut,platter} is to constrain the analytical form
of the functional as well as the values of its couplings from microscopic many-body perturbation
theory with realistic two- and three-nucleon forces. Within this framework, contributions to the
energy are given in terms of density-matrices convoluted with finite-range interaction kernels.
The resulting expression for the energy is then non-local, and in order for mean-field
calculations to be numerically feasible in heavy open-shell nuclei, approximations must be
developed in which only local densities and currents enter.

The density-matrix expansion provides a means to remove the non-local character of the
exchange contribution to the energy by expanding it in the form of a generalized Skyrme functional
with density-dependent couplings. For many years the density-matrix expansion of Negele and
Vautherin \cite{negele} has been explored for this purpose, but more recently an improved
density matrix expansion has been suggested \cite{dmeimprov} that improves the description
of spin-unsaturated nuclei. In fact, the phase-space averaging techniques in Ref.\
\cite{dmeimprov} were shown to provide a consistent expansion of both the spin-independent
part as well as the spin-dependent part of the density-matrix when employed with schematic
finite-range central, spin-orbit, and tensor interactions. First attempts \cite{microefun} at employing
chiral nuclear forces in the improved density matrix expansion were carried out at order N$^2$LO
in the chiral NN potential. There it was suggested that the density-dependent couplings associated
with one- and two-pion-exchange processes should be added to a standard Skyrme functional with
re-adjusted parameters. Later it was found \cite{stoitsov} that such an energy density functional
exhibits a small yet systematic reduction of the $\chi^2$ deviation in comparison to traditional Skyrme
functionals.

In this section we focus on recent calculations \cite{efun,isofun} of the nuclear energy density
functional with improved chiral two- and three-nucleon interactions. For the two-body interaction
the N$^3$LO chiral potential with momentum-space cutoff of $414$\,MeV is used
\cite{machleidtreview,n3low}. In addition we employ the N$^2$LO chiral three-nucleon force \cite{coraggio14}
whose short-range contributions, proportional to the low-energy constants $c_D$ and $c_E$, are
fit to the triton binding energy and lifetime. In addition to the energy per particle of infinite
homogeneous nuclear matter, the energy density functional includes as well strength
functions associated with the $(\vec \nabla \rho)^2$ surface term and the spin-orbit coupling.
In phenomenological Skyrme parameterizations, these strength functions are treated as
constants, whereas a microscopic approach including the finite-range character of the
pion-exchange terms naturally gives rise to density-dependent coupling strengths.

\subsubsection{\it Density-matrix expansion}

The construction of a microscopic nuclear energy density functional starts from
the density-matrix
\be
\rho(\vec r_1 \sigma_1 \tau_1; \vec r_2 \sigma_2 \tau_2) = \sum_\alpha \Psi_\alpha^*(\vec r_2 \sigma_2 \tau_2)
\Psi_\alpha(\vec r_1 \sigma_1 \tau_1)
\ee
given as a sum over the energy eigenfunctions $\Psi_\alpha$ associated with occupied
orbitals of the non-relativistic many-body Fermi system. It can be expanded \cite{dmeimprov}
in center-of-mass $\vec r$ and relative coordinates $\vec a$ as follows
\begin{eqnarray}
\sum_{\alpha}\Psi_\alpha( \vec r -\vec a/2)\Psi_\alpha^
\dagger(\vec r +\vec a/2) &=& {3 \rho\over a k_f}\, j_1(a k_f)-{a \over 2k_f}
\,j_1(a k_f) \bigg[ \tau - {3\over 5} \rho k_f^2 - {1\over 4} \vec \nabla^2
\rho \bigg] \nonumber \\ && + {3i \over 2a k_f} \,j_1(a k_f)\, \vec \sigma
\cdot (\vec a \times \vec J\,) + \dots\,,
\label{dme}
\end{eqnarray}
where $j_1(x) = (\sin x - x \cos x)/x^2$ is the spherical Bessel function. The terms on the
right-hand side of Eq.\ (\ref{dme}) are the nucleon density $\rho(\vec r\,)
=2k_f^3(\vec r\,)/3\pi^2 =  \sum_\alpha \Psi^\dagger_\alpha(\vec r\,) \Psi_\alpha(\vec r\,)$
with $k_f(\vec r\,)$ the local Fermi momentum,
the kinetic energy density $\tau(\vec r\,) =  \sum_\alpha \vec \nabla
\Psi^\dagger_\alpha (\vec r\,) \cdot \vec \nabla \Psi_\alpha(\vec r\,)$, and the
spin-orbit density $ \vec J(\vec r\,) = i \sum_\alpha \vec \Psi^\dagger_\alpha(\vec r\,)
\vec \sigma \times \vec \nabla \Psi_\alpha(\vec r\,)$.

Nuclear interactions derived within chiral effective field theory are naturally given in
momentum space. The Fourier transform (with respect to $\vec a$ and $\vec r$) of
the expanded density-matrix in Eq.\ (\ref{dme}) therefore provides the appropriate tool
for an efficient calculation of the nuclear energy density functional. This Fourier transform
has the structure
\begin{eqnarray}
\Gamma(\vec p,\vec q\,)& =& \int d^3 r \, e^{-i \vec q \cdot
\vec r}\,\bigg\{ \theta(k_f-|\vec p\,|) +{\pi^2 \over 4k_f^4}\Big[k_f\,\delta'
(k_f-|\vec p\,|)-2 \delta(k_f-|\vec p\,|)\Big] \nonumber \\ && \times \bigg(
\tau - {3\over 5} \rho k_f^2 - {1\over 4} \vec \nabla^2 \rho \bigg) -{3\pi^2
\over 4k_f^4}\,\delta(k_f-|\vec p\,|) \, \vec \sigma \cdot (\vec p \times
\vec J\,)  \bigg\}\,,
\label{gamma}
\end{eqnarray}
and generalizes the concept of ``medium-insertion'' to inhomogeneous many-body systems
characterized by the time-reversal-even densities $\rho(\vec r\,)$,  $\tau(\vec r\,)$ and
$ \vec J(\vec r\,)$. In this sense $\Gamma(\vec p,\vec q\,)$ extends the
step-function-like momentum distribution $\theta(k_f-|\vec p\,|)$, appropriate for infinite nuclear matter,
to inhomogeneous many-body systems. Note that the
delta-function $\delta(k_f-|\vec p\,|)$ occurring in Eq.\ (\ref{gamma}) gives weight to the
nucleon-nucleon interaction in the vicinity of the local
Fermi momentum $|\vec p\,|=k_f(\vec r\,)$ only.

Up to second order in spatial gradients the
energy density functional relevant for $N=Z$ even-even nuclei has the form
\begin{eqnarray}
{\cal E}[\rho,\tau,\vec J\,] &=& \rho\,\bar E(\rho)+\bigg[\tau-
{3\over 5} \rho k_f^2\bigg] \bigg[{1\over 2M_N}-{k_f^2 \over 4M_N^3}+F_\tau(\rho)
\bigg] \nonumber \\ && + (\vec \nabla \rho)^2\, F_\nabla(\rho)+  \vec \nabla
\rho \cdot\vec J\, F_{so}(\rho)+ \vec J\,^2 \, F_J(\rho)\, .
\label{edf}
\end{eqnarray}
In this equation $\bar E(\rho)$ is the energy per particle of symmetric nuclear
matter at the local nucleon density. The strength function $F_\tau(\rho)$ is related to the
nuclear single-particle potential $U(p,k_f)$ by
\begin{equation}
F_\tau(\rho) = {1 \over 2k_f} {\partial U(p,k_f) \over \partial p}
\Big|_{p=k_f} = -{k_f \over 3\pi^2} f_1(k_f)\,
\label{edfmf}
\end{equation}
and can be associated with an effective density-dependent nucleon mass $M_N^*(\rho)$.
The rightmost term in Eq.\ (\ref{edfmf}) relates the effective mass to the
spin- and isospin-independent $p$-wave Landau parameter $f_1(k_f)$ of Section
\ref{flsnm}. The improved density-matrix expansion therefore leads to the same
expression for the effective nucleon mass as established in Fermi liquid theory for
quasiparticles on the Fermi surface. In contrast, for the original density matrix expansion
\cite{negele} this close relationship does not hold generally, that is $U(p,k_f)$
deviates from a simple quadratic $p$-dependence.

The strength function $F_\nabla(\rho)$ is associated with the energy due to surface gradients
and can be written as the sum of two terms:
\begin{equation}
F_\nabla(\rho) = {1\over 4}\, {\partial F_\tau(\rho) \over  \partial \rho} +F_d(\rho) \,,
\end{equation}
where $F_d(\rho)$ contains all contributions for which the $(\vec \nabla \rho
)^2$-factor originates from the momentum dependence of the interaction. Only the nuclear
matter component $\theta(k_f-|\vec p\,|)$ of the density-matrix expansion enters
into the calculation of the strength function $F_d(\rho)$. The last two terms in Eq.\ (\ref{edf})
represent the spin-orbit interaction and quadratic spin-orbit interaction in nuclei. The latter
gives rise to an additional spin-orbit single-particle potential proportional to $\vec J$, whereas
the regular spin-orbit potential is proportional to the density-gradient $\vec \nabla \rho$.

\subsubsection{\it Two- and three-body contributions}

We now discuss the two- and three-body contributions to the density-dependent
strength functions entering the nuclear energy density functional ${\cal E}[
\rho,\tau,\vec J\,]$. We employ the chiral NN potential \cite{machleidtreview,n3low} with a
sharp cutoff at the scale $\Lambda= 414\,$MeV. This value of the cutoff coincides with the
resolution scale below which NN potentials become nearly model independent and exhibit
improved convergence properties in many-body perturbation theory. We note that in the
Hartree-Fock approximation it is not necessary to include the effects of the regulating function,
since the nuclear interactions are probed only at relatively small momenta
$|\vec p_{1,2}|\leq k_f\leq 285\,$MeV. The regulator function together with its associated cutoff
scale $\Lambda$ become relevant only at high orders in perturbation theory. The finite-range
part of the N$^3$LO chiral NN potential arising from one- and two-pion exchange processes
has the general form
\begin{eqnarray}
V_{NN}^{(\pi)} &=& V_C(q) + \vec \tau_1 \cdot \vec \tau_2\, W_C(q) +
\big[V_S(q) + \vec \tau_1 \cdot \vec \tau_2\, W_S(q)\big] \, \vec \sigma_1 \cdot
\vec\sigma_2 +\big[V_T(q) + \vec \tau_1 \cdot \vec\tau_2\,
W_T(q) \big]\,\vec \sigma_1 \cdot \vec q \,\, \vec\sigma_2 \cdot \vec q
\nonumber \\ && +\big[V_{SO}(q) + \vec \tau_1 \cdot \vec\tau_2\,W_{SO}(q)\big]\,
i (\vec \sigma_1+\vec\sigma_2)\cdot (\vec q \times \vec p\,) \,.
\end{eqnarray}
A special feature is that all scalar functions $V_C,\dots, W_{SO}$ depend on the momentum transfer
$q$ alone. Moreover, quadratic spin-orbit components proportional to $\vec \sigma_1\cdot
(\vec q \times \vec p\,)\,\vec \sigma_2\cdot (\vec q \times \vec p\,)$ are absent in pion-exchange
processes up to N$^3$LO.
In the Hartree-Fock approximation the pion-exchange terms lead to the following
two-body contributions to the energy density functional:
\begin{equation}
\bar E(\rho) = {\rho\over 2} V_C(0) - {3\rho\over 2} \int_0^1
\!\!dx\, x^2(1-x)^2(2+x) \Big[V_C(q)+3W_C(q) +3V_S(q)+9W_S(q) +q^2V_T(q) +3q^2W_T(q)\Big] \,,
\end{equation}
\begin{equation} F_\tau(\rho) = {k_f\over 2\pi^2}\int_0^1\!\!dx(x-2x^3)\Big[
V_C(q)+3W_C(q) +3V_S(q)+9W_S(q) +q^2V_T(q) +3q^2W_T(q)\Big] \,, \end{equation}
\begin{equation} F_d(\rho) = {1\over 4} V_C''(0)\,,  \label{fdeq}
\end{equation}
\begin{equation} F_{so}(\rho) = {1\over 2}V_{SO}(0)+ \int_0^1\!\!dx\,x^3\Big[
V_{SO}(q)+3W_{SO}(q)\Big] \,, \end{equation}
\begin{equation} F_J(\rho) = {3\over 8k_f^2} \int_0^1\!\!dx\Big\{(2x^3-x)\Big[
V_C(q)+3W_C(q) -V_S(q)-3W_S(q)\Big] -x^3 q^2\Big[V_T(q) +3W_T(q)\Big]\Big\}
\,, \end{equation}
where $q = 2x k_f$ and in Eq.\ (\ref{fdeq}) the double-prime denotes
the second derivative with respect to $q$. These terms are obtained from the density
matrix expansion employing the product of two medium-insertions $\Gamma(\vec p_1,\vec q\,)\,
\Gamma(\vec p_2, -\vec q\,)$). In addition to the finite-range parts of the NN potential arising
from explicit pion exchange, the two-body zero-range contact forces contribute to the
density-dependent strength functions in the energy density functional as follows:
\begin{equation}
\bar E(\rho) = {3\rho\over 8}(C_S-C_T)+{3 \rho k_f^2 \over 20}
(C_2-C_1-3C_3-C_6)+{9\rho k_f^4\over 140}(D_2-4D_1-12D_5-4D_{11})\,,
\end{equation}
\begin{equation}
F_\tau(\rho) = {\rho \over 4} (C_2-C_1-3C_3-C_6)+{\rho k_f^2\over
 4}(D_2-4D_1-12D_5-4D_{11})\,,
\end{equation}
\begin{equation}
F_d(\rho) = {1\over 32} (16C_1-C_2-3C_4-C_7)+{k_f^2\over 48}
(9D_3+6D_4-9D_7-6D_8-3D_{12}-3D_{13}-2D_{15})\,,
\end{equation}
\begin{equation}
F_{so}(\rho) = {3\over 8} C_5+{k_f^2\over 6}(2D_9+D_{10})\,,
\end{equation}
\begin{equation}
F_J(\rho) = {1\over 16} (2C_1-2C_3-2C_4-4C_6+C_7)+{k_f^2\over 32}
(16D_1-16D_5-4D_6-24D_{11}+D_{14})\,.
\end{equation}

\begin{figure}
\begin{center}
\includegraphics[scale=0.90,clip]{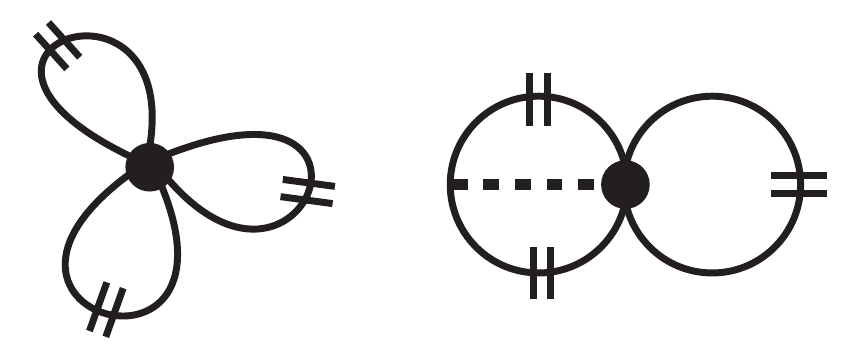}\hspace{.1in}
\includegraphics[scale=0.90,clip]{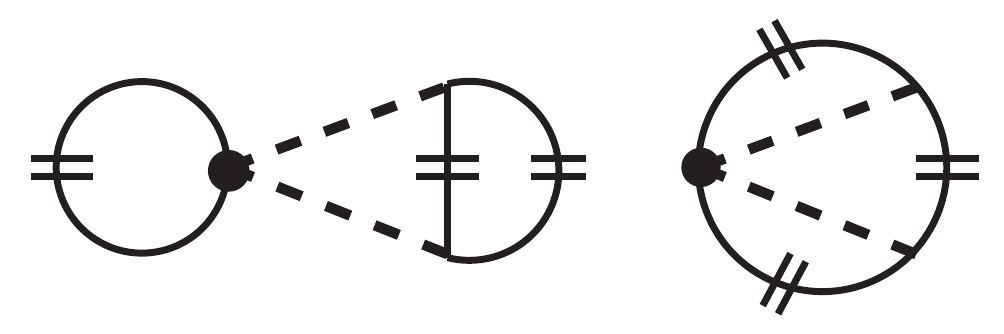}
\end{center}
\vspace{-.6cm}
\caption{Three-body diagrams related to the contact, $1\pi$-exchange,
and $2\pi$-exchange parts of the chiral three-nucleon force. The short double-line
represents the medium insertion $\Gamma(\vec p, \vec q\,)$ for inhomogeneous nuclear matter.}
\label{efunfig1}
\end{figure}

Finally, we consider the three-body contributions to the nuclear energy density
functional. The leading N$^2$LO chiral three-nucleon force consists of the well known contact,
$1\pi$-exchange and $2\pi$-exchange components. To simplify
calculation of the three-body correlations in inhomogeneous nuclear systems, one can assume
\cite{efun} that the relevant product of density-matrices is written in the factorized form
$\Gamma(\vec p_1,\vec q_1)\,\Gamma(\vec p_2,\vec q_2)\,
\Gamma(\vec p_3,-\vec q_1-\vec q_2)$. This ansatz respects the infinite nuclear matter limit,
but it involves approximations compared to more detailed treatments explored in Ref.\
\cite{platter}. Diagrammatic contributions from the chiral three-nucleon force are shown in
Fig.\ \ref{efunfig1} and explicit expressions can be found in Ref.\ \cite{efun}.
The three-body contact-term proportional to $c_E$ (see leftmost diagram in Fig.\ \ref{efunfig1})
gives rise to a contribution to the energy per particle that is quadratic in the density. Since
the contact term is momentum independent, it gives vanishing contributions to all other strength
functions. The $1\pi$-exchange component of the chiral three-nucleon interaction involves the
low-energy constant $c_D$, and with three inhomogeneous medium insertions (see the
second diagram in Fig.\ \ref{efunfig1}) one obtains contributions to all strength functions except
the spin-orbit \cite{efun}.

In the third diagram of Fig.\ \ref{efunfig1}, we show the three-body Hartree contribution arising
from the $2\pi$-exchange term. All terms in the energy density functional Eq.\ (\ref{edf}) are
generated, including a strong spin-orbit interaction \cite{efun} that is the dominant
part of the three-body contribution to $F_{so}(\rho)$. This was originally suggested by Fujita
and Miyazawa \cite{fujita} in the context of intermediate-state $\Delta(1232)$ excitation.
Although the two-pion exchange three-body force has components proportional to
$c_1$, $c_3$ and $c_4$, only the isoscalar $c_1$ and $c_3$ parts give nonvanishing Hartree terms.
Finally, we consider the three-body $2\pi$-exchange Fock contribution, shown as the fourth diagram
in Fig.\ \ref{efunfig1}. This term generates for symmetric nuclear matter non-vanishing
contributions from both the isoscalar and isovector $\pi \pi N N$
contact vertices, the latter proportional to the low-energy constant $c_4$.
The various contributions to the energy density strength functions involve Fermi sphere integrals
that cannot be evaluated in closed analytical form. The rather lengthy expressions can be found in
Ref.\ \cite{efun}.

\begin{figure}
\begin{center}
\includegraphics[scale=0.36,angle=270]{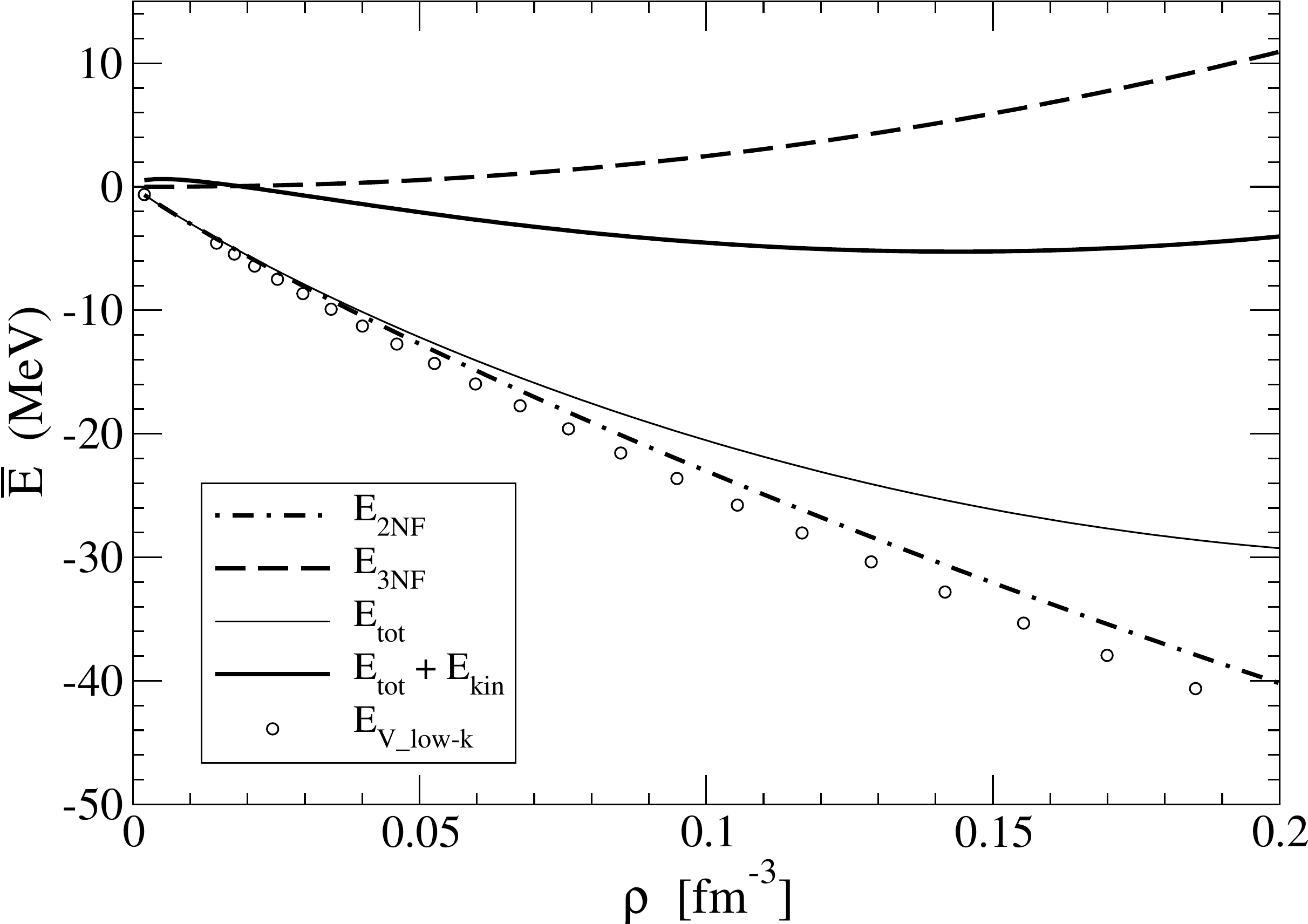}
\end{center}
\vspace{-.4cm}
\caption{Energy per particle $\bar E(\rho)$ of symmetric nuclear matter from chiral two- and three-nucleon forces.}
\label{ebar}
\end{figure}

We now discuss the results for the nuclear energy density functional obtained in the
first-order Hartree-Fock approximation. The contributions to the energy per particle
for densities up to $\rho = 0.2\,$fm$^{-3}$ are shown in Fig.\ \ref{ebar}. The
dash-dotted line gives the attractive two-body contribution, while the dashed line
denotes the repulsive three-body contribution. We compare as well the results from the
N$^3$LO chiral nucleon-nucleon interaction (with $\Lambda = 414$\,MeV) to those from
the universal low-momentum NN potential $V_{\rm low-k}$ obtained from renormalization
group methods \cite{vlowk} (shown by the dashed--double-dotted line). We observe that the
low-momentum chiral potential N$^3$LO potential in fact fairly accurately
reproduces the results from $V_{\rm low-k}$. The full line shows the sum of the two- and
three-body contributions, which indeed exhibits saturation of nuclear matter but with too
little binding when the kinetic energy $\bar E_{\rm kin}(\rho) = 3k_f^2/10M_N-3k_f^4/56M_N^3$,
is included. This particular feature of the Hartree-Fock energy per particle with
renormalization-group evolved low-momentum interactions
has been observed in previous studies \cite{bogner05,hebeler11}. A much
improved description of the nuclear matter equation of state is achieved by
including the two- and three-body interactions to second or third \cite{coraggio14} order.

\vskip 0.5cm

\begin{figure}
\begin{center}
\includegraphics[scale=0.36,angle=270]{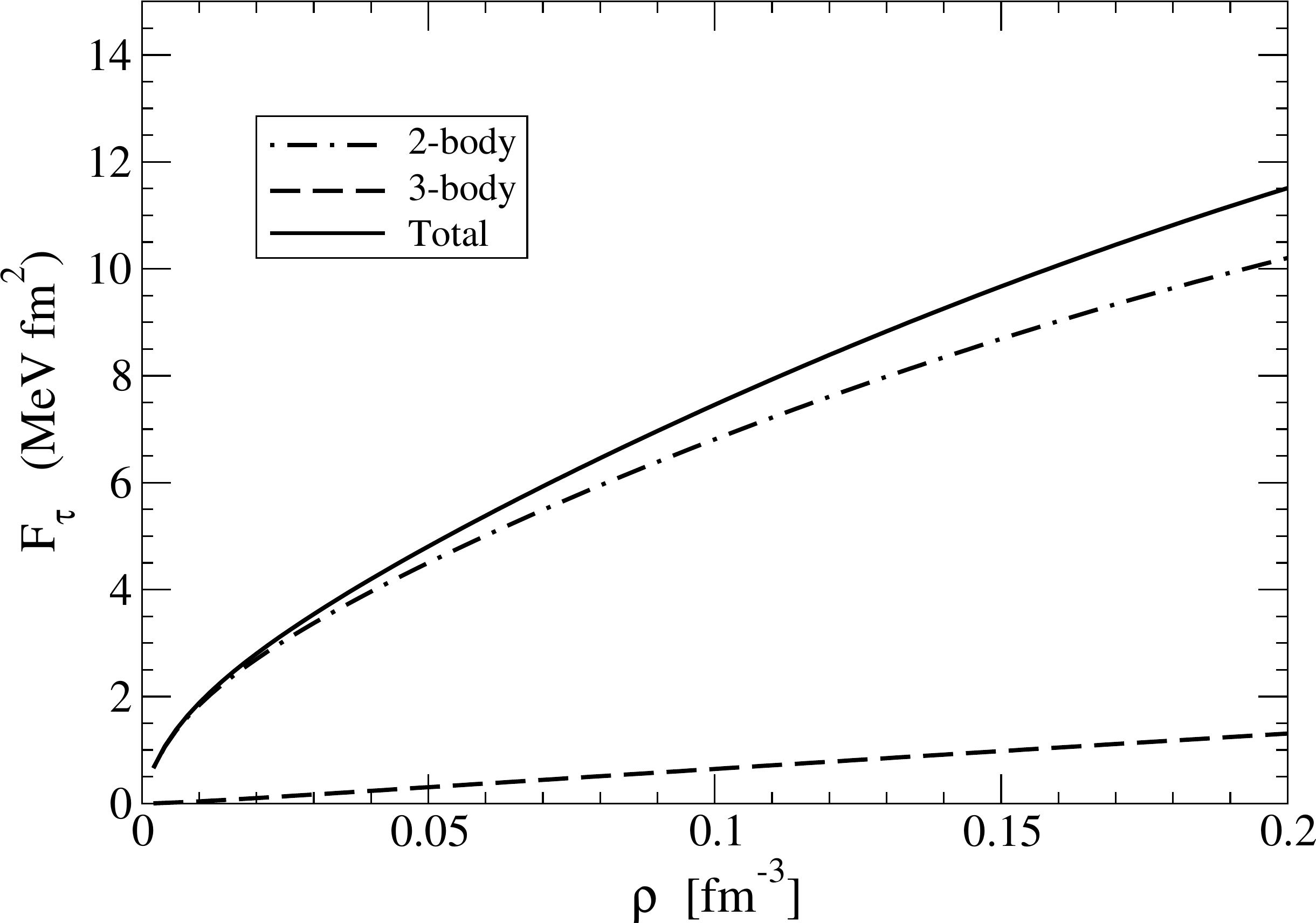}\hspace{.1in}
\includegraphics[scale=0.36,angle=270]{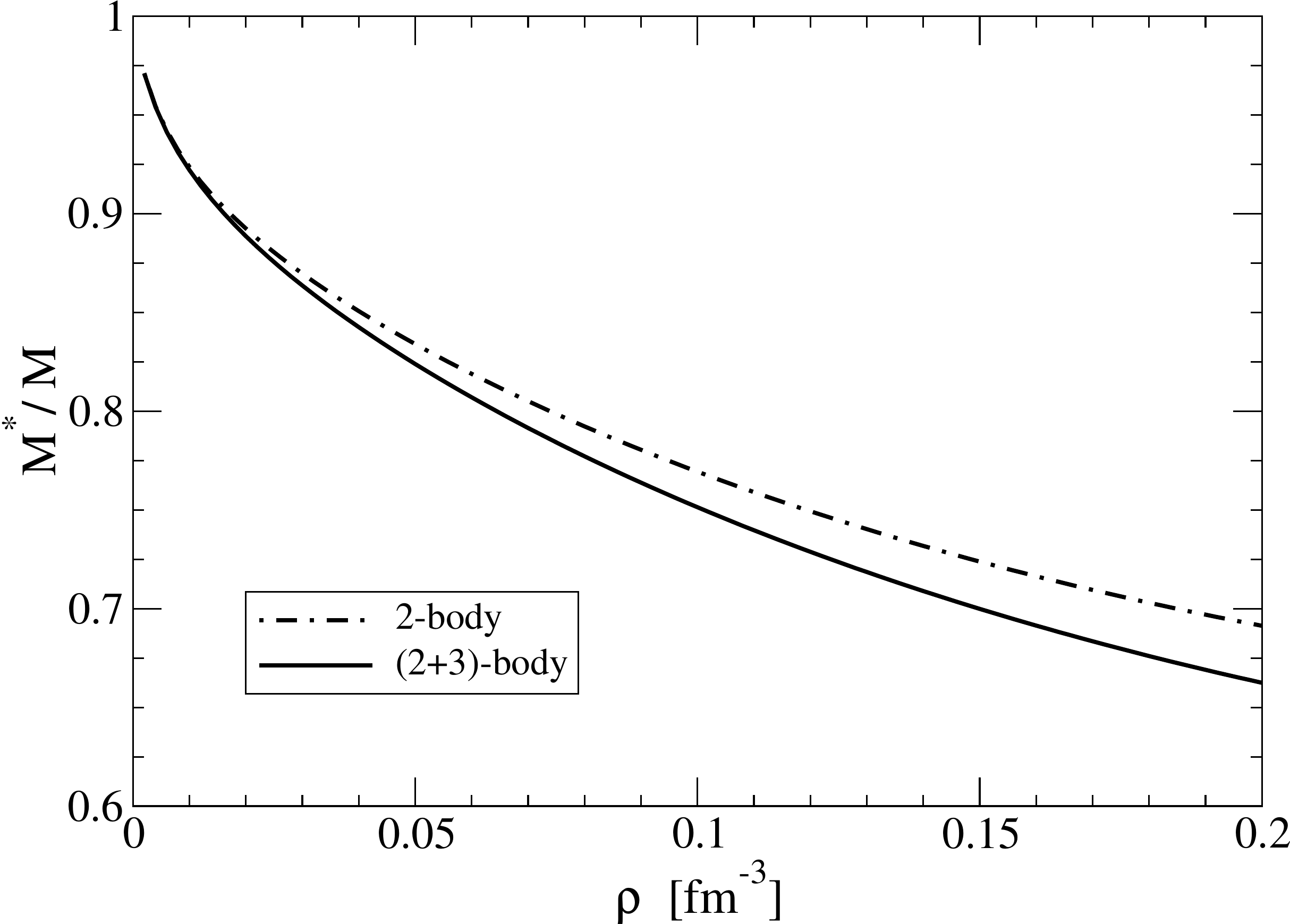}
\end{center}
\vspace{-.5cm}
\caption{Contributions to the strength function $F_\tau(\rho)$ and effective mass $M_N^*(\rho)$ as a function of the nuclear
density $\rho$.}\label{ftau}
\end{figure}


\begin{figure}
\begin{center}
\includegraphics[scale=0.36,angle=270]{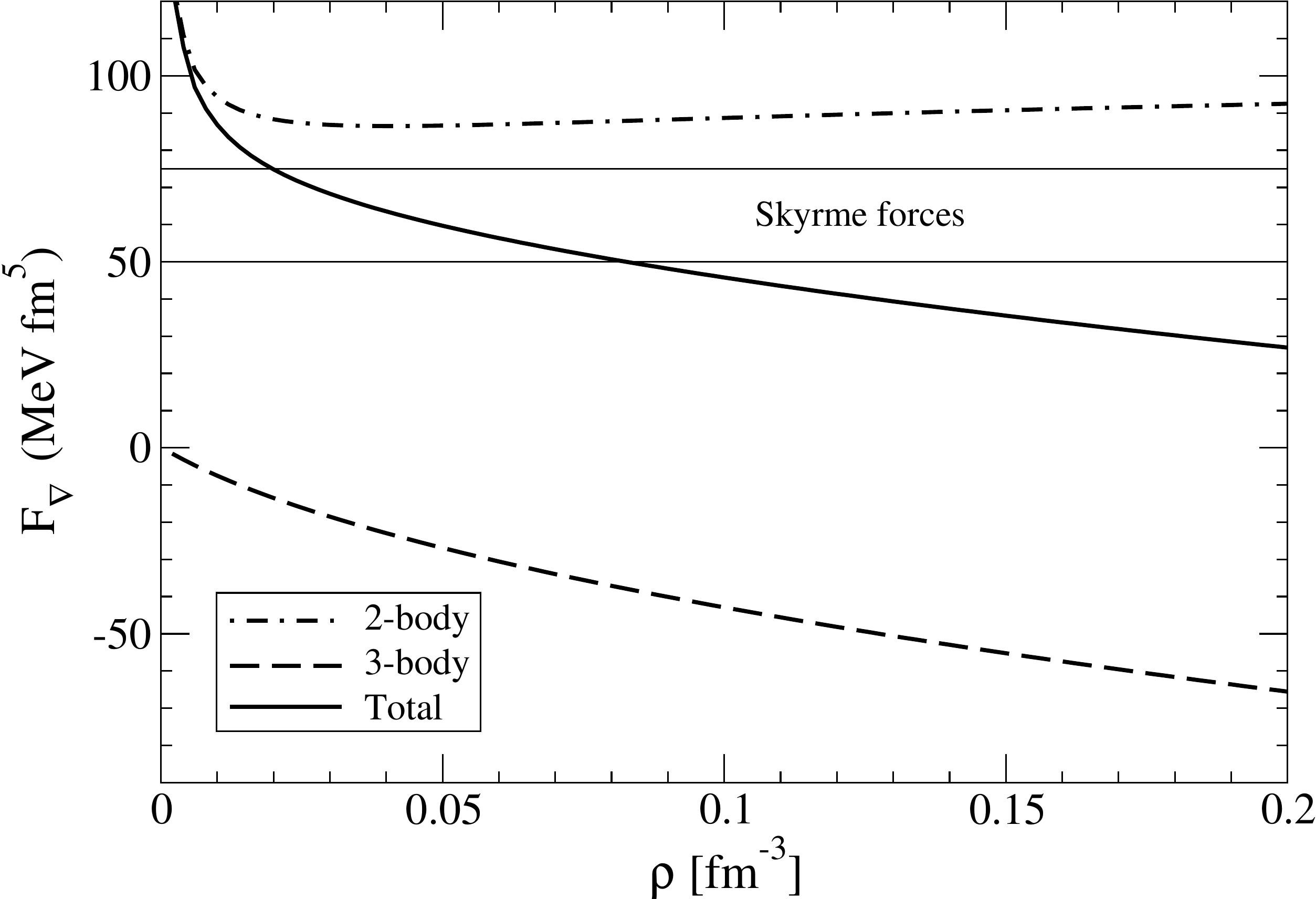}
\end{center}
\vspace{-.5cm}
\caption{Surface-gradient strength coupling as a function of the nuclear density.} 
\label{fgrad}
\end{figure}

\begin{figure}
\begin{center}
\includegraphics[scale=0.36,angle=270]{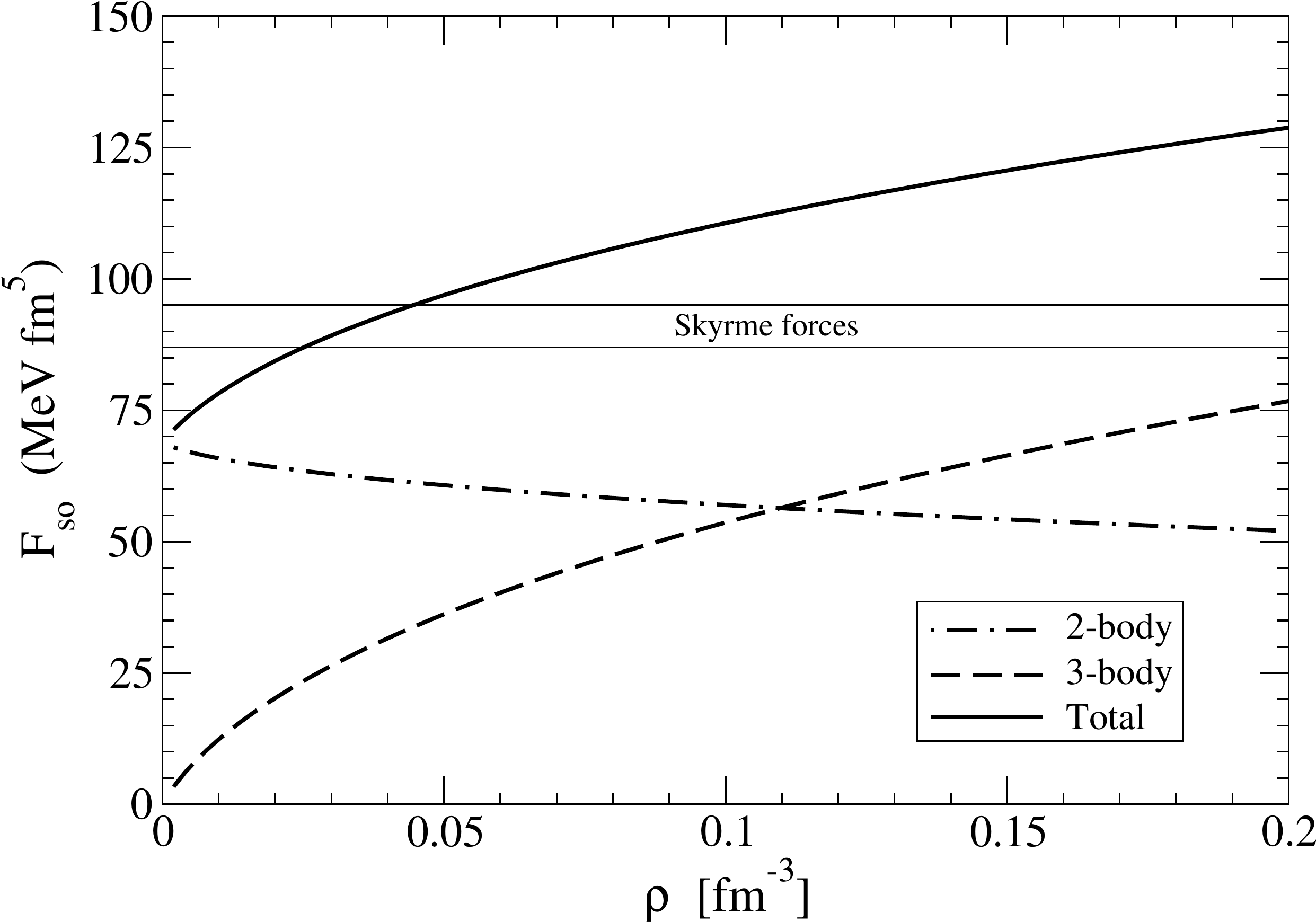}\hspace{.05in}
\includegraphics[scale=0.36,angle=270]{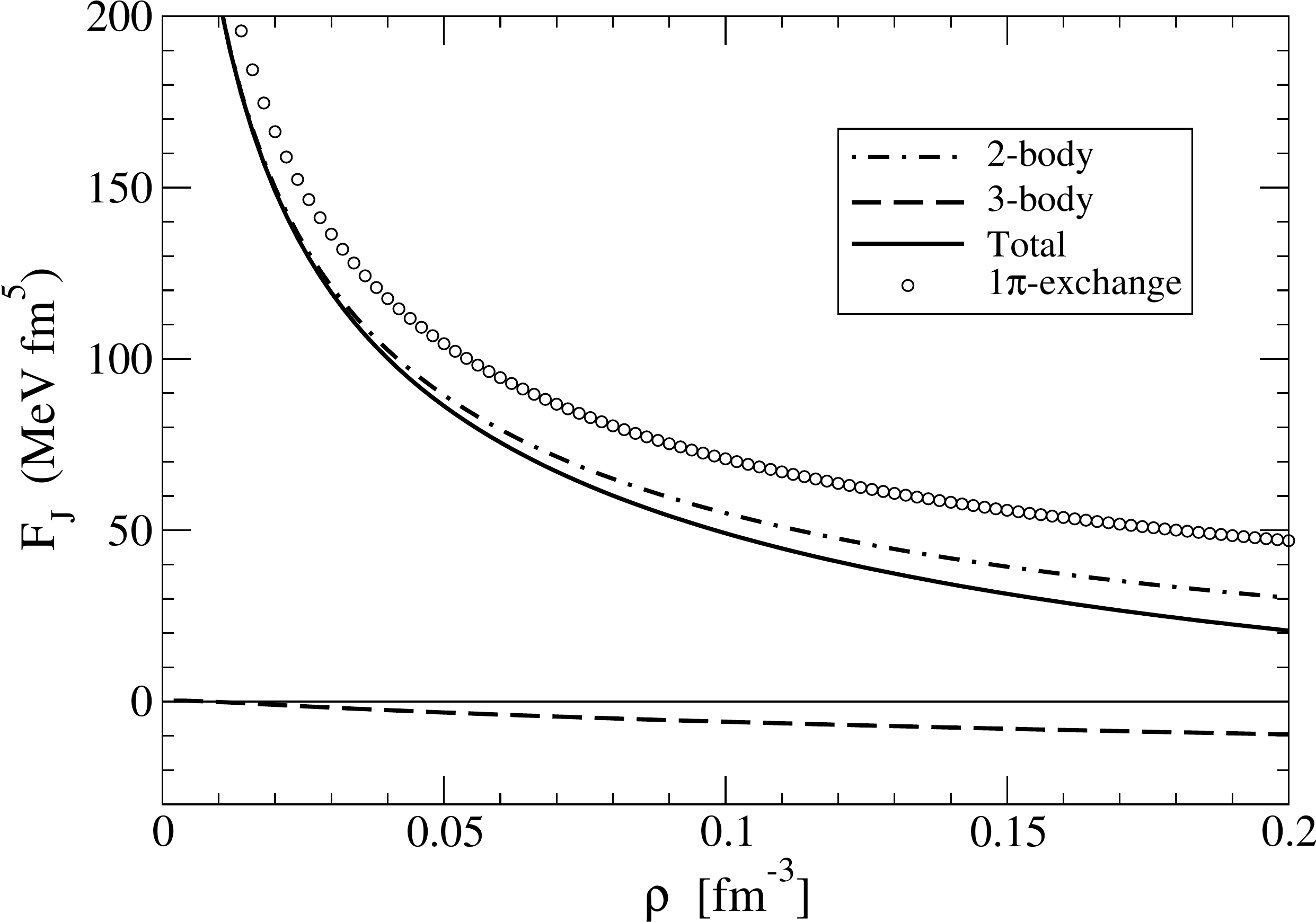}
\end{center}
\vspace{-.5cm}
\caption{Spin-orbit and quadratic spin-orbit strengths as a function of the density.}
\label{fso}
\end{figure}


In Fig.\ \ref{ftau} we show the individual contributions to the strength function $F_\tau(\rho)$
associated with the kinetic energy term in the density functional. Here also we
see that the contribution from the chiral low-momentum potential agrees fairly well with that
from $V_{\rm low-k}$. The three-body part increases nearly linearly with the density, and
comes out relatively small compared to the two-body contribution. At nuclear matter saturation
density the three-body force gives a correction of about $20\%$. As mentioned earlier, the
bracketed term that multiplies the kinetic energy density $\tau(\vec r\,)$ in the energy density
functional can be interpreted as the reciprocal of a density-dependent effective nucleon mass:
\begin{equation}
M_N^*(\rho) = M_N\bigg[ 1 -{k_f^2 \over 2M_N^2}+2M_N\, F_\tau(\rho)\bigg]^{-1}\,.
\end{equation}
The small correction term $-k_f^2/2M_N^2$ accounts for the small relativistic increase in the mass.
In Fig.\ \ref{ftau} we show explicitly the ratio $M_N^*(\rho)/M_N$ of the effective nucleon mass
to the free-space nucleon mass as a function of the density. At the Hartree-Fock level, the
effective mass is reduced significantly in comparison to the free-space
mass, and the ratio reaches the value  $M_N^*(\rho_0) \simeq 0.7 M_N$ at nuclear
matter saturation density $\rho_0=0.16\,$fm$^{-3}$. This is consistent with the range of values
$0.7 < M_N^*(\rho_0)/M_N < 1$ in phenomenological Skyrme forces
\cite{sly,pearson}. However, second-order perturbative corrections (mainly from two-body forces)
enhance the effective mass at the Fermi surface \cite{fermiliq,holt13b}.

In Fig.\ \ref{fgrad} we plot the strength function $F_\nabla(\rho)$ of the squared
surface gradient term as a function of the nuclear density. As the density decreases to
very low values, the gradient coupling strength from the two-body contribution increases
strongly. However, for higher densities the two-body contributions are nearly independent of
the density. The three-body contribution (dashed line) is sizable and strongly density dependent.
It reduces the two-body (positive) contribution such that the total value of $F_\nabla(\rho)$
decreases with increasing density. To compare with phenomenology we show
the band spanned by parameterized Skyrme forces \cite{sly,pearson}. The Hartree-Fock
result from realistic chiral two- and three-body forces is somewhat too small at densities
close to $\rho_0/2 =0.08\,$fm$^{-3}$, where the surface energy in finite nuclei gains
most of its weight. Iterated $1\pi$-exchange contributions to $F_\nabla(\rho)$ have been
considered in Ref.\ \cite{efunold} with the conclusion that a second-order treatment
will increase the values of this strength function.

Of particular interest is the spin-orbit coupling strength $F_{so}(\rho)$, and the different
contributions from two- and three-body terms shown in Fig.\ \ref{fso}. The coupling strength
from two-body forces is dominated by the short-distance dynamics, encoded in the
low-energy constant $3C_5/8$. Three-body forces strongly enhance the spin-orbit interaction,
the most significant contribution arises in the Hartree term proportional to the low-energy constant
$c_3 = -3.2\,$GeV$^{-1}$. With this particular value of $c_3$, the three-body spin-orbit strength is slightly larger
than that proposed by Fujita and Miyazawa \cite{fujita}, where the $\Delta(1232)$-excitation
mechanism would correspond to $c_3^{(\Delta)} = -g_A^2/2
\Delta \simeq -2.9\,$GeV$^{-1}$ ($\Delta = 293\,$MeV is the delta-nucleon mass splitting).
At around half saturation density, the total Hartree-Fock contribution exceeds the empirical
spin-orbit coupling strength $F_{so}^{\rm (emp)}(\rho) \simeq 90\,$MeV\,fm$^{5}$ \cite{sly,
pearson}. In Ref.\ \cite{efunold} it was suggested that the $1\pi$-exchange tensor force at
second-order generates such a spin-orbit coupling, which could reduce the strength of the
spin-orbit coupling $F_{so}(\rho)$ to a value close to the phenomenological one.

Finally we show in Fig.\ \ref{fso} the strength function $F_J(\rho)$ of the squared
spin-orbit coupling, which receives only a very small contribution from three-body forces.
The two-body contribution is strongly density-dependent, and at quite low densities
($\rho < 0.05$\,fm$^{-3}$) it reaches quite large values. As in the case of the gradient
coupling strength $F_\nabla(\rho)$, the strong density dependence in $F_J(\rho)$
originates largely from the $1\pi$-exchange contribution, which we show separately in
Fig.\ \ref{fso} by the dashed-double-dotted line. We emphasize that the $\vec J^{\,2}$
term in the energy density functional represents non-local Fock contributions from
tensor forces, etc., and therefore it is not surprising that such an outstanding $1\pi$-exchange
contribution to the strength function $F_J(\rho)$ exists.

\subsubsection{\it Matching to QCD}\label{matching2qcd}

It is perhaps appropriate to inject here a few words on the Wilsonian marching that we will elaborate on in Section \ref{hadrons-in-matter} that has to do with the ``ultraviolet (UV) completion" of the effective theory to QCD. ChEFT discussed thus far has no Wisonian matching to QCD. Without it, an essential part of Brown-Rho scaling, defined below, could be missing. It will be argued that with vector mesons suitably incorporated as in $hls$EFT, the Wilsonian matching will endow the parameters of ChEFT with certain density dependence associated with the (density-dependent) quark condensate $\la\bar{q}q\ra$, an ``intrinsic density dependence" inherited from QCD at the matching scale denoted $\Lambda_M$. Part of this scaling is included in the density-dependent ChEFT discussed above obtained by integrating out many-body forces that figure at higher chiral order, e.g., the contact three-body force, but some are not, as we will see in Section \ref{hadrons-in-matter}.






\section{Nuclear chiral thermodynamics}
\label{nct}
\subsection{Nuclear phase diagram and liquid-gas transition}

The thermodynamic description of nuclear and neutron matter arises in applications related to
heavy-ion collisions and astrophysics. Previous chapters have developed the framework of
chiral effective field theory applied to the nuclear many-body problem at zero temperature. Here
we review recent progress in extending this framework to finite temperatures. The ordering
scheme introduced in Sections \ref{inm} and \ref{dfm} for the nuclear energy density in terms
of the number of medium insertions is generalized and applied to the free energy density $F(\rho,T)$
as a function of density $\rho$ and temperature $T$. It can be written generally as the sum of
convolution integrals:
\begin{eqnarray}
\label{convolution}
\rho \, \bar{F}(\rho, T) &=& 4 \int \limits_0^\infty dp \, p \, {\cal K}_1(p) \, n(p)
+ \int \limits_0^\infty dp_1 \int \limits_0^\infty dp_2 \, {\cal K}_2(p_1,p_2) \, n(p_1) \, n(p_2)
\nonumber \\ && + \int \limits_0^\infty dp_1 \int \limits_0^\infty dp_2 \int \limits_0^\infty dp_3 \,
{\cal K}_3(p_1,p_2,p_3) \, n(p_1) \, n(p_2) \, n(p_3) + \rho \, \bar {\cal A}(\rho,T) \ ,
\end{eqnarray}
where we have introduced the notation
\begin{equation}\label{density}
	n(p) = {p \over 2\pi^2} \bigg[ 1 + \exp{\frac{p^2 /2 M_N -\tilde{\mu}}{T}} \bigg]^{-1},
\end{equation}
representing the density of nucleon states in momentum space. It is a product of the
temperature-dependent Fermi-Dirac distribution function and a kinematical prefactor $p/2\pi^2$.
We first discuss the equation of state of isospin-symmetric nuclear matter and focus on the
physics of the liquid-gas phase transition, experimentally linked \cite{gross,pocho,natowitz2}
to the plateau observed in the caloric curve of the nuclear fragments in nucleus-nucleus collisions.
The calculations are then extended to isospin-asymmetric nuclear matter, focusing on the change of
thermodynamic properties with varying proton-neutron ratio. Of particular interest is the asymmetry
free energy and its dependence on density and temperature as well as the validity of
the parabolic approximation to the free energy as a function of the isospin asymmetry
$\delta = (\rho_n-\rho_p)/ (\rho_n+\rho_p)$. When possible we compare the results from in-medium chiral
perturbation theory to recent calculations \cite{wellenhofer14,wellenhofer15} 
of the thermodynamic equation of state from
microscopic nuclear two- and three-body forces defined at the resolution scale $\Lambda = 414$\,MeV \cite{coraggio13}.

The one-body kernel $ {\cal K}_1$ in Eq.\ (\ref{convolution})
\begin{equation}
{\cal K}_1(p) = M_N +\tilde \mu- {p^2\over 3M_N}- {p^4\over 8M_N^3} \,.
\label{K1}
\end{equation}
represents the contribution of a free nucleon gas. Contributions to the two- and three-body
kernels $ {\cal K}_2$ and $ {\cal K}_3$, encoding the effects of interactions, include one-pion exchange,
iterated one-pion exchange and two-pion exchange (with virtual $\Delta$-isobar
excitation). Explicit expressions for these terms can be found in Ref.\ \cite{FKW2012}. The anomalous contribution
$\bar{\cal A}(\rho,T)$, which is not present in the zero-temperature theory, first arises from the second-order
$1\pi$-exchange Fock term and results from the smoothing of the Fermi surface at nonzero
temperature. For the temperatures discussed here, the effect of the anomalous term is small.
The effective ``chemical potential''  $\tilde{\mu}$ in the finite-temperature distribution function $n(p)$ is related
to the density by
\begin{equation}
	\rho = 4 \int\limits_0^\infty dp \, p \, n(p) \,.
\end{equation}
Once these contributions to the free energy density are given, the pressure can be computed from the
standard thermodynamic relation
\begin{equation} \label{pressure}
P(\rho, T) = \rho^2 \, \frac{\partial \bar{F}(\rho, T)}{\partial \rho} \, .
\end{equation}

\begin{figure}[tbp]
\center
\includegraphics[scale=0.71,clip]{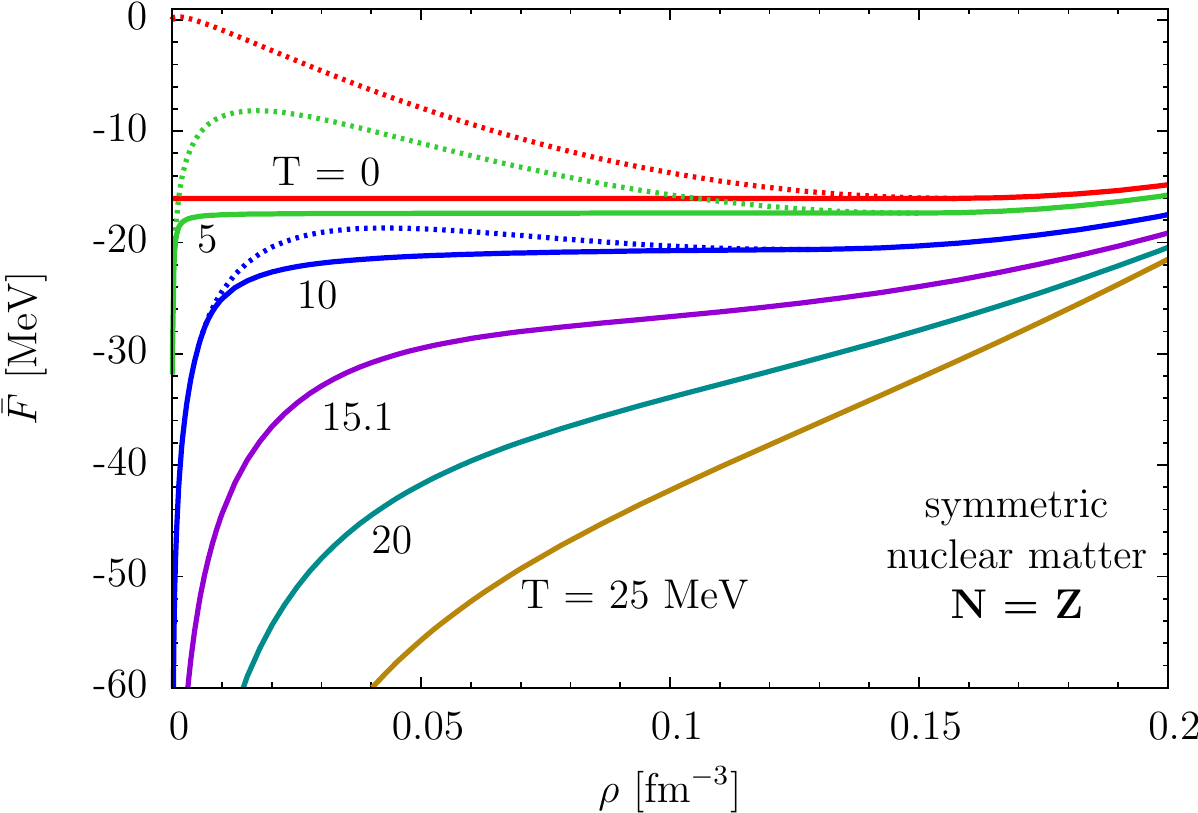} \hspace{.2in}
\includegraphics[width=3.28in]{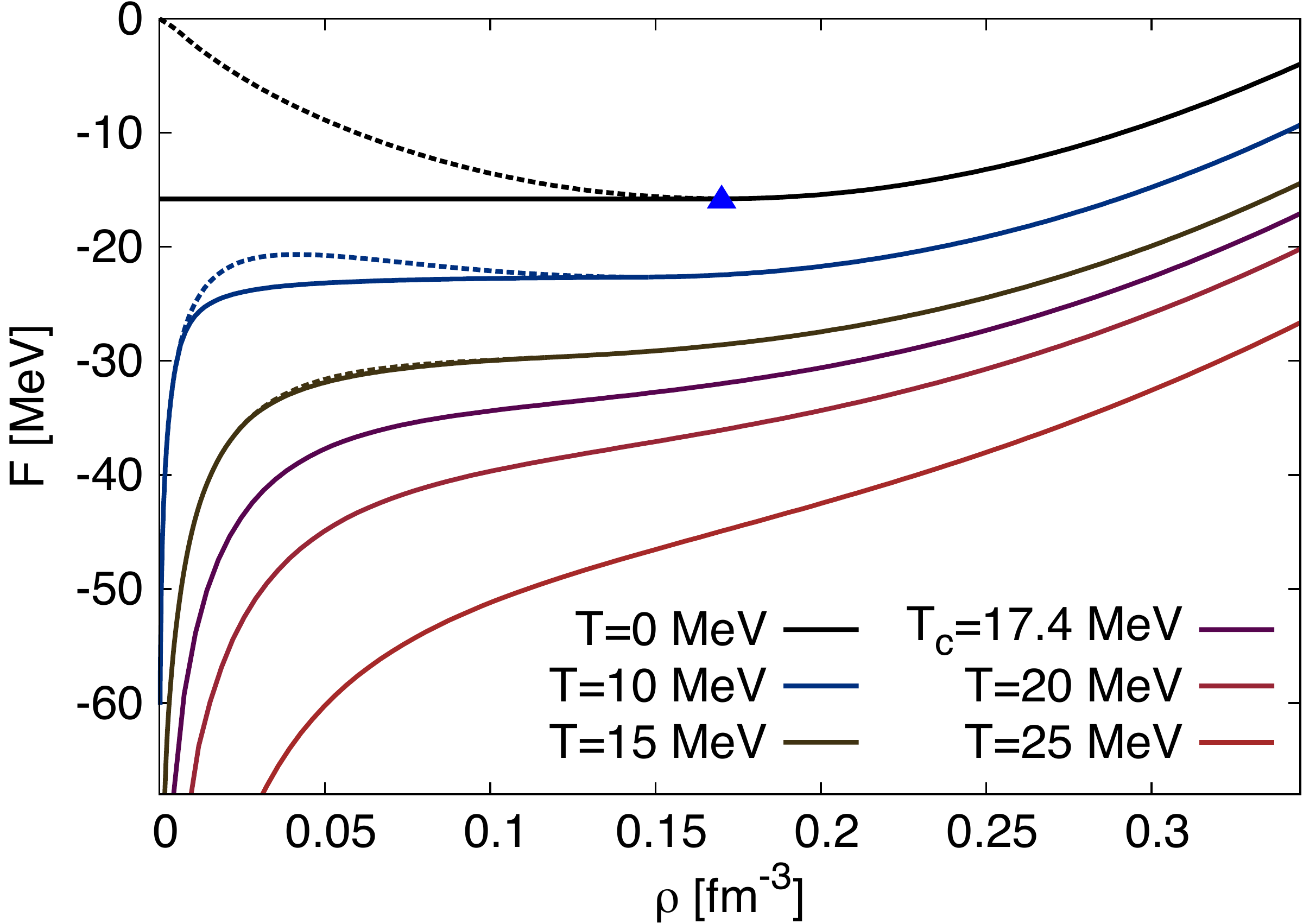}
\caption{Free energy per particle of symmetric nuclear matter as a function of the
temperature and nucleon density $\rho$. Both the physical (solid line) and nonphysical
(dotted line) free energy in the liquid-gas coexistence region are shown. Left panel: in-medium
chiral perturbation theory calculations. Right panel: calculations from high-precision two- and
three-body chiral nuclear forces \cite{wellenhofer14}.}
\label{fenergy50}
\end{figure}

In Fig.\ \ref{fenergy50} we show the free energy per particle $ \bar{F} (\rho, T)$ calculated
as a function of the density $\rho$ for a sequence of temperatures up to 25 MeV. The left panel
shows the results from chiral perturbation theory, while the right panel is based on high-precision
two- and three-body chiral nuclear forces \cite{wellenhofer14}. The
input for the interaction kernels ${\cal K}_n$ in Eq.\ (\ref{convolution}) contain contributions
up to three loop order, including the effects of intermediate-state $\Delta$ excitations.
Explicit expressions are given in Ref.\ \cite{FKW2012}. In the first-order liquid-gas transition
region, the calculated equation of state exhibits non-physical behavior, shown by the dotted
lines. The Maxwell construction is then used to obtain the physical equation of state, shown
by the solid lines. At zero temperature the free energy is equal to the internal energy of the
system with saturation point located at $ \bar{E}_0 = -16.0 $ MeV and $ \rho_0 = 0.16\,$.
As the temperature increases, the free energy develops a singularity at $ \rho = 0 $,
a feature present in numerous many-body calculations \cite{akmal,horowitz}. Qualitatively
the results from Refs.~\cite{FKW2012} and \cite{wellenhofer14} are similar, but in-medium
chiral perturbation theory gives a greater free energy per particle for most temperatures
and densities.

\begin{figure}[tbp]
\center
\includegraphics[scale=0.7,clip]{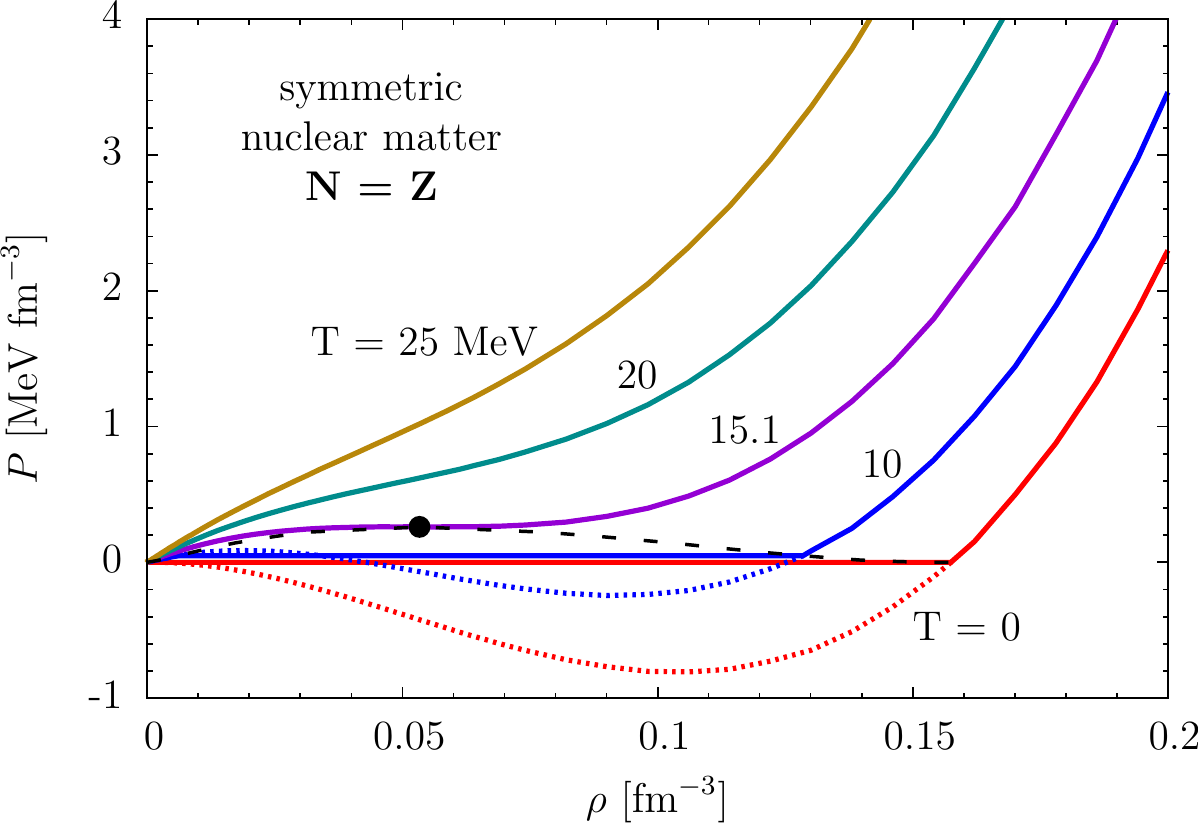}\hspace{.2in}
\includegraphics[width=2.8in]{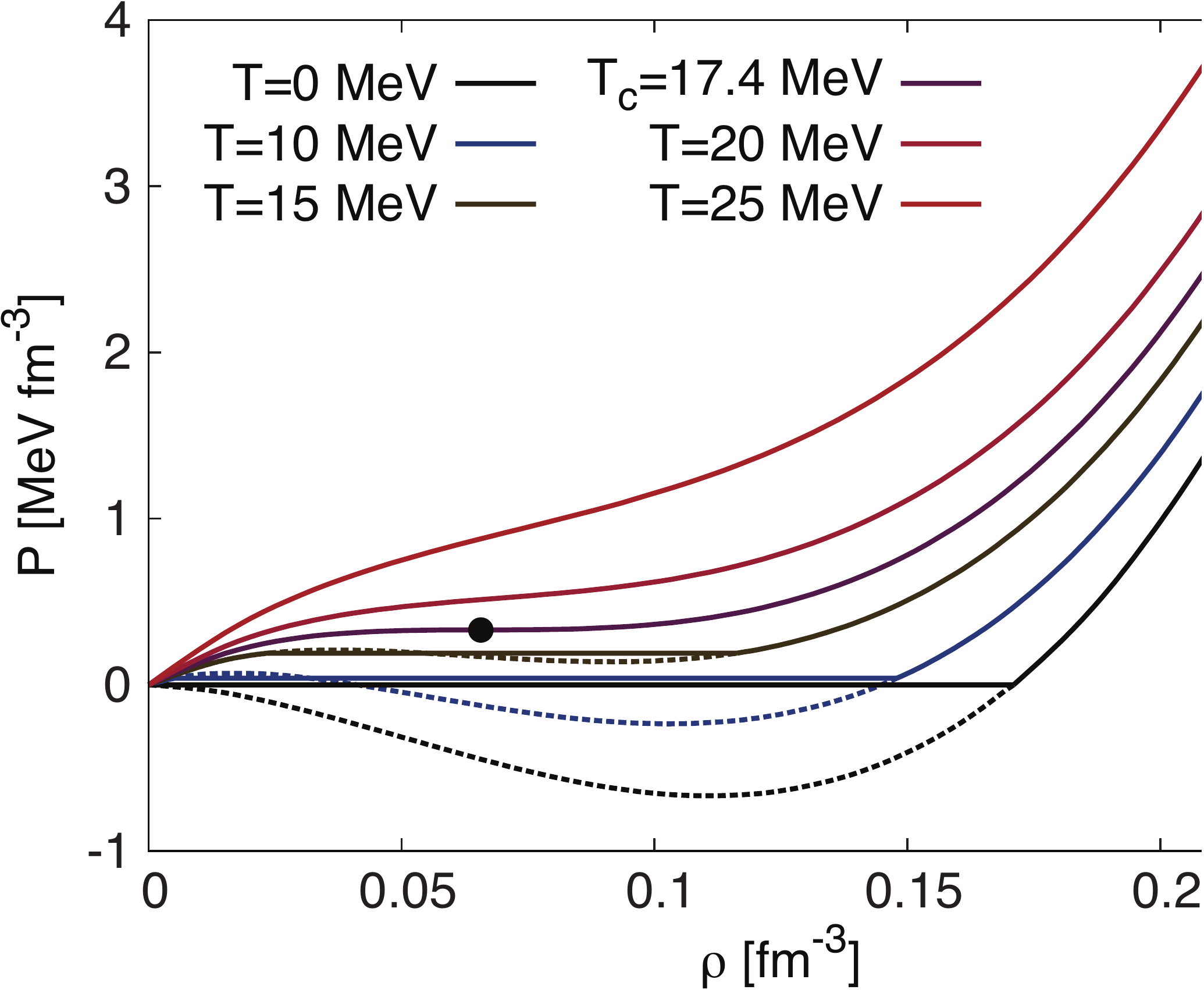}
\caption{Pressure isotherms as a function of temperature and density for symmetric nuclear
matter. The dotted lines indicate the non-physical behavior of the isotherms in the
phase transition region, while the physical pressure (represented by solid lines) is calculated
from the Maxwell construction. The boundary of the coexistence region is shown by the dashed
line, and the dot denotes the critical point at temperature $T_c \simeq 15.1$ MeV and
density $\rho_c \simeq \rho_0/3$. Left and right panels as in Fig.~\ref{fenergy50}.}
\label{eos50}
\end{figure}

\begin{figure}
\includegraphics[scale=0.72,clip]{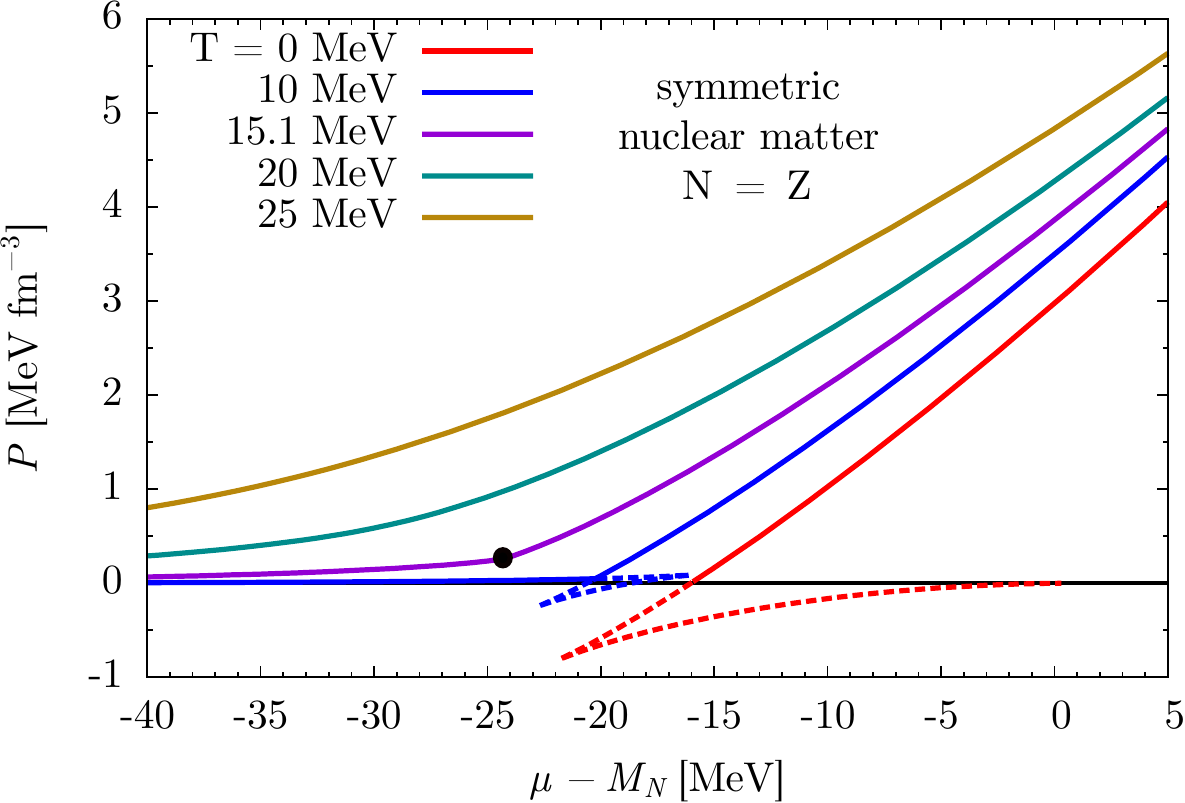}  \hspace{.2in}
\includegraphics[width=3.2in]{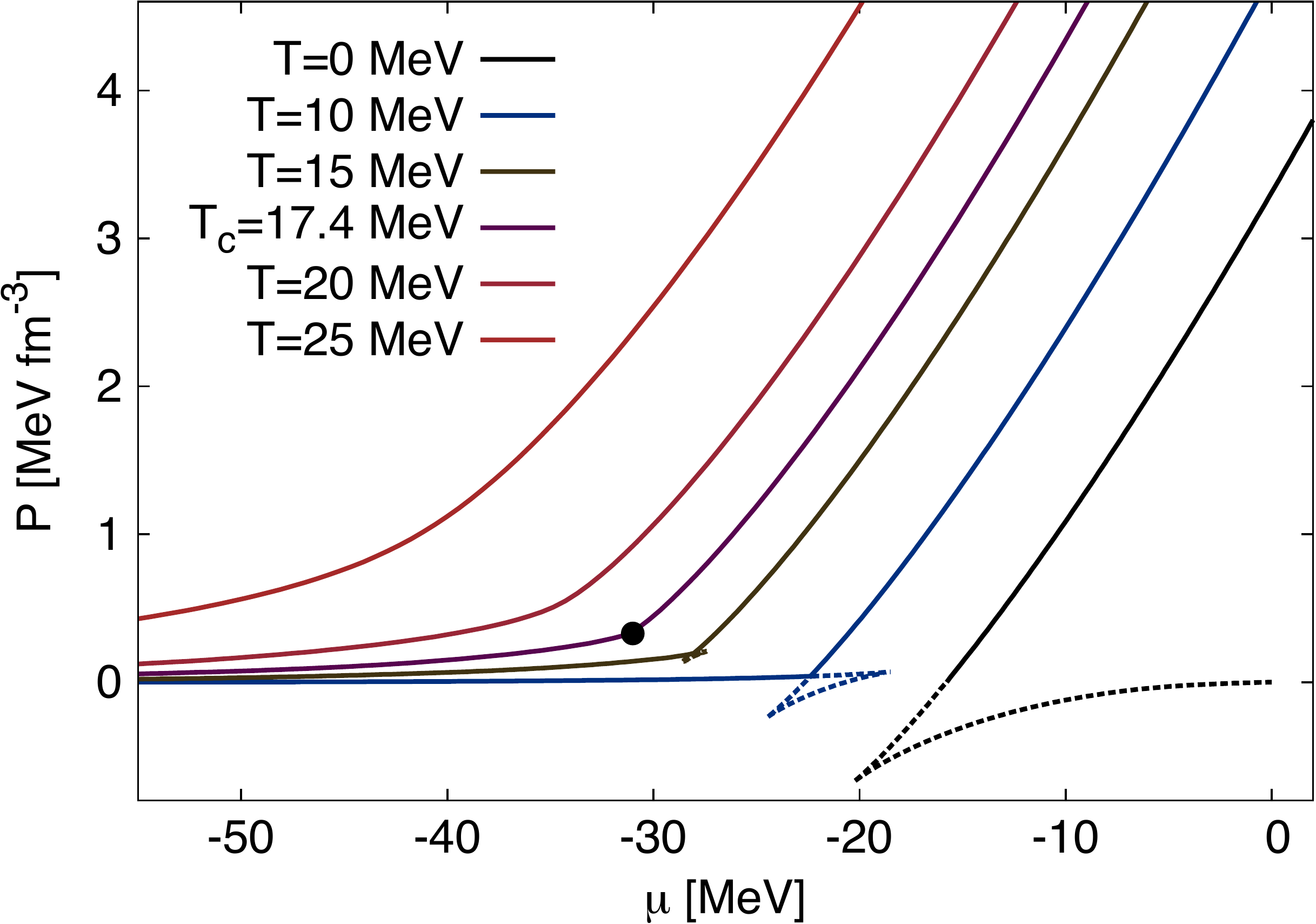}
\caption{Dependence of the pressure isotherms on the nucleon chemical potential $\mu$ for
symmetric nuclear matter. The non-physical behavior of the
equation of state in the liquid-gas coexistence region for temperatures below $T_c$ is denoted with
the dotted lines. The physical pressure and chemical potential in the coexistence region are constant
and determined from the Maxwell construction. The critical point is denoted with the large dot.
Left and right panels as in Fig.~\ref{fenergy50}.}
 \label{pchem50}
 \end{figure}

In Fig.\ \ref{eos50} we show the pressure $P(\rho, T)$ isotherms as a function of the density.
The results are qualitatively similar to that of a Van der Waals gas with its generic first-order
liquid-gas phase transition. In the present case chiral nucleon-nucleon dynamics generates
intermediate-range attractive interactions (most notably from $2\pi$-exchange with intermediate
$\Delta$ excitations), which as discussed in Section \ref{delt} exhibit a characteristic
$e^{-2\,m_\pi r} / r^6$ behavior. This mechanism provides nearly half of the attraction required
to bind symmetric nuclear matter at zero temperature. As in the case of nuclear matter saturation,
the liquid-gas phase transition results from a fine-tuned balance between intermediate-range
attraction and short-range repulsion, the latter coming from short-distance dynamics
unresolved at the relevant nuclear Fermi momenta $k_f$. Comparing in-medium chiral perturbation
theory results (left panel of Fig.\ \ref{eos50}) to those from microscopic forces fit to NN phase shift
data (right panel of Fig.\ \ref{eos50}), we see that in-medium chiral perturbation theory gives in
general a larger pressure for a given temperature and density.
In Fig.\ \ref{pchem50} we show the pressure as a function of the nucleon chemical potential
(including the free nucleon mass),
\begin{equation}
\mu = M_N + \left( 1+\rho\, \frac{\partial}{\partial \rho} \right) \bar{F}(\rho, T)  \,.
\end{equation}
As in previous figures, the non-physical part of the pressure is shown in the dotted curves of
Fig.\ \ref{eos50} and manifested in the double-valued behavior of $P$ for temperatures below the
critical temperature. In the coexistence region, both the physical pressure and chemical potential are
constant and given according to the Maxwell construction. The temperature at which the pressure
becomes single-valued is then identified as the critical temperature $T_c$.
From Figs.\ \ref{eos50} and \ref{pchem50} we see that the critical temperature of the liquid-gas phase
transition lies at $T_c \simeq 15.1$ MeV. For high-precision chiral two- and three-body forces (results shown in
the right panels of Figs.\ \ref{eos50} and \ref{pchem50}), the critical
temperature is a little higher $T_c \simeq 17.4$ MeV. Empirically the critical temperature
can be deduced from fission or multi-fragmentation measurements. The available data for both
indicate a value of $T_c \simeq 15-20$\,MeV \cite{karnaukhov}. Theoretical studies
employing phenomenological Skyrme interactions find similar values \cite{sauer}.

\begin{figure}[htbp]
\center
\includegraphics[scale=0.37,clip]{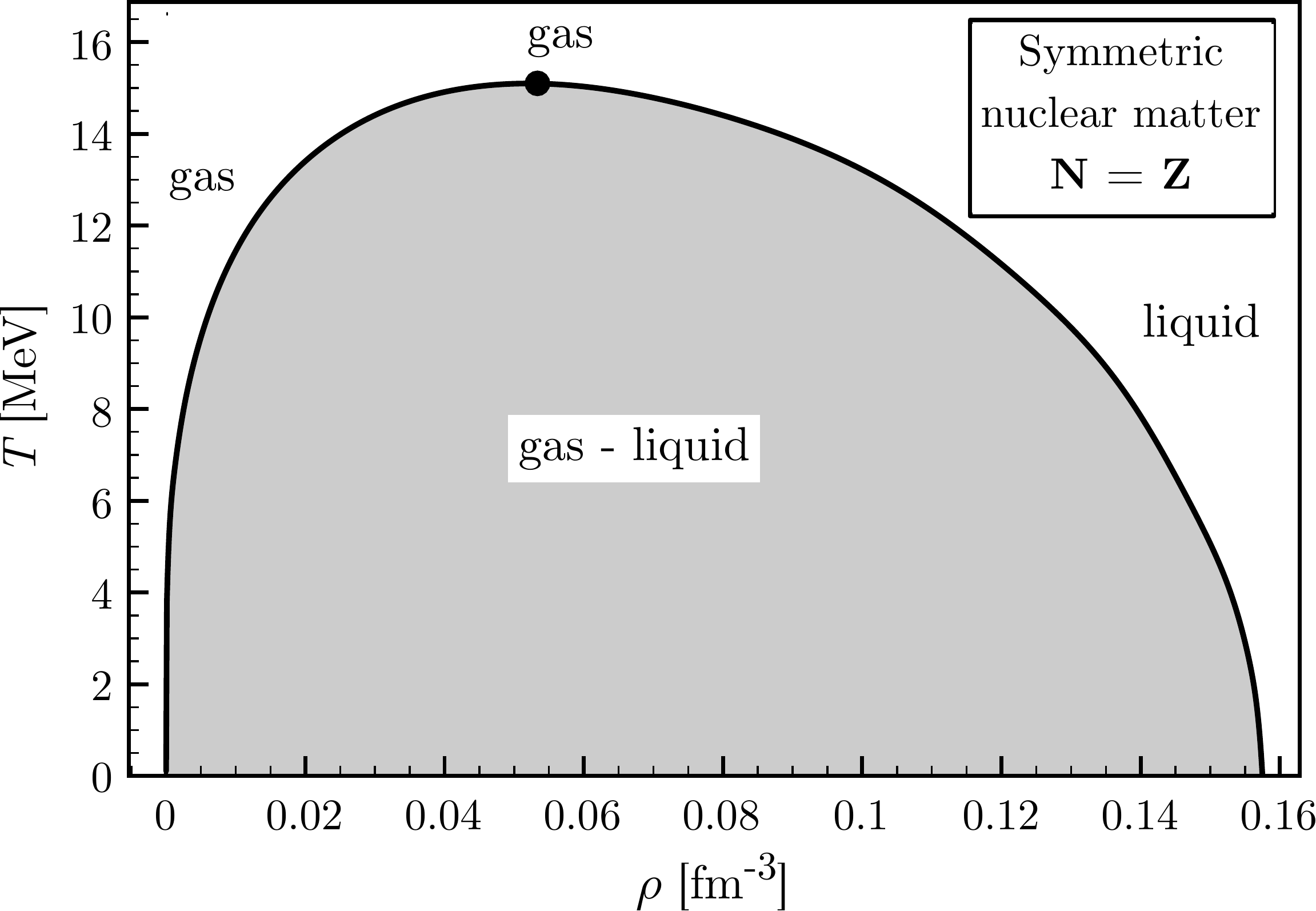}\hspace{.2in}
\includegraphics[width=3.4in]{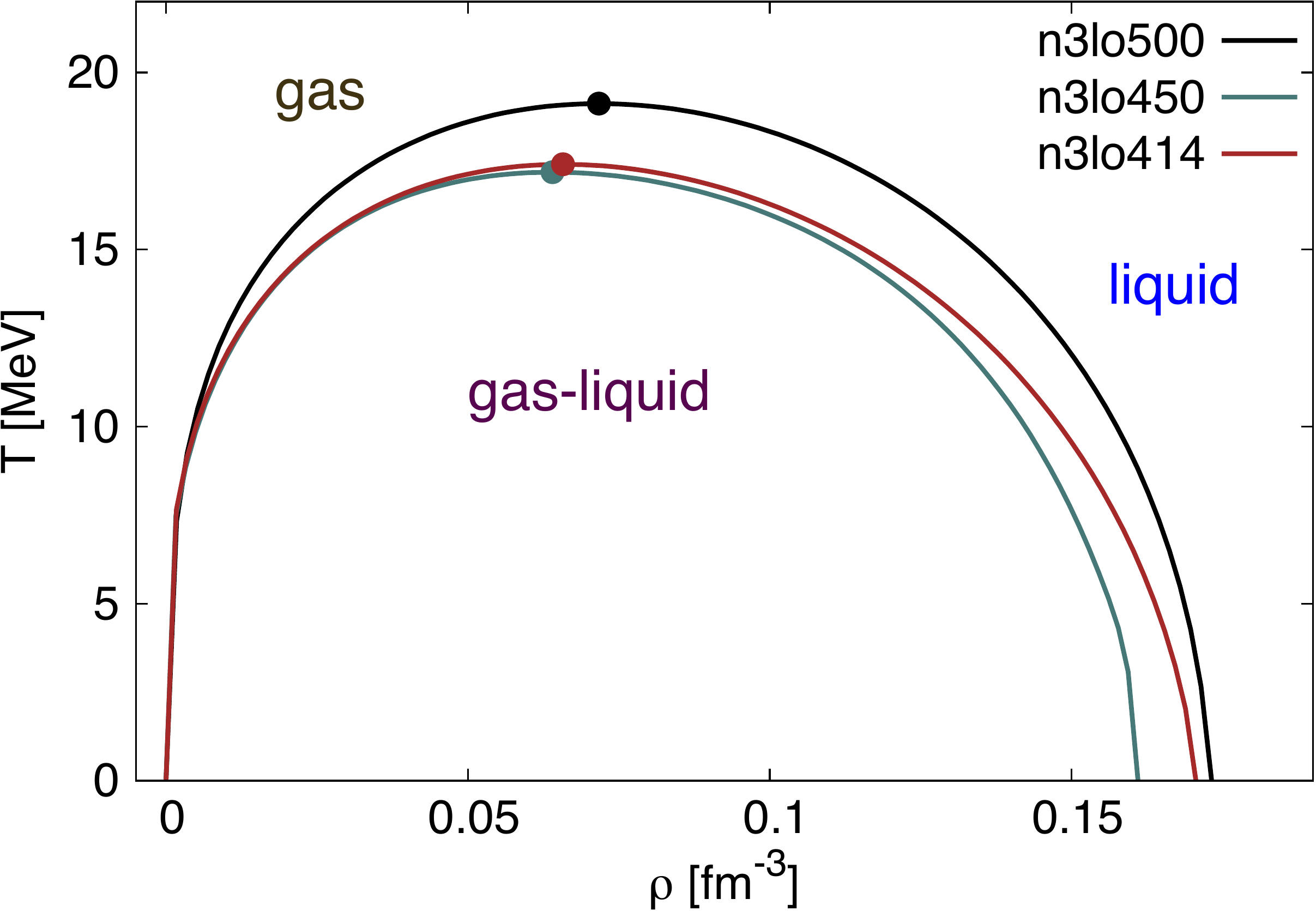}
\vspace{-.1in}
 \caption{The $T-\rho$ phase diagram of isospin-symmetric nuclear matter. The solid line denotes the
first-order phase transition boundary, and the critical point is indicated by the dot. Left and right panels
as in Fig.~\ref{fenergy50}.}
 \label{phasdiag}
 \end{figure}

The $T-\rho$ phase diagram, shown in Fig.\ \ref{phasdiag}, summarizes the features of the
liquid-gas phase transition. The solid line denotes the first-order transition region, which ends at the
critical point ($T_c=15.1$\,MeV) denoted with the dot. The associated values of the critical
pressure, baryon chemical potential, and density are found to be
$ P_c \simeq 0.261\,$ MeV\,fm$^{-3}$, $ \mu_c \simeq 914.7 $ MeV, and
$ \rho_c \simeq 0.053\, {\rm fm}^{-3}$. The liquid-gas coexistence region is found to extend over a
large range of densities up to $\rho_0 = 0.16$\,fm$^{-3}$, the equilibrium density marking the onset of the
Fermi liquid phase.
At $T=0$ the third law of thermodynamics constrains several features of
the $T-\rho$ diagram. In particular, at $T=0$ the boundary of the phase coexistence
region has an infinite slope and the chemical potential is given by the total energy per
particle at the saturation point, namely, $ \mu = M_N + \bar{E}_0 \simeq 923 $ MeV.

\subsection{Isospin-asymmetric nuclear matter}

\begin{figure}[tbp]
\center
\includegraphics[scale=0.7,clip]{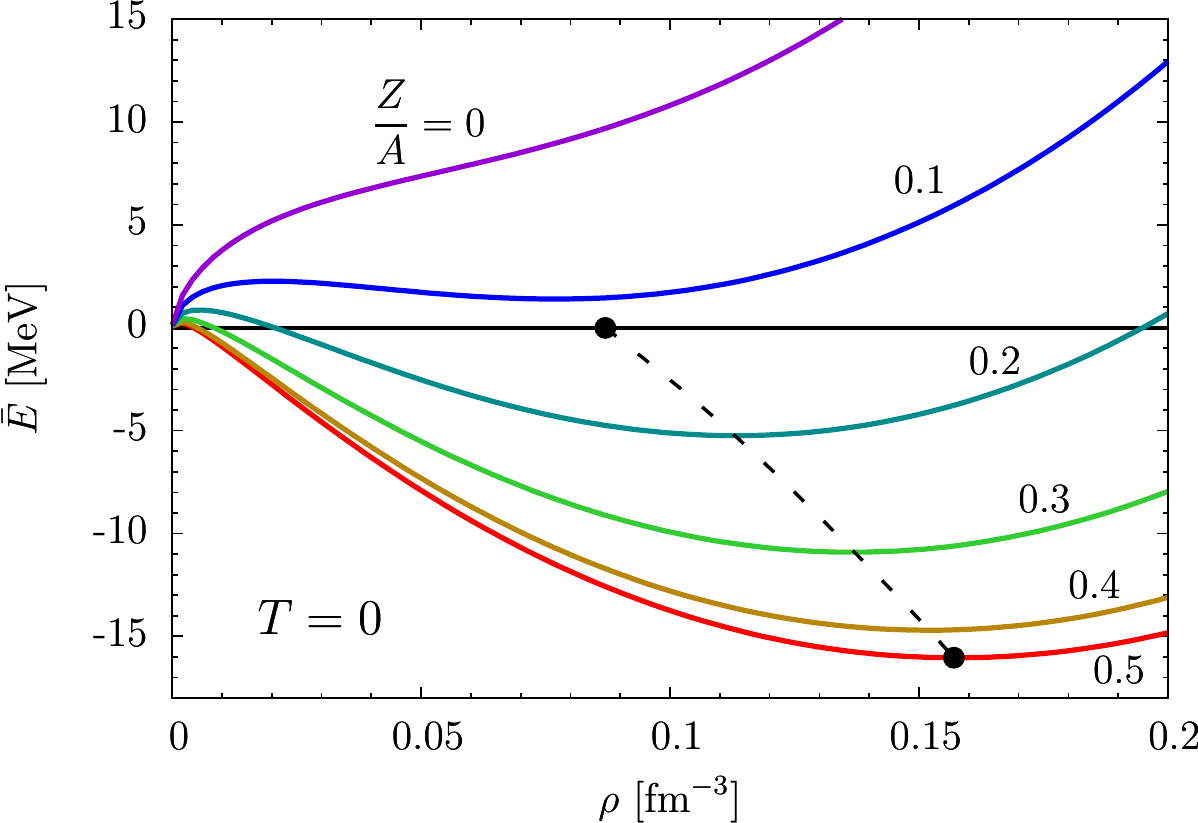}\hspace{.2in}
\includegraphics[scale=0.67,clip]{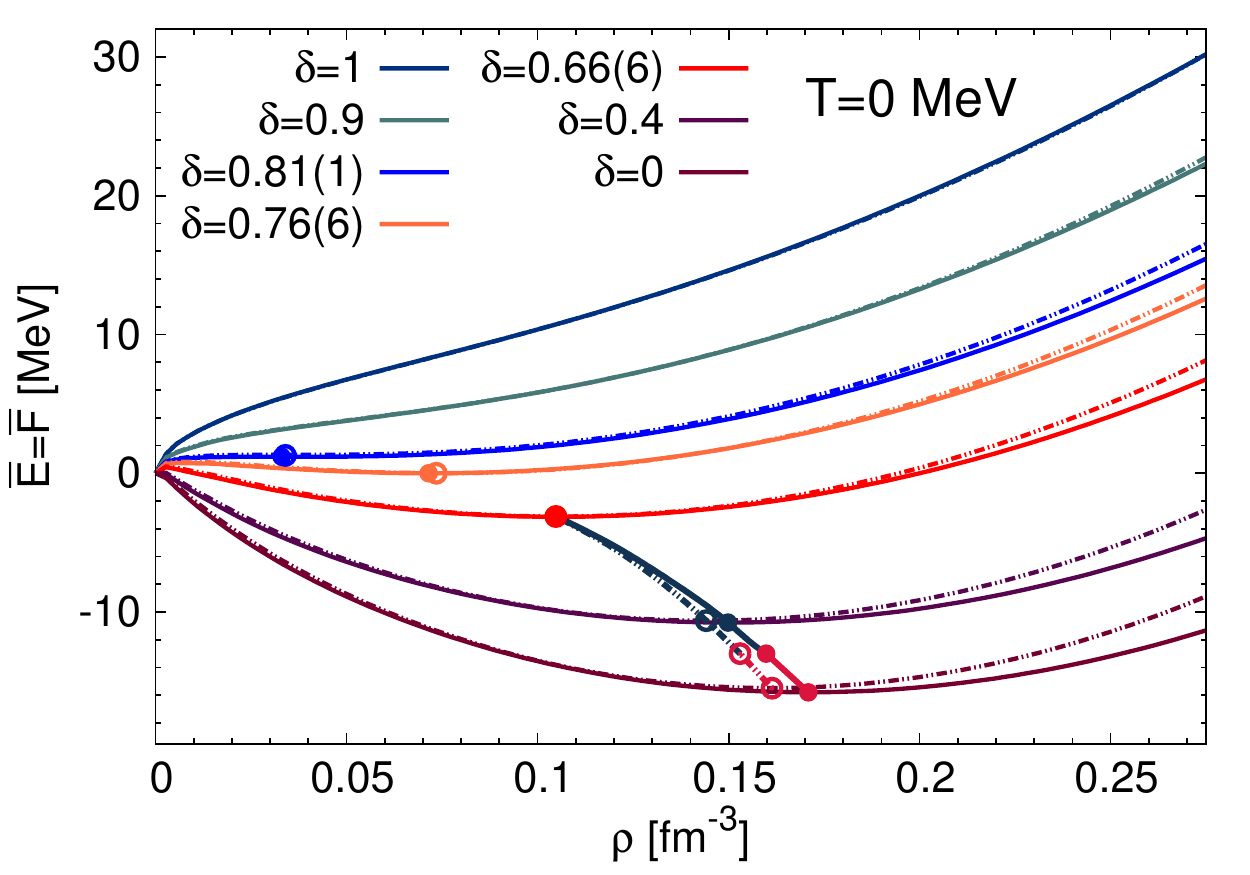}
\caption{Evolution of the zero-temperature equation of state for nuclear matter with varying
proton fraction $x_p$. The dashed line denotes the change in the saturation point.}
\label{satpoint}
\end{figure}

We now consider the extension of the above formalism to isospin-asymmetric nuclear matter and
pure neutron matter as a special limiting case. In comparison to symmetric nuclear matter the only
changes involve isospin factors and the introduction of separate proton and neutron thermal
occupation probabilities $d_p(p_j)$ and $d_n(p_j)$. The properties
of asymmetric matter with proton fraction $0.5 > x_p > 0$ are governed by the detailed isospin
dependence of the chiral nuclear interaction, arising almost entirely from one- and two-pion exchange
processes once the relevant nucleon-nucleon contact terms are fit to the empirical isospin
asymmetry energy $A(\rho_0,T=0) \simeq 34$\,MeV at zero temperature and saturation density.
The evolution of the saturation point, defined as the minimum of the energy per nucleon at
$T = 0$, with varying proton fraction is shown in Fig.\ \ref{satpoint}. As the proton fraction is reduced
starting from $x_p = 0.5$ where $ \bar{E}_0 \simeq -16 $ MeV at $ \rho_0 \simeq 0.157\,
{\rm fm}^{-3} $, the binding energy per particle continuously reduces in magnitude with
decreasing $ x_p $. For a proton fraction of $ x_p < 0.12$, neutron-rich matter is unbound at
zero temperature. This is in excellent agreement with the prediction \cite{wellenhofer15} from
microscopic many-body perturbation theory with chiral low-momentum interactions, which gives
the same value of the proton fraction: $x_p = \frac{1}{2}(1-\delta) < 0.12$.

\begin{figure}[tbp]
\center
\includegraphics[scale=0.4,clip]{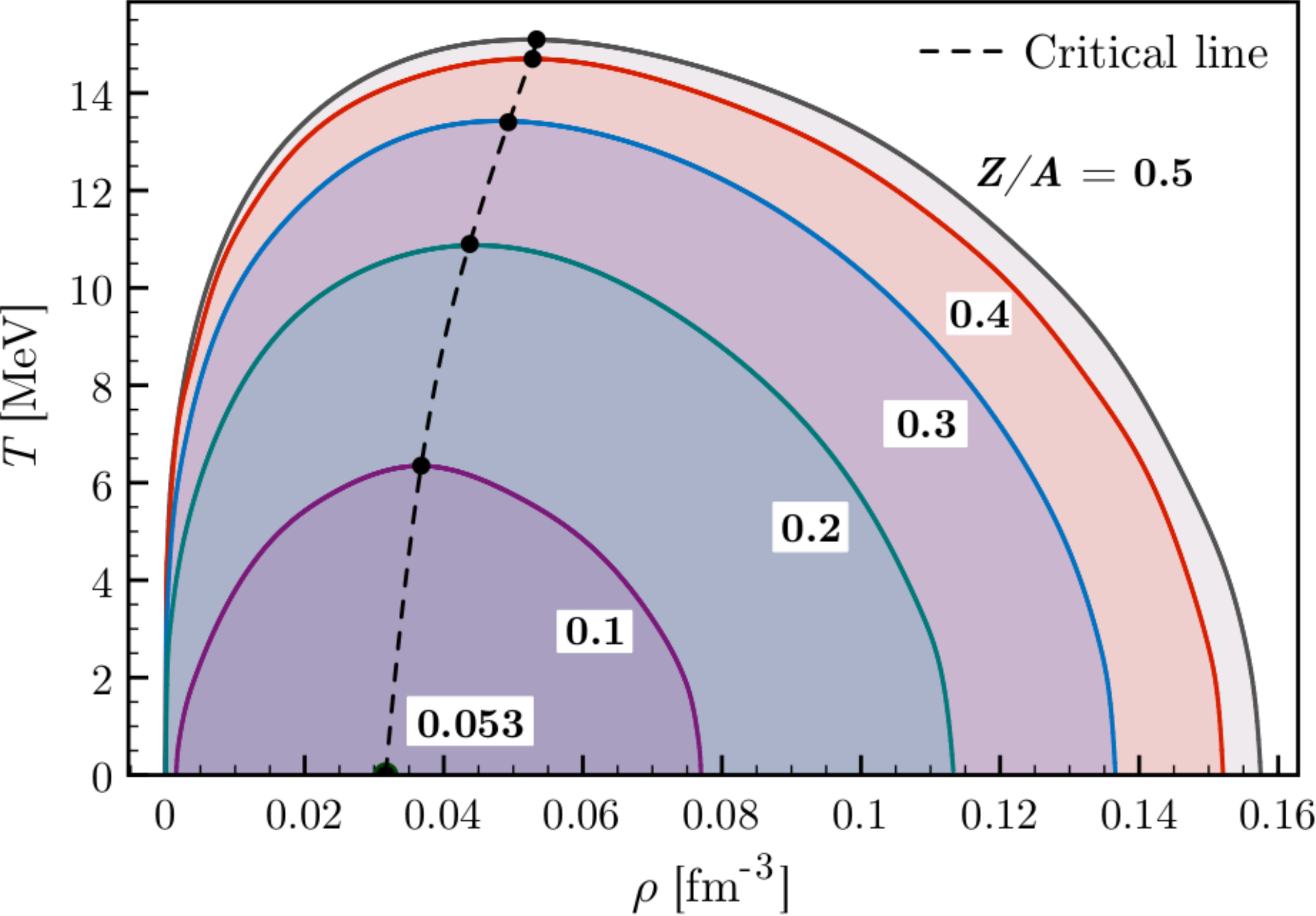}
\caption{Evolution of the $T-\rho$ phase diagram with changing proton fraction $x_p$. The dashed line
denotes the trajectory of the critical point.}
\label{phaseasy}
\end{figure}

In Fig.\ \ref{phaseasy} we show the $T-\rho$ phase diagram as a function of the proton fraction $x_p$.
The dashed line shows the trajectory of the liquid-gas phase transition critical point, which vanishes
for proton fractions below $x_p \simeq 0.05$. At this critical value of the proton fraction,
the coexistence region is a single point (at nonzero pressure), indicating the absence of a
liquid-gas phase transition. Neutron-rich matter
can therefore exist in a liquid phase only for $x_p \geq 0.05$.
From Fig.\ \ref{phaseasy} we see that when the proton fraction is above $x_p \simeq 0.1$,
the liquid-gas coexistence region starts at $\rho=0$.
The behavior of the coexistence region changes qualitatively for the proton fraction $ x_p = 0.12$,
which is the value at which the zero-temperature matter is not self-bound.
From Fig.\ \ref{satpoint} we see that for $x_p \leq 0.12$ although the absolute minimum of the energy
per particle is located at $\rho = 0$ there still exists a local minimum in $ \bar{E}(\rho, x_p) $
at nonzero density. This implies that neutron-rich nuclear matter is in the gaseous phase at very low
density and zero temperature for values of the proton fraction $x_p \leq 0.12$. As the density increases, the
neutron-rich matter then enters the liquid-gas coexistence region. Interestingly for values of the proton fraction
$ 0.05 \leq x_p \leq 0.12$, nuclear matter is not self-bound but nevertheless has a
liquid-gas phase transition. The discussion so far has neglected the effects of clustering at low
densities, which arises from an interplay of the short-range attractive nuclear force and long-range
repulsive Coulomb force. Relativistic mean field studies \cite{typel} that included the effects of light nuclei
(deuteron, triton, and alpha), however, resulted in only modest changes in the $T-\mu$ phase diagram
at low densities. The critical point, for instance was found to change by less than $ 1 \% $ in $ \mu_c $ and
less than $ 10 \% $ in $ T_c $.

Expanding the free energy per particle in powers of the nuclear asymmetry parameter,
$\delta = (\rho_n - \rho_p) / \rho = 1-2x_p$, around $\delta=0$ yields
 \begin{equation}
\bar{F}(\rho_p,\rho_n, T) = \bar{F}(\rho, T) + A(\rho, T)\, \delta^2 +
{\cal O} (\delta^4) \, .
\label{afe}
\end{equation}
Here $A(\rho, T)$ defines the asymmetry free energy. Ignoring isospin-symmetry
breaking effects, the expansion in Eq.\ (\ref{afe}) includes only even
powers of $ \delta$ and nuclear matter is invariant under the interchange of protons and
neutrons.
The parabolic approximation, keeping only the leading $ \delta^2 $ term in Eq.\ (\ref{afe}),
has been shown \cite{FKW2012} to work exceptionally well across a large range of densities.
At $ T = 0 $ the quadratic dependence holds very well up to large values of the isospin
asymmetry parameter, but for higher temperatures (e.g., $ T = 20\,$MeV) and low densities,
higher-order terms are needed to describe the dependence on the isospin asymmetry.

Previous model-dependent determinations \cite{blaizot,seeger,vretenar,cao} of the isospin
asymmetry $A(\rho, T)$ parameter at saturation density have obtained values in the range
$33 - 37$\,MeV. In the present chiral perturbation theory description, the contact terms of the
isospin-dependent part of the nucleon-nucleon interaction have been fit to produce
$A(\rho_0, T = 0) \simeq 34.0 $\,MeV at saturation density. The nuclear asymmetry energy at
a lower density, $ \rho = 0.1 $ fm$^{-3} $, was estimated in Ref.\ \cite{cao} to lie between 21
and 23 MeV. The chiral effective field theory prediction, $ A(\rho = 0.1\, {\rm fm}^{-3}, T = 0)
\simeq 23.9 $ MeV, is slightly above the empirically estimated value.
Moreover, in the vicinity of the saturation density $\rho_0$, the asymmetry energy at zero
temperature can be Taylor expanded:
\begin{equation}
A(\rho) = A(\rho_0) + L\,\frac{\rho - \rho_0}{3\,\rho_0} + \frac{K_{as}}{2}
\left(  \frac{\rho - \rho_0}{3\,\rho_0} \right)^2 + \dots  \end{equation}
The coefficients $L$ and $K_{as}$ of the linear and quadratic terms are found in the present
case to be $L \simeq 90$ MeV and $ K_{as} \simeq 153 $ MeV. This value of
$L$ is consistent with empirical constraints \cite{baoanli}, which give $ L = 88 \pm 25 $ MeV .

For small values of the isospin asymmetry $\delta$, the saturation density is related to the
nuclear compression modulus ${\cal K}(\delta)$ and the slope $L$ of the
asymmetry parameter through $\rho_0(\delta) = \rho_0 (1 - 3L\,\delta^2/{\cal K} )$.
The isospin dependence of the nuclear compression modulus ${\cal K}(\delta)$ is then often
expressed as
\begin{equation}
{\cal K}(\delta) = {\cal K} + {\cal K}_\tau \delta^2 + {\cal O}(\delta^4) \ , \qquad {\cal K}_\tau= K_{as} - 6\,L \ ,
\end{equation}
where $ {\cal K}_\tau $ is referred to as the isobaric compressibility. In the present framework we find
$ {\cal K}_\tau = -388 $ MeV, which should be compared to the empirical value $ {\cal K}_\tau = -550 \pm 100 $ MeV
obtained from measurements of giant monopole resonances in even-mass-number
isotopes \cite{li}.



\subsection{Thermodynamics of the chiral condensate}

The scalar quark condensate $\langle \bar{q}q \rangle$, introduced in Chapter \ref{intro}
as the expectation value of the scalar quark density, plays a fundamental role as an order
parameter for the spontaneous breaking of chiral symmetry in QCD. Chiral symmetry is
expected to be restored in hot and/or dense matter, and therefore the
evolution of $\langle \bar{q}q \rangle$ with increasing temperature and density is key to
mapping out the boundary of the chiral restoration transition in the phase diagram of
QCD. In-medium chiral perturbation theory provides a starting point for such
a study \cite{condpap}, since the quark-mass dependence (or equivalently, the pion mass
dependence) of the free energy density is accessible and can be linked to the thermodynamics
of the chiral condensate, which we now discuss in detail.

The two- and three-body kernels, ${\cal K}_2$ and ${\cal K}_3$ introduced in the previous section,
encode all one- and two-pion exchange processes up to three-loop order for the
free energy density. Explicit expressions can be found in
Refs.\ \cite{nucmatt,fkw2,FKW2012}. Let us recall that after fixing the unresolved short-distance
dynamics, the free energy density computed from these interaction kernels
provides a realistic equation of state up to about twice saturation density: $\rho
\leq 2\,\rho_0$. In the present context, the relevant question is how these interaction
kernels depend on the light quark mass, $m_q$, through pion
propagators and pion loops.

From the Feynman-Hellmann theorem, an exact relation can be established
between the temperature- and density-dependent quark condensate
$\langle \bar q q\rangle(\rho,T)$ and the derivative of the free energy
density with respect to the light quark mass $m_q$. Making use of the Gell-Mann-Oakes-Renner
relation in Eq. (\ref{eq:GOR}), one obtains the ratio of the in-medium to vacuum
quark condensate:
\begin{equation}
{\langle \bar q q\rangle(\rho,T)\over  \langle 0|\bar q q|0
\rangle} = 1 - {\rho \over f_\pi^2} {\partial \bar F(\rho,T) \over \partial
m_\pi^2} \,,\end{equation}
where the derivative of the free energy density is
taken at fixed $T$ and $\rho$. The vacuum scalar quark condensate
$\langle 0|\bar q q|0\rangle$ and the pion decay constant $f_\pi$ are both taken in the
chiral limit. Similarly, only the leading linear term in the expansion of
the squared pion mass in terms of the quark mass is considered.

The quark mass dependence of the one-body kernel ${\cal K}_1$ in Eq.\ (\ref{K1})
is implicit through its dependence on the nucleon mass $M_N$. Imposing the
condition $\partial \rho/\partial M_N=0$ leads to the
dependence of the effective one-body chemical potential $\tilde \mu$
on the nucleon mass $M_N$:
\begin{equation}
{\partial \tilde \mu \over \partial M_N }= {3 \rho \over 2M_N
\Omega_0''} \,, \qquad   \Omega_0''= -4M_N  \int_0^\infty dp\,
{n(p) \over p}\,.
\label{dudm}
\end{equation}
The nucleon sigma term $\sigma_N = \langle N|m_q \bar q q |N\rangle= m_\pi^2 \,
\partial M_N/\partial m_\pi^2$ relates variations in the nucleon mass to
that of the pion mass, and from Eq.\ (\ref{dudm}) the
expression for the derivative of the one-body kernel with respect to the pion mass
can be written
\begin{equation} {\partial {\cal K}_1 \over \partial m_\pi^2} = {\sigma_N
\over m_\pi^2} \bigg\{ 1+ {3 \rho \over 2M_N \Omega_0''} +{p^2 \over 3M_N^2}
+{3p^4 \over 8M_N^4} \bigg\}\,. \label{derk1}\end{equation}
In the zero temperature limit, the terms in Eq.\ (\ref{derk1}) give rise to a linear
decrease of the chiral condensate with the nucleon density. At the physical
pion mass, the empirical value \cite{GLS91} of the nucleon sigma term is
$\sigma_N =(45\pm 8)\,$MeV. Recently, lattice QCD calculations of the quark mass
dependence of baryon masses and accurate chiral extrapolations \cite{sigmaterm}
have found smaller values of the nucleon sigma term that are however consistent
with the empirical value to within the experimental uncertainty. In the present discussion
based on the work in Ref.\ \cite{condpap}, the central value of $\sigma_N = 45$\,MeV
has been employed.

The two-pion exchange mechanisms arising from the chiral symmetry breaking
$\pi \pi N N$ contact-vertex proportional to $c_1m_\pi^2$ have been neglected so
far in calculations of the equation of state because they are almost negligible.
However, when differentiating with respect to $m_\pi^2$ in the calculation of the
in-medium condensate, such contributions turn out to be of similar importance as the
other interaction terms. Therefore, they have been included in the calculations
discussed here, and the relevant two- and three-body kernels
${\cal K}_{2,3}^{(c_1)}$ can be found in Ref.\ \cite{condpap}.
At high temperatures it is also necessary to consider effects from thermal pions.
The pressure of thermal pions gives rise, through its $m_\pi^2$-derivative, to a
reduction of the $T$-dependent in-medium condensate. In the two-loop approximation
one finds the following shift of the chiral condensate ratio due to the presence
of a pionic heat bath \cite{gerber,toublan,pipit}:
\begin{equation}
{\delta\langle \bar q q\rangle(T)\over  \langle 0|\bar q q|0
\rangle} =-{3m_\pi^2 \over (2\pi f_\pi)^2} H_3\Big({m_\pi
\over T}\Big) \bigg\{1+ {m_\pi^2 \over 8\pi^2 f_\pi^2}
\bigg[H_3\Big({m_\pi \over T}\Big)-  H_1\Big({m_\pi \over T}\Big) + {2-3
\bar{\ell}_3 \over 8} \bigg] \bigg\}\,,
\end{equation}
where the functions $H_{1,3}(m_\pi/T)$ represent integrals over the
Bose distribution of thermal pions:
\begin{eqnarray}
 H_1(y) = \int_y^\infty dx\, {1 \over \sqrt{x^2-y^2}
(e^x-1)}\,,\qquad\qquad H_3(y) = y^{-2} \int_y^\infty dx\,
{\sqrt{x^2-y^2} \over e^x-1}\,.
\end{eqnarray}
The three-loop calculation of the free energy density is expected be reliable up
temperatures on the order of $T\sim 100$ MeV and densities up to about $\rho \sim 2\rho_0$,
where the hot and dense hadronic medium remains in the spontaneously-broken
chiral symmetry phase.

\begin{figure}
\begin{center}
\includegraphics[width=10cm]{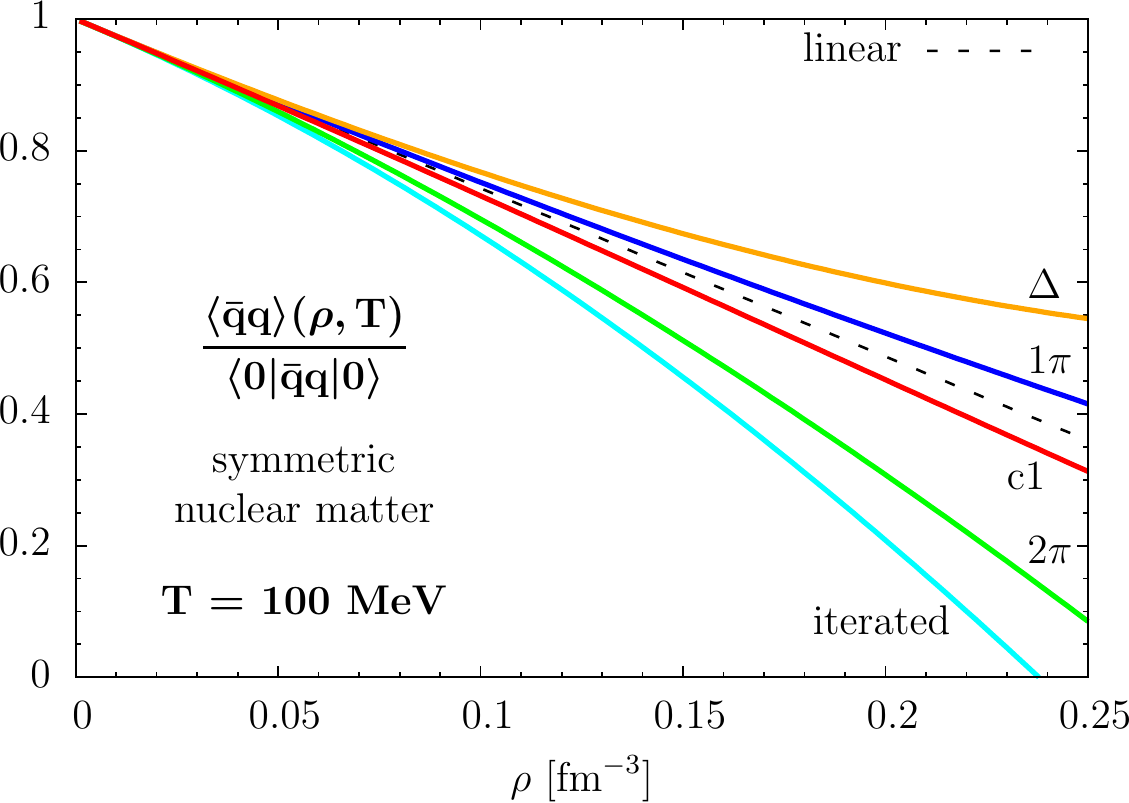}
\end{center}
\vspace{-.3in}
\caption{Density dependence of the chiral condensate at a fixed temperature of $T = 100$
MeV. The full density dependence is built by starting from the linear free Fermi gas approximation,
and then the following interaction contributions are successively
added: the one-pion exchange Fock contribution, iterated one-pion exchange,
irreducible two-pion exchange, two-pion exchange processes with intermediate $\Delta$ excitations,
and the two-pion exchange contribution proportional to the low-energy constant $c_1$. Here
we have omitted the contribution from thermal pions.}
\label{condfig2}
\end{figure}

In Fig.\ \ref{condfig2} we show the ratio of the in-medium chiral condensate to the vacuum condensate
as a function of density for the temperature $T = 100$ MeV. Interaction contributions are successively
summed together, starting from the linear free-gas approximation. First, the one-pion exchange contribution
followed by the iterated $1\pi$- and irreducible $2\pi$-exchange processes are considered. Taken alone
these terms would result in the system becoming unstable not far above nuclear matter saturation density. Of
particular importance is the pion-mass dependence of the terms with virtual $\Delta(1232)$ excitations, which
strongly delay the tendency towards chiral symmetry restoration at larger densities. Including finally the effects
from two-pion exchange contributions proportional to $c_1$, we see that the chiral condensate at $T=100$ MeV
is not far from that of a free Fermi gas. However, this behavior clearly results from a subtle cancellation between
attractive and repulsive correlations and their detailed pion-mass dependences. It is interesting to note that in
the chiral limit $(m_\pi \rightarrow 0)$ the instability seen from iterated $1\pi$ and irreducible $2\pi$ would have
appeared even at much lower densities. This emphasizes once more the important role played by $\Delta(1232)$
excitations and also highlights the significance of explicit chiral symmetry breaking.

\begin{figure}[t]
\begin{center}
\includegraphics[width=10cm]{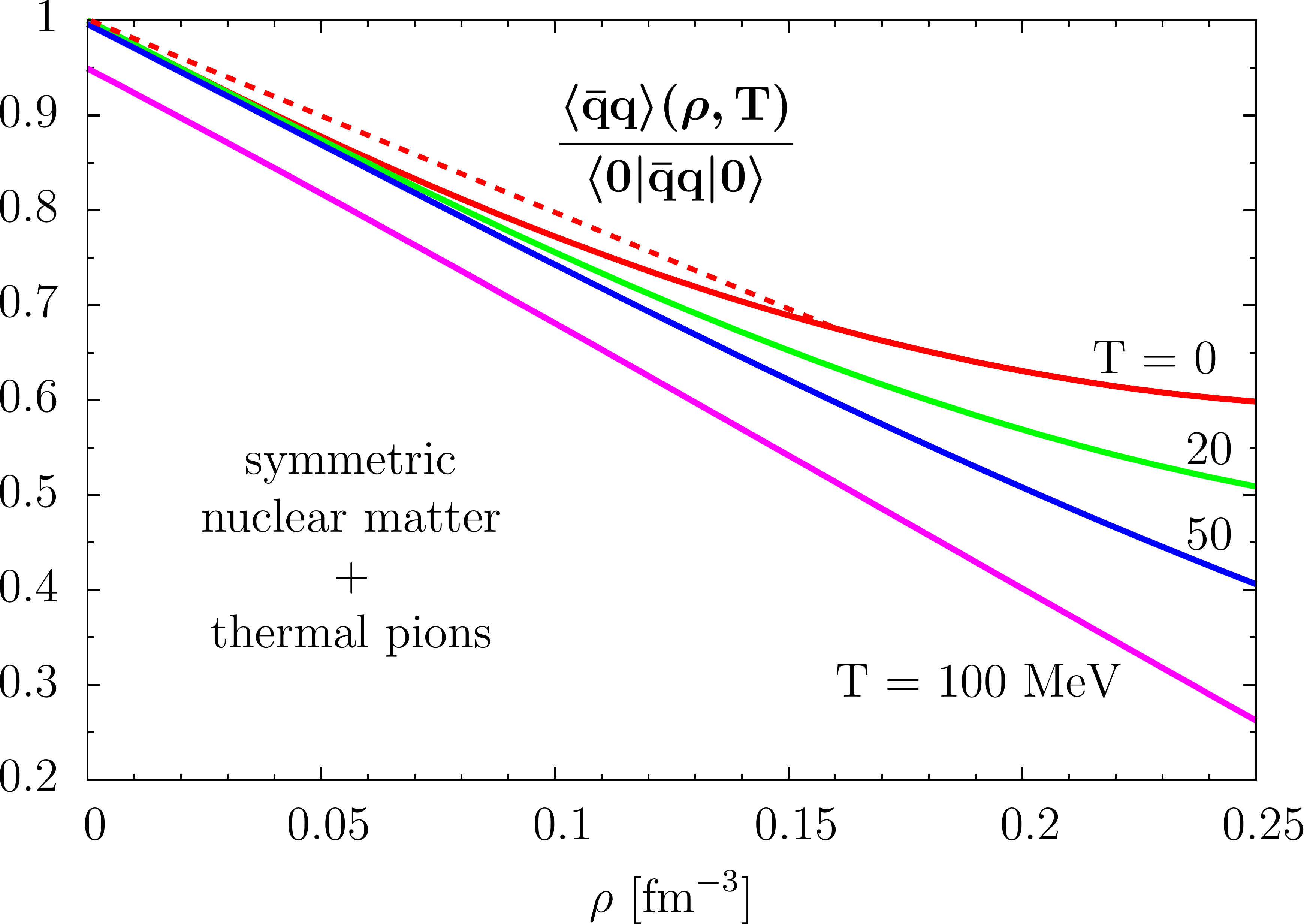}
\end{center}
\vspace{-.2in}
\caption{Ratio of the in-medium chiral condensate to the vacuum condensate as a function
of the baryon density and temperature in isospin-symmetric nuclear matter.
Thermal pion effects are included, and the dashed line at $T \simeq 0$ results from
the Maxwell construction in the nuclear liquid-gas coexistence region.}
\label{condfig3}
\end{figure}

\begin{figure}[htb]
\begin{center}
\includegraphics[width=10cm]{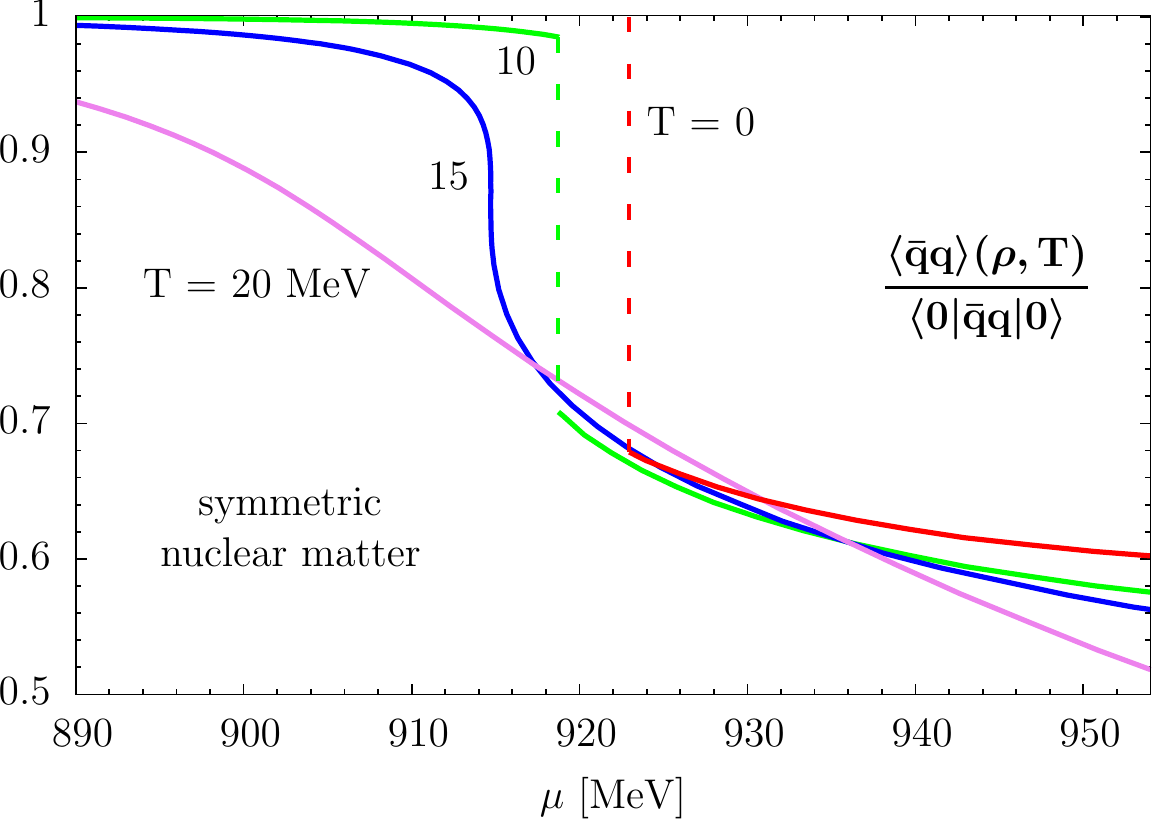}
\end{center}
\vspace{-.2in}
\caption{Ratio of the chiral condensate to its vacuum value as function
of baryon chemical potential in symmetric nuclear matter at low
temperatures within the liquid-gas phase coexistence region.}
\label{condfig4}
\end{figure}

In Fig.\ \ref{condfig3} we show the dependence of the chiral condensate on the
density $\rho$ and temperature $T$. All nuclear correlation effects discussed previously
are included as well as the small additional shift from thermal pions, which can be seen
only at the highest temperature $T=100$\,MeV considered here. The actual crossover
transition can be extracted from lattice QCD simulations at zero baryon density and has the value
$T \sim 170$ MeV \cite{lattice}. At zero temperature, the decrease in the chiral condensate above
nuclear matter saturation density is suppressed due to three-body correlations
in ${\cal K}_3$ that grow faster than ${\cal K}_2$ as the density increases. However, as the
temperature increases, the relative influence of ${\cal K}_3$ to ${\cal K}_2$ is reduced and
at $T=100$ MeV there is a much stronger tendency toward chiral symmetry restoration at large
baryon densities. The solid line for the $T=0$ curve in Fig.\ \ref{condfig3} does not account for the
liquid-gas coexistence region. Any first-order phase transition is expected to appear also in other order
parameters, and the red dashed line (based on the usual Maxwell construction) accounts for this effect.
When the chiral condensate is plotted as a function of the baryon chemical potential
\begin{equation}
\mu = M_N +\bigg( 1 + \rho {\partial \over\partial \rho}\bigg)
\bar F(\rho,T) \,.
\end{equation}
this feature becomes much more pronounced.
We show in Fig.\ \ref{condfig4} the discontinuity in the chiral condensate associated with
the first-order liquid-gas transition at $T$ smaller $T_c \simeq 15$\,MeV.
Another consequence of the first-order liquid-gas phase transition is
that the frequently advocated ``low-density theorem'' must be modified:
\begin{equation}
{\langle \bar q q\rangle(\rho)\over  \langle 0|\bar q q|0
\rangle} = 1 - {\widetilde\sigma_N \over m_\pi^2f_\pi^2} \rho  \, ,
\label{ldtmod}
\end{equation}
where the effective nucleon sigma term $\widetilde\sigma_N\simeq 36\,$MeV
has been introduced and represents the quark mass dependence of the sum
$M_N+\bar E_0$, with $\bar E_0 \simeq -16\,$MeV the energy per particle of
nuclear matter at the saturation density.
The usual version of Eq.\ (\ref{ldtmod}) contains the nucleon sigma term $\sigma_N$
in vacuum and treats low-density nuclear matter as a non-interacting Fermi gas.

The results discussed in this section set important nuclear physics constraints on
the QCD equation of state at baryon densities and temperatures of
relevance in relativistic heavy-ion collisions. In particular, the present analysis provides
no indication for chiral symmetry restoration at temperatures $T \leq
100\,$MeV and baryon densities at least up to about $\rho \simeq 2\rho_0$.
The prominent role played by intermediate $\Delta(1232)$-isobar excitations
together with Pauli blocking in the nuclear medium play a central role in stabilizing the
scalar quark condensate at and beyond nuclear matter saturation density.

\subsection{Functional renormalization group approach to nuclear and neutron matter}

Chiral effective field theory applied to the nuclear many-body problem is based on a well organized
perturbative hierarchy of in-medium forces driven by one- and two-pion exchange
dynamics at long and
intermediate distances, and supplemented by short-distance contact interactions. The
calculations include three-body forces which play an increasingly important role
when matter is compressed to densities beyond that of equilibrium nuclear matter. While
ChEFT as a perturbative scheme is remarkably successful within its range of applicability,
questions about convergence in extrapolations to higher baryon densities remain. It is therefore
important to examine such issues by comparison with models that do allow for a non-perturbative
treatment of pionic fluctuations in the nuclear medium.

Here we consider a chiral model of interacting mesons and nucleons \cite{Berges2000,FW2012} combined with functional renormalization group (FRG) methods \cite{Drews2013,Drews2014}, as an instructive example of such an approach. Its starting point is a linear sigma model in which an isospin doublet of nucleons,
$\Psi_N = (p,n)^\top$, is coupled to pion $(\boldsymbol\pi)$ and scalar $(\sigma)$
fields, plus short distance dynamics generated schematically by isoscalar and isovector
vector bosons (``$\omega$" and ``$\boldsymbol\rho$"):
\begin{eqnarray}\label{eq:NMLagrangian}
\mathcal L=\bar{\Psi}_N\left[ i\gamma_\mu\partial^\mu - g(\sigma+i\gamma_5\,\boldsymbol\tau\cdot\boldsymbol\pi)-\gamma_\mu(g_\omega\, \omega^\mu+g_\rho\boldsymbol\tau\cdot\boldsymbol\rho^\mu)\right]\Psi_N +
\mathcal L_\chi(\sigma,\boldsymbol\pi) + \mathcal L_V(\omega,\boldsymbol\rho)~,
\end{eqnarray}
\begin{equation}
\mathcal L_\chi(\sigma,\boldsymbol\pi) = \frac 12\left(\partial_\mu\sigma\,\partial^\mu\sigma+\partial_\mu\boldsymbol\pi\cdot\partial^\mu\boldsymbol\pi\right)
- {\cal U}(\sigma,\boldsymbol\pi) ~,
\end{equation}
\begin{equation}
\mathcal L_V(\omega,\boldsymbol\rho) = \frac 12m_V^2\left(\omega_\mu\,\omega^\mu+ \boldsymbol\rho_\mu\cdot\boldsymbol\rho^\mu\right) -\frac 14\left( F^{(\omega)}_{\mu\nu}F^{(\omega)\mu\nu} - \boldsymbol F^{(\rho)}_{\mu\nu}\cdot\boldsymbol F^{(\rho)\mu\nu}\right) ~,
\label{NMLagrangian}
\end{equation}
where the potential ${\cal U}(\sigma,\boldsymbol\pi) = {\cal U}_0(\chi)- m_\pi^2 f_\pi(\sigma-f_\pi)$ has a piece, ${\cal U}_0$, given as a fourth-order polynomial in the chiral invariant field $\chi = {1\over 2}(\sigma^2 + \boldsymbol\pi^2)$, and an explicit symmetry breaking term proportional to the quark mass, $m_q \sim m_\pi^2$. The input pion mass and decay constant are fixed at their physical values in vacuum,
$m_\pi = 140$ MeV and $f_\pi = 92$ MeV. The Goldstone boson field $\boldsymbol\pi$ is light (massless in the chiral limit) and subject to fluctuations beyond the mean-field approximation. These fluctuations include the one- and two-pion exchange mechanisms plus emergent three-body forces as they appear prominently in the chiral EFT treatment. The non-linear structure of ${\cal U}(\sigma,\boldsymbol\pi)$ is even richer in content as it potentially generates iterations to all orders of such mechanisms, together with multi-nucleon interactions, when solving the FRG equations.

The massive vector fields introduced here have their correspondence in the contact terms
of chiral EFT, representing short-distance dynamics. These $\omega$ and $\rho$ bosons\footnote{The most natural way to introduce vector-meson fields in chiral effective field theories is to resort to hidden local symmetry as described in Section \ref{HLS}.}
will be treated as mean fields, fluctuations being suppressed by their large masses. As a consequence,
only the combinations $G_\omega = g_\omega^2/m_V^2$ and $G_\rho = g_\rho^2/m_V^2$
appear as effective coupling strength parameters of dimension $(length)^2$.

The treatment of the equilibrium thermodynamics based on the model Lagrangian in Eq.\ (\ref{eq:NMLagrangian}) involves again the following standard steps. First, the action in Minkowski space, $S_{\mathrm M} = \int d^4x \,{\cal L}$, is rewritten in Euclidean space-time with $t\equiv x^0 \rightarrow -i\tau$. The time integral $\int dt$ is replaced by $-i\int_0^\beta d\tau$ with the inverse temperature \mbox{$\beta = 1/T$}. Periodic (antiperiodic) boundary conditions apply to bosonic (fermionic) fields. A non-zero baryon chemical potential, $\mu$, is defined by adding the term $\Delta S_{\mathrm E}(\mu) = -\beta\mu B$ to the Euclidean action $S_{\mathrm E}$, with the baryon number $B = \int d^3x \,\psi^\dagger \psi$.

Next, an effective
action $\Gamma_k$ is introduced that depends on a renormalization scale, $k$. Starting from a microscopic action, $\Gamma_{k=\Lambda}$, defined at an ultraviolet renormalization scale $\Lambda$, an interpolation is performed to reach the full quantum effective action, $\Gamma_{\mathrm{eff}}=\Gamma_{k=0}$. As the scale $k$ is lowered, the renormalization group flow of $\Gamma_k$ is determined by Wetterich's equation~\cite{Wetterich1993},
\begin{eqnarray}\label{eq:Wetterich}
k\,\frac{\partial\Gamma_k}{\partial k}=
		\frac 12 \mathrm{Tr}\frac{k\,\frac{\partial R_k}{\partial k}}{\Gamma_k^{(2)}+R_k}~\,,
\end{eqnarray}
where $R_k$ is a regulator function (the one commonly used is $R_k(p^2) = (k^2 - p^2)\,\theta(k^2 - p^2)$)
and $\Gamma_k^{(2)} = \delta^2\Gamma_k/\delta \phi^2$ is the full inverse propagator ($\phi$ generically denotes all fluctuating fields involved).
Using the derivative expansion to leading order, ignoring possible
anomalous dimensions and restricting the (heavy) vector boson fields to their expectation
values, the effective action takes the form
\begin{eqnarray}\label{eq:effaction}
\Gamma_k = \int d^4x\left\{\bar{\Psi}_N\left[ i\gamma_\mu\partial^\mu - g(\sigma+i\gamma_5\,\boldsymbol\tau\cdot\boldsymbol\pi)\right]\Psi_N- \Psi^\dagger(g_\omega\, \omega^0+g_\rho\boldsymbol\tau\cdot\boldsymbol\rho^0)\Psi_N + \mathcal L_\chi(\sigma,\boldsymbol\pi)- U_k(\sigma,\boldsymbol\pi,\omega_0,\rho_0^3)\right\}~,
\end{eqnarray}
where the $k$-dependence of the action is now relegated entirely to the chirally symmetric
part, $U_k(\chi)$, of the effective potential:
\begin{equation}
U_k = U_k(\chi) - m_\pi^2\,f_\pi^2(\sigma - f_\pi) -
{1\over 2}m_V^2\left[\omega_0^2 + (\rho_0^3)^2\right]~.
\end{equation}
Since only potential differences matter, a conveniently subtracted potential is introduced:
\begin{equation}
\bar{U}_k(\chi;T,\mu) = U_k(\chi; T,\mu)  - U_k(\chi; T=0,\mu=\mu_0)~,
\end{equation}
where the subtraction point is fixed at temperature $T = 0$ and at the chemical potential
$\mu_0 = M_N - 16$ MeV $= 923$ MeV corresponding to equilibrium nuclear matter.
Returning to the flow equation (\ref{eq:Wetterich}), the equation to be solved is now:
\begin{equation}
{\partial \bar{U}_k(\chi;T,\mu)\over\partial k} = f_k(T,\mu) - f_k(0,\mu_0)~,
\end{equation}
\begin{eqnarray}
\mathrm{with}~~~~~~~~f_k(T,\mu) = {k^4\over 12\pi^2}\left[{3(1+2n_{\mathrm{B}}(E_\pi))\over E_\pi}
+ {1+2n_{\mathrm{B}}(E_\sigma)\over E_\sigma} - 4\sum_{i=p,n}{1 - n_{\mathrm{F}}(E_N,\mu_i^{\mathrm{eff}})\over E_N}\right]~.
\end{eqnarray}
Here we have dropped negligibly small anti-nucleon terms while nucleons are kept as fluctuating degrees of freedom, taking into account important low-energy particle-hole excitations around the nucleon
Fermi surface. Furthermore,
\begin{eqnarray}
E_N^2 = k^2 + 2g^2\chi~,~~~~E_\pi^2 = k^2 + {\partial U_k\over\partial\chi}~,~~~~
E_\sigma^2 = k^2 + {\partial U_k\over\partial\chi} + 2\chi{\partial^2 U_k\over\partial \chi^2}~,\nonumber\\
n_\mathrm{B}(E) = {1\over e^{E/T} - 1}~,~~~~~n_\mathrm{F}(E,\mu) = {1\over e^{(E-\mu)/T} + 1}~,~~~~~
\mu_{p,n}^{\mathrm{eff}} = \mu_{p,n} - \left[g_\omega\,\omega_0(k) \pm g_\rho\,\rho_0^3(k)\right]~.
\end{eqnarray}
The $k$-dependent vector mean fields are computed self-consistently by minimizing $U_k$ for each
scale $k$.

\begin{figure}[t]
\begin{center}
\includegraphics[height=6cm,angle=0]{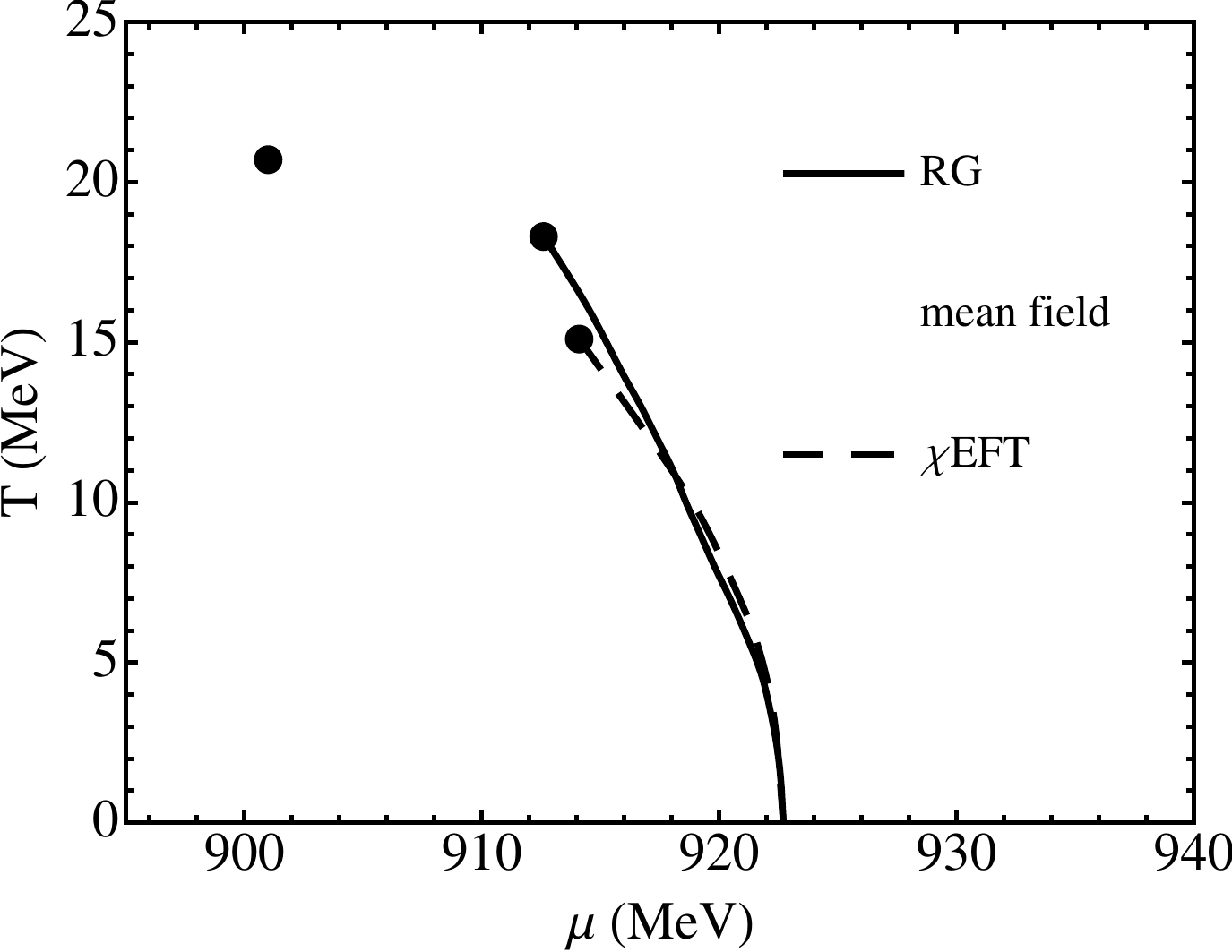}
\end{center}
\vspace{-.5cm}
\caption{Phase diagram of the liquid-gas transition in nuclear matter (from Ref.\,\cite{Drews2013}). Dotted curve: mean-field result of the chiral meson-nucleon model. Solid curve: FRG calculation including mesonic fluctuations and nucleonic particle-hole excitations. Dashed curve: in-medium chiral effective field theory calculation of Ref.\,\cite{FKW2012}.}
\label{fig:phasediagram}
\end{figure}

Once the effective potential $U_{k\rightarrow 0}$ is found by solving the FRG equations, all
thermodynamic quantities (pressure, energy density, proton and neutron densities) can be
calculated. The input parameters
of the potential are determined such that empirical nuclear matter properties (saturation density
binding energy, surface tension, compression modulus and symmetry energy) are reproduced
\cite{Drews2013,Drews2014}. Interesting physics questions to be studied are then the following:
how does the phase diagram for the nuclear liquid-gas transition look like in comparison to the
previously discussed results from chiral effective field theory calculations? What is the role of
fluctuations beyond the mean-field approximation in this context? And what does this
non-perturbative approach tell us about the density and temperature dependence of the chiral
condensate?

Examples of results are presented in the following. The phase diagram of symmetric
nuclear matter is shown in Fig.~\ref{fig:phasediagram}. It is instructive to see the effect of fluctuations
as compared to the mean-field approximation. It is also quite remarkable that the fully non-perturbative
FRG calculation based on the chiral meson-nucleon (ChMN) model produces a phase diagram close to
the one found in the three-loop calculation using in-medium chiral perturbation theory \cite{FKW2012}.

\begin{figure}[t]
\begin{center}
\includegraphics[height=6cm,angle=0]{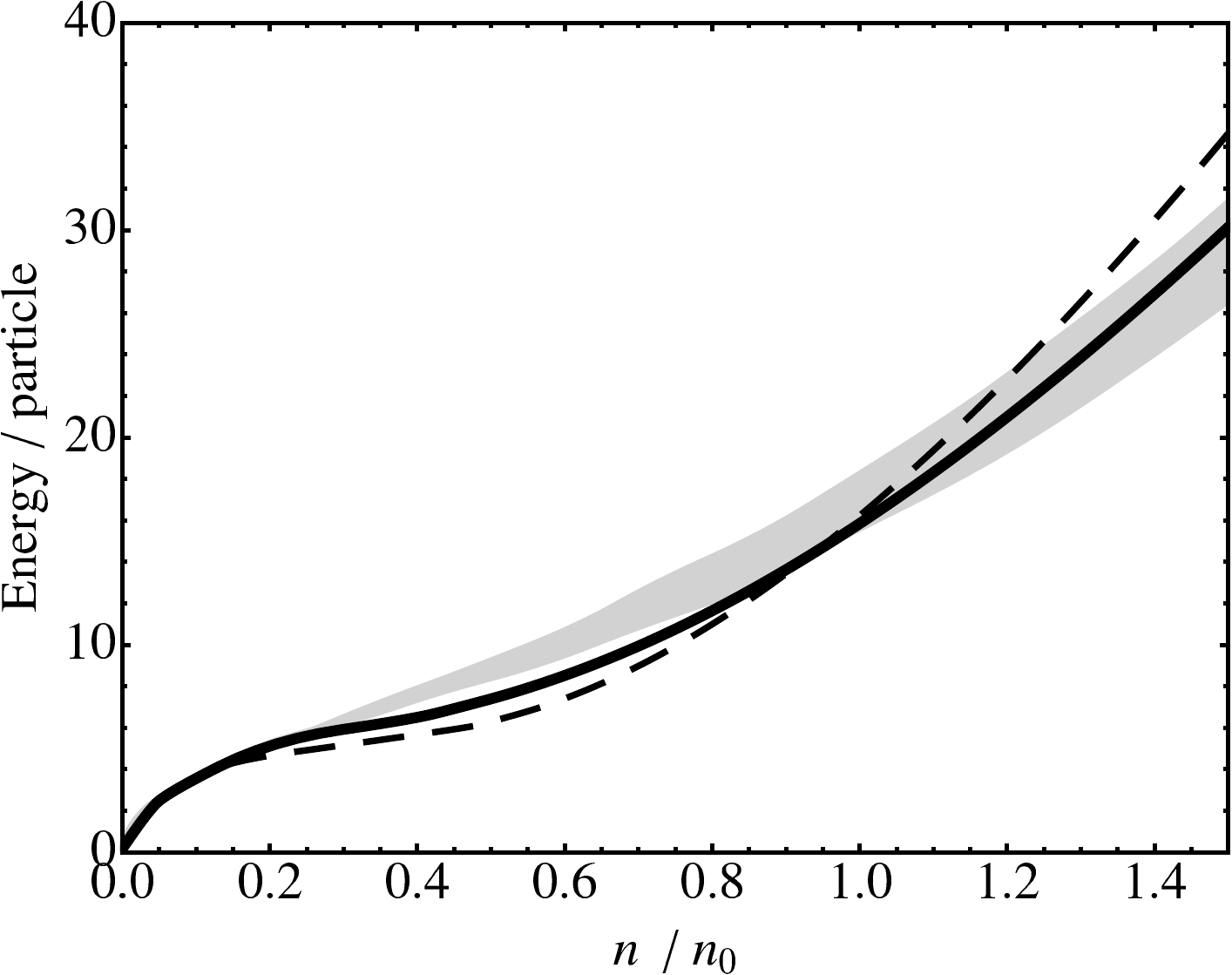}
\end{center}
\vspace{-.5cm}
\caption{Energy per particle of pure neutron matter as a function of density (in units of normal nuclear matter density, $n_0 = 0.16$ fm$^{-3}$) calculated using the FRG approach \cite{Drews2014}. The full FRG result (solid curve) is compared to the mean-field approximation (dashed line) and to the result from a chiral Fermi liquid theory calculation (grey band) \cite{holt13}. Here and in what follows, except otherwise noted, $n$ will replace $\rho$ for baryonic density.}
\label{fig:nmatter}
\end{figure}

Fig.~\ref{fig:nmatter} presents the energy per particle of neutron matter at zero temperature as a
function of density \cite{Drews2014}. The isovector-vector coupling strength $G_\rho$ has been
set to reproduce the symmetry energy, $E_{sym} = 32$ MeV. The mean-field result encounters the typical problem familiar from relativistic mean field models: $E/N$ comes out too small at low densities, while
at large densities, the equation-of-state is too stiff compared to realistic many-body calculations. However,
fluctuations generated by the FRG method greatly improve the behavior. There is now much better agreement
with results both from chiral Fermi liquid theory \cite{holt13} and from Quantum Monte Carlo (QMC) calculations \cite{GCRSW:2014}.

\begin{figure}[t]
\begin{center}
\includegraphics[height=5.8cm,angle=0]{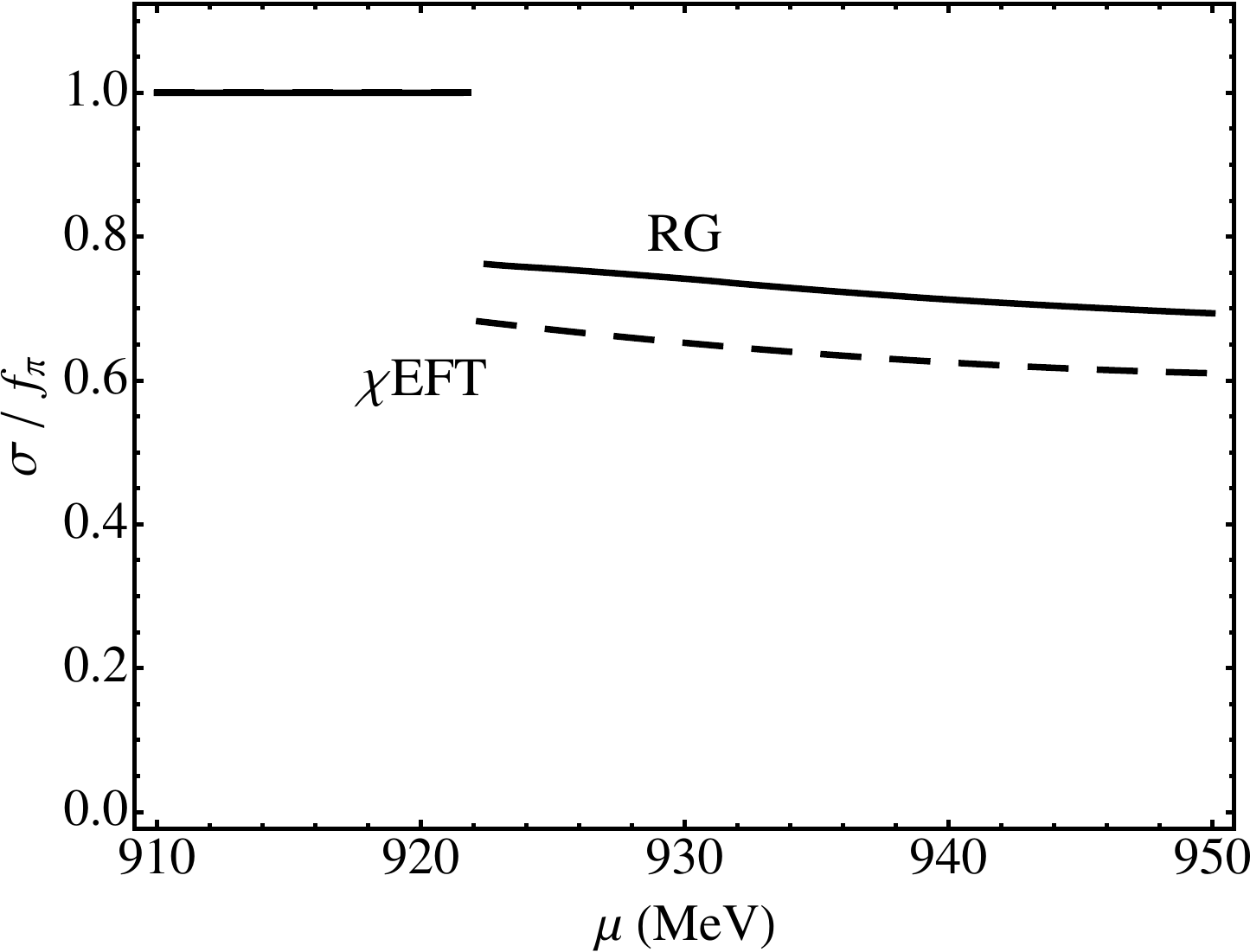}
\includegraphics[height=6cm,angle=0]{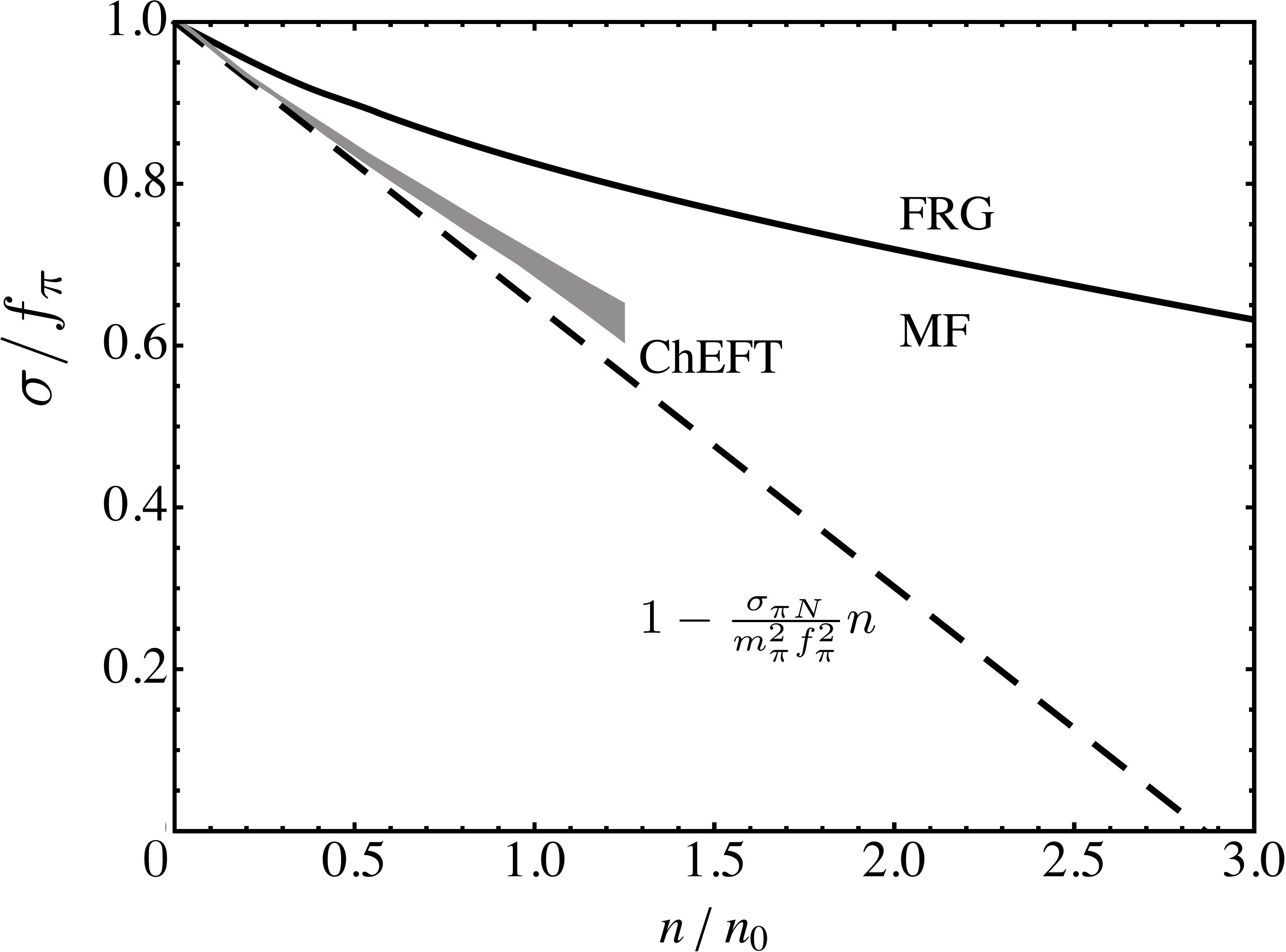}
\end{center}
\vspace{-.5cm}
\caption{Chiral order parameter (sigma field in units of the pion decay constant, $f_\pi = 92$ MeV) at zero temperature. Dependence on baryon chemical potential in symmetric nuclear matter (left) or on density in pure neutron matter (right). Left: Comparison of full FRG calculation \cite{Drews2013} with result from
3-loop chiral EFT \cite{FKW2012}. Right: leding-order linear behavior (dashed curve) in comparison to full FRG \cite{Drews2014}. Dotted curve: mean-field approximation using the chiral meson-baryon model.
Grey band: chiral EFT result \cite{Krueger2013}.}
\label{fig:condensate}
\end{figure}

A further point of interest is the behavior of the chiral condensate as a function of baryon density or
chemical potential, represented here by the expectation value of the $\sigma$ field.
Fig.~\ref{fig:condensate} shows results for this quantity in different situations: first, for symmetric nuclear
matter at zero temperature. In this case the liquid-gas transition and its coexistence region leaves its
prominent mark also in the chiral order parameter, as already described in the previous sections of the
present chapter. At chemical potentials $\mu > \mu_0 = 923$ MeV, corresponding to densities above that of normal nuclear matter,
the results derived from the non-perturbative FRG approach show again a close correspondence with those from a ChEFT in-medium perturbative calculation \cite{condpap}:
fluctuations involving Pauli principle effects and
many-body forces stabilize the chiral condensate against restoration of chiral symmetry at too low densities,
rendering approaches that work with nucleon and pion degrees of freedom valid up to $\sim 3$ three times the
density of normal nuclear matter. A qualitatively similar and even more pronounced feature is seen in neutron matter. The leading-order linearly decreasing tendency of the condensate with increasing density, controlled by the pion-nucleon sigma term $\sigma_{\pi N}$, is again stabilized. Deviations from linear behavior set in already at low densities, while the calculated value of the sigma term, $\sigma_{\pi N} = 44$ MeV, turns out to be consistent with the empirical $\sigma_{\pi N} = 45 \pm 5$ MeV. Fluctuations treated with the FRG method evidently play an important role in providing the stabilization mechanisms.

These chiral FRG calculations have been further extended to isospin-asymmetric nuclear matter \cite{DW:2015}. The computed evolution with changing proton fraction of the nuclear liquid-gas phase transition turns out to be remarkably similar to the pattern that emerged from the thermodynamics based on in-medium chiral perturbation theory, shown in Fig.~\ref{phaseasy}. One of the significant results of this extended FRG calculation concerns again the behavior of the $\sigma$ field as a function of density, as it evolves from symmetric via asymmetric matter to pure neutron matter at zero temperature. In compressed neutron matter, the mean-field calculation would suggest the appearance of a chiral first-order phase transition at about $n\sim 3 n_0$. However, the full FRG treatment incorporating fluctuations stabilizes the Nambu-Goldstone phase of spontaneously broken chiral symmetry up to densities beyond $5 n_0$, indicating that hadronic degrees of freedom may persist even in the deep interior of neutron stars. 


In Section \ref{half-skyrmion}, the same quantity as $\sigma/f_\pi$ of the right panel of Fig.~\ref{fig:condensate} is studied in $hls$EFT model. There the result differs from $\sim 2n_0$, indicating a novel phenomenon may be taking place due to a topology change that is not captured in ChEFT. This matter, which raises the puzzle as to where the proton mass comes from, will be addressed in the conclusion section.


\section{Hadrons in matter and chiral symmetry restoration: conceptual aspects}\label{hadrons-in-matter}
\label{him}
We have shown that the chiral effective field theory with only Goldstone bosons, with the baryon field introduced in a chirally symmetric way as a matter field, with the parameters of the {\em bare} chiral Lagrangian fixed in the vacuum, is found to be stunningly successful and provides a striking example of the working of Weinberg's folk theorem in nuclear physics. The question that is raised next is how far in density can the theory be pushed in addressing the state of compressed baryonic matter believed to be present inside large-mass compact stars. This requires going beyond the regime where experiments are available and play a crucial role in guiding the theory.

In extending an effective field theory to higher scales, in energy or density, one meets a stumbling block, namely that unless one has an ``ultraviolet (UV) completion," there is no unique way to make the {\em correct and unique} extrapolation. One prominent example in nuclear physics is the symmetry energy in the EoS for compact stars. Guided by experiments, various models that satisfy required symmetry as well as dynamical constraints can arrive at results that agree with each other and with experiments up to nuclear matter density. However when extrapolated beyond nuclear matter density to where there are no experimental data to guide the theory, the predictions turn out to vary extremely wildly, going all the way from very soft to very hard EoS. Lacking model-independent constraints from QCD, the EoS, especially,  the symmetry energy, is totally unknown in the density regime appropriate for compact stars.

What one does with ChEFT is that one starts with a theory constrained at the IR by low-energy theorems, i.e., soft-pion theorems etc., and go up to higher scale in the spirit of the Weinberg theorem. It works stunningly well at the pion mass scale at low density as well as in nuclear matter, up to $n_0$ or even to $\sim 2n_0$. A striking evidence for this is seen in the unprecedented precision in exchange current processes such as the thermal neutron capture and the solar proton fusion. But if the mass scale reaches that of vector mesons, it is very unlikely that ChEFT will continue to work without significant modifications. At low density, the vector-meson mass is much greater than the pion mass, thus no problem. However if the $\rho$ mass could be tweaked to come down to that of the pion, then extending ChEFT to that regime becomes highly problematic, if not impossible. In this section, we discuss how such a situation can arise at high density and/or high temperature, with a focus on high density. In order to address this possibility, we argue that vector mesons and a scalar dilaton need to be introduced, the former in hidden local symmetry and the latter in spontaneously broken conformal invariance. We describe in this section the effective Lagrangian -- called $hls$EFT to be distinguished from ChEFT treated above -- that incorporates the pion, the vector mesons and the dilaton consistently with chiral symmetry -- to which spontaneous breaking of scale symmetry is intimately connected.

One important aspect of hidden local symmetry is that $hls$EFT can be matched to QCD in terms of correlators at a matching scale $\Lambda_M\sim \Lambda_\chi$. The matching provides a sort of UV completion to a UV fixed point, called the ``vector manifestation (VM) fixed point," at which the vector meson mass vanishes. This endows the parameters of $hls$EFT with the dependence on the quark and gluon condensates with specific properties manifested as the chiral restoration point is reached. The condensates, particularly the quark condensate $\la\bar{q}q\ra$, depend on the vacuum, and in a dense medium on the density. This renders the parameters of the ``bare" $hls$EFT {\it necessarily} density-dependent. We will identify this {\it intrinsic density dependence} with BR scaling. At very low energy/density scales where the vector mesons do not intervene, one can simply integrate them out. Then we will wind up with a ChEFT with the parameters {\it intrinsically} density-dependent, inherited from the matching, even if vector mesons are explicitly present. We might call this Lagrangian ChEFT$^*$. It is found that this simplified $hls$EFT, ChEFT$^*$,  in an appropriate approximation works well in nuclei.

In what we have discussed so far, ChEFT is found to work superbly well, except for some exceptions, up to, and slightly above, nuclear matter density without the  ``intrinsic" density dependence identified with BR scaling that is inherited from QCD. Does this mean that BR scaling is absent or is it simply buried in the parameters of ChEFT? This situation is not unlike the case of meson-exchange currents in nuclei. Until the soft-pion current was identified and exhibited as discussed in Section \ref{MEC}, the general belief in some circles of nuclear physics was that there was no need for meson-exchange currents. The observables could be explained simply with a bit of fiddling with the many parameters present in the models.  We will see that the situation is somewhat similar with BR scaling.

\subsection{Defining Brown-Rho (BR) scaling}\label{BRscaling}

What is referred to in the literature as Brown-Rho scaling (BR scaling for short)~\cite{br91}  was originally formulated with the Skyrme Lagrangian that consists of the current algebra term quadratic in derivatives and the quartic Skyrme term. Then implementing a scalar dilaton field $\chi$ and relying on certain current algebra relations such as, among others, KSRF relation for the vector meson mass (both $\rho$ and $\omega$ assuming $U(2)$ flavor symmetry), GMOR relation, Goldberger-Treiman relation etc., how the light-quark hadrons behave in nuclear medium reflecting the {\em intrinsic} change of vacuum structure, signaling how chiral symmetry is manifested in nuclear matter and compressed baryonic matter was predicted. What was at issue there was the in-medium properties of light-quark hadrons involved in nuclear physics such as the nucleon, scalar meson and vector mesons, specially the $\rho$ meson. Much attention was focused on the $\rho$ meson, because it figured importantly in nuclear tensor forces which affect a variety of nuclear processes, including the saturation phenomenon, and because its properties could be explored in experimental activities at both heavy-ion and electron accelerators. Although there was a flurry of experimental work searching for the works of chiral symmetry in hadronic matter under extreme conditions, a large number of these experiments misinterpreted the implication of the scaling given in Ref.\ \cite{br91}, leading to a wide confusion in the field. In this chapter we will clarify some of these issues.

With the standard chiral Lagrangians\footnote{By ``standard chiral Lagrangian" we mean the Lagrangian with pions only without other meson degrees of freedom, namely what's referred to as ChEFT.}, other mesons such as the vector and scalar mesons are to be generated dynamically through multi-pion exchange processes involving quantum loops. Such an effective theory is likely to break down at the scale at which such mesons figure in the dynamics explicitly. Now suppose, as we will conjecture, that the dynamically generated $\rho$ meson mass decreases in the approach to chiral restoration. In this case it should be treated as a light degree of freedom on equal footing with the pion. The $\rho$ meson should then not be ignored or integrated out. In fact this was the underlying reason for formulating BR scaling. In what follows, we therefore will develop BR scaling with the vector mesons taken as explicit degrees of freedom.

Flavor $SU(N_f)$ is not a gauge symmetry in QCD. Hence vector mesons with such flavor symmetry can be introduced in a variety of different ways as effective phenomenological fields, all giving equivalent physics at low energy in free space. In dealing with the medium changes to the vacuum structure, however, the most natural way to introduce the vector mesons as dynamical fields is to have them be endowed with hidden local symmetry~\cite{HLS,HY:PR}. A natural interpretation is to consider this symmetry as ``emergent" and not intrinsic to QCD. We will find that this is the only way to arrive at  the situation where the pion and the $\rho$ can become degenerate, a scenario that is encoded in BR scaling, thereby distinguishing HLS theory from other phenomenological models without gauge symmetry.

It should be understood how the quark condensate, the order parameter of chiral symmetry, enters into BR scaling. In fact, as we will stress throughout this chapter, the BR scaling that figures in physical observables does not exhibit transparently the property of the quark condensate. In the HLS framework that we will adopt, the Wilsonian matching of the vector and axial-vector correlators of the ``bare" HLS Lagrangian and the operator product expansion (OPE) of QCD at a matching scale $\Lambda$ makes the bare parameters of HLS {\em dependent} on the quark condensate (as well as other quantities) of QCD. The effective HLS Lagrangian is evolved through the Wilsonian renormalization group from $\Lambda$ to the energy scale where physics is studied. The resulting Lagrangian will then {\em necessarily} depend on the quark condensate, which itself depends on the density of the baryonic system. This density dependence of the Lagrangian is called the ``intrinsic density dependence" to  distinguish it from other sources -- such as many-body interactions -- of density dependence.  Now it remains to show how BR scaling is obtained from this HLS Lagrangian. This will be clarified below.

There are two ways that HLS can figure in the strong interactions of nuclear physics. One is that it appears as an ``emergent" symmetry -- bottom up -- and the other a ``reduced symmetry" from a higher energy scale -- top down.

\subsection{Hidden local symmetry (HLS)}\label{HLS}

\subsubsection{\it HLS as ``emergent symmetry"}

We will consider the two-flavor chiral symmetry $SU(2)_L\times SU(2)_R$ for simplicity\footnote{In what follows in this section, unless otherwise noted, we will ignore quark masses and take the chiral limit.}. It is straightforward to go to the three-flavor case considered later.
Start with the current algebra term of the non-linear sigma model applicable at near zero energy scale,
\be
{\cal L}=\frac{f_\pi^2}{4} \Tr\ \del_\mu U^\dagger \del^\mu U +\cdots.
\ee
Given the $SU(2)_L\times SU(2)_R$  chiral symmetry, the chiral field $U$ can be written as a product of left (L) and right (R) fields,
\be
U =e^{2i\pi/f_\pi} = \xi_L^\dagger \xi_R\label{U}
\ee
transforming
\begin{equation}
U \to g_L U g_R^\dagger\,
\end{equation}
where $g_{L,R} \in \left[ SU(2)_{L,R}\right]_{\rm global}$.  Since the $U$ field is given by the product, there is an inherent redundancy since one can always insert $h^\dagger (x)h (x)=1$ without changing anything,
\be
U= \xi_L^\dagger [h^\dagger (x) h(x)] \xi_R\label{emergent}
\ee
and have the invariance under the local transformation
\begin{equation}
\xi_{L,R} \to h\,\xi_{L,R}\,g_{L,R}^\dagger.
\end{equation}
One can elevate this redundancy to a local gauge symmetry by introducing a gauge field\footnote{Later we will include the $\omega$ field in a $U(2)$ symmetric way.} $V_\mu$ that transforms as
\be
V_\mu \to h\,(V_\mu\ + i\partial_\mu)  h^\dagger\, .
\ee
As it stands, $h$ is a totally arbitrary unitary matrix function. If one chooses it to be
\be
h=h(\pi(x), g_L, g_R) \in \left[ SU(2)_{L+R}\right]_{\rm local},
\ee
then the corresponding local theory is gauge equivalent to the non-linear sigma model~\cite{HLS}. If one parameterizes
\begin{eqnarray}
\xi_{L,R}= e^{i\sigma/{f_{\sigma}}}e^{\mp i\pi/{f_\pi}},
\end{eqnarray}
then the non-linear sigma model is obtained by the unitary gauge $\sigma=0$.

Making the vector field dynamical by introducing the kinetic energy term, the HLS Lagrangian in the chiral limit takes the form
\begin{eqnarray}
{\mathcal L}_M
 = - \frac{1}{2g^2}\mbox{tr}\left[ V_{\mu\nu}V^{\mu\nu} \right] + f_\pi^2\mbox{tr}\left[ \hat{\alpha}_{\perp\mu}   \hat{\alpha}_{\perp}^{\mu} \right]
+ f_{\sigma}^2\mbox{tr}\left[ \hat{\alpha}_{\parallel\mu}
  \hat{\alpha}_{\parallel}^{\mu} \right],
\label{lagmeson}
\end{eqnarray}
with $V^{\mu\nu} = \partial^\mu V^\nu - \partial^\nu V^\mu - i\left[ V^\mu, V^\nu \right]$. In terms of the isovector $\rho$ field, $V_\mu=\frac 12 \vec{\rho}\cdot\vec{\tau}$. Here the Maurer-Cartan 1-forms
\begin{eqnarray}
\hat{\alpha}_{\perp }^{\mu}
&=& \frac{1}{2i}\left[ D^\mu\xi_R \cdot \xi_R^{\dagger}
{}- D^\mu\xi_L \cdot \xi_L^{\dagger} \right]\,,
\nonumber\\
\hat{\alpha}_{\parallel}^{\mu}
&=& \frac{1}{2i}\left[ D^\mu\xi_R \cdot \xi_R^{\dagger}
{}+ D^\mu\xi_L \cdot \xi_L^{\dagger} \right]\,,
\end{eqnarray}
given with the covariant derivatives of $\xi_{L,R}$
\begin{eqnarray}
D^\mu \xi_{L,R}
 = (\partial^\mu - iV^\mu) \xi_{L,R}
\end{eqnarray}
transform homogeneously,
\begin{equation}
\hat{\alpha}_{\perp,\parallel}^\mu
\to  h\, \hat{\alpha}_{\perp,\parallel}^\mu\, h^\dagger\, ,
\end{equation}
 so the Lagrangian Eq.\ (\ref{lagmeson}) is manifestly locally gauge invariant. The gauge field so generated is a ``hidden" local field emerging from strong interactions.

Now given that one can put any number of redundancies in writing the $U$ field, one can then generate an infinite number of hidden local fields. By a suitable construction, one can elevate the theory of infinite number of gauge fields to a gauge theory in one dimension higher, i.e., ``deconstructing"  5D in our case. The resulting dynamics can be encoded in a 5D Yang-Mills action in a background of curved space-time metric with a dilaton~\cite{son-stephanov}.

The recent application of string theory to QCD showed that  a similar 5D YM action in a background with certain space-time metrics can arise from gravity-gauge dualities in string theory~\cite{SS2}. The resulting 5D action is of the same form as the dimensionally deconstructed one of Ref.\ \cite{son-stephanov}, the main difference being in the curvature metric, encoding different UV completions. We will have more to say about this development later. It suffices to mention here that  the 5D YM action in the limit of large $N_c$ and large 't Hooft constant $\lambda=N_c g_s^2$ (where $g_s$ is the QCD gauge coupling), gives rise, when Klein-Kaluza-reduced to 4D, to an infinite tower of hidden gauge fields in the presence of Goldstone fields, i.e., the pions. In this article, unless otherwise noted, HLS will typically refer to the lowest-lying vector fields $\rho$ and $\omega$ when $U(2)$ symmetry is assumed or the $\rho$ field alone is considered in specific cases. The role of the lowest-lying axial field $a_1$ will be considered toward the end of this section.

\subsubsection{\it HLS and fractionization}
\label{fraction}

The emergence of local gauge symmetry in the form Eq.\ (\ref{emergent}) hints at the possible fractionization of internal symmetries in the strong interactions. Indeed, when skrymions of the HLS Lagrangian are put on a crystal lattice to simulate dense baryonic matter in terms of skyrmions (to be discussed in Chapter \ref{skyrm}), the single skyrmion in medium fractionizes into two half skyrmions at a certain density denoted $n_{1/2}$ lying higher than the normal nuclear matter density $n_0$. Such fractionization phenomena are fairly well-known to take place in condensed matter physics, albeit at lower dimensions (see e.g., Ref.~\cite{multifacet}). For instance if one writes an electron field $\psi$ as a product of two local fields, one bosonic $b(x)$ and the other fermionic $f(x)$ as $\psi (x)=b(x)f(x)$, introducing local gauge redundancy, then the electron can fractionize into two quasiparticles -- holons and spinons -- which independently carry the flavor and the spin of the electron. Such phenomena are experimentally observed in a variety of condensed matter systems, exhibiting essential roles of emergent gauge fields. In fact, a condensed matter case resembling the half-skyrmion structure is the ``deconfined quantum critical phenomenon"~\cite{senthil} for the phase transition from the N\'eel antiferromagnet to the VBS (valence bond solid) phase -- with a broken symmetry different from that of N\'eel -- in 2 spatial dimensions. When the N\'eel field $\hat{n}$ is written in the CP$^1$ parametrization as $\hat{n}=z^\dagger\vec{\sigma}z$, the $U(1)$  gauge field that arises from the redundancy plays a crucial role in mediating the fractionization of the skyrmion in the N\'eel phase into two half skyrmions at the critical point where the phase transition to the VBS state takes place. Here there is no local order parameter field that describes the phase transition, so the transition does not belong to the standard Landau-Ginzburg-Wilson paradigm.

In the case of skyrmions in dense matter, up to date, there is no experimental verification of the fractionization phenomenon. In fact, the appearance of half-skyrmions is seen in the Skyrme model itself with no gauge fields present, so one cannot say whether the emergent gauge fields are essential or not, and if essential, what their role is. In the 5D YM theory with the infinite number of 4D HLS fields mentioned above, the baryon is an instanton and it is known that the instanton in dense medium fractionizes into two dyons~\cite{RSZ}. This indicates the robust nature of emerging half-soliton quasiparticles.  We will see in Chapter \ref{skyrm} that by putting  skyrmions on an FCC crystal the half-skyrmions appearing at a density $n_{1/2}\sim 2 n_0$ can have a drastic effect on nuclear tensor forces and hence on the EoS of dense compact-star matter.

\subsection{Relativistic mean field and BR scaling}\label{RMFDilaton}

The degree of freedom that does not find a natural habitat in the HLS framework is the scalar excitation associated with the broken scale invariance tied to the trace anomaly of QCD. The trace anomaly is, in the chiral limit, given by the gluon condensate. The scalar that would be involved would then be mostly, if not all, gluonic and is most likely heavy, so would be expected not to enter in low-energy dynamics. Therefore low-mass scalars observed in nature at low energy must be generated in pionic interactions. This means that at low energy, a dilaton, the pseudo-Goldstone of scale symmetry, can figure only if it is a strongly mixed state of  quarkyonic and gluonic components.  These two components were identified in \cite{BR:DD} with ``soft" glue and  ``hard" glue, the former participating in the spontaneous breaking and the latter in the explicit breaking of scale invariance.  Whether this identification is correct or not is not known at the moment and remains a problem to be resolved. This highly intricate structure of scalar in the strong interactions
shares its mysterious nature with other areas of physics, say, particle physics with the Higgs boson and cosmology with the dilaton (associated with conformal invariance), among others.

Now in nuclear physics, the Walecka-type mean-field theory which seems to work surprisingly well requires a chiral-singlet scalar field that lies low at a density commensurate with that of the equilibrium nuclear matter, with a mass in the vicinity of that of vector mesons.  This indicates that in medium, there must be an intricate interplay between the spontaneous breaking and explicit breaking of scale symmetry. A priori, the dilaton does not belong to any visible multiplet with the other mesons ($\pi$, $\rho$, $\omega$ ...) we are concerned with. However, following the ``mended symmetry" argument of  Weinberg~\cite{weinberg-mended}, we will assume that the dilaton\footnote{It is denoted $\epsilon$ and identified as $\sigma$ of the linear sigma model by Weinberg. How the chiral scalar $\chi$ transforms to the chiral four-vector $\sigma$ is described in \cite{vankolck,SLPR}.} which we denote as $\chi$ joins the $\pi$, $\rho$ and $a_1$ as chiral restoration is approached~\cite{BR:DD}. This assumption leads to the in-medium scaling relation known as Brown-Rho scaling to be discussed later in this section.

A unified approach to dense baryonic matter would be to have baryons appear as solitons and treat both baryons and mesons on the same footing with the single HLS Lagrangian of the form in Eq.~(\ref{lagmeson}). This approach is at an early stage of development, the recent preliminary results of which will be discussed in Chapter \ref{skyrm}.
Here we write an HLS Lagrangian in which the baryon fields are put in explicitly in place of the skyrmion in conformity with the Weinberg (folk) theorem. We may call it baryon-HLS (BHLS for short). In nuclear dynamics a scalar excitation which in principle could be generated by quantum loop effects (involving pions and intermediate-state $\Delta$ excitations)\footnote{This is similar to the scalar $\sigma$ discussed in ChEFT in Section \ref{delta-sigma}.} from the BHLS plays an absolutely crucial role. Without it, there can be no binding of nucleons into nuclei. In principle, one could do high chiral-order perturbation theory calculations with BHLS as one does in chiral perturbation theory with the standard chiral Lagrangian discussed in Chapter \ref{nceft}. Here we will bypass this complication and resort to a mean field theory using the given BHLS Lagrangian.

The underlying assumption we make is that nuclear matter is at the Fermi-liquid fixed point in the sense of Landau Fermi liquid theory and that the relativistic mean field approximation with a Lagrangian possessing the full relevant degrees of freedom and symmetries corresponds to doing Landau Fermi liquid fixed point theory. We will return  to this matter in more detail below.
Coupling baryon fields to the HLS Lagrangian in Eq.~(\ref{lagmeson}), the preservation of hidden local symmetry is straightforward. What is subtle is how to introduce a requisite scalar field into the Lagrangian to do the mean field calculation. There is no well established way to do this in the framework of hidden local symmetry. Here, we will follow the procedure adopted in Ref.~\cite{br91}. The key observation exploited there is that a chiral scalar field figures in the trace anomaly of QCD, and this scalar associated with the spontaneous breaking of scale symmetry involving a dilaton is tied intricately to the spontaneous breaking of chiral symmetry.

\subsubsection{\it Trace anomaly: ``Soft" and ``hard glue"}\label{glue}
The trace anomaly is given entirely by the gluon condensate $G^2\equiv \la\Tr \, G_{\mu\nu}G^{\mu\nu}\ra$ when the quark masses are ignored.  This gluon condensate is non-zero all the way to the asymptotic limit of QCD where the gauge coupling $g_s$ vanishes. Thus it remains nonzero even when the quark condensate vanishes at chiral restoration (in the chiral limit). This condensate can be ``measured" as a function of temperature in lattice QCD. It is found (see the discussion in Ref.~\cite{BR:DD}) that as the temperature nears the critical temperature $T_c$ for chiral restoration, the gluon condensate drops rapidly within a small temperature window and then stays non-zero above the critical temperature. About half of the zero-temperature condensate remains unmelted above $T_c$. This suggests the separation of the condensate into $G^2 (T)=G^2_{soft} (T) +G^2_{hard} (T)$ such that $G^2_{soft} (T_c)\approx 0$ and $G^2_{hard} (T_c)\approx \frac 12 G^2 (0)$. There is no lattice information on how the gluon condensate behaves in a baryonic medium. However, it seems reasonable to suppose that the behavior should be similar in density, at least qualitatively if not quantitatively. In what follows, we will assume that $G^2_{soft}$ goes to zero at the chiral restoration density $n_c$ while leaving $G^2_{hard}$ non-zero as in the finite-temperature case.

In Ref.~\cite{br91}, it is this behavior of  $G^2_{soft}$ that was proposed to be tied to the behavior of hadron masses in medium. Since the trace of the energy-momentum tensor is tied to scale symmetry, the vanishing of $G^2_{soft}$ could be associated with the restoration of spontaneously broken scale symmetry and can therefore be phrased in terms of a dilaton field, denoted as $\chi$, whose condensate signals the manifestation of scale symmetry.
As is well-known, the spontaneous breaking of scale symmetry is not possible in the absence of an explicit breaking. The potential is flat without explicit breaking~\cite{freund-nambu}. One natural candidate for the source of explicit scale symmetry breaking in QCD is the non-vanishing gluon condensate above the chiral critical point. It is not known how this produces the precise potential that triggers the spontaneous breaking. What is known is that if the explicit breaking is a small perturbation, then one can develop a Coleman-Weinberg type potential used in the literature~\cite{conformalon}. For our purpose the precise form is not needed. We will just denote it as $V(\chi)$.

Now the procedure for introducing the scalar we need is as follows. Under the scale transformation $x\rightarrow \lambda x$, the fields involved transform with canonical dimension, i.e., the boson fields $\phi (x)\rightarrow \lambda \phi (\lambda x)$ and the fermion fields $\psi (x)\rightarrow \lambda^{3/2}\psi (\lambda x)$. Write the action as the sum of terms
\be
S=\int d^4x \sum_i c_i {\cal O}_i \, ,
\ee
where $c_i$ is a constant and ${\cal O}_i$ is an object consisting of a product of field operators, derivatives etc. of total scale dimension $d_i$. Then the simple way of incorporating non-linearly realized scale invariance is to add a field $\chi (x)$ that serves as a ``conformal compensator"~\cite{conformalon}, also called ``conformalon" for short,
\be
S_{SI}=\int d^4x \sum_i \left (\frac{\chi}{f_\chi} \right)^{4-d_i}c_i {\cal O}_i\, ,
\label{scalaraction}
\ee
where $f_\chi$ is the scalar condensate in the zero-density vacuum, $f_\chi=\la\chi\ra|_{n=0}$. This action is scale-invariant by construction in the chiral limit. For explicit chiral symmetry breaking, i.e., the quark mass term that gives the pion a mass, the same prescription applies with the  quark mass matrix replaced by a spurion field with $d=1$. Because of the spurion, scale symmetry is {\em explicitly} broken by the quark mass term.
The scale invariance in the action of Eq.~(\ref{scalaraction}) is broken spontaneously by the potential $V(\chi)$ which has the minimum at nonzero condensate $\la\chi\ra=f_\chi\neq 0$. The scale symmetry is restored as $G_{soft}\to 0$ if the potential is driven (e.g., by temperature or density) so that  $\frac{d}{d\la\chi\ra}V(\la\chi\ra)|_{\la\chi\ra=0}\to 0$.

Now imagine that the Lagrangian so constructed is embedded in a medium characterized by the density $n$. Since the scale symmetry is spontaneously broken, the $\chi$ field condenses in medium with the VEV, $\la\chi\ra_n =f_\chi^*$. Expand $\chi$ around the VEV, $\chi=f_\chi^* + s$, denoting the fluctuating scalar dilaton field by $s$. The Lagrangian, when expanded, takes the form
\be
&&{\mathcal L}
= {\mathcal L}_N + {\mathcal L}_M
{}+ {\mathcal L}_\chi\,,\label{Efflag}
\\
&&{\mathcal L}_N
= \bar{N}(i\gamma_\mu D^\mu
{}- \Phi m_N + hs)N
{}+ g_A \bar{N}\not{\hat{\alpha}}_\perp\gamma_5 N
{}+ g_V \bar{N}\not{\hat{\alpha}}_\parallel N\ +\cdots,,
\\
&&{\mathcal L}_M
= - \frac{1}{2g^2}\mbox{tr}\left[ V_{\mu\nu} V^{\mu\nu} \right]\ + \Phi^2 m_V^2  \mbox{tr}V^2+\cdots,,
\\
&&{\mathcal L}_\chi
= \frac{1}{2}\partial_\mu s \cdot \partial^\mu s
{}+ \frac 12 \Phi^2 m_s^2 s^2 +\cdots
\ee
where $N$ represents the nucleon field and the ellipsis stands for higher derivative and higher $s$ (``dilaton") fields. We have added the dilaton kinetic energy term to make it dynamical. It follows then that we can write the mass parameters of Eq.~(\ref{Efflag}) that depend on density as
\be
m_N^*/m_N\approx m_V^*/m_V\approx m_s^*/m_s\approx \Phi ,
\label{BR}
\ee
with $\Phi(n)=f_\chi^*/f_\chi=\la\chi\ra^*/\la\chi\ra$. From the pionic part of the Lagrangian, one finds that the pion decay constant is also given by
\be
f_\pi^*/f_\pi\approx \Phi.\label{fstar}
\ee
The relations given in Eqs.~(\ref{BR}) and (\ref{fstar}) define BR scaling as it was first written down in Ref.~\cite{br91}. As stressed there this is a relation connecting the parameters of the Lagrangian to the spontaneous breaking of scale invariance which is valid in the mean-field approximation, applicable in the vicinity of the Fermi-liquid fixed point, i.e., normal nuclear matter.

{\it Note added}: After this review article was completed and uploaded on arXiv, a conceptually superior and potentially more systematic approach than what is described above to the {\it locking} of chiral symmetry to scale symmetry, exploited for nuclear matter as well as for dense matter, appeared in the literature. The idea put forward by Crewther and Tunstall~\cite{CT} is based on the conjectured presence, appealing though as-yet unconfirmed by lattice calculations, of an infrared(IR) fixed-point for three flavors in the QCD beta function for the gauge coupling constant, which, together with the quark masses,  dictates the explicit scale symmetry breaking. This new scenario does not affect the discussion made in this paper up to nuclear matter density, entirely consistent with the Brown-Rho scaling in its 1991 form. However it could very well have a profound impact on the dense compact star matter treated below. The new development coming from this new approach vis-a-vis with what's treated in this and other sections is discussed in \cite{paengetal}.

\subsubsection{\it Relativistic mean field theory and Landau Fermi-lquid theory}\label{rmf-landau}
The Lagrangian in Eq.~(\ref{Efflag}) is essentially Walecka's relativistic mean field theory Lagrangian (including the isovector fields $\pi$ and $\rho$) used for nuclei and nuclear matter~\cite{walecka}, except that the parameters vary in a specific way with the density, given in Eq.~(\ref{BR}). To confront nature with this Lagrangian, one may notice that the mean field approximation of this type corresponds~\cite{matsui} to Landau Fermi liquid theory.  The reason why nuclear matter is well described by Fermi-liquid theory is because nuclear matter at the equilibrium density $n_0$ is at the Fermi liquid fixed point~\cite{shankar94}.
In fact, it has been shown~\cite{song} that the Lagrangian in Eq.~(\ref{Efflag}) with a suitable scaling defined in Eq.~(\ref{BR}) with an $\omega$ field for $V$ can describe nuclear matter quantitatively well, even better than Walecka's linear mean field model, which gives too high a compression modulus.\footnote{To improve on the linear RMF model, usually higher-dimension field operators are added with the relevant constraints such as chiral symmetry, naturalness etc.\ taken into account. At low densities BR scaling does away with such high dimension operators, simplifying the matter a great deal~\cite{RMF}.}
\vskip 0.5cm
\begin{figure}[th]
\begin{center}
\includegraphics[width=10cm,angle=-90]{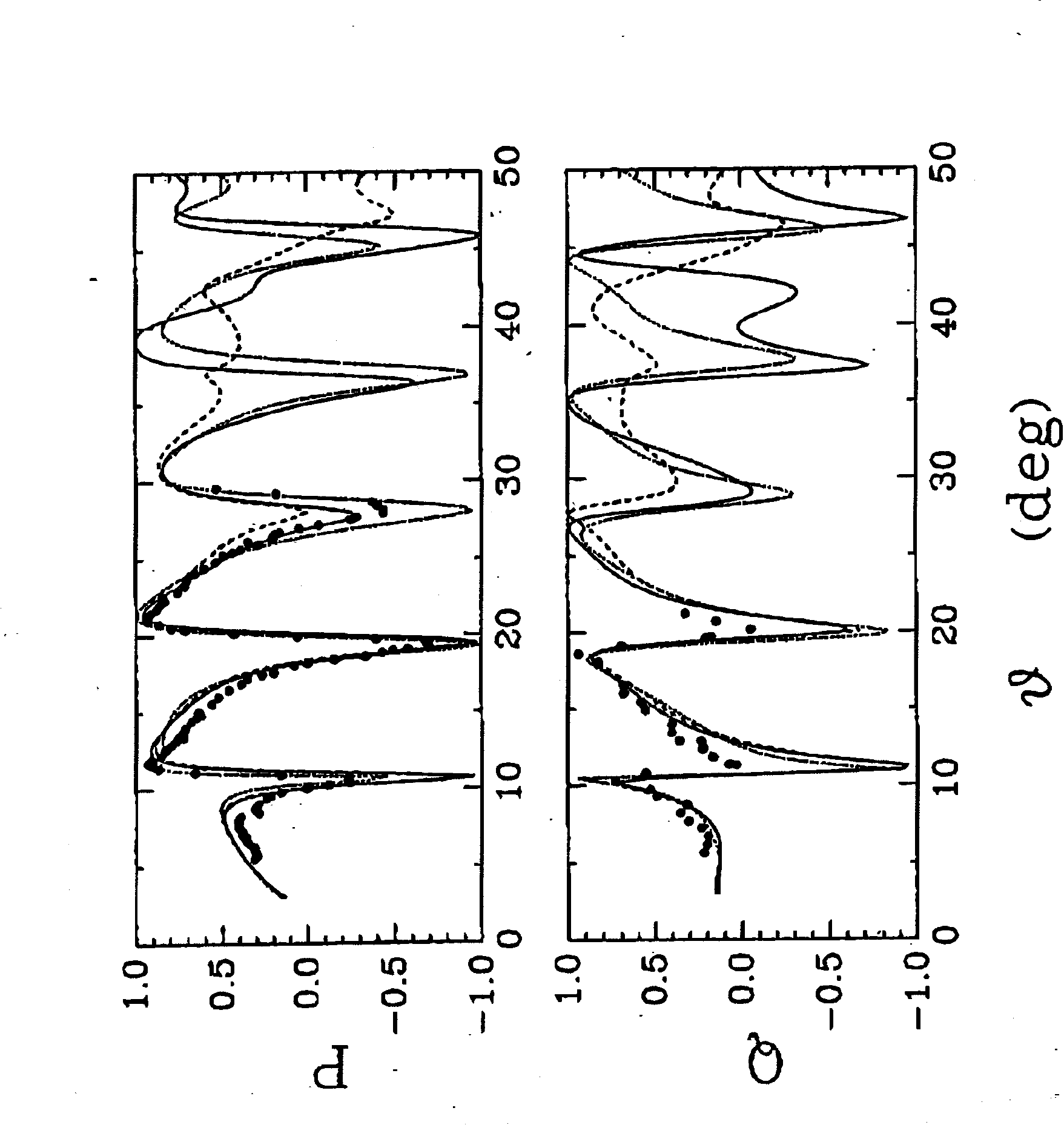}
 \caption{The spin observables for
500 MeV protons scattered elastically from $^{40}$Ca. The dashed line is for the case where the $\omega$ and the dilaton $s$ scale as in Eq.~(\ref{BR}) but the $\rho$ is unscaled. Note that it starts deviating strongly from the others beyond about 30 degrees.}
\label{fig2}
\end{center}
\end{figure}

One of the immediate questions raised when the scaling relation was put forward was whether it would upset the good agreement with experiments of some of the observables obtained without the scaling. A highly nontrivial consistency check of the BR-scaling mean-field model was the spin observables for 500 MeV protons scattered elastically from $^{40}$Ca~\cite{hintz}. The successful fit of the spin observables $P,Q$ up to the angle $\sim\theta=30$ degrees stands as a benchmark for Dirac phenomenology anchored on the mean field theory without BR. It has been shown that the Lagrangian in Eq.~(\ref{Efflag}) with the mass scaling Eq.~(\ref{BR}), suitably parameterized, not only fits the differential cross section up to 35 degrees measured, but also works as well as the Walecka model for the measured spin observables. This is shown in Fig.~\ref{fig2} taken from Ref.~\cite{hintz}. Note that if the $\rho$ meson is unscaled while other mesons are, the predicted spin observables differ strongly from the BR-scaled prediction.

An issue that has raised controversies regarding the notion of scaling is how it is related to the property of chiral symmetry in nuclear and dense medium.  As seen in Eq.~(\ref{fstar}), the in-medium pion decay constant is controlled by the dilaton condensate, which as it stands, is not explicitly connected to the quark condensate, the presumed order parameter of chiral symmetry. This apparent disconnection was not correctly interpreted in Ref.~\cite{br91} and led to a wide confusion among those looking for the signal -- in terms of BR scaling -- for chiral restoration at high temperature or density. This issue is related to the next question of how far in density the mean-field approximation can be applied. Although one can do chiral perturbation theory with the Lagrangian given in Eq.~(\ref{Efflag}) and calculate loop corrections to the mean-field approximation, this would not provide an answer to the question of whether nonperturbative phenomena -- such as meson condensation, a Fermi-liquid--non-Fermi-liquid transition etc. -- intervene at higher density. We will see indeed that possible topological phase changes can take place at some high density and make the mean-field approximation break down. This would then invalidate the scaling relation since it is anchored on the Fermi-liquid structure.

\subsection{Vector manifestation and $m_\rho^*$ as the order parameter for chiral symmetry}\label{vector-manisfestation}
The most important point to recognize in what follows is what is called ``vector manifestation" (VM for short) at chiral restoration predicted by HLS theory~\cite{HY:PR}. It has been shown that regardless of what happens in between, when matched \`a la Wilsonian to QCD in terms of correlators at some scale $\Lambda_M$ above the $\rho$ mass\footnote{This amounts to doing a sort of UV completion of the effective theory to QCD, which is facilitated when a gauge symmetry -- here HLS -- is present.}, the $\rho$ mass must go to zero as the chiral transition point is approached:
\be
m_\rho^*/m_\rho\to g_V^*/g_V\propto (\la\bar{q}q\ra^*/ \la\bar{q}q\ra)\to 0.\label{vm}
\ee
This is an unambiguous prediction of hidden local symmetry provided that the notion of local fields makes sense in the vicinity of chiral restoration.
{\it Note that it is here that the scaling of mass makes contact with the scaling of the quark condensate.} This relation is established in HLS without other fields, such as $a_1$, dilaton $\chi$ etc. As explained below, arguments based on ``mended symmetries" suggest that very near the chiral restoration point, the relation in Eq.~(\ref{vm}) could still hold in the presence of $a_1$ as well as a scalar $\sigma$ -- which is related to the dilaton $\chi$ mentioned below.

It is significant that Adami and Brown proposed even before the VM of HLS was discovered that  it is the $\rho$ mass that is the bona-fide signal for chiral transition, independently of how the $\rho$ mass is related to the pion decay constant~\cite{adami-brown}. Combined with the argument of HLS and the role of the dilaton condensate in BR scaling, this would suggest that what happens near nuclear matter -- where measurements are made -- which is far away from the vector manifestation (VM) fixed point, cannot give a bona-fide signal for what is called ``partial restoration" of chiral symmetry that has been searched for in the literature. It will be seen in the skyrmion crystal description (Chapter \ref{skyrm}) that the effective pion decay constant in medium $f_\pi^*$ {\it does not} signal the property of chiral symmetry in medium when density exceeds $\sim n_0$.
BR scaling {\it per se} is seen in a variety of processes in nuclear medium as we will discuss, but {\it when interpreted in terms of HLS -- or equivalently of BR}, there are no direct signals for the behavior of the quark condensate itself, except perhaps at the chiral restoration point.

\subsection{Dilaton and mean field approximation}\label{dilaton}
There is an intricate relation between the structure of the dilaton and the validity of the relativistic mean field approximation in Eqs.~(\ref{Efflag}) and (\ref{BR}).  The scalar $\chi$ associated with the trace anomaly is a chiral scalar. On the other hand, what is observed in nature at low energy is the $f_0(500)$ which is most likely a complicated mixture of gluonium and quarkonium components in quark-gluon description.  The scalar needed in the mean-field theory for nuclear matter cannot have a substantial component of $q\bar{q}$ as one would expect in the linear sigma model. In fact the $\sigma$ of the sigma model would make nuclear matter collapse. It is instead a chiral-singlet scalar that would play the role of attraction in a relativistic mean-field description.

In the regime of long-wavelength excitations, that is, at low energy/momentum, the would-be chiral partner of the pion, denoted $\sigma$ as in the linear sigma model, should be high-lying as is known in the current algebras embodying the nonlinear realization of chiral symmetry. But as one goes to higher energy/momentum scales, presumably also in higher density, certain algebraic consequences of chiral symmetry constrained by, for example, Regge asymptotic behavior, require that the scalar be more or less degenerate with the pion, thereby joining the chiral four-vector~\cite{vankolck}. One can think of this behavior of the scalar meson as a level-crossing between the $s$ dominantly chiral scalar at low density and the $\sigma$ dominantly the fourth component of the chiral four vector\footnote{Such a level crossing is discussed in Refs.~\cite{level-crossing,SLPR} in terms of the Fock-space expansion in leading order in $N_c$. The $N_c$ dependence exploited here may be connected to the $N_c$ dependence of the various Fock spaces at different energy scales. Weinberg has argued that the lowest-lying tetraquark configuration for small $N_c$ in Ref.~\cite{level-crossing} is not suppressed in the large $N_c$ limit.}. One way to understand such a level crossing is to say that at low density, the lowest-lying scalar is the $s$ of multi-quark-multi-antiquark components and at high density, it is the $\sigma$ of a quark-anti-quark configuration.

It is not known whether by higher-order loop calculations that bring in corrections in $1/N_c$ one can produce the level-crossing. Given the number of parameters involved, this is a daunting task. One can, however, re-parameterize the fields and take certain (singular) limits to arrive at the configuration where the $\sigma$ arises with the nonlinear sigma model, as was done in Ref.~\cite{vankolck}. We do this with the dilaton-baryon-HLS model. One introduces a new field
\begin{eqnarray}
\Sigma
&=& U\chi {\kappa} = \xi_L^\dagger\xi_R \chi {\kappa}
= \sigma + i\vec{\tau}\cdot\vec{\pi}\,,
\\
{\mathcal N}
&=& \frac{1}{2}\left[ \left( \xi_R^\dagger + \xi_L^\dagger \right)
{}+ \gamma_5\left( \xi_R^\dagger - \xi_L^\dagger \right) \right] N\,,
\end{eqnarray}
with the Pauli matrices $\vec{\tau}$ in the isospin space and $\kappa=(f_\pi/f_\chi)$.
The linearized Lagrangian includes terms that generate singularities, i.e., have
negative powers of $\la\mbox{tr}\left[ \Sigma\Sigma^\dagger\right]\ra\to 0$ as chiral symmetry is restored. The singular terms are eliminated if one takes the limits\footnote{We are assuming $U(2)$ flavor symmetry for $V=(\rho,\omega)$. In medium, this symmetry may not hold~\cite{plrs}. In what follows we will be focusing on $V=\rho$.}
\be
\kappa=g_V=g_A=1,
\label{DLFP}
\ee
where $g_A$ is the axial-vector coupling constant and $g_V$ is the induced coupling constant defined in the effective vector-nucleon coupling as $g_{VNN}=g(g_V-1)$ with $g$ the hidden gauge coupling constant. Thus as one approaches chiral restoration, one gets
\be
(g_{VNN}, g_A-1, m_N)\to (0,0,0).
\ee
This is what is referred to as the ``dilaton limit fixed point (DLFP)". In this limit, the vector mesons, becoming massless, decouple from the nucleon -- though not from the quartet scalars -- and the nucleon and $\pi$ and $\sigma$ are arranged into the Gell-Mann-L\'evy linear sigma model.

We can now return to the question: To what density can the relativistic mean field theory so far developed be applied? This is a wide-open question that will be the subject of future research, e.g., at such accelerators as FAIR in Darmstadt. Theoretically this is a challenging task. It is practically impossible to perform systematic quantum calculations with the effective HLS Lagrangian at densities much higher than the saturation density. As mentioned in Section \ref{fraction} and discussed in Chapter \ref{skyrm}, if there is a phase transition of the type that is indicated by the skyrmion fractionizing into half-skyrmions at some density $n_{1/2}$ with a possible Fermi-liquid--non-Fermi-liquid transition, then the mean field approximation must be breaking down. Chiral perturbation theory calculations -- even if possible -- are unlikely to access such a phase change. Furthermore if the DLFP is approached, then the effective $\rho$NN coupling will go to zero, independently of the hidden gauge coupling. This means that the vector mean field will vanish and the relativistic mean field approach used for the EoS of dense matter will break down. This point will be discussed later in connection with the symmetry energy that figures in the EoS of compact stars.

\subsection{Mended symmetries: DLFP and VM}\label{mended}
HLS in the presence~\cite{HKR} or absence~\cite{HY:PR} of baryons with the assumed matching of the correlators to QCD at the matching scale predicts that at the chiral transition, it is the hidden gauge coupling $g\sim \la\bar{q}q\ra\to 0$ making the $\rho$ mass go to zero as $m_\rho\sim g\to 0$, the VM fixed point. The multiplet structure here is that the pion and the longitudinal $\rho$ come together while the scalar joins the $a_1$ meson.  On the other hand, the movement toward the dilaton limit fixed point discussed above with the $\rho$NN coupling $g_{VNN}$ going to zero while $g\neq 0$, involves the scalar joining the triplet of pions. Matching with QCD at a suitable matching scale should give a UV fixed point from which the IR flow will be on a different trajectory from the VM UV fixed point of HLS.

The question that is raised is: How does one go over from the flow trajectory on which the DLFP lies to the VM fixed point? Since the multiplet structures are different involving different symmetries, one the Gell-Mann-L\'evy sigma model and the other HLS, it looks that there has to be a phase transition going from the DLFP to the VM fixed point. This question is suggested to be tied to the issue of ``mended symmetries"~\cite{weinberg-mended}.

Recall that the behavior of the $\rho$ meson mass, given in Eq.~(\ref{vm}), in approaching chiral restoration is obtained in HLS with the $\rho$ as the only vector meson degree of freedom. One can ask whether this concept of the vector manifestation survives when other degrees of freedom are included. For instance, what about the $a_1$ meson? In the presence of $\rho$ and $\pi$, the matching of the HLS and QCD correlators yields simple and unique relations consistent with various sum rules such as Weinberg's. However, in the presence of the $a_1$ meson, this simplicity is lost. What happens to these degrees of freedom as the chiral restoration is approached? This question was addressed by two groups~\cite{harada-sasaki,hidaka} for $N_f=2$ for chiral $SU(N_f)_L\times SU(N_f)_R$ symmetry in a generalized HLS model, called ``GHLS ".  The fields included in their analyses were the quartets that figure Weinberg's mended symmetries, i.e., $\rho$, $a_1$, $\pi$ and $\sigma$. There was a perfect agreement between the two groups.

By matching the GHLS and QCD correlators at the matching scale $\Lambda$ and by doing the Wilsonian RG, three fixed points were found at chiral restoration: the first that corresponds to the Gell-Mann-L\'evy linear sigma model (or Ginzburg-Landau, GL for short) with the $\pi$ and $\sigma$ coming together and likewise for $\rho$ with $a_1$, the  second that corresponds to the VM with the longitudinal $\rho$ and $\pi$ coming together and  likewise for the longitudinal $a_1$ with $\sigma$, and the third which is a mixed hybrid type. The first two are IR fixed points, but the third is not, and is arrived at only by fine-tuning. In all cases, both $\rho$ and $a_1$ become massless at the restoration point, since the GHLS gauge coupling goes to zero.

The analysis does not indicate which fixed point is arrived at in which condition. For instance, it cannot answer the question raised above, namely, whether and how the density triggers the transition from the dilaton-limit fixed point (in GL mode) to the VM fixed point. There may be a variety of different flows possible depending on probes and conditions. However, a highly plausible possibility of what might happen at the chiral restoration point is offered by Weinberg's mended symmetries~\cite{weinberg-mended}. It has been shown using a colinear current algebra for Goldstone bosons of a broken symmetry group that under certain assumptions in the context of QCD such as large $N_c$ and the high-energy behavior of exclusive inelastic scattering amplitude, the quartet mesons, $\rho$, $a_1$, $\pi$ and $\sigma$, become massless at a second-order phase transition  at which the chiral symmetry becomes unbroken~\cite{weinberg-mended2}. This implies the emergence of massless hidden gauge bosons $\rho$ and $a_1$.

What this scenario of mended symmetries is pointing to is that the approach to the chiral phase transition in dense matter we are concerned with is most likely complex and intricate at the same time. The dilaton-limit fixed point and the VM fixed point discussed above could be just the tip of the iceberg of what might lie in the hadronic phase diagram going toward dense matter relevant to compact stars. It also implies a variety of scenarios in the response of vector mesons to photons as has been explored in relativistic heavy-ion collisions (which will be briefly mentioned below.)

\subsection{``Seeing" BR scaling}
Since BR scaling Eq.~(\ref{BR}) is {\em defined} in the specific framework adopted, namely, $hls$EFT in the mean field approximation or equivalently for Fermi-liquid system of nucleons in matter, whether it is viable or not has yet to be checked by experiments at densities above that of nuclear matter. It is of course crucial that theoretical predictions are made within the framework so defined. This is the point somewhat loosely interpreted or misunderstood in the works so far done in the field, both theoretically and experimentally, purporting to verify the notion of the scaling relation, in particular in the context of chiral symmetry. In confronting experimental data with the scaling given by the VEV of ``soft" dilaton $\chi$, $f_\chi^*$, and interpreting it as a manifestation of chiral symmetry properties of the ``vacuum" that is supposed to be changing at different densities, it has been overlooked that there could be a highly nontrivial relation between $f_\chi^*$ (which enters in BR scaling) and the quark condensate $\la\bar{q}q\ra^*$ (which enters in the bare HLS Lagrangian via Wilsonian matching with QCD).

What can be said with confidence is that the scaling in the mean-field theory framework has generally met with success in finite nuclei, far away from the critical restoration point.
Up to nuclear matter density -- and perhaps up to the density $n_{1/2}$ where a half-skyrmion phase appears in the skyrmion crystal description of nuclear matter described in Chapter \ref{skyrm}, the mean-field treatment of Eq.~(\ref{Efflag}) with the scaling Eq.~(\ref{BR}) should work. The mass scaling then can be given in terms of the scaling of the pion decay constant $f_\pi^*$ for which information is available from experiments and chiral perturbation theory. We discuss a few cases here.

\subsubsection{\it Dileptons}
Since the $\rho$ meson in the HLS framework has the property suggested by the vector manifestation and the photon couples to the pion via vector dominance (VD), it was thought, understandably, that dilepton productions in heavy ion collisions could cleanly map out, more or less unencumbered by background processes,  the in-medium behavior of the $\rho$ meson, in particular at high temperature. If the VM were to hold near the critical point\footnote{The argument based on mended symmetries suggests that some of the VM properties (such as the vanishing of the gauge coupling) apply more generally.}, then near chiral restoration, dileptons would be emitted from nearly zero-mass $\rho$ mesons, with the width strongly suppressed due to the vanishing vector coupling.

It turns out that this expectation, which at a first glance appears to be sound, is not actually founded  in the framework of HLS. Because of the intricate structure of HLS theory with its matching to QCD and renormalization group flows, some of the dominant processes operative in the medium-free vacuum, such as vector dominance, do not hold in medium.\footnote{This is also the case in GHLS.} Furthermore, certain processes not present in the vacuum -- what might be classed as ``mundane" nuclear many-body effects -- produce medium-dependent {\em background}. As a consequence, even if the BR-scaled vector mesons were present in the process, their coupling to the dileptons would be swamped by the background nuclear effects so that ``seeing" the $\rho$ mesons carrying the pristine imprint on BR scaling or the VM would be like finding a ``needle in a haystack"~\cite{needle}. So far the experiments performed in heavy-ion collisions have failed to  zero in on the scaling property or more generally on the chiral symmetry property of dense or hot matter.

In connection with relativistic heavy ion collisions, a much more trivial explanation of why the $\rho$-meson-mass scaling is not ``seen" in the experiments may be that the particle (or quasi-particle) description of the degrees of freedom breaks down near the chiral transition. This would mean that the local field description of the degrees of freedom implicity assumed with HLS makes no sense there, a rather uninteresting possibility. At present this issue is left unresolved.

\subsubsection{\it Anomalous orbital gyromagnetic ratio $\delta g_l$} \label{orbital}
\vskip 0.5cm
The Migdal (isovector) anomalous orbital gyromagnetic ratio $\delta g_l$ in heavy nuclei can be obtained from the Lagrangian Eq.~(\ref{Efflag}) in the mean field approximation, including a Fock term from the pion exchange~\cite{friman-rho}:
\be
\delta g_l=\frac 16 \frac{F_1^\prime -F_1}{1+F_1/3}\tau_3=\frac 49[\Phi^{-1} -1-\frac 12 \tilde{F}_1^\pi]\tau_3\label{deltagl}
\ee
where $F_1$ and $F_1^\prime$ are the Landau parameters introduced in Section \ref{flsnm} and $\tilde F_1^\pi = (M_N/M_N^*)F_1^\pi$ is the pion Fock-term contribution precisely given by chiral symmetry for any density. Since $\Phi$ is related to the in-medium pion decay constant, it can be obtained from experiments in the vicinity of the saturation density. Let's evaluate Eq.~(\ref{deltagl}) at nuclear matter density $n_0 = 0.16$ fm$^{-3}$. Evaluating the pion contribution to the Landau parameters, we have $\frac 13 \bar{F}_1^\prime =-0.153$, and from experiments in deeply bound pionic-nuclear systems~\cite{kienle-yamazaki}, we have
\be
(f^*_\pi (n_0)/f_\pi)^2\approx 0.64.
\ee
Eq.~(\ref{deltagl}) gives at $n\approx n_0$
\be
\delta g_l\approx 0.21\tau_3.\label{friman-rho}
\ee
This is in strikingly good agreement with experiments in Pb nuclei, which give $\delta g_l^{proton}=0.23\pm 0.03$.

The difference between Eq.~(\ref{friman-rho}) obtained here and Eq.~(\ref{chpt-deltagl}) is an interesting object. The former is from $hls$EFT and the latter from ChEFT. The former is obtained in RMF with BR scaling with a correction from the pion-exchange Fock term and the latter is the standard chiral perturbation approach as defined in the framework of ChEFT. Both are in the framework of Landau Fermi liquid theory. The question then is what makes the difference? The answer to this question, not known at this moment, should be most illuminating.

{\it The lesson from this result, underlining what was stated at the beginning of this section, is that the effective BR scaling {\it determined in nuclear matter}, representing  scale symmetry spontaneously broken at that density, encapsulates the Landau fixed-point parameters that capture nonperturbative strong interaction dynamics of nuclear matter. The concept makes sense in the vicinity of normal nuclear matter, the Fermi-liquid fixed point. This clearly indicates that BR scaling, directly connected to the soft dilaton condensate, is {\em not  a simple or explicit} signal for a fundamental vacuum property associated with the quark condensate inherited from Wilsonian matching with QCD at the scale $\Lambda_M$ although it is in principle included in the ``measured" BR scaling. The fundamental link between scale or conformal symmetry and chiral symmetry needs to be worked out to give a correct interpretation between what is observed in experiments and the chiral structure of matter at high density.}
\subsubsection{\it Axial charge transitions in nuclei}
As discussed in Section \ref{MEC}, the axial charge operator in the weak current receives a potentially large contribution from a soft-pion dominated exchange current. This contribution dominates the meson exchange weak current with corrections appearing at next-to-next-to-leading order. A simple consideration shows that the one-pion exchange axial charge operator appears with $1/f_\pi$ which in medium will be enhanced by the factor $f_\pi/f_\pi^*$~\cite{KRaxial}, so that the total matrix element divided by the matrix element of the single-particle operator is enhanced in medium by the factor
\be
\epsilon_{MEC}=\frac{1}{\Phi} (1+\Delta_a),
\label{axial}
\ee
where $\Delta_a$ is the ratio of the matrix element of the ``bare" two-body axial-charge operator over that of the one-body axial-charge operator. Because of the soft-pion exchange involved~\cite{KDR}, $\Delta_a$ is fairly accurately calculable with little dependence on density and is found to be $\Delta_a \approx 0.6$, a big value for a meson-exchange correction. Now evaluating at $n\approx n_0$ for which $\Phi \approx 0.8$, we have
\be
\epsilon_{MEC} (n_0)\approx 2.
\ee
This prediction is confirmed by experiments in heavy nuclei~\cite{warburton}:
\be
\epsilon^{Pb}_{MEC}= 1.9-2.0.
\ee
This is one of the strongest manifestations of soft-pion effects in nuclear processes. With an accurate treatment of nuclear dynamics with ChEFT$^\prime$ (ChEFT with BR scaling correctly implemented), it would provide another ``smoking gun" signal for chiral symmetry in nuclei.

\subsubsection{\it Nuclear tensor forces}

From early on, it was the special property of the nuclear tensor forces that motivated the proposal of BR scaling~\cite{BRtensor}. In baryon-HLS (BHLS), the nuclear tensor forces are dominated by the one-pion exchange and one-$\rho$ exchange in the form
\begin{eqnarray}
V_M^T(r)&&= S_M\frac{f_{NM}^2}{4\pi}m_M \vec \tau_1 \cdot \vec \tau_2 S_{12}\nonumber\\
&& \left(
 \left[ \frac{1}{(m_M r)^3} + \frac{1}{(m_M r)^2}
+ \frac{1}{3 m_Mr} \right] e^{-m_M r}\right),
\label{tenforce}
\end{eqnarray}
where $M=\pi, \rho$, $S_{\rho(\pi)}=+1(-1)$. Note that it is the sign difference between the two contributions that is crucial.

\begin{figure}
\begin{center}
\includegraphics[height=7cm]{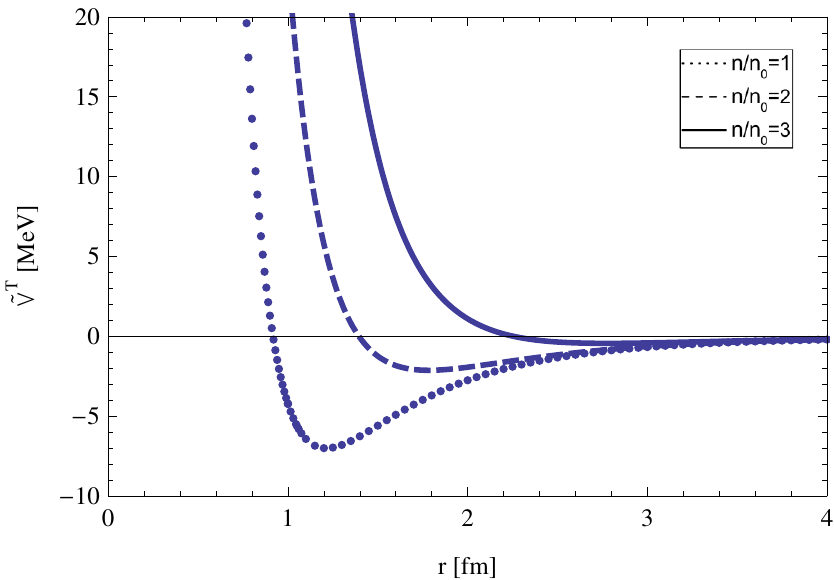}
\caption{Schematic form of the sum of $\pi$ and $\rho$ tensor forces (in units of MeV) for 
densities $n/n_0$ =1 (dotted), 2 (dashed) and 3 (solid) with $m_\rho^*/m_\rho \approx  
1-0.15 n/n_0$ and $R\approx 1$ for all $n$ {\it in the absence of topology change}.}
\label{tensor}
\end{center}
\end{figure}

In applying Eq.~(\ref{tenforce}) in medium, it is reasonable to assume that the pion tensor is more or less unaffected by the density for the range of density we are concerned with. So we leave it unchanged. As for the in-medium $\rho$ meson, apart from its scaling mass $m_\rho^*$, we have to take into account the scaling of $f_{N\rho}$ in medium. Written in terms of the parameters of the baryon BHLS Lagrangian, the scaling figures in the ratio
\be
R\equiv \frac{f_{N\rho}^*}{f_{N\rho}}\approx \frac{g_{\rho NN}^*}{g_{\rho NN}}\frac{m_\rho^*}{m_\rho}\frac{m_N}{m_N^*}\,.
\ee
Now using the scaling relation in Eq.~(\ref{BR}), which is to hold up to nuclear matter density $n_0$, and the observation that the hidden gauge coupling $g_V$ does not scale up to $n_0$,  we get\footnote{In Chapter \ref{skyrm}, we will suggest that this scaling holds up to $n_{1/2} \gsim 2 n_0$ where $n_{1/2}$ is the density at which skyrmions turn to half-skyrmions with topology change. At $n_{1/2}$ there takes place a drastic modification of the scaling in the net tensor force, due principally to the change in the ratio $R$.}
\be
R&\approx& 1.\label{R}
\ee
 For a  simple parametrization, the net tensor force behaves as shown schematically in  Fig.~\ref{tensor}: As the density increases, the net tensor force strength effective in nuclear processes -- with the short-range correlations cutting off short-distance interactions -- gets weaker and disappears nearly completely at $n\gsim 2n_0$.

\subsubsection{\it Revisiting C14 dating}
The stiffening of the nucleon-nucleon spin-isospin interaction in medium was already noted in Ref.~\cite{BRtensor} and has since been observed in a variety of nuclear response functions. More recently, this mechanism of decreasing tensor force in the vicinity of nuclear matter density was invoked to explain the drastic suppression of the Gamow-Teller matrix element of the beta decay in $^{14}$C, accounting for its long lifetime exploited in the C14 dating~\cite{holt-BR}. Although a decreasing tensor force does inhibit the beta decay of $^{14}$C, the effect is enhanced with the introduction of additional short-range repulsion via the scaled $\omega$ meson mass, also included in the calculation of Ref.~\cite{holt-BR}. Later work examined the suppression of the Gamow-Teller matrix element in the context of the microscopic shell model with chiral two- and three-nucleon forces \cite{holt09}. There it was shown that although three-body forces increase the strength of the effective tensor force, the repulsive three-nucleon contact term has sufficient strength to significantly reduce the Gamow-Teller transition strength.

We note that these seemingly different mechanisms are not incompatible. In fact, they are logically consistent. As emphasized in various places in this review, the BR scaling that reflects the ``intrinsic density dependence" is defined at the scale at which QCD and HLS correlators are matched, i.e., at $\Lambda$. On the other hand, applied to nuclear matter, the scaling parameter is numerically fitted at the Fermi-liquid fixed point that is captured in the mean-field approximation of the BHLS Lagrangian. In effective field theories anchored on the Fermi-liquid fixed point, 3-body forces are ``irrelevant" in the renormalization-group sense, with their effect subsumed in the parameters of the two-body forces with coefficients that are ``marginal." Now when the effect of BR scaling is calculated in \cite{holt-BR}, the scaling should contain both the {\it intrinsic} density dependence coming from the matching of the HLS/QCD correlators {\em and} the induced one inherited from integrating-out of higher nuclear excitation modes. In Ref.~\cite{holtGEBFS} it was suggested that in this way the chiral three-nucleon contact term and the BR scaled $\omega$ meson mass may be linked. In fact, one way of seeing this is that the BR scaling inherited from matching to QCD at the matching scale $\Lambda_M \sim 1$ GeV encodes what is integrated out above that scale. Now the $\omega$ mass is comparable to the matching scale, so the three-nucleon contact interaction attributable to the exchanges of mesons whose masses are of the order of $\Lambda_M$ or greater must be encoded in BR scaling. A similar mechanism is found to be operative in the saturation property of nuclear matter as discussed in the conclusion section.
An interesting observation that follows from the above reasoning is that the effect of the {\it long-range} 3-body force that contributes importantly to the oxygen neutron drip line found in \cite{otsuka10} is {\em not} captured in BR scaling.
\subsubsection{\it Shell evolution}

There seems to be a way to check whether or  not the stiffening of the tensor force predicted by BR scaling -- as represented by  Fig.~\ref{tensor} -- is viable and this is most cleanly seen in the shell evolution in exotic nuclei. In a series of papers, Otsuka and his co-wokers have shown, rather convincingly, that  in the ``monopole" matrix element of the two-body interaction between two single-particle states labeled $j$ and $j^\prime$ and total two-particle isospin $T$ defined by
\be
V_{j,j^\prime}^T=
\frac{\sum_J (2J+1)\la jj^\prime|V|jj^\prime\ra_{JT}}{\sum_J(2J+1)},
\ee
the strength of the ``bare" tensor force remains, most surprisingly, unrenormalized by many-body long-range and short-range correlations~\cite{otsuka}. Thus the single-particle shell evolution
\be
\Delta\epsilon_p (j)=\frac 12(V_{jj^\prime}^{T=0} +V_{jj^\prime}^{T=1})n_n(j^\prime),
\ee
where  $\Delta\epsilon_p (j)$ represents the change of the single-particle energy of protons in the state $j$ when  $n_n(j^\prime)$ neutrons occupy the state $j^\prime$ will single out the strength of the tensor force. It is not at all obvious why the tensor force strength is free of many-body renormalization, but if it is, then this would imply that the BR scaling that encodes the density dependent quark condensate in the ``bare" parameters of the effective Lagrangian -- for which a renormalization-group flow analysis indicates scale independence --  could be ``seen" in the shell evolution, unscathed by nuclear correlations. Such a feat will require a highly sophisticated precision calculation of the matrix element, sampling different density regimes -- perhaps a daunting task, but worthwhile investigating.

It will be seen in Chapter \ref{skyrm} that the topology change implied in the dense skyrmion crystal induces a drastic modification in the behavior of the tensor force once the density reaches $n_{1/2}$. Such a high density cannot be reached in ordinary nuclear systems but could be relevant in compact stars.

\subsection{Meson condensation in dense matter}\label{mesoncondensation}
One more place where effective field theory anchored on chiral symmetry plays a significant role is in describing the possibility of pseudo-Goldstone boson condensation in compact-star matter. Early work on this was focused on pion condensation, particularly p-wave pion condensation. Meson condensations can be driven by attractive meson-nuclear interactions. The s-wave pion condensation did not attract attention, since there is no strong attraction in the s-wave channel. The p-wave channel, however, has the well-known attraction that produces the $\Delta(1232)$ resonance, and this was exploited by A.B. Migdal and others to predict p-wave pion condensation at a density more or less relevant to compact stars~\cite{rho-wilkinson}. It was, however, quickly recognized that there is a strong short-range spin-isospin repulsion associated with the Landau-Migdal $g_0^\prime$ parameter that leads, in chiral models, to a critical density, going roughly like $n_c^\pi\sim 1/\sqrt{({g_A^*}^2-1)}$~\cite{baym}.  It was observed in medium and heavy nuclei that $g_A^* \sim 1$. This suggested that the critical density would become much too high to be relevant for compact stars. This is perhaps too simplistic an argument, but it gives a rough but qualitatively correct answer. It should be admitted that this phenomenon is not yet completely ruled out\footnote{As will be shown in Chapter \ref{skyrm} and mentioned below, dense matter described with skyrmions put on crystal lattice undergoes a topological change from skyrmions of unit baryon charge to half-skyrmions of half-baryon charge. What happens then is that the $\rho$ tensor force gets killed and the pion tensor takes over at increasing density. In this case, through the enhanced pion tensor, the p-wave pion-nuclear interaction can become strongly attractive and the pions can get condensed into a  pionic crystal, leading, for instance, as discussed a long time ago by Pandharipande and Smith~\cite{pandha}, to a possible neutron-rich $\pi^0$-condensed solid at high density in compact stars. Such a pion-condensed crystal is expected in the large $N_c$ limit.},  but so far there is no evidence for it, although there have been extensive experimental searches. We will not enter it here.

On the other hand, conditions are met for kaons to condense in the s-wave channel~\cite{kaplan-nelson,BTKR}. This has an interesting ramification on compact-star physics in view of the recent observations, rather well-established, of $\sim 2$ solar-mass stars and the Brown-Bethe scenario of light-mass black holes due to kaon condensation. This matter will most likely be treated in the astrophysics section of this volume by other authors, so we will not go into that aspect of the matter.

Here we will give a brief discussion on kaon condensation proper. It concerns not only its effect on the equation of state for compact stars but also a possible break-down of the Fermi liquid state of baryonic matter, the corner stone of baryonic phase structure. In order to make the discussion as general as possible, consider a (complex) scalar boson field $\phi$ interacting with a fermion $\psi$ on the Fermi surface in a Fermi liquid in 4D. The partition function in Euclidean space is taken as
\begin{equation}
Z = \int [d\phi][d\phi^\ast][d\psi][d\bar{\psi}] { e}^{-\tilde{S}^E}
\end{equation} with the action in momentum space
\begin{eqnarray}
\tilde{S}^E&=& \tilde{S}_\psi^E + \tilde{S}_\phi^E +\tilde{S}_{\psi\phi}^E\ \label{S2}\\
\tilde{S}_\psi^E &=&\int  \frac{d\epsilon  d^3 \vec{k}}{(2\pi)^4}
\,\bar{\psi}_\sigma \left\{ -i\epsilon +(e(\vec{k})-e_F)\right\} \psi_\sigma\nonumber\\
&& - \int \prod_{i=1}^{4}\left( \frac{d\epsilon_i  d^3 \vec{k}_i}{(2\pi)^4}\right) \lambda \bar{\psi}_\sigma \bar{\psi}_{\sigma^\prime} \psi_\sigma\psi_{\sigma^\prime} \delta^4(\epsilon,\,\vec{k})\label{fermion2}\\
\tilde{S}_\phi^E &=& \int \frac{d\omega d^3\vec{q}}{(2\pi)^4} \{\phi^*(\omega^2 +q^2)\phi +m_\phi^2\phi^*\phi+\cdots\} \label{boson2}\\
\tilde{S}_{\psi\phi}^E &=&-\int \prod_{i=1}^{2}\left(\frac{d\epsilon_i d^3\vec{k}_i}{(2\pi)^4}\right) \left(\frac{d\omega_i d^3 \vec{q}_i}{(2\pi)^4}\right) \, h\,\phi^*\phi \bar{\psi}_\sigma\psi_\sigma \delta^4(\omega,\,\epsilon,\,\vec{q},\,\vec{k})\,. \label{couple2}
\end{eqnarray}
For kaon condensation, the field $\phi$ can be identified with the negatively-charged kaon field $K^-$ and the fermion field $\psi$ with the proton and neutron doublet in Fermi liquid. The four-Fermi interaction in Eq.~(\ref{fermion2}) is the marginal quasiparticle (Landau-Migdal) interaction in the sense of the renormalization group equation~\cite{shankar94}.  The baryonic matter is at the Fermi-liquid fixed point in the large $N$ limit where $N\propto k_F$ where $k_F$ is the Fermi momentum. The boson-fermion coupling, Eq.~(\ref{couple2}), can be thought of as arising from a chiral Lagrangian. There the constant $h$ can be identified  with the $KN$ sigma term, $h\propto \Sigma_{KN}/f^2$ with $f\sim f_\pi$ or with the Weinberg-Tomozawa term, $h\propto q_0/f^2$.

Recalling that the Fermi liquid fixed-point theory can be identified with the relativistic mean field theory of an $hls$EFT, we can also do the mean field approximation to kaons interacting with the quasiparticle nucleons. This gives essentially what was obtained by Kaplan and Nelson~\cite{kaplan-nelson}, which in ordinary (symmetric) nuclear matter happens at the critical density $n_c\approx m_K^2 f^2/\Sigma_{KN}\sim 6 n_0$.
In neutron-rich compact-star matter in beta equilibrium, one simple way of seeing what happens is to view kaon condensation as electrons with the chemical potential $\mu_e$ decaying to kaons with the in-medium mass $m_K^*$ when $\mu_e=m_K^*$~\cite{BTKR}. This makes kaons condense at a lower density than in the absence of electron chemical potential, typically at $n_c\sim (3-4)n_0$.

That kaons may condense at a density not too far above $n_0$ has raised several serious issues. First of all, the kaon condensation can substantially soften the EoS of compact-star matter, leading to the Brown-Bethe scenario of light-mass black-hole formation mentioned above. The second issue is that the softened EoS would be at odds with the observed massive neutron stars~\cite{demorest,antoniadis}, prompting to conclude that kaon condensation is ruled out in the
star matter.
These are, however, a controversial matter: the treatment of kaon condensation by which the conclusions are reached is far from reliable and trustworthy.
It has been shown in a (perturbative) renormalization-group analysis using an action of the form of Eq.~(\ref{S2}) in beta equilibrium that kaons {\em most likely} condense within a certain window of parameter space compatible with chiral symmetry of QCD~\cite{LRS}. Whether the parameter space obtained is strictly compatible with nature is hard to judge at the moment for various reasons given below.

A recent observation that is crucially pertinent to the issue is that treating kaon condensation in baryonic matter in a Fermi liquid may be invalid if the skyrmion--half-skyrmion phase transition discussed in Chapter \ref{skyrm} takes place at a density commensurate with that of kaon condensation. The topology change could induce the breakdown of Fermi liquid structure of the nucleonic matter. The mean field treatment with the action in Eq.~(\ref{S2}) would then be invalid.
Furthermore, a refined renormalization group analysis of the action in Eq.~(\ref{S2}) reveals further subtleties. In Ref.~\cite{LRS} the kaon-nucleon interaction was taken to be irrelevant. This is consistent with the Goldstone-nucleon derivative coupling as the Weinberg-Tomozawa term is. It is irrelevant in the RG sense. However, depending on the Fermi surface structure, the $\Sigma_{KN}$  term can be relevant. It is known in correlated condensed matter systems that even marginal boson-Fermi-liquid system can result in the breakdown of the Fermi-liquid structure when the boson becomes massless. The situation would be a lot more drastic if the coupling were relevant.

Now whether the sigma term in our case is relevant or not will crucially depend on where the effective action of Eq.~(\ref{couple2}) comes from, i.e., HLS or GHLS in the meson sector. The two new effects, one in the kaon sector and the other in the nucleon sector, combined together could bring a totally new perspective on how strangeness figures in nuclear matter and in compact-star matter. It is suggested in Chapter \ref{skyrm} that kaon-condensed matter in a non-Fermi liquid state (or half-skyrmion state) could be ``dual" to a region of cross-over from baryonic matter to strong-coupled strange quark matter, lending itself to a Cheshire Cat mechanism. This is an open problem.



\section{Implications for nuclear astrophysics}
\label{dnmai}

\subsection{Neutron matter equation of state}
\label{nmeos}

The equation of state of strongly-interacting neutron-rich matter arises in many different
astrophysical contexts, including neutron-star structure, the dynamics of core-collapse supernovae,
binary neutron-star coalescence, and $r$-process nucleosynthesis. The low-density
equation of state is governed to a large extent by universal features associated with large
scattering-length physics in the BCS regime \cite{gandolfi,gezerlis}, while at intermediate
densities in the vicinity of saturated nuclear matter, microscopic chiral nuclear forces with resolution scales
between $\Lambda = 400-500$\,MeV have been employed successfully in perturbative and
nonperturbative calculations of the equation of state \cite{tews12,coraggio13,carbone14,
gezerlis13,roggero14,wlazlowski14}. At yet higher densities beyond that of saturated nuclear matter, the calculations of the energy per particle become increasingly model dependent, and reliable extrapolation schemes that account for constraints coming from neutron star observations and general causality arguments provide a path forward \cite{hebeler10b} although the density regime relevant to the interior of compact stars is more or less unconstrained.

\begin{figure}
\centering{
\includegraphics[width=8.5cm] {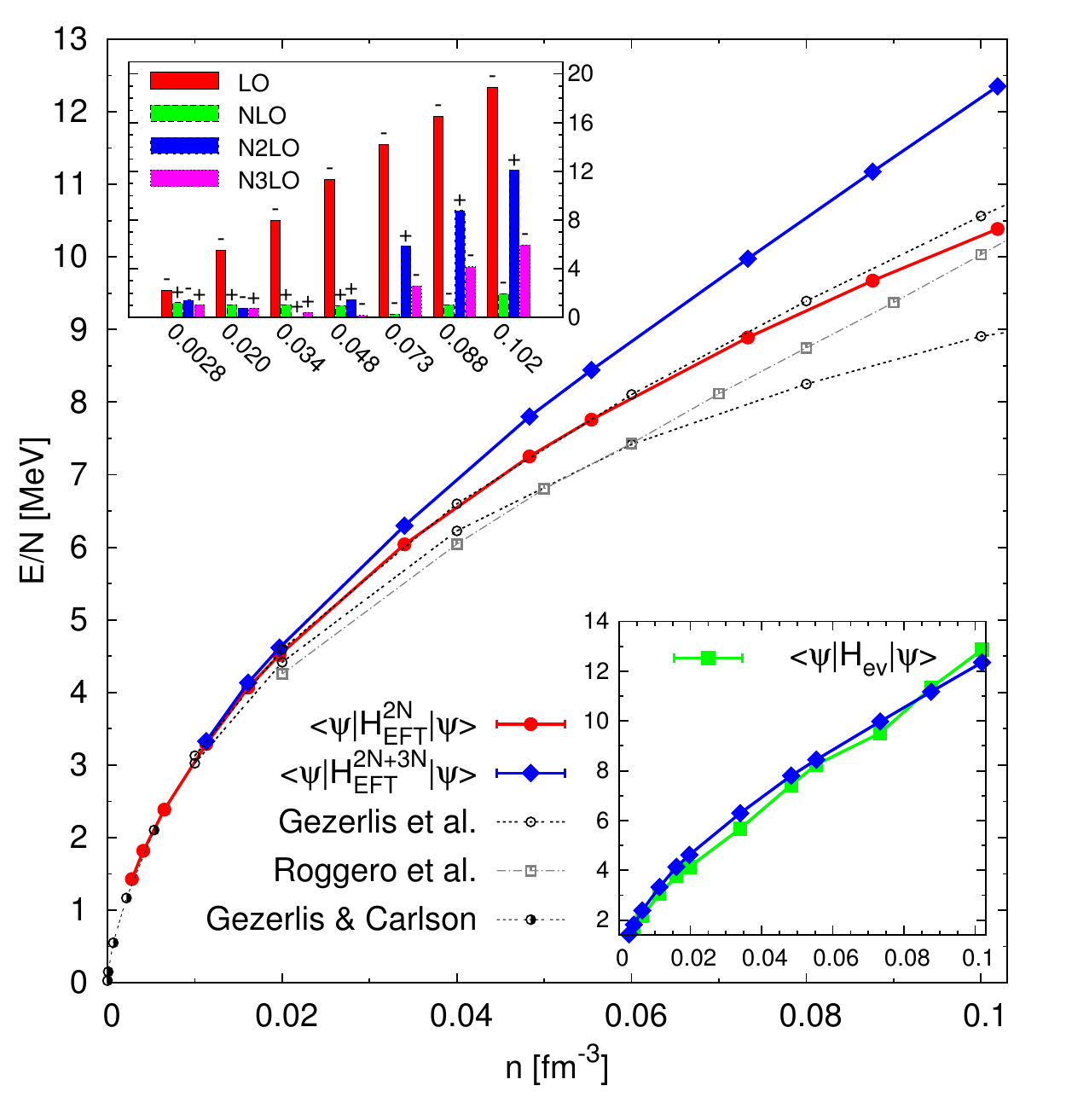} }
\caption{The energy per particle of neutron matter as a function of the neutron density $\rho_n$.
The dash-dotted line is a reproduction of the results from the variational many-body calculation
of ref.\ \cite{akmal}.}
\label{gabriel}
\end{figure}

The results of these studies with different chiral effective field theory forces, however, exhibit striking consistency, particularly at low densities. In the left panel of Fig.~\ref{gabriel} we show benchmark calculations of the neutron matter equation of state at low to moderate densities from several recent quantum Monte Carlo (QMC) studies \cite{gezerlis13,roggero14,wlazlowski14} employing chiral nuclear potentials at different resolution scales. The red \cite{wlazlowski14},
dotted \cite{gezerlis13}, and dashed-dotted \cite{roggero14} lines include two-body forces
only, while the blue curve \cite{wlazlowski14} includes as well the N$^2$LO chiral three-nucleon force. The upper-left inset in Fig.~\ref{gabriel} shows the contributions to the energy per particle
from different orders of the chiral nuclear force. In particular, the large contributions at N$^2$LO
reflect the presence of intermediate-state $\Delta$ degrees of freedom. In a chiral effective
field theory approach with explicit $\Delta$s, such effects would be promoted to NLO.
A notable feature of the equations of state shown in Fig.~\ref{gabriel} is that compared relativistic
mean field calculations (see e.g.\ Ref.~\cite{shen11}), chiral nuclear forces lead to a stiffer equation
of state at low densities and a softer one at high densities. At very low densities the results in
Ref.~\cite{wlazlowski14} are shown to match very well onto QMC results obtained with an effective
interaction that includes only the physics of the neutron-neutron scattering length and effective
range \cite{gezerlis08}.

Analytical calculations of the low-density equation of state of neutron matter have recently been
performed in the particle-particle hole-hole ladder approximation \cite{resum} using an effective
interaction proportional to the neutron-neutron scattering length $a$. The in-medium loop function
with zero, one, and two medium insertions is taken as the
starting point to generate a series of contributions to the energy per particle proportional to
$(ak_f)^n$. Due to the special nature of the momentum-independent contact interaction, all
multi-loop diagrams factorize, allowing their sum to be written in the form of a power of the
in-medium loop. The resulting series can then be resummed to all orders with the result
\begin{equation}
\bar E(k_f)^{(\rm lad)}= -{24k_f^2 \over \pi M} \int\limits_0^1 \!ds\, s^2
\!\!\int\limits_0^{\sqrt{1-s^2}}  \!\!d\kappa \,
\kappa \, \arctan { a k_f\, I(s,\kappa) \over 1+ \pi^{-1}ak_f \,R(s,\kappa)} \,,
\label{ladseries}
\end{equation}
where the arctangent function refers to the normal branch with odd parity taking on values in the
interval $[-\pi/2,\pi/2]$.
The expansion of the resummed energy per particle $\bar E(k_f)^{(\rm lad)}$ in powers of $a k_f$
has been tested against known low-order results based on the traditional particle-hole counting
scheme \cite{furnstahl,steele}. Up to fourth power in the scattering length $a^4$, where comparisons
can be made, the results are in perfect agreement. Moreover, since only two-dimensional integrals
are involved in the expression for the energy per particle Eq.~(\ref{ladseries}), the relevant coefficients
can be calculated with very high numerical precision \cite{resum}.

\begin{figure}
\centering{
\includegraphics[width=10cm] {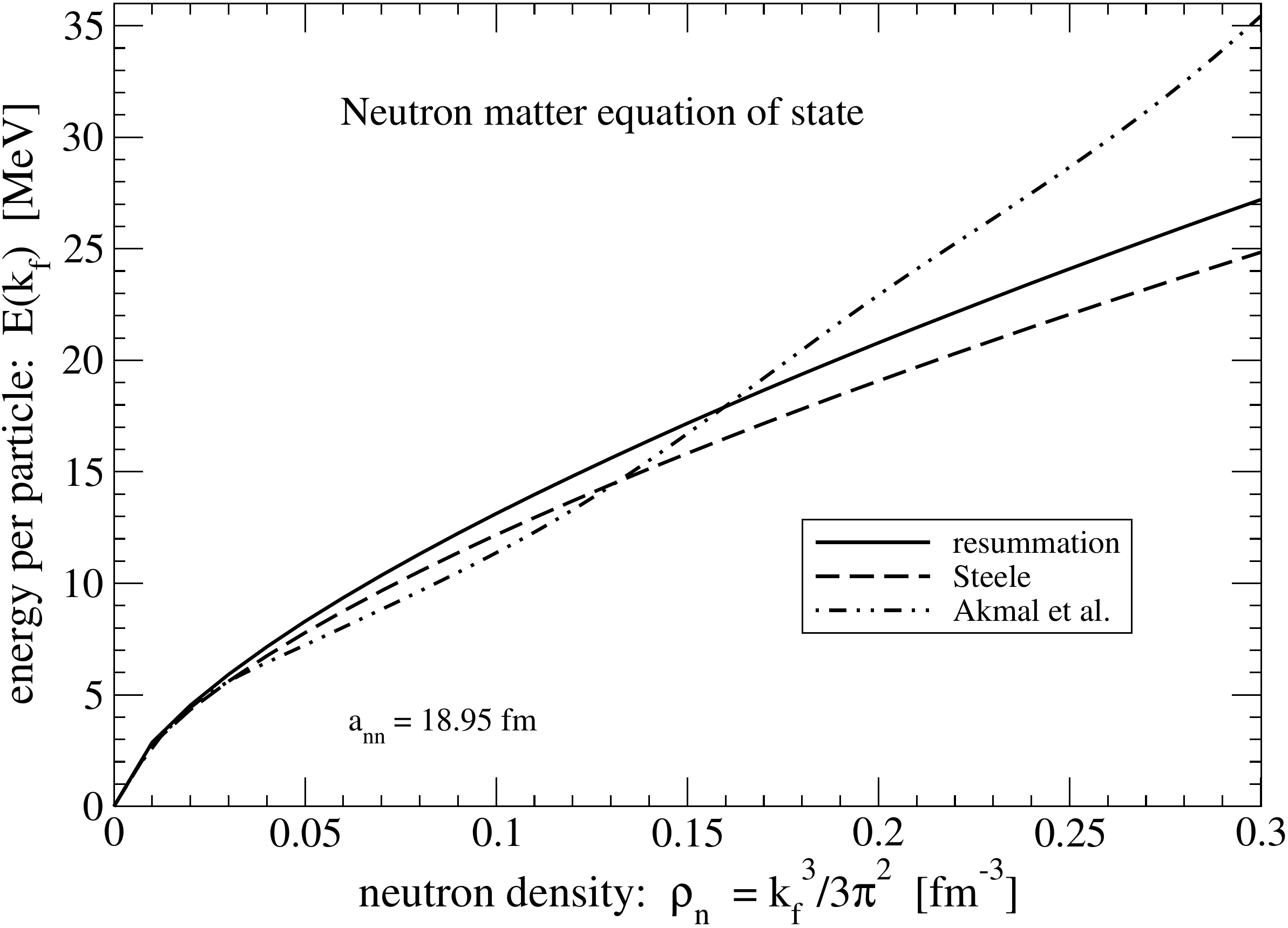} }
\caption{The energy per particle of neutron matter as a function of the neutron density $\rho_n$.
The dash-dotted line is a reproduction of the results from the variational many-body calculation
of ref.\ \cite{akmal}.}
\label{neutronmatter}
\end{figure}

The results for the energy per particle in Eq.\ (\ref{ladseries}) for the exact resummed in-medium
ladder diagrams holds for all values of the scattering length $a$. Substituting the very large
neutron-neutron scattering length $a_{nn} = (18.95 \pm  0.40)\,$fm \cite{chen} then provides
an approximation to the neutron matter equation of state, valid in particular at low densities where
only $S$-wave interactions are relevant. In Fig.\ \ref{neutronmatter} we show the resulting
energy per particle of neutron matter as a function of the density in the approximation
$\bar E(k_f)^{(\rm lad)}+ 3k_f^2/10M$ \cite{resum}. The solid line represents the treatment
given in Eq.\ (\ref{ladseries}) with $a=a_{nn}=18.95$\,fm for the
scattering length and $M=M_n = 939.57\,$MeV for the fermion mass. The
dash-dotted line comes from sophisticated variational many-body calculations \cite{akmal}
employing Argonne and Urbana two- and three-body forces. The dashed line corresponds to
the results of Steele's suggestion \cite{steele} for a simple geometrical series, $\bar E(k_f)^{(St)} =
-a k_f^3[3M(\pi+2a k_f)]^{-1}$. There is good agreement among the different results up to rather
larger neutron densities$\rho_n \simeq 0.2\,$fm$^{-3}$, where the dimensionless parameter
$a_{nn} k_f \simeq 34$. Beyond nuclear matter saturation density, repulsive effects from three-body
forces (treated Ref.\ \cite{akmal}) begin to play a significant role.

\subsection{Chiral Fermi liquid description of neutron matter}
\label{flt}

The framework of Fermi liquid theory, outlined in Section \ref{flsnm} to study the
bulk properties of symmetric nuclear matter, can be applied to pure neutron matter,
where the quasiparticle interaction determines the vector and axial-vector
response functions governing neutrino and anti-neutrino propagation \cite{iwamoto82,
navarro99,lykasov08,bacca09} as well as the response of the matter to
strong magnetic fields \cite{haensel82,olsson04,pethick09,garcia09}. Previous
calculations with phenomenological nuclear potentials have predicted the instability of
neutron matter to a ferromagnetic phase transition at several times nuclear matter
saturation density \cite{garcia09, fantoni01,rios05}, while microscopic forces have not
confirmed this qualitative feature. The magnetic susceptibility of neutron matter as well as
neutrino absorption, emission and elastic scattering rates depend in part on the noncentral
components of the quasiparticle interaction \cite{haensel75,schwenk04}
as discussed in Refs.\ \cite{olsson04,lykasov08,bacca09}.
Recently, the full quasiparticle interaction in the low-energy, long-wavelength limit
has been computed for pure neutron within the framework of chiral
effective field theory, including three-nucleon forces \cite{holt13,davesne14}.

The most general form for the quasiparticle interaction in neutron matter respecting
all strong interaction symmetries can be written \cite{schwenk04}:
\begin{eqnarray}
{\cal F}(\vec p_1, \vec p_2\,) &=& f(\vec p_1, \vec p_2\,) + g(\vec p_1,
\vec p_2\,) \vec \sigma_1 \cdot \vec \sigma_2 + h (\vec p_1, \vec p_2\,)
S_{12}(\hat q) + k (\vec p_1, \vec p_2\,) S_{12}(\hat P) \nonumber \\
&& +l (\vec p_1, \vec p_2\,) (\vec \sigma_1 \times \vec \sigma_2)\cdot
(\hat q \times \hat P),
\label{qpi}
\end{eqnarray}
where $\vec q = \vec p_1 - \vec p_2$ is the momentum transfer in the
exchange channel, $\vec P = \vec p_1 + \vec p_2$ is the two-particle center-of-mass
momentum, and $S_{12}(\hat v)$ defines the usual tensor
operator $S_{12}(\hat v) = 3 \vec \sigma_1 \cdot \hat v\, \vec
\sigma_2 \cdot \hat v -\vec \sigma_1 \cdot\vec \sigma_2$.
The interaction in Eq.\ (\ref{qpi}) is invariant
under parity, time-reversal, and particle exchange.
The presence of the medium induces terms that break Galilean
invariance and that depend explicitly on the center of mass momentum vector $\vec P$; namely,
$S_{12}(\hat P)$ and $A_{12}(\hat q, \hat P) = (\vec \sigma_1 \times \vec \sigma_2)\cdot
(\hat q \times \hat P)$.
As outlined in Ref.\ \cite{holt13}, linear combinations of the spin-space matrix elements of
the quasiparticle interaction can be used to extract the various scalar functions in Eq.\ (\ref{qpi}).

\begin{table}
\begin{center}
\begin{tabular}{|c||c|c|c||c|c|c||c|c|c||}\hline
\multicolumn{1}{|c||}{} & \multicolumn{9}{c||}{Chiral 2NF + 3NF \hspace{.3in}($k_f=1.7$\,fm$^{-1}$)} \\ \hline
\multicolumn{1}{|c||}{$L$} & \multicolumn{3}{c||}{0} & \multicolumn{3}{c||}{1} &
\multicolumn{3}{c||}{2} \\ \hline
\multicolumn{1}{|c||}{} & \multicolumn{1}{c|}{$V_{2N}^{(1)}$} & \multicolumn{1}{c|}{$V_{2N}^{(2)}$} &
\multicolumn{1}{c||}{$V_{3N}^{(1)}$} & \multicolumn{1}{c|}{$V_{2N}^{(1)}$} &
\multicolumn{1}{c|}{$V_{2N}^{(2)}$} & \multicolumn{1}{c||}{$V_{3N}^{(1)}$} &
\multicolumn{1}{c|}{$V_{2N}^{(1)}$} & \multicolumn{1}{c|}{$V_{2N}^{(2)}$} &
\multicolumn{1}{c||}{$V_{3N}^{(1)}$} \\ \hline
$f$ [fm$^2$] & $-$0.700 & 0.069 & 1.319 & $-$1.025 & 1.197 & $-$0.037 & $-$0.230 & $-$0.500 & $-$0.293  \\ \hline
$g$ [fm$^2$] & 1.053 & 0.293 & $-$0.283 & 0.613 & 0.159 & $-$0.364 &  0.337 & $-$0.089 & 0.043  \\ \hline
$h$ [fm$^2$] & 0.270 & $-$0.212 & 0.075 & 0.060 & 0.106 & 0.164 & $-$0.040 & $-$0.143 & $-$0.087  \\ \hline
$k$ [fm$^2$] & 0 & $-$0.156 & 0 & 0 & 0.085 & 0 & 0 & 0.063 & 0  \\ \hline
$l$ [fm$^2$] & 0 & 0.135 & $-$0.168 & 0 & $-$0.031 & $-$0.134 & 0 & $-$0.279 & 0.083  \\ \hline
\end{tabular}
\caption{Fermi liquid parameters for the particle-hole interaction in
neutron matter at a density corresponding to the Fermi momentum $k_f=1.7$\,fm$^{-1}$.
The low-energy constants of the N$^2$LO chiral three-body force are chosen to
be $c_1 =-0.81\,$GeV$^{-1}$ and $c_3=-3.2\,$GeV$^{-1}$.}
\label{flpkf}
\end{center}
\end{table}

We show in Table \ref{flpkf} the individual contributions from two- and three-nucleon
forces to the $L=0,1,2$ Landau parameters at various orders
in perturbation theory for a density corresponding to $k_f = 1.7$\,fm$^{-1}$. The first-order
contribution from two-nucleon forces is nothing more than a kinematically constrained version
of the antisymmetrized free-space interaction and therefore is trivially independent of the total
quasiparticle momentum
$\vec P$. Second-order contributions generate all of the noncentral interactions, which
from Table \ref{flpkf} are seen to be smaller in overall magnitude than the central components.
An especially noteworthy feature is that the exchange tensor interaction (already present in
the free-space nucleon-nucleon interaction) is largely reduced by Pauli-blocking effects.
The center-of-mass tensor particle-hole interaction arises solely from Pauli-blocking and core
polarization terms coming in at second-order in perturbation theory with two-body forces, while
the cross-vector interaction $A_{12}(\hat q, \hat P)$ is generated both from
two- and three-body forces.


The leading-order contribution from two-body forces results in a value of the compression
modulus ${\cal K}$ of neutron matter at $\rho_0$ that is
unphysically small (though still free from instabilities).  Auxiliary-field diffusion Monte Carlo
simulations with realistic two- and three-nucleon forces find that at
$\rho=0.16$\,fm$^{-3}$, ${\cal K} \simeq 520$\,MeV \cite{fantoni01}. As in the case
of symmetric nuclear matter, the chiral three-body force is quite
repulsive and gives rise to a compression modulus ${\cal K} = 550$\,MeV. The $f_1$ Landau
parameter governs the quasiparticle effective mass at the Fermi surface and receives
only a small contribution from three-body forces. Strong polarization effects at
second order, however, increase the effective mass from $M^*/M_N = 0.82$ to $M^*/M_N = 1.04$.

\begin{figure}
\begin{center}
\includegraphics[width=10cm,angle=270]{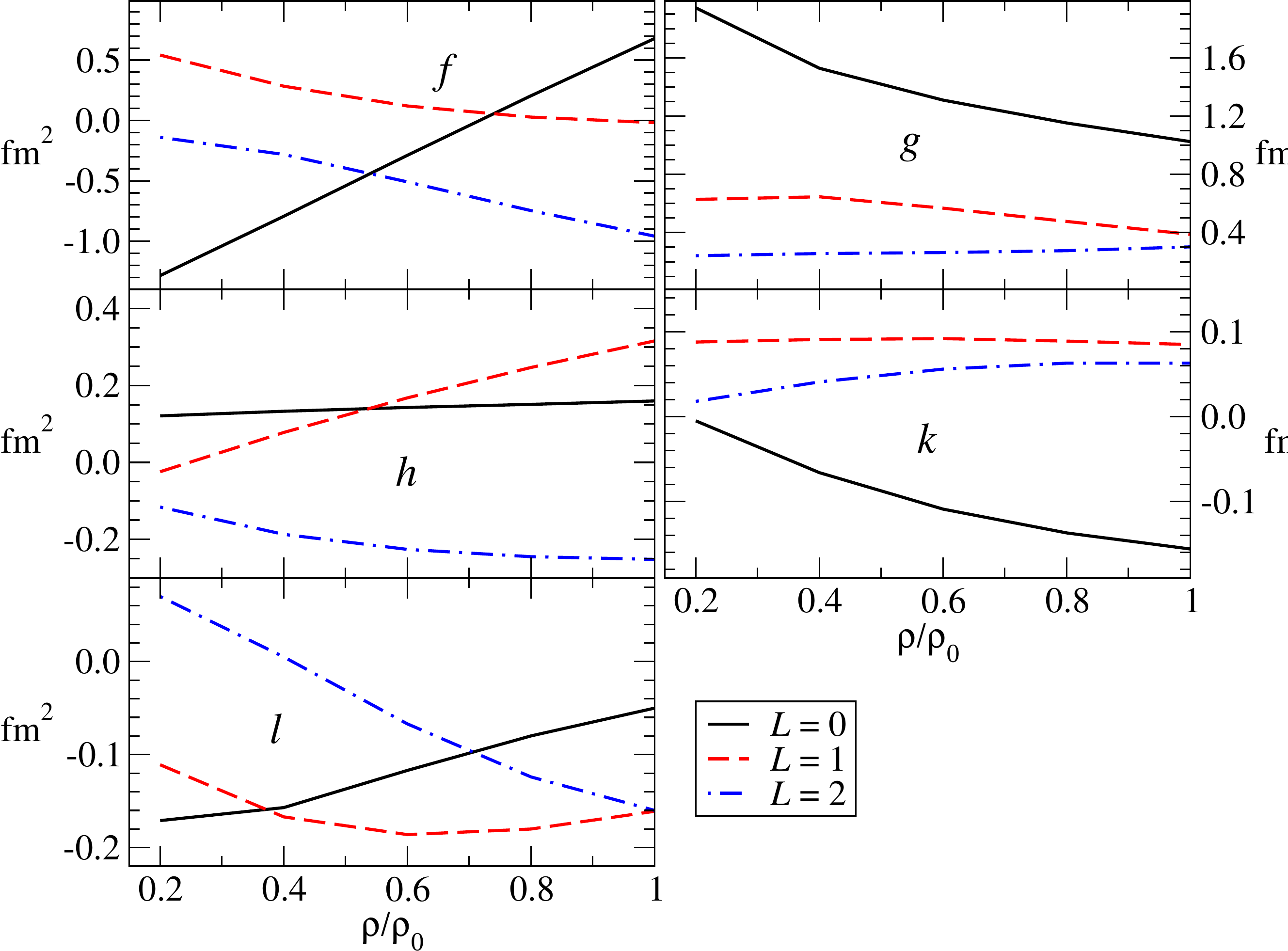}
\end{center}
\vspace{-.5cm}
\caption{(Color online) Density-dependent Fermi liquid parameters including first- and second-order
contributions from the chiral N$^3$LO nucleon-nucleon potential of ref.\ \cite{entem03}
as well as the N$^2$LO chiral three-nucleon force to leading order.}
\label{ddflp}
\end{figure}

Neglecting contributions from the noncentral particle-hole interactions, the
magnetic susceptibility is related to the Landau parameter $G_0$ by
\be
\chi = \mu_n^2 \frac{N_0}{1+G_0},
\label{susc}
\ee
where $\mu_n=-1.913$ is the neutron magnetic moment in units of
the nuclear magneton and $G_0 = N_0 g_0$. Noncentral interactions result in
effective charges (magnetic moments) that are not scalars under rotation. The spin
susceptibility then involves both the longitudinal and transverse components of the
magnetic moment \cite{haensel82,olsson04}. As seen in Fig.\ \ref{ddflp}, the Landau
parameter $g_0$ decreases with the nuclear density but there is no evidence for a
spin-instability close to nuclear matter saturation density.

\begin{figure}
\begin{center}
\includegraphics[scale=0.6,clip]{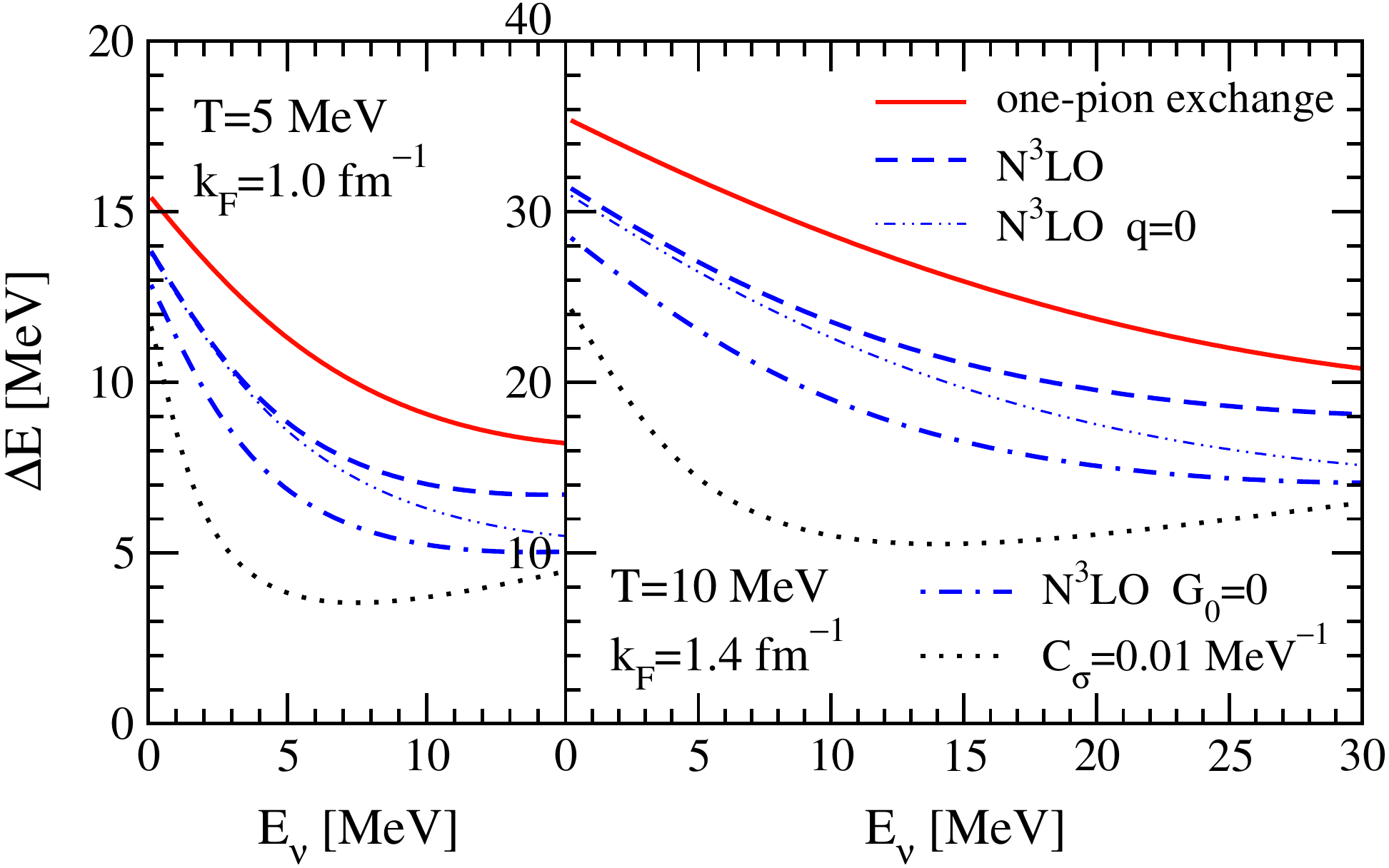}
\end{center}
\vspace{-.5cm}
\caption{The rms energy transfer in neutrino scattering from neutron matter as a function
of the temperature, density, and neutrino energy. Results are computed with chiral two-body
forces at different orders in the chiral power counting. Figure
reproduced from ref.\ \cite{bacca09}.
}
\label{etrans}
\end{figure}

Core-collapse supernovae release nearly all of their gravitational potential energy
through the emission of neutrinos. The temporal and spectral aspects of the
neutrino signal coming from core-collapse supernovae are sensitive to strong interaction
effects in the neutrino decoupling region of the proto-neutron star, called the neutrinosphere.
This region consists of warm ($T = 5 - 10$ MeV) and low-density ($n \simeq \rho_0/10$)
neutron-rich matter, and an accurate understanding of neutrino scattering, production, and
annihilation rates in the neutrinosphere is accessible within linear response theory.
At these densities, three-nucleon forces are expected to be negligible provided that the
densities and temperatures are large enough to prevent the formation of bound states.
In this partially degenerate regime, Landau's theory of Fermi liquids is
applicable and has been used \cite{bacca09} to compute the spin-spin response function
of neutron matter employing chiral two-body interactions at N$^3$LO. Assessing the role
of nonperturbative many-body correlations arising from the physics of large scattering lengths
in the neutrino-sphere is also being investigated \cite{bartl14,rrapaj14}.
Within Fermi liquid theory, the response functions are computed
by solving the quasiparticle transport equation. The relaxation rate that is used to
approximate the collision integral in the transport equation is reduced with the inclusion
of interactions beyond one-pion exchange. This results in a reduction of the mean-square
energy loss from neutrino scattering, as shown in Fig.\ \ref{etrans}.

\subsection{Constraints from neutron star observations}

The investigation of compressed baryonic matter is one of the persistently important themes in the physics of strongly interacting many-body systems. While high-energy heavy-ion collisions probe the transition from
the hadronic phase to deconfined quark-gluon matter at high temperatures and low baryon chemical potentials, the access to ``cold" and dense baryonic matter comes primarily through observations of neutron stars \cite{Lattimer:2010uk} in which central core densities several times the density of normal nuclear matter can be reached. Due to the difficulty in lattice QCD for dense matter, there are no trustful theoretical tools available.

Two remarkable examples of massive neutron stars have recently emerged. One of those is the radio pulsar J1614--2230 with a mass $M = (1.97\pm0.04) M_\odot$ \cite{demorest}. It is special because of the high accuracy of its mass determination made possible by the particular edge-on configuration (an inclination angle of almost $90^\circ$) of the binary system consisting of the pulsar and a white dwarf. Given this configuration, a pronounced Shapiro-delay signal of the neutron star's pulses could be detected. A second neutron star has been found with a comparable, accurately determined mass (J0348+0432 with $M = (2.01\pm0.04) M_\odot$) \cite{antoniadis}, further strengthening the case. While the empirical restrictions on neutron star radii are less stringent than those on the mass, the quest for a stiff EoS at high baryon densities nonetheless persists as a common theme throughout such investigations.
The established existence of two-solar-mass neutron stars rules out many equations of state (EoS) that are too soft to stabilize such stars against gravitational collapse. However, some selected equations of state based entirely on conventional nuclear degrees of freedom are able to develop a sufficiently high pressure so that the condition to reach $2 M_\odot$ can be satisfied  \cite{AP:1997, akmal, Engvik:1996}.

A considerable amount of work has been devoted to establishing the acceptable range of the pressure as function of density or energy density, $P({\cal E})$, given the new astrophysical constraints \cite{Steiner:2010fz,Steiner:2012xt,LS:2014a,LS:2014,HW:2014,Ozel:2009,Ozel:2010}. In this subsection\footnote{In this subsection, we revert to denoting baryonic density by $\rho$.} we summarize the conditions that any EoS of strongly interacting baryonic matter should fulfill in view of the recent neutron star observations. In the spirit of the present review, we also display an EoS with a firm foundation in the (chiral) symmetry breaking pattern of low-energy QCD. The modeling of this EoS and its extrapolation to neutron star core densities is guided by in-medium ChEFT. In this subsection, we rely on the developments reported in previous sections, namely that ChEFT has been applied successfully to the nuclear many-body problem and its thermodynamics, for symmetric nuclear matter, pure neutron matter and varying proton fractions Z/A between these extremes. In fact, reproducing the known properties of normal nuclear matter is a mandatory prerequisite for the construction of any realistic EoS, together with the requirement of consistency with the most advanced many-body computations of pure neutron matter (see, e.g., Refs. \cite{GCRSW:2014,roggero14}).

Traditionally, a primary source of information is the mass-radius relation of the star calculated using the Tolman-Oppenheimer-Volkov (TOV) equation \cite{tolman,oppenheimer},
with a given equation of state $P({\cal E})$ as input:
\begin{equation}
{dP(r)\over dr} = -{{\cal G}\over r^2}\left[{\cal E}(r) + P(r)\right]{M(r) + 4\pi r^3P(r)\over 1-2{\cal G}M(r)/r}~.
\end{equation}
Here ${\cal G}$ is the gravitational constant, $r$ is the radial coordinate, ${\cal E}(r)$ and $P(r)$ are the local energy density and pressure. Furthermore, $M(r)$ is the mass inside a sphere of radius $r$ determined
by
\begin{equation}
{dM(r)\over dr} = 4\pi r^2{\cal E}(r)~.
\end{equation}
The total mass, integrated up to the radius $R$ of the neutron star, is
\begin{equation}
M \equiv M(R) = 4\pi \int_0^R dr\,r^2{\cal E}(r)~.
\end{equation}

Neutron star matter interpolates between the extremes of isospin-symmetric nuclear matter and pure neutron matter. The fraction of protons added to the neutron sea is controlled by beta equilibrium. The passage from $N = Z$ matter to neutron-rich matter as it emerges in the core of the star is driven by detailed properties of the isospin-dependent part of the nuclear interaction. These isospin-dependent forces also determine the evolution of the nuclear liquid-gas transition from isospin-symmetric matter towards the disappearance of this phase transition around $Z/N \simeq 0.05$. Such properties of the phase diagram of highly asymmetric nuclear matter, elaborated in Section 5, provide further guidance and constraints for extrapolating the EoS into the high-density regime encountered in the cores of neutron stars.

Solving the TOV equation requires the knowledge of the EoS in the entire neutron star, including the low-density crust region at its surface. The outer crust is associated with densities $\varrho\lesssim\varrho_d$ below the neutron-drip point, $\varrho_d\approx 10^{-3}\,\varrho_0$ in units of nuclear saturation density, $\varrho_0=0.16\,{\rm fm}^{-3}$. In the transition region to a uniform nuclear medium (in the density range $0.2\,\varrho_0\lesssim\varrho\lesssim0.5\,\varrho_0$) extended clusters of so-called ``pasta" phases  might be formed. In order to describe this multifacet structure of the neutron star's crust we can use the empirical Skyrme-Lyon (SLy) EoS \cite{Douchin:2001sv} for the low-density region.

At a density of about ${1\over 2}\,\varrho_0$ the nuclei dissolve and turn into a uniform medium of neutrons with a small admixture of protons in the outer core region of the neutron star. In order to interpolate between regions from lower densities up to around $\varrho_0$, we adopt the ChEFT-based EoS determined in Refs.~\cite{FKW2012, holt13} (FKW). The FKW EoS is matched to the SLy EoS at their intersection point, $\epsilon_0\approx 118\,{\rm MeV}/{\rm fm}^3$ corresponding to a density $\varrho \approx 0.75\, \varrho_0$.
\begin{figure}[htbp]
\begin{center}
 \includegraphics[width=.55\textwidth]{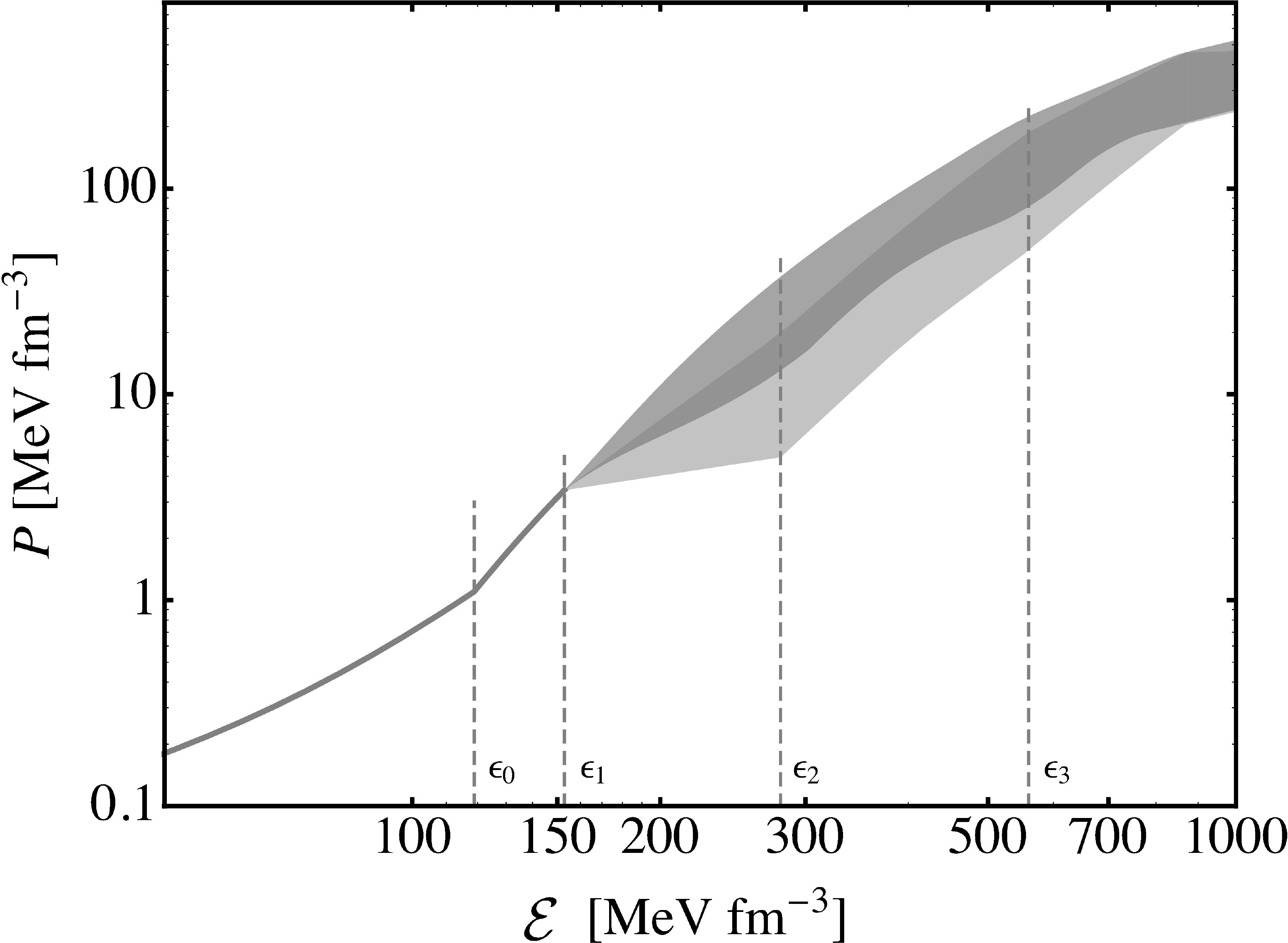}
\end{center}
\vspace{-.5cm}
\caption{Allowed regions for the equation of state $P(\cal E)$ as dictated by neutron star observables \cite{HW:2014}. The upper (dark grey) area takes into account the limitations as given by Tr\"umper \cite{Truemper2013} and constraints from causality. The lower (light grey) band uses, in addition to the two-solar-masses constraint, a permitted radius window $11.0$--$12.5\,{\rm km}$ from \cite{Steiner:2010fz,Steiner:2012xt,LS:2014a}. For energy densities smaller than $\epsilon_1$ and $\epsilon_0$ the FKW and SLy EoS, respectively, are used. The matching points $\epsilon_1,\epsilon_2,\epsilon_3$ of the polytropes in Eq.\ (\ref{eq:polytropes}) are also shown in the figure.
}
\label{constraintseos}
\end{figure}

The extrapolation to the high-density domain of the equation of state is parametrized using three polytropes fitted sequentially to one another (in a way similar to the procedure pursued in Refs.~\cite{hebeler10b,Hebeler:2013nza}): $P=K_i\,\varrho^{\Gamma_i}$, $i =1,2,3$. The energy density for each of the branches is
	\begin{equation}
			{\cal E}=a_i\left({P\over K_i}\right)^{1/\Gamma_i}+{1\over\Gamma_i-1}\,P~~~~~ (i = 1,2,3)\,,
\label{eq:polytropes}
	\end{equation}
where the $a_i$ are constants determined by the continuity of ${\cal E}(P)$.  It turns out that three polytropes are sufficient in order to represent a large variety of models for dense matter. We use the FKW EoS up to an energy density $\epsilon_1=153\,{\rm MeV}/{\rm fm}^3$ corresponding to nuclear saturation density. The three polytropes are then introduced in the ranges between the $\epsilon_i$ as shown in Fig.\ \ref{constraintseos}. The parameters $\Gamma_i$ and $K_i$ are fixed such that the equation of state is continuous at the matching points. The constraints from neutron star masses and radii then translate into a limited band area of $P(\cal E)$. Any acceptable EoS must lie within this belt.
These emerging allowed corridors are consistent with results reported in Ref.~\cite{Hebeler:2013nza}.

State-of-the-art EoS's computed using advanced quantum Monte Carlo methods \cite{GCRSW:2014}, as well as the time-honored EoS resulting from a variational many-body calculation \cite{akmal} (APR),
both  pass the test of being within the allowed $P(\cal E)$ region, once three-nucleon forces are included and the nuclear symmetry energy is constrained around $E_{sym} \simeq 33$ MeV. Notably, these equations of state work with ``conventional" (baryon and meson) degrees of freedom.

Consider now the EoS at zero temperature derived from nuclear chiral effective field theory as described in previous sections. We recall that this EoS is generated using in-medium chiral perturbation theory to three-loop order in the energy density. It includes explicitly one- and two-pion exchange dynamics and three-body forces in the presence of the nuclear medium, together with re-summed contact terms. The energy density is written as:
\begin{equation}
	{\cal E}(\varrho,x_p)=\varrho\left[M_N+\bar E(\varrho,x_p)\right]\,,
	\end{equation}
with the energy per nucleon, $\bar{E} = E/A$, given as a function of the density $\varrho = \varrho_n + \varrho_p$ and the proton fraction, $x_p = \varrho_p/\varrho$. The expansion of $\bar{E}$ provided by in-medium chiral effective field theory is actually in powers of the Fermi momentum, i.e. in fractional powers of the density $\varrho$. The nucleon mass is taken as the average of neutron and
proton masses, $M_N=\frac{1}{2}\left(M_n+M_p\right)$. It is useful to write the energy per nucleon as an expression to second order in the asymmetry parameter, $\delta=(\varrho_n-\varrho_p)/\varrho$, given the small proton fraction $x_p$ encountered in the neutron star interior. With the calculated energies per nucleon for symmetric nuclear matter, $\bar{E}_{SM}$, and pure neutron matter, $\bar{E}_{NM}$, and the symmetry energy,
$S(\varrho) = \bar{E}_{NM}(\varrho) - \bar{E}_{SM}(\varrho)$:
\begin{eqnarray}\label{easymexpansion}
	\bar{E} = \bar{E}_{SM}(\varrho)+S(\varrho)(1-2 x_p)^2
			=\left(1-2 x_p\right)^2\bar E_{NM}(\varrho) + 4 x_p(1-x_p)\,\bar E_{SM}(\varrho)\,.
	\end{eqnarray}
The ChEFT calculation of the symmetry energy at nuclear saturation density, $\varrho_0=0.16$ fm$^{-3}$, gives
\begin{equation}
		S_{\rm ChEFT}(\varrho_0)=33.5\,{\rm MeV}\,,
	\end{equation}
compatible with empirically deduced values.
Beta equilibrium involving electrons and muons, $n\leftrightarrow p+e^-+\bar\nu_e$ and $n\leftrightarrow p+\mu^-+\bar\nu_\mu$, together with charge neutrality imply:
	\begin{eqnarray}
			\varrho_p = \varrho_e +\varrho_\mu\,,~~~~
			\mu_n= \mu_p+\mu_e\,,~~~~\mu_e=\mu_\mu\,.\label{cheftbeta}
		\end{eqnarray}
where the neutron and proton chemical potentials are given by
\begin{equation}
\mu_{n,p} = \left({\partial{\cal E}\over\partial\varrho_{n,p}}\right)_V~.
\end{equation}
The lepton charge densities, $\varrho_e,\varrho_\mu$, and the corresponding chemical potentials, $\mu_e,\mu_\mu$, are again assumed to be those of a free Fermi gas of electrons and muons.

\begin{figure}[htbp]
\begin{center}
 \includegraphics[width=.55\textwidth]{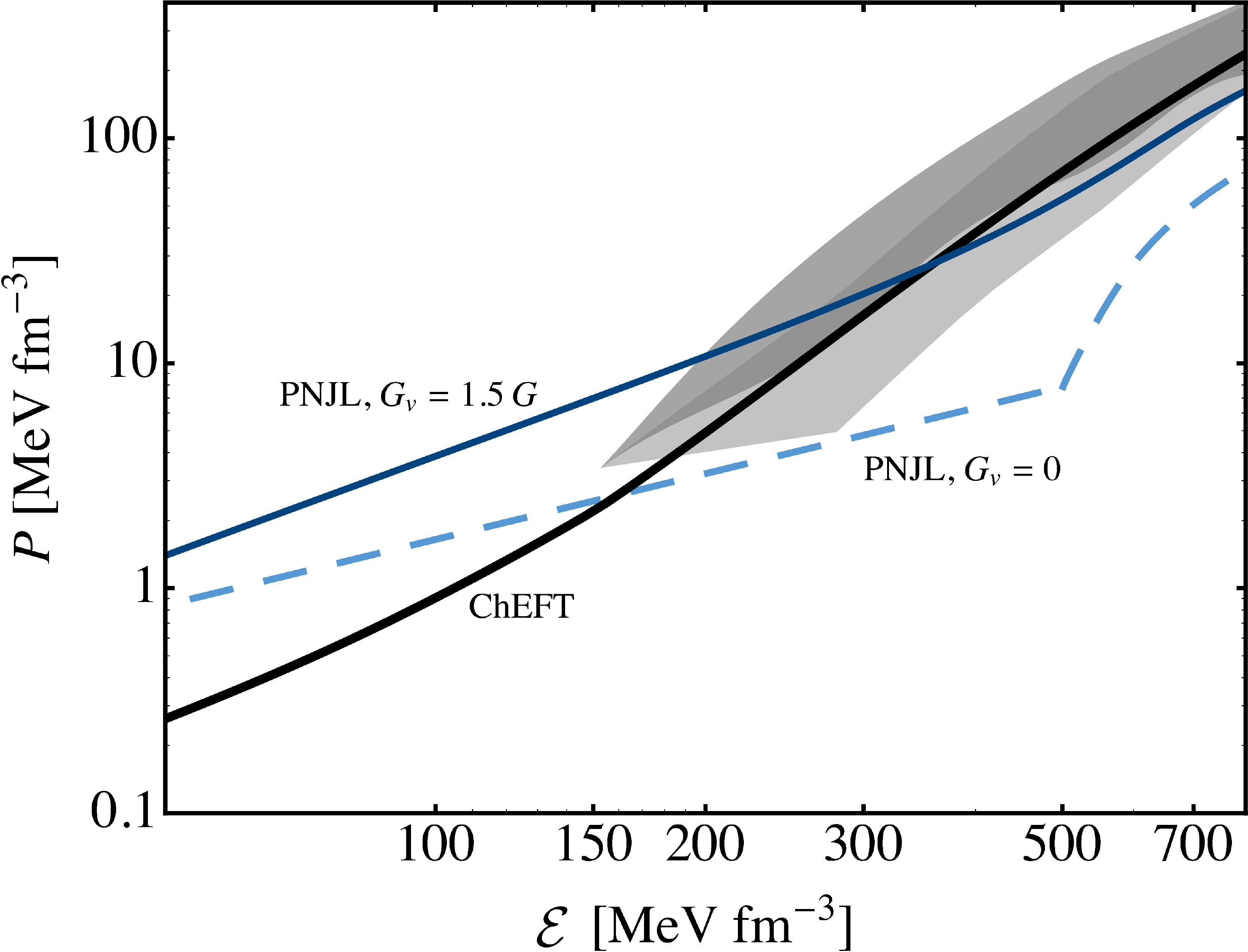}
\end{center}
\vspace{-.5cm}
\caption{Equations of state of neutron star matter at $T = 0$ (taken from Ref.\ \cite{HW:2014}). The black solid line displays the EoS derived from in-medium chiral effective field theory as described in previous sections.  The blue dashed and solid lines show results using a PNJL chiral quark model for different vector coupling strengths $G_v$ as indicated in the figure. The grey bands represent the constraints from neutron star observables (see Fig.\ \ref{constraintseos}). }
\label{EoS}
\end{figure}

Incorporating the conditions of charge neutrality and beta equilibrium the equation of state $P(\cal E)$, applicable for neutron star matter in beta equilibrium at zero temperature, is derived using
\begin{equation}
		P = -{\cal E}+\sum_i\mu_i\varrho_i\,~~~~ (i = n,p;e,\mu)~.
	\end{equation}
At very low densities this EoS based on ChEFT is matched again to the SLy EoS as in Fig.\,\ref{constraintseos}. The complete result is shown by the solid black curve in Fig.\,\ref{EoS}. The ChEFT equation of state turns out to satisfy the astrophysical constraints over the whole range of relevant energy densities. The exact microscopic treatment of the Pauli principle acting on the in-medium pion-exchange processes and the repulsive three-nucleon correlations provide the required stiffness of the EoS in the dense medium to support $2 M_\odot$ neutron stars.

Also shown in Fig.\ \ref{EoS} are two examples of EoS derived from a chiral quark model (the Polyakov - Nambu - Jona-Lasinio model), again respecting charge neutrality and beta equilibrium, but now at the level of quarks with dynamically generated constituent masses \cite{HW:2014}. Two versions are shown that differ in the
strength, $G_v$, of an assumed isoscalar-vector contact interaction between quarks. In the standard scenario
with only chiral pseudoscalar and scalar NJL interactions and $G_v = 0$, this model features a first-order chiral phase transition and is evidently nowhere close to the allowed belt. This EoS is far too soft to
support a $2 M_\odot$ neutron star. Only if a strongly repulsive vector interaction between quarks is introduced (here with $G_v = 1.5\,G$ where $G$ is the pseudoscalar-scalar coupling strength) is it possible
to make such a dynamical quark model compatible with the necessary constraints, at least at sufficiently high baryon densities. Hybrid matter with a region in which baryonic and quark matter coexist would still be an option, but in practical calculations subject to thermodynamic consistency conditions \cite{HW:2014}, such a quark-hadron coexistence region turns out to be restricted to very high densities that can be barely reached in the center of the neutron star.

\begin{figure}[htbp]
\begin{center}
 \includegraphics[width=.55\textwidth]{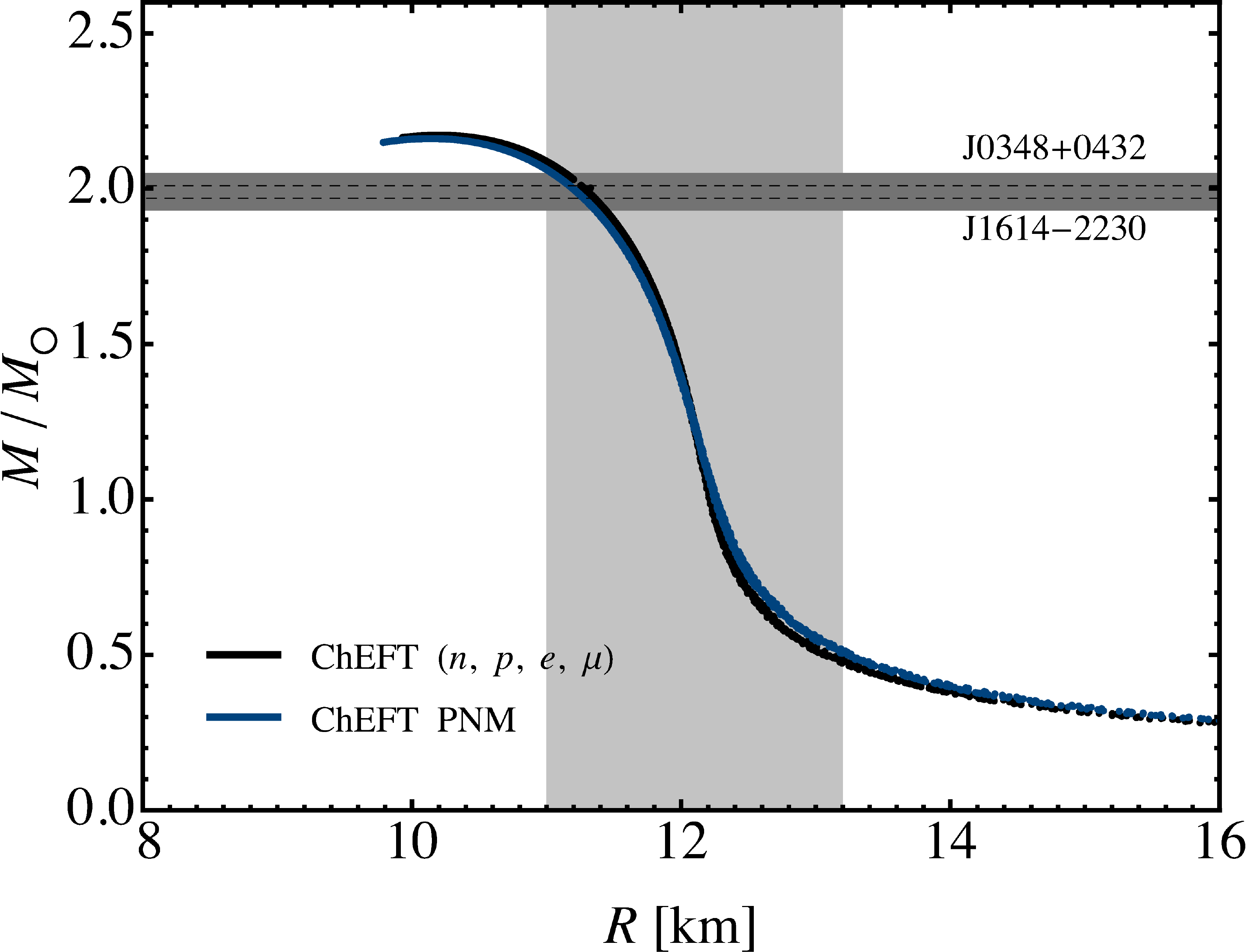}
\end{center}
\vspace{-.5cm}
\caption{Mass-radius relation computed with the ChEFT equation of state for neutron stars including beta equilibrium. Stable neutron stars can exist up to the maximum of this curve. The hardly distinguishable EoS for pure neutron matter (PNM) is also shown for reference. The horizontal band indicates the masses of the pulsars J1614--2230 and J0348+0432. The lighter grey band corresponds to the radius range deduced in Ref.\ \cite{Steiner:2010fz}.
}
\label{cheftmassradiusfig}
\end{figure}

\begin{figure}[htbp]
\begin{center}
 \includegraphics[width=.45\textwidth]{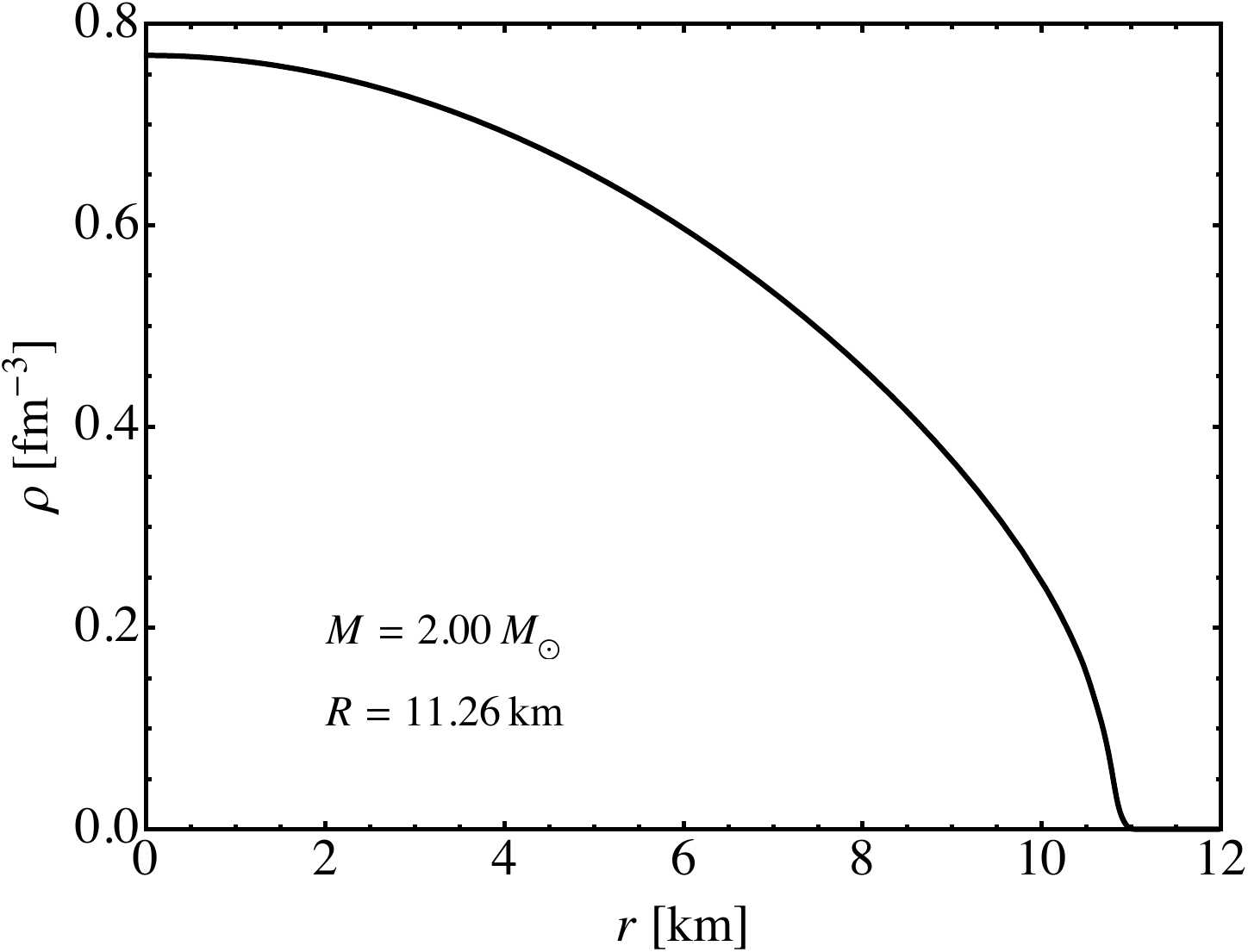}
\end{center}
\vspace{-.5cm}
\caption{Density profile in the interior of a neutron star with mass $M=2\,M_\odot$ and resulting radius of about $R=11\,{\rm km}$. Note that the central density does not exceed $\varrho_c\sim4.8\,\varrho_0$.
}
\label{chefttsmnsfig}
\end{figure}

In fact, the requirement of a very stiff equation of state that produces sufficient pressure to stabilize a $2 M_\odot$ neutron star has the (almost model-independent) implication that the central core density of
the star cannot exceed about five times normal nuclear matter density. Following the calculated mass-radius relation in Fig.\ \ref{cheftmassradiusfig} up to the two-solar-mass range, the density profile of such a $2 M_\odot$ neutron star is shown in Fig.\ \ref{chefttsmnsfig}. With a radius of about 11 km the density in the center of the star reaches not more than $4.8\,\varrho_0$. The high pressure (more than 150 MeV fm$^{-3}$) at these densities tends to keep the baryons apart.

Concerns might be raised at this point about how far ChEFT calculations can be extrapolated into the high-density regime. At $\varrho_n \sim 3\,\varrho_0$ the neutron Fermi momentum, $k_F^n \sim 2.4$ fm$^{-1}$, continues to be appreciably smaller than the chiral symmetry breaking scale of order $4\pi f_\pi \sim$ 1 GeV, rendering the ChEFT expansion in powers of $x = k_F/4\pi f_\pi$ still meaningful (note that $x\sim 0.5$ even at densities as high as $\varrho_n \sim 5\,\varrho_0$). The sensitivity to convergence issues in the chiral expansion of the energy per particle starts at order $x^4$ and then involves even higher powers of $x$. The only exception to this scheme is the case of reducible two-nucleon processes such as iterated one-pion exchange (dominated by the in-medium second-order tensor force), for which the relative scaling factor is $M_N\,k_F/(4\pi f_\pi)^2$. Such diagrams are calculated exactly up to three-loop order in the energy density. Elaborating further on questions of convergence, it is instructive to compare the (perturbative) ChEFT expansion in the nuclear medium with calculations that start from a chiral meson-nucleon Lagrangian based on a linear sigma model plus short-distance interactions, combined with a (non-perturbative) functional renormalization group (FRG) approach \cite{Drews2013,Drews2014}. The latter takes into account leading subclasses of in-medium pionic fluctuations and nucleonic particle-hole excitations to all orders. The close similarity of those ChEFT and FRG results, both for symmetric nuclear matter \cite{Drews2013} and for neutron matter \cite{Drews2014} holds up to at least three times the density of nuclear matter.

A necessary condition for the applicability of in-medium ChEFT is that the medium persists in the hadronic phase of QCD with spontaneously broken chiral symmetry. Investigations of the in-medium chiral condensate at zero temperature \cite{condpap,Drews2013,Drews2014} do indeed show a stabilization of the density-dependent condensate $\langle\bar{q}q\rangle(\varrho, T = 0)$, shifting the transition to chiral symmetry restoration far beyond three times $\varrho_0$.

\begin{figure}[htbp]
\begin{center}
 \includegraphics[width=.45\textwidth]{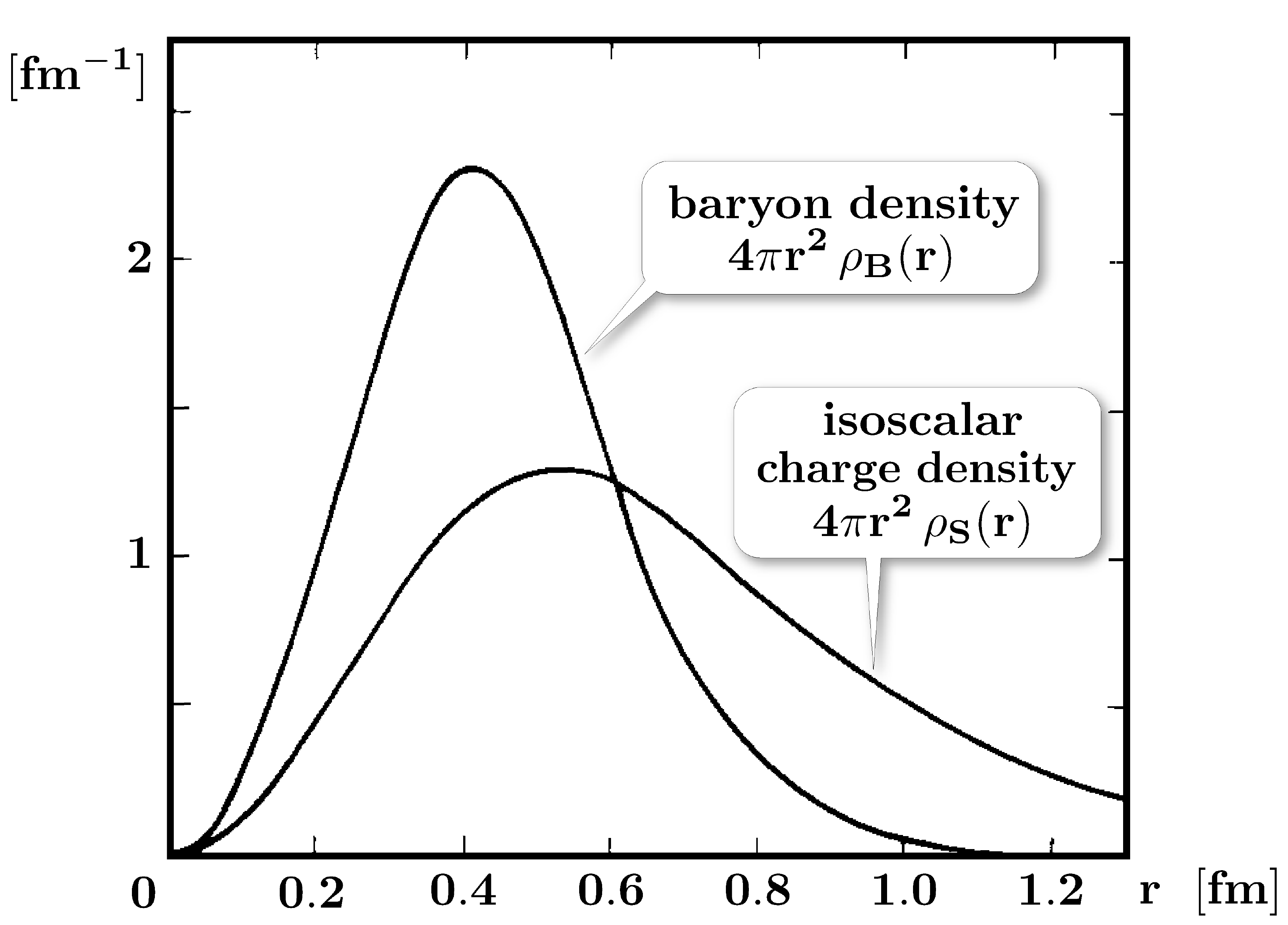}
\end{center}
\vspace{-.5cm}
\caption{Profiles of the baryon density and isoscalar charge density of the nucleon, each multiplied
by $4\pi r^2$, resulting from a chiral soliton model \cite{KMW:1987}.}
\label{nucleondensities}
\end{figure}

The in-medium ChEFT approach relies on the assumption that the proper baryonic degrees of freedom are nucleons (rather than liberated quarks) even in compressed baryonic matter. In this context the following qualitative picture may be useful for orientation.
Models based on chiral symmetry describe the nucleon as a compact valence quark core (typically with a radius of about 1/2 fm), surrounded by a mesonic cloud \cite{BRW}. The meson cloud determines most of the empirical proton rms charge radius of 0.87 fm. For the neutron the picture of core and cloud is analogous except that the electric charges of quark core and meson cloud now add up to form the overall neutral object. A typical example of such a picture of the nucleon is shown in Fig.\ \ref{nucleondensities}, with a baryonic core rms radius $\langle r^2 \rangle_B^{1/2} \simeq 0.5$ fm and an isoscalar charge rms radius
$\langle r^2 \rangle_{E,isoscalar}^{1/2} \simeq 0.8$ fm. Now, even at densities as high as $\varrho \sim 5\,\varrho_0$, the average distance between two neutrons is still about 1 fm, hence the baryonic cores do not yet overlap appreciably at such densities. The pionic field surrounding the baryonic sources is of course expected to be highly inhomogeneous and polarized in compressed matter, but this effect is properly dealt with in chiral EFT. It is therefore perhaps not so surprising that an EoS based entirely on nucleons and pionic degrees of freedom works well for neutron stars, once the repulsive correlations and mechanisms for generating stiffness and high pressure are properly incorporated. This picture is justified also in $hls$EFT where a topology change affecting the tensor forces intervenes to stiffen the EoS above $\sim 2n_0$ as will be shown below.



\section {Skyrmions, dense matter and compact stars}
\label{skyrm}

\subsection{Skyrmions on a crystal lattice}\label{half-skyrmion}

Given an effective Lagrangian arrived at from QCD in some trustful limits such as large $N_c$ limit, chiral limit etc., there are several ways of approaching nuclear matter and dense matter. In previous chapters an approach anchored on a renormalization group strategy has been discussed using an effective chiral Lagrangian in which baryons and pions  are the relevant degrees of freedom. Here one typically resorts to a ``double-decimation technique" that involves two scales, one a chiral scale $\Lambda_\chi\sim (0.7-1) $ GeV and another a Fermi momentum scale $\Lambda_{k_F}\sim k_F\sim 230$ MeV. Another approach is to add vector-meson (and other heavy-meson) degrees of freedom to the above (i.e., $hls$EFT)  and do a relativistic mean-field approximation. Underlying both approaches is the assumption that nuclear matter is a Fermi liquid.

In this section we discuss an alternative approach that starts with an effective Lagrangian containing meson fields only, i.e., the HLS Lagrangian, considered to result from QCD in the large $N_c$ limit. Since, at present, the only known technique available to access dense matter from QCD is the large $N_c$ limit, we will adopt this line of reasoning and push it as far as one can go in that limit.  It provides a justification for ChEFT applicable to $\sim n_0$ and slightly above. There are hints, however, that the large $N_c$ limit could be unreliable for certain properties of nuclear matter -- and most likely dense matter -- so the approach as it stands can only be taken as an initial step toward a more realistic approach.

In the large $N_c$ limit, baryons are solitons (that is, skyrmions) in chiral Lagrangians. It has been established in that limit, the skyrmion description and the non-relativistic quark model give the same results for baryons.  As described in Chapter \ref{him}, in holographic QCD that results from string theory, baryons are instantons in a 5D Yang-Mills Lagrangian in a warped space, which KK reduced to 4D are skyrmions in an effective  Lagrangian with the infinite tower of vector mesons in hidden gauge symmetry. The basic idea of approaching dense matter in this way is to start with the crystalline matter given in the large $N_c$ limit and study what happens to the crystal as the density increases. This is done by putting skyrmions in a crystal structure~\cite{klebanov}, typically on FCC which is favored energetically, and decreasing the crystal size. Given the large $N_c$ approximation, certain correlation effects sensitive to $1/N_c$ corrections will be missed. However, there are certain robust features associated with the topology involved, and these we would like to extract from the skyrmion crystal model.

The first calculations of dense skyrmion matter (reviewed in Refs.~\cite{CNDII,park-vento}) were done with the Skyrme model (with pion fields only) by looking at an array of skyrmions with neighboring skyrmions relatively rotated in isospin by $\pi$ so as to maximize their attraction at large separation. More recently, this treatment has been extended to the HLS Lagrangian, with the vector mesons suitably incorporated so as to account for short-distance interactions~\cite{ma-etal} and a dilaton to account for the trace anomaly of QCD. What emerged from such studies is that there is a phase change from the state made up of skyrmions of unit baryon number to a state in which the single skyrmion fractionizes into two half-skyrmions. A simple understanding of this phase change can be gained in a model of a skyrmion put on a hypersphere $S_3$~\cite{manton}. It is found in such a model~\cite{jackson} that there is a phase transition as the density is increased from a system of skyrmions with no preferred symmetry to a regular lattice of half-skyrmions. This transition, which will be seen both in terms of skyrmions as well as in terms of instantons, turns out to be quite robust, independent of the degrees of freedom involved, such as vector mesons, scalars etc. The  resulting half-skyrmion state can be identified as a state of ``enhanced symmetry" with alternating signs of quark condensates $\Sigma\sim \la\bar{q}q\ra$, which average to zero in a unit cell~\cite{goldhaber-manton}. Locally, however, the condensate is not zero, so the system is expected to support a chiral density oscillation. Furthermore, the effective pion decay constant in medium $f_\pi^*$ does not vanish even when the quark condensate $\Sigma \sim \la\bar{q}q\ra$ is zero, therefore $\Sigma$ is not an order parameter for chiral symmetry restoration. We will see that this will have an important implication on the EoS of dense matter.

\subsection{Skyrmion--half-skyrmion phase change}\label{nucleonmass}
In the large $N_c$ limit and at high density, the soliton -- skyrmion or instanton -- crystal is an energetically favored state. There must, therefore, be some sort of phase transition from skyrmions at low density to half-skyrmions at some higher density -- that we shall denote as $n_{1/2}$. The transition is topological with no well-defined local field order parameters. It is not known whether such a transition exists -- and if it exists, how it manifests itself -- in the continuum chiral symmetry approach described in previous chapters. It could be non-perturbative, perhaps inaccessible in chiral perturbation theory.

The density at which the half-skyrmion appears must be higher than normal nuclear matter density $n_0$. Were it below or near $n_0$, it would have been observed in normal nuclei and nuclear matter. Although it is not feasible at present to pin it down, for standard parameters of the Lagrangian, $n_{1/2}$ comes out at $\gsim 2n_0$. Up to $n_{1/2}$ the skyrmion crystal should -- modulo $1/N_c$ contributions -- reproduce more or less chiral perturbation theory results, so there should be no surprise or new phenomena here. However there are striking novel phenomena predicted at and above $n_{1/2}$, and this we would like to describe here. In what follows, we continue with the assumption that the predicted effects are not a mere artifact of the crystalline structure or equivalently of the large $N_c$ approximation and try to extract model-independent information from the results. The robust feature of the phase transition\footnote{We shall call this a ``phase transition" although there are no obvious order parameters signalling the change of symmetries. The half-skyrmion phase resembles the pseudo-gap phase in high-$T$ superconductivity, but we think it is more like a quarkyonic phase in which quarks are confined while hadrons are present, although the quark condensate is zero. The half-skyrmion phase is endowed with a higher symmetry not present in QCD, i.e., emergent symmetry.} allows us to make certain qualitative predictions that are unique in this approach.

One of the most prominent observations emerging from the analysis is that the vector mesons play an extremely important role in dense matter. With the vector mesons included, it is essential to consider HLS Lagrangian up to $\O (p^4)$. To that order, there are a large number of $\O(p^4)$ terms, eleven in the normal component of the Lagrangian and three in the anomalous component, referred to as homogeneous Wess-Zumino (hWZ) terms. In the previous treatments with vector mesons included,  the vector mesons were included in the normal component to $\O(p^2)$ and only one $\O(p^4)$ term proportional to $\omega^\mu B_\mu$ with $B_\mu$ the baryon current from the hWZ action was taken into account. We shall call this HLS$_{min} (\pi,\rho,\omega)$ model. The HLS to $\O(p^4)$ with no hWZ terms gets no contribution from the $\omega$ meson\footnote{This is because the $\omega$ couples to $\pi$ and $\rho$ only through the hWZ terms.}. This will be referred to as HLS$(\pi,\rho)$. The Lagrangian that contains {\em all} $\O(p^4)$ terms will be denoted simply as HLS$(\pi,\rho,\omega)$. In what follows, unless otherwise specified, HLS will refer to this full $\O(p^4)$ Lagrangian.

At first sight, it would appear unfeasible, if not impossible, to make any sensible calculation of the properties of elementary skyrmions, not to mention that of nuclear matter, with the HLS$(\pi,\rho,\omega)$ that carries 14 parameters. Naively, at most only three parameters could be fixed empirically, so how can one do anything with this complicated Lagrangian?

At present, in QCD as a gauge theory, this is indeed not possible. However, given a 5D YM theory with a warped metric and (topological) Chern-Simons term, arrived at from string theory (holographic) or from dimensional deconstruction starting from the current algebras exploiting ``emergent" hidden local symmetry mentioned in Chapter \ref{him}, this can be done. Both approaches arrive at the same form, essentially governed by hidden local symmety, the only difference lying in the warp factor that represents the ultraviolet structure encapsulated in the 5D action. From such an action, one can then formally integrate out {\em all} of the infinite tower of hidden local fields {\em except} the lowest $\rho$ and $\omega$ with all the coefficients of the $\O(p^4)$ HLS Lagrangian fixed, via a ``master formula,"  in terms of two known constants $f_\pi$ and $m_\rho$ \cite{integrateout}. This may sound surprising, but there is actually no miracle here. The 5D action contains only two unknown constants inherited from the string theory construction and it is expanded up to four derivatives\footnote{For those familiar with string theory, the expansion is made from what is called the ``Dirac-Born-Infeld (DBI)" action.} so it is valid only up to $\O (p^4)$ in the sense of the chiral expansion. What is highly nontrivial is that the resulting action can be given entirely in the form of HLS with only the $\pi$, $\rho$ and $\omega$ fields. What this represents within the large $N_c$ -- and in holographic QCD, also large 't Hooft constant -- limit is the renormalization-group flow in the extra dimension (known as holographic direction and corresponds roughly to length scale in gauge theory) and hence in some sense properly accounts for the high-energy scale integrated out. It should, however, be mentioned that the ultraviolet completion is unknown in both cases -- namely, reduction from string theory and dimensional deconstruction -- and could very well be wrong. Thus physical properties sensitive to short-distance physics might not be trusted.

The detailed discussion -- which can be found in Ref.~\cite{ma-etal} -- is rather involved. We will skip it and give instead a succinct summary of what we consider to be robust in the application to dense matter.
\begin{figure}
\begin{center}
\includegraphics[width=7.2cm]{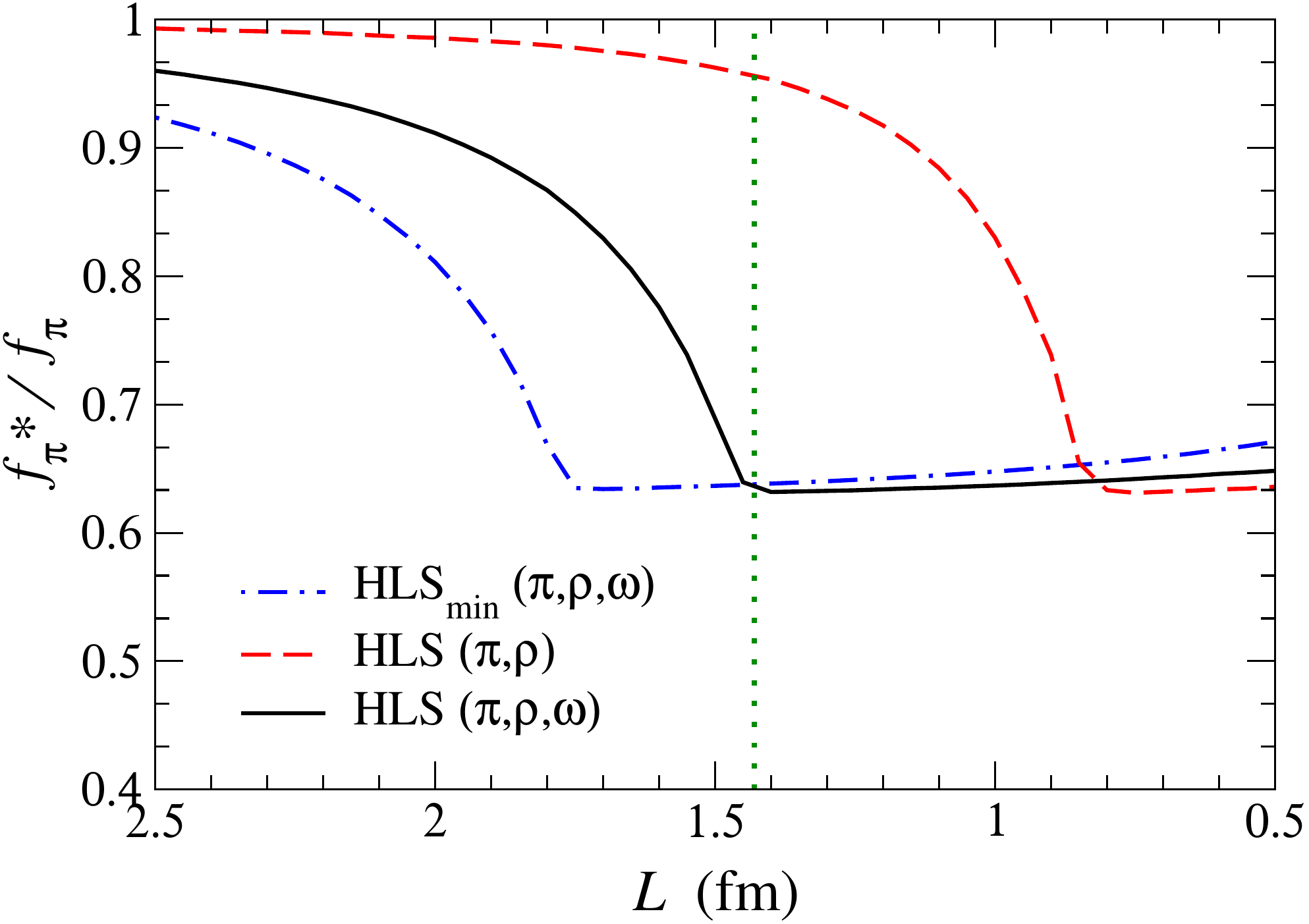}\hspace{.4in}
 \includegraphics[width=7.2cm]{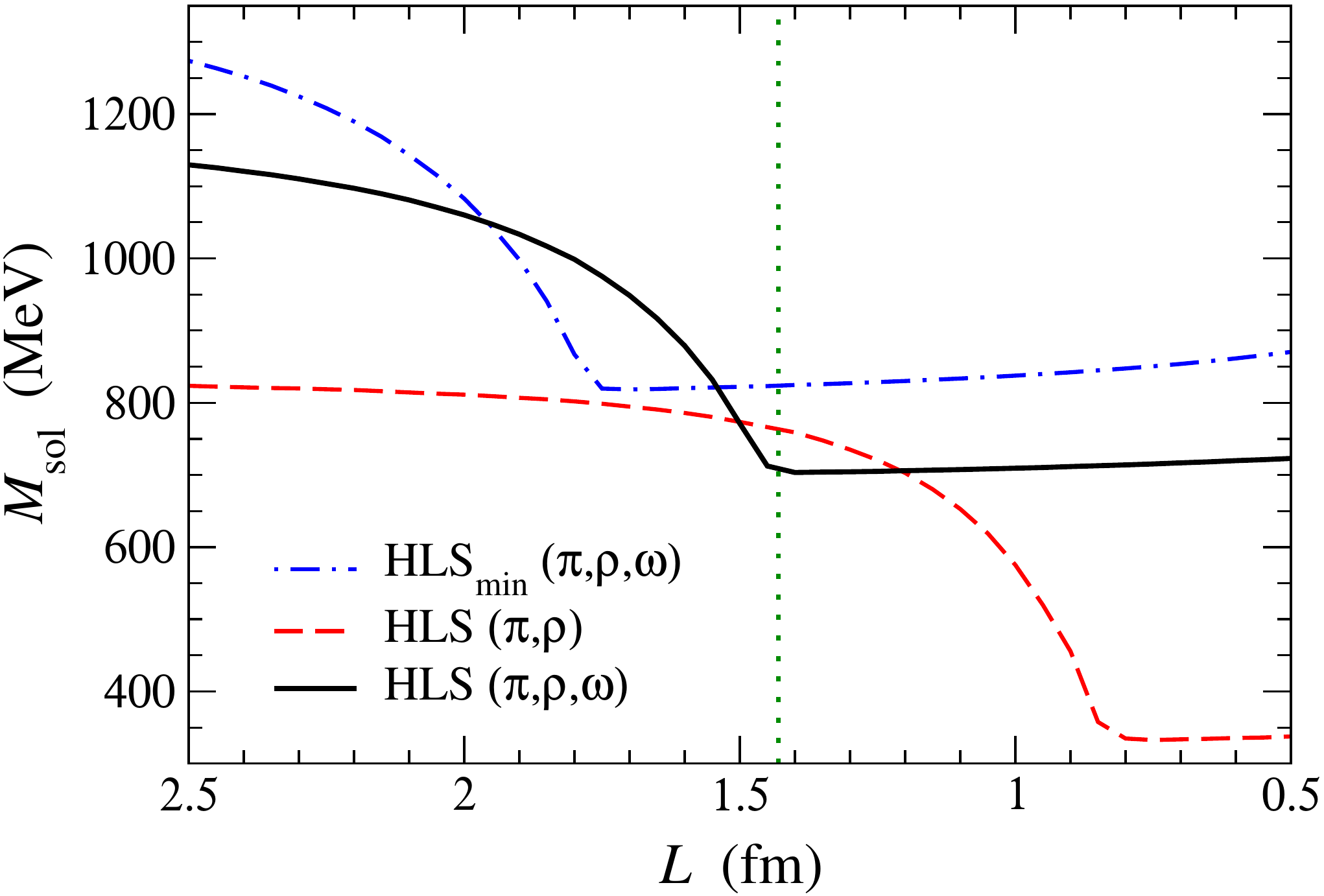}
\caption{\small In-medium pion decay constant (left panel) and nucleon mass (right panel) vs. the lattice size $L$ plotted such that density increases as $L$ decreases. The half-skyrmion phase appears at the density denoted $n_{1/2}$ at which the $\Delta$'s vanish. The notations for HLS's are given in the text.}\label{crystal}
\end{center}
\end{figure}
In Fig.~\ref{crystal} are plotted the in-medium pion decay constant $f_\pi^*$  and the effective soliton mass $M_{sol}$ (equivalent to the in-medium nucleon mass $m_N^*$ in the large $N_c$ limit) vs. the crystal lattice size $L$. Given the empirical free-space values of $f_\pi=92.4$ MeV and $m_\rho=775.5$ MeV, there are no free parameters. The results can be simply parameterized as
\be
f_\pi^* (n) &=& f_\pi^0+\Delta_{f_\pi}(n),\\
m_N^*(n) &=& m_N^0 +\Delta_m(n),
\ee
where $f_\pi^0$ and $m_N^0$ are constants and the $\Delta$'s depend on $\Sigma\sim \la\bar{q}q\ra$ which is supposed to drop as the density increases such that
\be
(\Delta_m, \Delta_{f_\pi})\rightarrow (0, 0) \ {\rm as} \ n\rightarrow n_{1/2},
\ee
where $n_{1/2}$ is the density at which the half-skyrmion phase appears.
The qualitatively striking features are:
\begin{itemize}
\item The half-skyrmion phase is present independently of the degrees of freedom involved as long as the pion field is present. This is because the pion carries topology. Thus one finds the transition in {\em all} skyrmion crystals, i.e., in two extreme cases: the Skyrme model (with the pion field only) and the instanton model (with the pion and the infinite tower of vector mesons).
\item The half-skyrmion onset density $n_{1/2}$ -- and as will be detailed below, various qualitative features of nuclear and dense matter -- are extremely sensitive to the presence of the $\omega$ meson. With the full HLS, i.e., HLS$(\pi,\rho,\omega)$, $n_{1/2}$ comes out reasonable: $n_{1/2}\sim (1.5-2)n_0$. With HLS$(\pi,\rho)$ that has no $\O(p^4)$ terms, hence no $\omega$ contribution, however, the transition occurs at a much higher density, more than 8 times what one expects with the full HLS.
\item At $n_{1/2}$, $f_\pi^*/f_\pi\approx m_N^*/m_N\approx 0.6-0.7$. One can understand the close relation between the ratio of $f_\pi^*$ and that of $m_N^*$ by the large $N_c$ relation that holds in dense skyrmion matter, i.e., $m_N^*\sim \sqrt{N_c} f_\pi^*$.
\item It is significant that in this model a large portion of the nucleon mass, contrary to the common lore anchored on the familiar Nambu-Goldstone mechanism, such as ChEFT -- which works well up to near $n_0$, is {\em not} associated with the spontaneous breaking of chiral symmetry. This is reminiscent of the parity-doublet model of the nucleon~\cite{detar} which is endowed with a large chiral-invariant mass $m_0/m_N\sim \O(1)$. A large $m_0$ in QCD would raise the question of where most of the nucleon mass comes from. However what we have here may be basically different from the parity-doublet structure in QCD in that the half-skyrmion structure is an ``enhanced symmetry," most likely not an intrinsic one in QCD.
\end{itemize}

\subsection{Role of $\rho$ and $\omega$ in nuclear structure}

There is an intriguing interplay between the isovector vector meson $\rho$ and the isoscalar vector meson $\omega$. In all our discussions in this section and elsewhere, we assume that the flavor $U(2)$ symmetry between the two mesons that more or less holds in the vacuum holds also in medium. But it has not been adequately verified that this assumption is valid in a dense medium. In fact, a (perturbative) renormalization group analysis with BHLS (HLS with baryon fields incorporated), indicates that this symmetry may break down in medium at least at one-loop order~\cite{plrs}. This caveat, however, will not affect what is discussed below

In the previous section, we saw that the $\omega$ plays a key role in making dense skyrmion matter be in qualitative agreement with nature. In the holographic description of the baryon as an instanton in 5D, it is the $U(1)$ vector which represents the $\omega$ in 4D and is essential for the soliton size, but it is suppressed in the large 't Hooft constant $\lambda=N_c g_c^2$ limit. In that limit, the Chern-Simons action that carries the $U(1)$ gauge field that provides the coupling to the $SU(2)$ gauge fields does not contribute. This is equivalent to the decoupling of the $\omega$ from the pion and $\rho$ meson in HLS$(\pi,\rho)$ in the absence of the hWZ terms. In the large $N_c$ and large $\lambda$ limit, the metric becomes flat in the 5D YM action, and hence the baryon(s) will be given by the Bogomol'nyi-Prasad-Sommerfield (BPS) soliton. Then a system with the baryon number $B$, say, a nucleus with $B$ nucleons, will saturate the BPS bound, i.e., a system with zero binding energy.

\subsubsection{\it BPS skyrmions and heavy nuclei}\label{bps}

\begin{figure}
\begin{center}
\includegraphics[width=10cm]{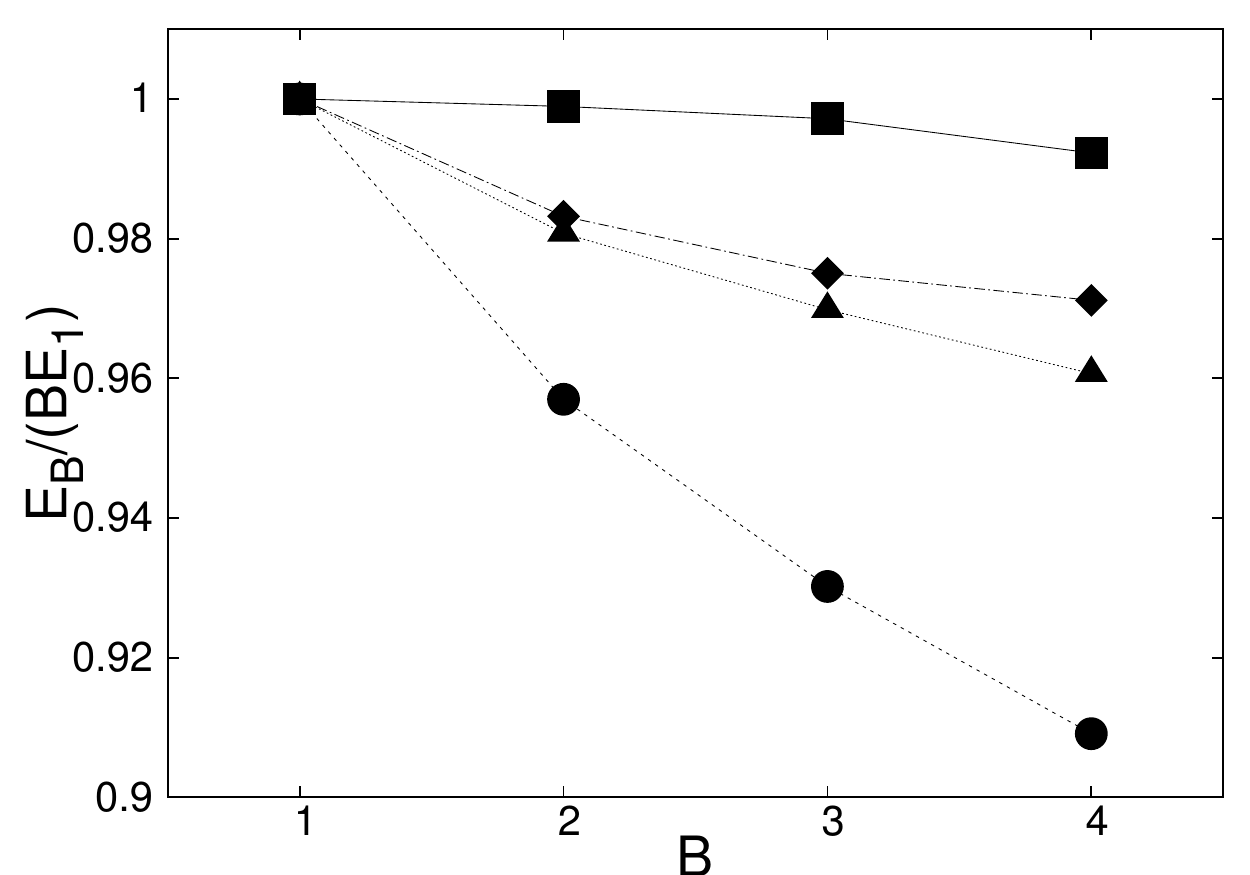}
\caption{The energy per baryon, in units of the single baryon energy,
for baryon numbers $B=1, 2, 3, 4$ taken from \cite{sutcliffe}. The notations are:
Squares = the experimental data, Circles = the energies in the Skyrme model (pions only),
Triangles = HLS$(\pi,\rho)$ and Diamonds = the energies in the theory that includes an $a_1$ to HLS$(\pi,\rho)$.
} \label{fig-sutcliffe}\end{center}\end{figure}

Now suppose that we compute the binding energy of a nucleus with $B$ nucleons, first with the Skyrme model (pion only) and then HLS$(\pi,\rho)$ -- i.e., HLS to $\O(p^2)$. The soliton obtained from the Skyrme model (with pions only) has the energy that exceeds the BPS bound, by more than 20\%, and the binding energy of the $B$-nucleon system will then come out to be much greater than the empirical value. As shown by Sutcliffe~\cite{sutcliffe}, the incorporation of the lowest $\rho$ meson lowers the binding energy by a large amount, more so for heavier nuclei,  going in the direction of empirical values. The inclusion of the lowest axial meson, $a_1$, improves the agreement even further. This feature is well illustrated  in Fig.~\ref{fig-sutcliffe} taken from Sutcliffe's analysis~\cite{sutcliffe}. This implies that as more vector and axial vector mesons are included, there is less binding energy in heavy nuclei in holographic QCD in the large $N_c$ and large $\lambda$ limit, eventually coming closer to what is observed in nature.

What is remarkable in this observation is that the lowest vector meson $\rho$ alone does almost all the work in reducing the discrepancy from the experimental value. This is very much like what happens in the isovector form factors in the holographic QCD model for both pion and nucleon where the $\rho$ contribution makes the largest effect in bringing the vector dominance to the isovector form factors -- saturated by the lowest three vector mesons -- as discussed in light of the Cheshire Cat phenomenon in Chapter \ref{cshs}.

That the BPS limit brings the binding energy of heavy nuclei closer to the experimental results poses a great puzzle in nuclear physics. In fact, it was shown in Refs.~\cite{bps-adam,bps-marleau} that a simple model that exploits the BPS structure of the topological baryon number density, proportional to $B_\mu B^\mu$ where $B_\mu$ is the baryon number current, can ``explain" remarkably well the binding energy of heavy nuclei. Note that the Lagrangian involved here has apparently nothing to do with chiral symmetry in the vacuum and also in light nuclear systems. This is opposite to the standard philosophy of the skyrmion which is anchored, albeit in some limit, on the gauge theory QCD: It captures  low-energy chiral dynamics in the vacuum and in light nuclear systems but meets difficulty in going to heavy nuclei. The BPS model works for heavy nuclei, but its connection to QCD, verifiable in the vacuum and in light nuclei,  is severed.

It has been shown in Ref.~\cite{ma-etal} that the presence of the $\omega$ meson drastically modifies the above scenario. As seen in Fig.~\ref{crystal}, this feature is most likely associated with the role of the $\omega$ meson on the appearance of a possible topology change manifested in the slyrmion--half-skyrmion transition. In the holographic model, the $\omega$ meson figures at the subleading order in $\lambda$. At this order, the metric also gets warped, so the BPS structure is inevitably lost. One can see what happens in HLS$(\pi,\rho,\omega)$. There, it is the hWZ terms that bring in the $\omega$ and obstruct the approach to the BPS seen in the large $N_c$ and large $\lambda$ limit of the holographic model. The small binding energy in heavy nuclei ``explained" by the BPS structure remains, therefore, totally unexplainable in an approach that starts with low-energy hadron dynamics based on chiral symmetry and predicts the robust presence of topology change in a dense medium. This calls for a profoundly revamped understanding of the half-skyrmion structure brought in by the skyrmion matter.

\subsection{Modified BR scaling and tensor forces}
The presence of the half-skyrmion structure at a density $n_{1/2}>n_0$ discussed above can have a drastic effect on the BR scaling discussed in Section \ref{him} when the density goes higher than that of saturated nuclear matter. In Section \ref{him} BR scaling was derived in RMF theory with $hls$HLS, so it was argued to be applicable in the vicinity of $n_0$ at which nuclear matter is at the Fermi-liquid fixed point. Considering that $n_{1/2}$ is not too high compared with $n_0$, we shall assume that the scaling holds up to $n_{1/2}$
\be
m_N^*/m_N\approx m_V^*/m_V\approx m_s^*/m_s\approx \Phi \ \ {\rm for}\ \ n<n_{1/2}.
\ee
Now the topology change at $n_{1/2}$ will modify the scaling at higher density. We see from Fig.~\ref{crystal}
\be
f_\pi^*/f_\pi\approx m_N^*/m_N\approx c \ \ {\rm for}\ \ n\geq n_{1/2},
\ee
where $c \sim 0.6$ is a constant. We cannot say what happens to this scaling as one approaches chiral restoration. It may go to zero, perhaps by a first-order phase transition. For our purpose, we won't need to go close to the chiral restoration point. The vector mesons will be governed not by the dilaton condensate but by the hidden gauge coupling $g$ as we know from matching the HLS correlators to the QCD ones, so
\be
m_\rho^*/m_\rho\to g^*/g\propto (\la\bar{q}q\ra^*/ \la\bar{q}q\ra).
\ee
In HLS, it will go to zero at the vector manifestation fixed point. As for the $\omega$ meson, we cannot say anything certain given that the flavor $U(2)$ most likely breaks down in a dense medium. In particular, there is no reason to expect that the $\omega$ meson will follow the $\rho$ to the vector manifestation fixed point. We will not need to know what happens to the $\omega$ for what follows.

\subsection{Symmetry energy and compact stars}\label{symmetry-star}
One of the most dramatic effects to set in that will have a strong influence on physical observables in nuclei and dense matter from the topology change at $n_{1/2}$ is on the nuclear tensor force. Recall from Chapter \ref{him} that in the HLS scheme, the tensor forces are given by the one-$\pi$ and one-$\rho$ exchanges between two nucleons entering with opposite signs.
%
There it was exploited that due to the cancelations between the two, the net tensor force stiffens up to $n_0$. We assume that the same feature holds beyond $n_0$, say, up to $n_{1/2}$. We will leave the pion tensor unaffected above $n_{1/2}$.  However the $\rho$ tensor undergoes an important modification from the topology change: the scaling of $f_{N\rho}$ gets drastically modified.   In the half-skyrmion phase, we have
\be
\frac{g_{\rho NN}^*}{g_{\rho NN}} \approx \frac{g^*}{g}, \ \ \frac{m_\rho^*}{m_\rho}\approx \frac{g^* f_\pi^*}{g f_\pi}, \ \ \frac{m_N}{m_N^*}\approx \frac{f_\pi^*}{f_\pi}.
\ee
Therefore the ratio $R$ scales as
\be
R\equiv \frac{f_{N\rho}^*}{f_{N\rho}}\approx \frac{g_{\rho NN}^*}{g_{\rho NN}}\frac{m_\rho^*}{m_\rho}\frac{m_N}{m_N^*}\approx (g^*/g)^2\approx (m_\rho^*/m_\rho)^2 .
\ee
This is in strong contrast to $R\approx 1$ for $n<n_{1/2}$.

\begin{figure}
\begin{center}
\includegraphics[height=7cm]{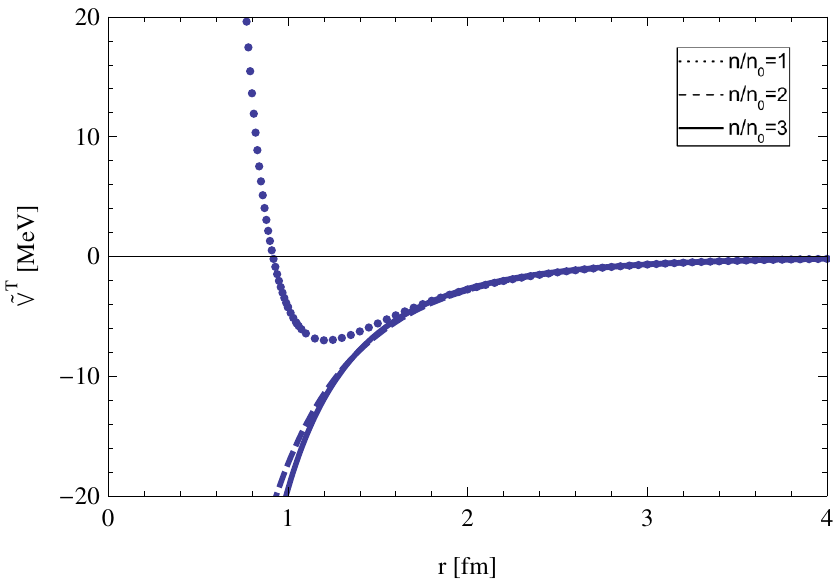}
\caption{Schematic form of the sum of $\pi$ and $\rho$ tensor forces (in units of MeV) {\it in the presence of topology change} for densities $n/n_0$ =1 (dotted), 2 (dashed) and 3 (solid) with  $m_\rho^*/m_\rho \approx 1-0.15 n/n_0$  with  $R\approx 1$ for $n<n_{1/2}$ and $\Phi\equiv m_\rho^*/m_\rho\approx 1-0.15 n/n_0$ and $R\approx \Phi^2$ for $n>n_{1/2}$, assuming $ n_0<n_{1/2}<2n_0$. }
\label{tensorforce2}
\end{center}
\end{figure}

The resulting tensor force is depicted in Fig.~\ref{tensorforce2}.
The effect of this changeover in $R$ can be best seen in the nuclear symmetry energy coefficient $S$ defined by
\be
E_{sym} (n,x)\equiv E(n,x)-E(n,x=1/2)=S(1-2x)^2+\O((1-2x)^3),
\ee
where $x$ is the proton fraction $x=n_p/n$ and $E$ is the energy per nucleon of the system. To have a rough idea, we can use the simple closure formula of Brown and Machleidt~\cite{brown-machleidt} for $S$:
\be
S\approx 12\la V_T^2\ra/\bar{E},
\label{BM}
\ee
where $\bar{E}\approx 200$ MeV is the excitation energy mediated by the tensor forces.
Although the precise scaling within the HLS framework\footnote{It should be stressed that the scaling in medium will depend on the effective theory used. Here we are adhering to the framework defined by the $hls$EFT Lagrangian defined in Chapter \ref{him}.} is not known, what happens is clear and unambiguous. The net tensor force decreases up to $n_{1/2}$ due to the cancelation of the two opposing tensor forces, an aspect highlighted in Chapter \ref{him}, and at $n_{1/2}$ as the contribution from the $\rho$ gets suppressed strongly with the pion tensor taking over, the tensor force will then increase as density increases above $n_{1/2}$. Eq.~(\ref{BM}) predicts a cusp at $n_{1/2}$. Remarkably this cusp is precisely reproduced ~\cite{LPR} by the Skyrme model (with pions only) where a neutron matter described on the crystal is rotationally quantized as suggested by Klebanov~\cite{klebanov}. This is a robust feature and is expected to be present in the HLS$(\pi,\rho,\omega)$ model. {\it This precise matching between the Skyrme model and the scaling deduced from the skyrmion--half-skyrmion transition provides a confirmation of the robustness of the skyrmion crystal formulation.}
The appearance of the cusp in the skyrmion crystal does not necessarily lend itself to a {\em direct} observation in experiments. So how does one go about checking that the cusp is not just an artifact of the approximations?

\begin{figure}
\begin{center}
\includegraphics[width=6.25cm,angle=270]{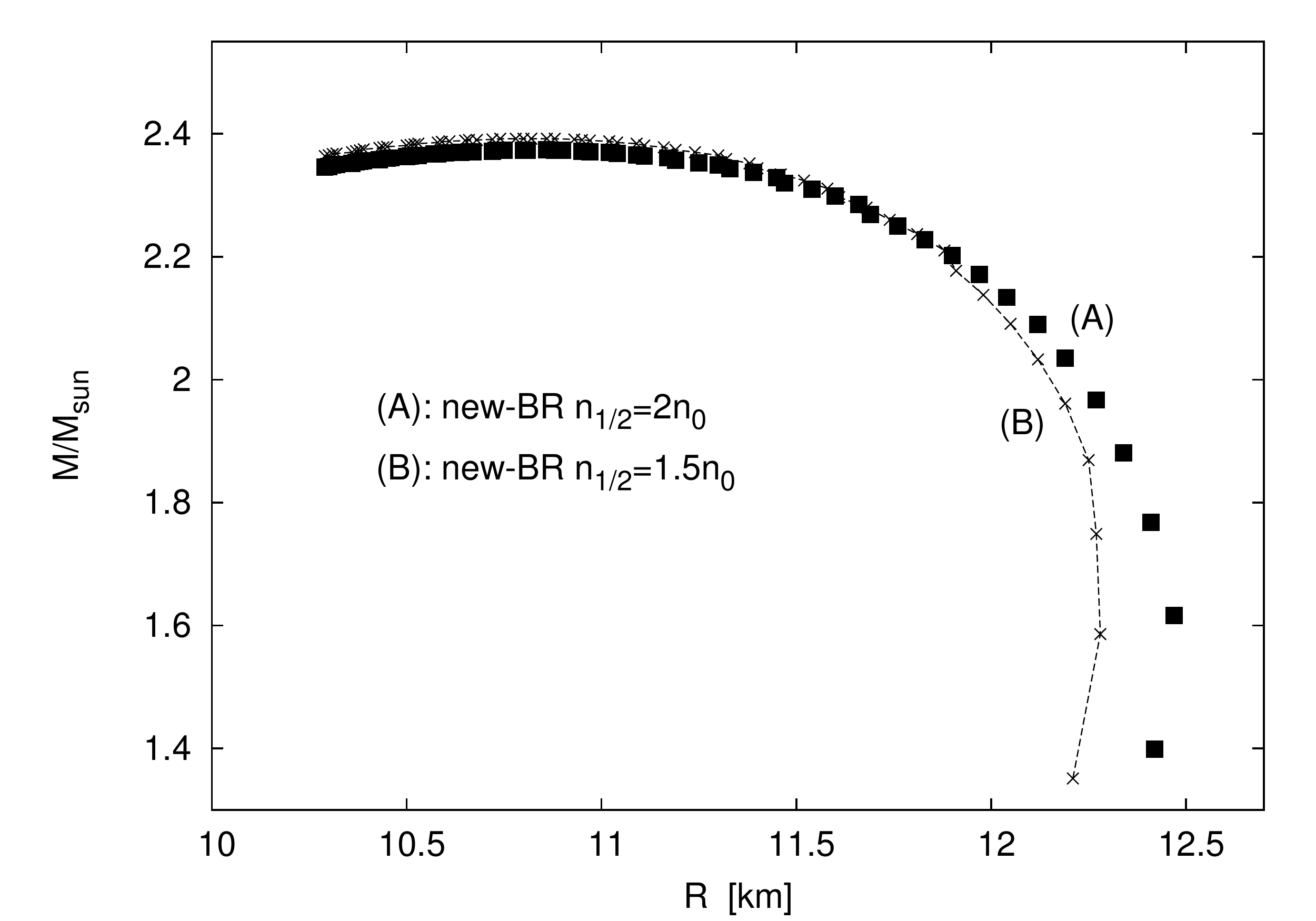}
\includegraphics[width=6.55cm,angle=270]{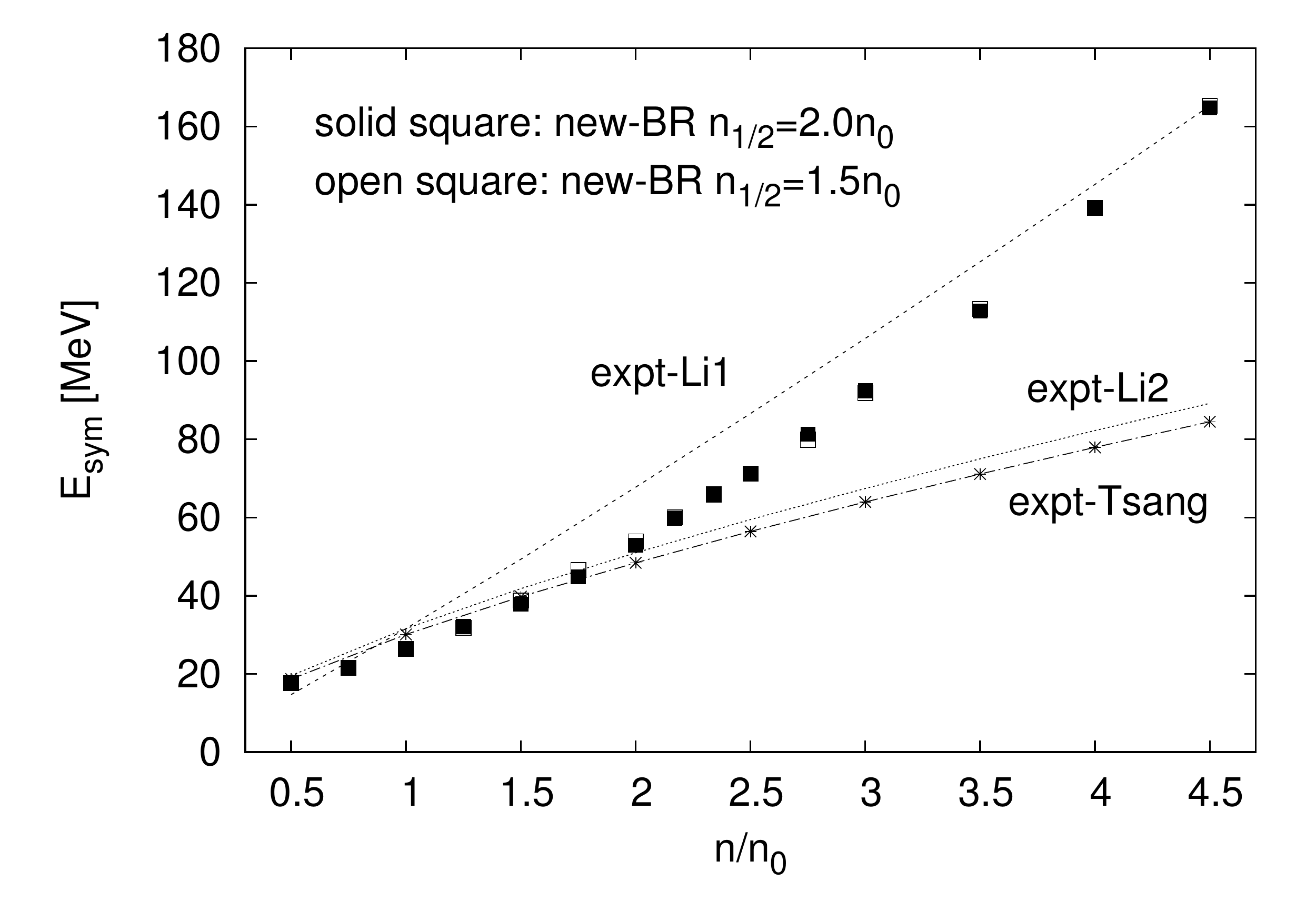}
\caption{Nuclear symmetry energy and the mass-radius relationship for neutrons stars from nuclear interaction models
incorporating new Brown-Rho scaling. Figures taken from Refs.~\cite{dong,kuo14}.}
\label{newbr}
\end{center}
\end{figure}

Coming from the rotational quantization of the soliton matter, it is easy to understand that the symmetry energy with the cusp is leading order in $N_c$.  To show that the cusp can be present in nature, one can take the procedure developed by Gerry Brown and his colleagues at Stony Brook~\cite{holt07}, which was to take an effective Lagrangian with suitable scaling parameters, do the ``double decimation" renormalization group analysis via $V_{low-k}$ and then take into account higher-order nuclear correlations. One such calculation employing a ring-diagram summation with $V_{low-k}$ derived from the HLS Lagrangian so obtained,  was applied to the EoS of compact stars~\cite{dong}. With a suitable scaling function that is consistent up to normal nuclear matter density but is an extrapolation  for $n\gsim n_{1/2}$, the maximum mass of the star is found to come out $\sim 2.4 M_\odot$, with radius $\sim 11$\,km and the maximum central density $\sim 5.5n_0$ (see left panel of Fig.~\ref{newbr}). These results are quite reasonable. One potentially important  distinctive feature of this approach is that while the symmetry energy is soft below $n_{1/2}$ -- which is consistent with nuclear matter at $n\lsim n_0$, it stiffens for $n>n_{1/2}$, just the right mechanism to accommodate the large mass stars being observed in nature (see right panel of Fig.~\ref{newbr}). The EoS so obtained falls in the allowed region of Fig.~\ref{constraintseos} constrained by neutron-star observations.

There is one caveat here that should be emphasized. In the same framework, as the density increases to the vicinity of $n_{1/2}$ or not too far above, kaons should condense~\cite{PKR}. Then the standard lore is that kaon condensation, as treated in mean field theory of the chiral Lagrangian, should soften the EoS to the extent that a star as massive as the accurately measured 1.97$M_\odot$ star~\cite{demorest} could not be supported. This could falsify the result of Ref.~\cite{dong}.

It is, however, not obvious that kaon condensation can be treated in the mean-field approximation with the $hls$EFT Lagrangian in the presence of the half-skyrmion phase. With the topology change, when kaons condense, the condensed mesons can induce a change in the structure of baryonic matter and alter importantly the EoS. For instance, the relativistic mean field approximation, usually employed with an effective Lagrangian for the EoS, could no longer be reliable. What happens in the condensation process is that strangeness enters, and the strange kaon-condensed matter could be in a state ``dual" to a strange quark matter. In fact, a similar stiffening of the EoS is observed in the description that involves the hadronic phase crossing over to strong-coupling quark matter~\cite{hatsuda}.


\section{Concluding remarks and outlooks}
It has been shown that the chiral effective field theory (ChEFT) of Goldstone bosons alone,
including nucleons as matter fields, works extremely well at low energy and low density. As the
low-energy realization of QCD in the sector of light up and down quarks, ChEFT
provides a successful framework for the description of both pion-pion and pion-nucleon
interactions. In modern low-energy nuclear physics, ChEFT allows for the systematic organization
of nuclear forces in powers of the ``small parameter" $Q/\Lambda_\chi$ (where $Q$ stands
generically for an external momentum or the pion mass and $\Lambda_\chi \sim 1$ GeV
is the scale of spontaneous chiral symmetry breaking). This conceptual framework, anchored in
Weinberg's theorem that addresses the general tenet of effective field theories, turns out to be
remarkably successful in nuclear physics, surely up to nuclear matter density $n_0$
and possibly beyond.

Much of the well known phenomenology associated with the nuclear many-body problem
can in the ChEFT framework be given a systematic foundation that incorporates
the spontaneous chiral symmetry breaking pattern of low-energy QCD. The small expansion
parameter that figures in ChEFT for baryonic matter is the Fermi momentum divided by a
characteristic energy/momentum scale involved, namely, $k_f/{\Lambda}$ where ${\Lambda}$
is the typical mass scale integrated out, e.g., the vector meson mass $\sim m_V\sim 0.7$ GeV.
The nuclear equation of state and the related phase diagram featuring the liquid-gas transition, the
nuclear mean field associated with the shell model potential for bound states and the
optical potential for scattering states, the quasiparticle approach given by Landau-Migdal
Fermi liquid theory and the nuclear energy density functional description of finite
systems -- all these important aspects of nuclear theory are well described as
shown in this report. The framework allows in addition the investigation of the
energy density as a function of the pion mass, which provides a path to study the behavior of
the chiral condensate at temperatures up to about 100 MeV and densities reaching two to
three times normal nuclear matter densities, indicating stability of this chiral order parameter
beyond previous expectations based merely on the Fermi gas limit.

The striking prediction of the pion-exchange current in the threshold radiative $np$
capture and the solar $pp$ fusion processes, calculated with an unprecedented precision
$< 1\%$, provide a ``smoking gun" signal for both the role of the soft pion in the
exchange currents {\it and most significantly} the working of QCD chiral symmetry in
multi-nucleon systems. Highly notable is the nontrivial parameter-free prediction for
the highest energy solar neutrinos, namely those produced in the solar $hep$ process,
which has defied explanation for decades and which has been declared by John Bahcall as
a major challenge. In the past, different model predictions for the associated $S$ factor
ranged over two orders of magnitude. The estimated theoretical error in the effective field
theory calculation comes out to be $\sim 15\%$ and could be greatly reduced with the
modern technology.

Compact stars can in principle achieve densities sufficiently high to fall outside the purview of
chiral effective field theory. As is well recognized, dense matter beyond about three times the
saturation density of nuclear matter is a largely unknown territory with a paucity of terrestrial
experiments and reliable theoretical tools. It therefore provides an arena for genuine theoretical
predictions. One way to approach this novel regime of strongly interacting matter is to elevate
ChEFT to a hidden local symmetry theory $hls$EFT that is matched to QCD at a scale $\Lambda_M$
within which both theories are valid. The $hls$EFT necessarily incorporates BR scaling, suitably
implemented with the mended symmetries, and is in principle equivalent to ChEFT at low energies
while enabling the extrapolation to higher energy scales through the inclusion of explicit mesons
heavier than the pion. That ChEFT works up to a specific density (denoted as $\bar{n} > n_0$)
in the Landau-Migdal framework can be viewed through the lens of $hls$EFT in the {\it mean
field approximation}, which is equivalent to Landau Fermi-liquid fixed point theory.

There is evidence that $hls$EFT and ChEFT are compatible up to $\sim n_0$. One case is the
mechanism for the long half-life of carbon-14. In ChEFT, it is the contact (short-range)
three-nucleon potential that is operative for the quenching of the Gamow-Teller matrix element
and in $hls$EFT it is the BR scaling whose density dependence is inherited from the Wilsonian
matching that suppresses the tensor force {\em and} increases the short-range repulsion. The
same mechanism is active in the stabilization of nuclear matter in ChEFT through the
three-nucleon force and in $hls$EFT through the decrease of the vector meson masses
\cite{kuo-brown-BR,holt07}. What is nontrivial in ChEFT for stable nuclear matter is the interplay
between terms of ${\cal O}(k_F^3)$ and ${\cal O} (k_F^4)$, the former attractive and the latter
repulsive, typically involving three-body forces. This would suggest that the ``low" $k_F$
expansion must breakdown at some high density. This observation follows also from
renormalization-group (RG) studies of Fermi liquids.

In the Wilsonian RG approach, nuclear matter is stable because it is a fixed point in the
effective field theory with Fermi surface. Now in Fermi liquid theory, the expansion is in
terms of $1/N$, where $N= k_F/(\Lambda - k_F)$, and the fixed point is reached in the limit
$N\rightarrow \infty$. Thus viewed in terms of the Wilsonian RG, the decimation toward the
Fermi surface -- i.e., ``second decimation"~\cite{BR:DD} starting from the chiral scale
$\Lambda$ involves the changeover from ``small" $k_F/\Lambda$ to ``small" $(\Lambda - k_F)/k_F$.
This changeover is present implicitly in $hls$EFT theory. Now taking $\Lambda$ in ChEFT
to be of the order of the mass of heavy mesons integrated out, say $\Lambda \sim 600$ MeV,
and $\bar{\Lambda} = \Lambda-k_F$ the cutoff for the Fermi surface decimation, one can
estimate that the changeover occurs at around $\sim 3n_0$, close to $\bar{n}$ found in the
ChEFT calculation. This is close to the density at which the skyrmion matter goes over to a
half-skyrmion matter as discussed in Section \ref{half-skyrmion}.

We end with some predictions made in $hls$EFT going beyond ChEFT. They will be
confirmed or falsified by future experiments. One of the most intriguing predictions of
$hls$EFT theory is the behavior of the nucleon mass as the density increases beyond a few
times $n_0$. It is predicted that the effective pion decay constant drops as the density increases
as it does in ChEFT, up to say, $n\equiv n_{1/2}\sim 2n_0$ and then stops dropping or if it
does at all, drops very slowly beyond that density. Now in the skyrmion crystal calculation
with $hls$EFT (Section \ref{nucleonmass}), the in-medium nucleon mass drops roughly the
same way as the in-medium pion decay constant: the nucleon mass stops decreasing at
$n_{1/2}$ with the spatially averaged quark condensate $\la\bar{q}q\ra$ vanishing (see
Fig.~\ref{crystal}). There is an indication in RG-implemented chiral perturbation theory that
the dropping slows down by non-perturbative effects (see the curve FRG in
Fig.~\ref{fig:condensate}). However it does not give rise to the flattening seen in
Fig.~\ref{crystal}. This (flattening) behavior is indicative of a medium-enhanced symmetry, i.e.,
parity-doubling, at the topology change in $hls$EFT. The substantially large, $\gsim 60\%$,
nucleon mass that does not vanish with the vanishing quark condensate cannot be accounted
for by the Nambu-Goldstone mechanism of mass generation.

Finally one can ask whether, in addressing the wide-open regime of high density, there can
be alternatives to the ``bottom-up" approach anchored on Weinberg's theorem. Within the
precisely formulated Harada-Yamawaki hidden local symmetry framework, with the crucial
assumption that local field theory applies in the vicinity of the chiral transition, the vector
manifestation (VM) fixed point is a well-established point around which quantum fluctuations
can be made. Given no experimental or theoretical check (such as lattice), the VM
fixed point cannot be more than a mere conjecture. However, certain explorations of hadronic
matter under extreme conditions, starting from the VM fixed point, do lead to interesting
novel observations, both at high temperature and at high density~\cite{BLR-review}.


\section{Acknowledgments}

Two of us (JWH and WW) would like to thank Norbert Kaiser for many useful discussions and help. The third (MR) acknowledges the invaluable influence and guidance of Masayasu Harada and Koichi Yamawaki on the development made in collaboration with Gerry Brown on the role and power of hidden local symmetry in nuclear physics. The work of JWH was supported in part by US DOE grant DE-FG02-97ER-41014, MR acknowledges support by the WCU project of Korean Ministry of
Education, Science and Technology (R33-2008-000-10087-0), and the work of WW was supported by BMBF and by the DFG Cluster of Excellence ``Origin and Structure of the Universe''.





\end{document}